\documentclass[12pt,fleqn]{article}

\usepackage{authblk}

\usepackage[top=3cm,bottom=3cm,left=3cm,right=3cm,headsep=10pt,a4paper]{geometry} % Page margins

\usepackage{graphicx} % Required for including pictures
\graphicspath{{Pictures/}} % Specifies the directory where pictures are stored

\usepackage{lipsum} % Inserts dummy text

\usepackage{tikz} % Required for drawing custom shapes

\usepackage[english]{babel} % English language/hyphenation

\usepackage{enumitem} % Customize lists
\setlist{nolistsep} % Reduce spacing between bullet points and numbered lists

\usepackage{booktabs} % Required for nicer horizontal rules in tables

\usepackage{xcolor} % Required for specifying colors by name
\definecolor{ocre}{RGB}{243,102,25} % Define the orange color used for highlighting throughout the book

\usepackage{avant} % Use the Avantgarde font for headings
\usepackage{mathptmx} % Use the Adobe Times Roman as the default text font together with math symbols from the Sym­bol, Chancery and Com­puter Modern fonts

\usepackage{microtype} % Slightly tweak font spacing for aesthetics
\usepackage[utf8]{inputenc} % Required for including letters with accents
\usepackage[T1]{fontenc} % Use 8-bit encoding that has 256 glyphs

\usepackage{csquotes}
\usepackage[style=phys,hyperref=true,backref=true,backend=biber]{biblatex}
\defbibheading{bibempty}{}

\usepackage{calc} % For simpler calculation - used for spacing the index letter headings correctly
\usepackage{makeidx} % Required to make an index
\makeindex % Tells LaTeX to create the files required for indexing

\usepackage{titletoc} % Required for manipulating the table of contents

\contentsmargin{0cm} % Removes the default margin

\titlecontents{part}[0cm]
{\addvspace{20pt}\centering\large\bfseries}
{}
{}
{}

\titlecontents{chapter}[1.25cm] % Indentation
{\addvspace{12pt}\large\bfseries} % Spacing and font options for chapters
{\color{ocre!60}\contentslabel[\Large\thecontentslabel]{1.25cm}\color{ocre}} % Chapter number
{\color{ocre}}  
{\color{ocre!60}\normalsize\;\titlerule*[.5pc]{.}\;\thecontentspage} % Page number

\titlecontents{section}[1cm] % Indentation
{\addvspace{1pt}\bfseries} % Spacing and font options for sections
{\contentslabel[\thecontentslabel]{1cm}} % Section number
{}
{\hfill\color{black}\thecontentspage} % Page number
[]

\titlecontents{subsection}[2cm] % Indentation
{\addvspace{0pt}\small} % Spacing and font options for subsections
{\contentslabel[\thecontentslabel]{1cm}} % Subsection number
{}
{\ \titlerule*[.5pc]{.}\;\thecontentspage} % Page number
[]

\titlecontents{subsubsection}[3cm]
{\addvspace{0pt}\small}
{\contentslabel[\thecontentslabel]{1cm}}
{}
{\ \titlerule*[.5pc]{.}\;\thecontentspage}
[]

\titlecontents{figure}[0em]
{\addvspace{-5pt}}
{\thecontentslabel\hspace*{1em}}
{}
{\ \titlerule*[.5pc]{.}\;\thecontentspage}
[]

\titlecontents{table}[0em]
{\addvspace{-5pt}}
{\thecontentslabel\hspace*{1em}}
{}
{\ \titlerule*[.5pc]{.}\;\thecontentspage}
[]

\titlecontents{lchapter}[0em] % Indenting
{\addvspace{15pt}\large\bfseries} % Spacing and font options for chapters
{\color{ocre}\contentslabel[\Large\thecontentslabel]{1.25cm}\color{ocre}} % Chapter number
{}  
{\color{ocre}\normalsize\bfseries\;\titlerule*[.5pc]{.}\;\thecontentspage} % Page number

\titlecontents{lsection}[0em] % Indenting
{\small} % Spacing and font options for sections
{\contentslabel[\thecontentslabel]{1.25cm}} % Section number
{}
{}

\titlecontents{lsubsection}[.5em] % Indentation
{\normalfont\footnotesize} % Font settings
{}
{}
{}

\usepackage{fancyhdr} % Required for header and footer configuration

\renewcommand{\sectionmark}[1]{\markright{\normalsize\thesection\hspace{5pt}#1}{}} % Section text font settings
\fancypagestyle{plain}{\fancyhead{}} % Style for when a plain pagestyle is specified

\makeatletter
\renewcommand{\cleardoublepage}{
\clearpage\ifodd\c@page\else
\hbox{}
\vspace*{\fill}
\thispagestyle{empty}
\newpage
\fi}

\usepackage{amsmath,amsfonts,amssymb,amsthm} % For math equations, theorems, symbols, etc

\newtheoremstyle{ocrenumbox}% % Theorem style name
{0pt}% Space above
{0pt}% Space below
{\normalfont}% % Body font
{}% Indent amount
{\small\bf\color{ocre}}% % Theorem head font
{\;}% Punctuation after theorem head
{0.25em}% Space after theorem head
{\small\color{ocre}\thmname{#1}\nobreakspace\thmnumber{\@ifnotempty{#1}{}\@upn{#2}}% Theorem text (e.g. Theorem 2.1)
\thmnote{\nobreakspace\the\thm@notefont\bfseries\color{black}---\nobreakspace#3.}} % Optional theorem note
\newtheoremstyle{blacknumex}% Theorem style name
{5pt}% Space above
{5pt}% Space below
{\normalfont}% Body font
{} % Indent amount
{\small\bf}% Theorem head font
{\;}% Punctuation after theorem head
{0.25em}% Space after theorem head
{\small{\tiny\ensuremath{\blacksquare}}\nobreakspace\thmname{#1}\nobreakspace\thmnumber{\@ifnotempty{#1}{}\@upn{#2}}% Theorem text (e.g. Theorem 2.1)
\thmnote{\nobreakspace\the\thm@notefont\bfseries---\nobreakspace#3.}}% Optional theorem note

\newtheoremstyle{blacknumbox} % Theorem style name
{0pt}% Space above
{0pt}% Space below
{\normalfont}% Body font
{}% Indent amount
{\small\bf}% Theorem head font
{\;}% Punctuation after theorem head
{0.25em}% Space after theorem head
{\small\thmname{#1}\nobreakspace\thmnumber{\@ifnotempty{#1}{}\@upn{#2}}% Theorem text (e.g. Theorem 2.1)
\thmnote{\nobreakspace\the\thm@notefont\bfseries---\nobreakspace#3.}}% Optional theorem note

\newtheoremstyle{ocrenum}% % Theorem style name
{5pt}% Space above
{5pt}% Space below
{\normalfont}% % Body font
{}% Indent amount
{\small\bf\color{ocre}}% % Theorem head font
{\;}% Punctuation after theorem head
{0.25em}% Space after theorem head
{\small\color{ocre}\thmname{#1}\nobreakspace\thmnumber{\@ifnotempty{#1}{}\@upn{#2}}% Theorem text (e.g. Theorem 2.1)
\thmnote{\nobreakspace\the\thm@notefont\bfseries\color{black}---\nobreakspace#3.}} % Optional theorem note
\makeatother

\newcounter{dummy} 
\numberwithin{dummy}{section}
\theoremstyle{ocrenumbox}
\newtheorem{theoremeT}[dummy]{Theorem}

\newtheorem{exerciseT}{Exercise}[section]
\theoremstyle{blacknumex}
\newtheorem{exampleT}{Example}[section]
\theoremstyle{blacknumbox}

\newtheorem{definitionT}{Definition}[section]
\newtheorem{corollaryT}[dummy]{Corollary}
\theoremstyle{ocrenum}

\RequirePackage[framemethod=default]{mdframed} % Required for creating the theorem, definition, exercise and corollary boxes

\newmdenv[skipabove=7pt,
skipbelow=7pt,
backgroundcolor=black!5,
linecolor=ocre,
innerleftmargin=5pt,
innerrightmargin=5pt,
innertopmargin=5pt,
leftmargin=0cm,
rightmargin=0cm,
innerbottommargin=5pt]{tBox}

\newmdenv[skipabove=7pt,
skipbelow=7pt,
rightline=false,
leftline=true,
topline=false,
bottomline=false,
backgroundcolor=ocre!10,
linecolor=ocre,
innerleftmargin=5pt,
innerrightmargin=5pt,
innertopmargin=5pt,
innerbottommargin=5pt,
leftmargin=0cm,
rightmargin=0cm,
linewidth=4pt]{eBox}	

\newmdenv[skipabove=7pt,
skipbelow=7pt,
rightline=false,
leftline=true,
topline=false,
bottomline=false,
linecolor=ocre,
innerleftmargin=5pt,
innerrightmargin=5pt,
innertopmargin=0pt,
leftmargin=0cm,
rightmargin=0cm,
linewidth=4pt,
innerbottommargin=0pt]{dBox}	

\newmdenv[skipabove=7pt,
skipbelow=7pt,
rightline=false,
leftline=true,
topline=false,
bottomline=false,
linecolor=gray,
backgroundcolor=black!5,
innerleftmargin=5pt,
innerrightmargin=5pt,
innertopmargin=5pt,
leftmargin=0cm,
rightmargin=0cm,
linewidth=4pt,
innerbottommargin=5pt]{cBox}

\newcounter{openq}

\newenvironment{openq}{%
\refstepcounter{openq}%
\par\vspace{10pt}\small
\begin{list}{}{%
\leftmargin=45pt
\rightmargin=25pt}%
\item\ignorespaces
\makebox[-2.5pt]{%
\begin{tikzpicture}[overlay]
\node[
  draw=ocre!60,
  line width=1pt,
  circle,
  fill=ocre!25,
  font=\bfseries,
  inner sep=2pt,
  outer sep=0pt
] at (-25pt,0pt){\textcolor{ocre}{OQ~\theopenq}};
\end{tikzpicture}}%
\advance\baselineskip -1pt%
}{%
\end{list}\vskip5pt%
}
\makeatletter
\renewcommand{\@seccntformat}[1]{\llap{\textcolor{ocre}{\csname the#1\endcsname}\hspace{1em}}}                    
\renewcommand{\section}{\@startsection{section}{1}{\z@}
{-4ex \@plus -1ex \@minus -.4ex}
{1ex \@plus.2ex }
{\normalfont\large\bfseries}}
\renewcommand{\subsection}{\@startsection {subsection}{2}{\z@}
{-3ex \@plus -0.1ex \@minus -.4ex}
{0.5ex \@plus.2ex }
{\normalfont\bfseries}}
\renewcommand{\subsubsection}{\@startsection {subsubsection}{3}{\z@}
{-2ex \@plus -0.1ex \@minus -.2ex}
{.2ex \@plus.2ex }
{\normalfont\small\bfseries}}                        
\renewcommand\paragraph{\@startsection{paragraph}{4}{\z@}
{-2ex \@plus-.2ex \@minus .2ex}
{.1ex}
{\normalfont\small\bfseries}}

\newcommand{\@mypartnumtocformat}[2]{%
\setlength\fboxsep{0pt}%
\noindent\colorbox{ocre!20}{\strut\parbox[c][.7cm]{\ecart}{\color{ocre!70}\Large\bfseries\centering#1}}\hskip\esp\colorbox{ocre!40}{\strut\parbox[c][.7cm]{\linewidth-\ecart-\esp}{\Large\centering#2}}}%
\newcommand{\@myparttocformat}[1]{%
\setlength\fboxsep{0pt}%
\noindent\colorbox{ocre!40}{\strut\parbox[c][.7cm]{\linewidth}{\Large\centering#1}}}%
\newlength\esp
\newlength\ecart
\def\@part[#1]#2{%
\ifnum \c@secnumdepth >-2\relax%
\refstepcounter{part}%
\addcontentsline{toc}{part}{\texorpdfstring{\protect\@mypartnumtocformat{\thepart}{#1}}{\partname~\thepart\ ---\ #1}}
\else%
\addcontentsline{toc}{part}{\texorpdfstring{\protect\@myparttocformat{#1}}{#1}}%
\fi%
\startcontents%
\markboth{}{}%
{\thispagestyle{empty}%
\begin{tikzpicture}[remember picture,overlay]%
\node at (current page.north west){\begin{tikzpicture}[remember picture,overlay]%	
\fill[ocre!20](0cm,0cm) rectangle (\paperwidth,-\paperheight);
\node[anchor=north] at (4cm,-3.25cm){\color{ocre!40}\fontsize{220}{100}\bfseries\@Roman\c@part}; 
\node[anchor=south east] at (\paperwidth-1cm,-\paperheight+1cm){\parbox[t][][t]{8.5cm}{
\printcontents{l}{0}{\setcounter{tocdepth}{1}}%
}};
\node[anchor=north east] at (\paperwidth-1.5cm,-3.25cm){\parbox[t][][t]{15cm}{\strut\raggedleft\color{white}\fontsize{30}{30}\bfseries#2}};
\end{tikzpicture}};
\end{tikzpicture}}%
\@endpart}
\def\@spart#1{%
\startcontents%
\phantomsection
{\thispagestyle{empty}%
\begin{tikzpicture}[remember picture,overlay]%
\node at (current page.north west){\begin{tikzpicture}[remember picture,overlay]%	
\fill[ocre!20](0cm,0cm) rectangle (\paperwidth,-\paperheight);
\node[anchor=north east] at (\paperwidth-1.5cm,-3.25cm){\parbox[t][][t]{15cm}{\strut\raggedleft\color{white}\fontsize{30}{30}\bfseries#1}};
\end{tikzpicture}};
\end{tikzpicture}}
\addcontentsline{toc}{part}{\texorpdfstring{%
\setlength\fboxsep{0pt}%
\noindent\protect\colorbox{ocre!40}{\strut\protect\parbox[c][.7cm]{\linewidth}{\Large\protect\centering #1\quad\mbox{}}}}{#1}}%
\@endpart}
\def\@endpart{\vfil\newpage
\if@twoside
\if@openright
\null
\thispagestyle{empty}%
\newpage
\fi
\fi
\if@tempswa
\twocolumn
\fi}

\newif\ifusechapterimage
\usechapterimagetrue
\newcommand{\thechapterimage}{}%
\def\@makechapterhead#1{%
{\parindent \z@ \raggedright \normalfont
\ifnum \c@secnumdepth >\m@ne
\if@mainmatter
\begin{tikzpicture}[remember picture,overlay]
\node at (current page.north west)
{\begin{tikzpicture}[remember picture,overlay]
\node[anchor=north west,inner sep=0pt] at (0,0) {\ifusechapterimage\includegraphics[width=\paperwidth]{\thechapterimage}\fi};
\draw[anchor=west] (\Gm@lmargin,-9cm) node [line width=2pt,rounded corners=15pt,draw=ocre,fill=white,fill opacity=0.5,inner sep=15pt]{\strut\makebox[22cm]{}};
\draw[anchor=west] (\Gm@lmargin+.3cm,-9cm) node {\huge\bfseries\color{black}\thechapter. #1\strut};
\end{tikzpicture}};
\end{tikzpicture}
\else
\begin{tikzpicture}[remember picture,overlay]
\node at (current page.north west)
{\begin{tikzpicture}[remember picture,overlay]
\node[anchor=north west,inner sep=0pt] at (0,0) {\ifusechapterimage\includegraphics[width=\paperwidth]{\thechapterimage}\fi};
\draw[anchor=west] (\Gm@lmargin,-9cm) node [line width=2pt,rounded corners=15pt,draw=ocre,fill=white,fill opacity=0.5,inner sep=15pt]{\strut\makebox[22cm]{}};
\draw[anchor=west] (\Gm@lmargin+.3cm,-9cm) node {\huge\bfseries\color{black}#1\strut};
\end{tikzpicture}};
\end{tikzpicture}
\fi\fi\par\vspace*{270\p@}}}

\def\@makeschapterhead#1{%
\begin{tikzpicture}[remember picture,overlay]
\node at (current page.north west)
{\begin{tikzpicture}[remember picture,overlay]
\node[anchor=north west,inner sep=0pt] at (0,0) {\ifusechapterimage\includegraphics[width=\paperwidth]{\thechapterimage}\fi};
\draw[anchor=west] (\Gm@lmargin,-9cm) node [line width=2pt,rounded corners=15pt,draw=ocre,fill=white,fill opacity=0.5,inner sep=15pt]{\strut\makebox[22cm]{}};
\draw[anchor=west] (\Gm@lmargin+.3cm,-9cm) node {\huge\bfseries\color{black}#1\strut};
\end{tikzpicture}};
\end{tikzpicture}
\par\vspace*{270\p@}}
\makeatother

\usepackage{hyperref}
\hypersetup{hidelinks,colorlinks=false,breaklinks=true,urlcolor= ocre,bookmarksopen=false,pdftitle={Title},pdfauthor={Author}}
\usepackage{bookmark}
\bookmarksetup{
open,
numbered,
addtohook={%
\ifnum\bookmarkget{level}=0 % chapter
\bookmarksetup{bold}%
\fi
\ifnum\bookmarkget{level}=-1 % part
\bookmarksetup{color=ocre,bold}%
\fi
}
}

\usepackage{wrapfig}

\usepackage[utf8]{inputenc}
\usepackage{amsmath,esint}
\usepackage{mathtools}
\usepackage[T1]{fontenc}
\usepackage{amsfonts}
\usepackage{mathrsfs}
\usepackage{siunitx}
\usepackage{graphicx}
\usepackage{bm}

\newcommand{\citeasnoun}[1]{Ref.~\cite{#1}}

\newcommand{\Figref}[1]{Figure~\ref{fig:#1}}
\newcommand{\figref}[1]{Fig.~\ref{fig:#1}}
\renewcommand{\eqref}[1]{Eq.~(\ref{#1})}
\newcommand{\Eqref}[1]{Equation~(\ref{#1})}

\newcommand{\secref}[1]{Sec.~\ref{sec:#1}}
\newcommand{\eqrefs}[2]{Eqs.~(\ref{#1},\ref{#2})}

\DeclareMathAlphabet{\CMcal}{OMS}{cmsy}{m}{n}
\newcommand{\Bl}{\CMcal{B}}
\newcommand{\Pvl}{\CMcal{P}}

\renewcommand{\Re}{\operatorname{Re}}
\renewcommand{\Im}{\operatorname{Im}}
\newcommand{\Tr}{\operatorname{Tr}}
\newcommand{\vect}[1]{\mathbf{#1}}
\newcommand{\Ev}{\vect{E}}
\newcommand{\Dv}{\vect{D}}
\newcommand{\Hv}{\vect{H}}
\newcommand{\Jv}{\vect{J}}
\newcommand{\Pv}{\vect{P}}
\newcommand{\Mv}{\vect{M}}
\newcommand{\Nv}{\vect{N}}
\newcommand{\xv}{\vect{x}}
\newcommand{\yv}{\vect{y}}
\newcommand{\zv}{\vect{z}}
\newcommand{\av}{\vect{a}}
\newcommand{\bv}{\vect{b}}
\newcommand{\pv}{\vect{p}}
\newcommand{\kv}{\vect{k}}
\newcommand{\kp}{\kv_{\parallel}}
\newcommand{\xp}{\xv_{\parallel}}
\newcommand{\ev}{\vect{e}}
\newcommand{\sv}{\vect{s}}
\newcommand{\iv}{\vect{i}}
\newcommand{\vv}{\vect{v}}
\newcommand*{\ep}[1]{\hat{\ev}_p(#1)}
\newcommand*{\qhat}{\hat{\vect{q}}}
\newcommand*{\nhat}{\hat{\vect{n}}}
\newcommand*{\dkk}{{\rm d}\kp}
\newcommand{\Einc}{\Ev_{\rm inc}}
\newcommand{\Hinc}{\Hv_{\rm inc}}

\newcommand{\TT}{\mathbb{T}}
\renewcommand{\AA}{A}
\newcommand{\CC}{C}
\newcommand{\DD}{\mathbb{D}}
\newcommand{\II}{\mathbb{I}}
\newcommand{\GG}{\mathbb{G}}
\newcommand{\QQ}{Q}
\newcommand{\XX}{\mathbb{X}}
\renewcommand{\SS}{\mathbb{S}}
\newcommand{\cout}{\vect{c}_{\rm out}}
\newcommand{\cin}{\vect{c}_{\rm in}}
\title{Fundamental limits in photonics and electromagnetics: a tutorial}
\author[1]{Owen D. Miller}
\author[2]{Francesco Monticone}
\affil[1]{Yale, owen.miller@yale.edu}
\affil[2]{Cornell, francesco.monticone@cornell.edu}

\DeclareRobustCommand{\hbar}{{\mathchar'26\mkern-9muh}}

\begin{document}

\usechapterimagefalse 
\pagestyle{fancy} 
\maketitle
\date{\vspace{-5ex}}

\begin{abstract}
Theoretical limits and physical bounds across many areas of science, mathematics, and technology---including Shannon's information-capacity limits, Bennett and Landauer's thermodynamic limits on computation, and G\"{o}del's incompleteness theorem in formal logic---serve as defining pillars of their fields. In photonics and electromagnetism, numerous physical bounds and constraints have likewise been uncovered over the past several decades. In this Tutorial, we first review the fundamental principles that constrain light-matter interactions, and then discuss limits at different hierarchical scales, from optical material responses to wave propagation, scattering, and related optical phenomena and functions, relevant to a wide range of applications. By providing a more unified treatment of these results and highlighting the many open questions that remain, our goal is to help readers rapidly get up to speed with the frontier of this research area and contribute to advancing the broader vision of a universal framework for fundamental limits of light interactions with matter.
\end{abstract}

\clearpage
\thispagestyle{empty}
\tableofcontents

\section*{Introduction}
\phantomsection
\addcontentsline{toc}{section}{\protect\numberline{I}Introduction}
\markboth{Introduction}{Introduction}
The laws of physics allow for vast, but not unlimited, engineering opportunities. Understanding the fundamental limits of what is possible has long been a crucial part of modern science and technology, often sparking breakthroughs in understanding and design approaches by distilling the fundamental aspects of a problem from an immense space of design choices and physical details. In photonics, experimental advances now enable wavelength-scale patterning of nanophotonic materials across cm-scale wafers, opening design spaces with structural degrees of freedom numbering from the hundreds to well past billions. While computational design methods have made tremendous progress over the past few decades, we argue that, to fully harness the vast opportunities offered by such high-dimensional design spaces, it is critical to establish, \emph{a priori}, the fundamental bounds and limits on what is achievable.

Identifying optimal bounds is a balancing act. One must relax the problem of interest, for instance by removing physics or practical constraints, to achieve typically conflicting goals: (1) simplify the mathematical structure to one whose global optimum can be found, and (2) retain the key (``binding'') constraints so that the resulting bounds are meaningful. For many decades, only aggregate quantities and/or highly simplified electromagnetic scenarios were amenable to bounds. Some of the most notable and longstanding examples include Abbe’s diffraction limit for imaging systems \cite{abbe1873beitrage} and the all-frequency, ray-optical Yablonovitch $4n^2$ limit for photovoltaic absorbers \cite{yablonovitch1982statistical}. In recent years, however, efforts to identify the extreme limits of light-matter interactions have evolved from such isolated results tailored for specific applications to a rapidly emerging field with a broader outlook and vision at the intersection of modern electromagnetism, applied optimization theory, and nanophotonic engineering.

In this Tutorial, we aim to organize a disparate (and rapidly growing) set of results within this nascent field, viewed through the lens of basic constraints of electrodynamics---causality, linearity, reciprocity, and conservation laws. We discuss the wide-ranging results that already provide bounds for many problems and applications of current interest, and we offer a more unified treatment of the general principles and approaches used to derive theoretical limits and physical bounds in electromagnetics and photonics. The general structure of the Tutorial, illustrated and summarized in Fig. 1, consists of six chapters, the first five of which cover a wide array of topics:
\begin{enumerate}
    \item \textbf{Fundamental principles.} We begin by discussing several fundamental principles, including linearity, causality, and energy conservation, that enable general insights into photonic systems independent of their specific implementation details. These principles and their mathematical implications---analyticity, positivity, symmetries---impose a certain degree of ``rigidity'' on the relevant response functions, from which one can derive fundamental limits on a wide range of optical phenomena and functions.
    \item \textbf{Bounds on constitutive material properties.} Chapter 2 discusses fundamental limits on the macroscopic electromagnetic response of continuous media, with an emphasis on the linear electric susceptibility/permittivity and refractive index. We examine how large or how small the real and imaginary parts of these quantities can fundamentally be, particularly in relation to bandwidth and dispersion constraints. We also briefly discuss limitations and tradeoffs associated with nonlinear optical susceptibilities, refractive-index modulation, magneto-electric coupling, and chirality.
    \item \textbf{Bounds on wavefront shaping.} Chapter 3 explores physical bounds on \emph{wavefront shaping}---the optimization of source distributions to control electromagnetic wave fields in scenarios with fixed scatterers, from free space to complex media. Most such bounds reduce, via the Rayleigh-quotient structure of quadratic objectives normalized by total source power, to extremal-eigenvalue computations of source-to-output operators. Within this framework, we discuss limits on free-space focusing (including the diffraction limit, super-oscillations, and spacetime focusing), focusing in complex media (where reciprocity, rather than time reversal, sets the underlying constraint), and other objectives including Wigner--Smith time delay, optical force and torque, communication channels, and optical-thickness bounds.
    \item \textbf{Bounds on electromagnetic scattering from global symmetries.} While Chapters 2 and 3 discuss fundamental limits on the electromagnetic response of materials and on the behavior of electromagnetic waves largely independently, Chapters 4 and 5 examine the limitations governing the interaction between waves and material bodies through the lens of electromagnetic scattering theory. Specifically, Chapter 4 presents physical bounds that arise from global symmetries and constraints of the scattering matrix, including time-reversal symmetry, reciprocity, and conservation laws.
    \item \textbf{Bounds on electromagnetic scattering from local constraints.} Chapter 5 ``zooms-in'' from a more abstract description in terms of scattering matrix bounded by global symmetries (Chapter 4) to actual physical structures whose response is limited by more ``local'' constraints. In this context, the typical problem setup specifies certain material properties (such as the available permittivity or electron density) and/or geometrical constraints (such as the size or shape of a bounding volume), while the optimal arrangement or structuring of matter needed to achieve a given objective remains unknown. One then seeks global bounds on the desired objective function (e.g., scattering or absorption efficiency, invisibility, light confinement and enhancement, time delay) given the available material properties and volume, or, conversely, limits and scaling laws on the minimum material or geometrical ``resources'' required to realize a specified function. An important theme of this chapter is whether, and to what extent, \emph{structuring} is beneficial for achieving a given goal, or whether a homogeneous system is already optimal. This also helps clarify when a \textbf{materials-first} approach (optimizing the choice of material) or a \textbf{structure-first} approach (optimizing the specific structure) is preferable in a given design problem.

\end{enumerate}

Throughout this Tutorial, we emphasize general principles that can be widely applied and identify \textbf{open questions} (labeled OQ~1, OQ~2, \dots) that we believe are particularly important and worth exploring further. In Chapter 6, we look ahead, predicting a continued flowering of approaches from a small set of fundamental constraints, and identifying new opportunities and next frontiers.

We have tried to keep the chapters largely self-contained, so that readers can sample them according to their interests; the main exception is Chapter 1, on which all later chapters build in some form. Chapter 2 draws mainly on causality and passivity (Secs.~\ref{sec:causality} and \ref{sec:passivity}), through the Kramers--Kronig relations and sum rules. Chapters 3 and 4 rest on the extremal-eigenvalue (Rayleigh-quotient) results of \secref{RQs} and on reciprocity (\secref{tr_symmetry}), with Chapter 4 also using the conservation laws of \secref{conslaws}. Chapter 5 combines causality, passivity, and the conservation laws (notably the optical theorem), and leverages some of the techniques developed for the material bounds of Chapter 2.

Given how broad the fields of electromagnetics and photonics are today and how wide-ranging the topic of fundamental limits can be, it is also worthwhile to briefly outline what the scope of this Tutorial is and what it does \emph{not} cover. Specifically, our focus is primarily on:
\begin{itemize}
    \item \textbf{Linear time-invariant (LTI) systems.} We primarily focus on systems with linear material properties and continuous time-translation invariance. In a few sections, we briefly discuss possible extensions to non-LTI systems, while also highlighting the challenges involved. Extending physical bounds and limits to broader classes of nonlinear and time-varying systems is, arguably, one of the most important directions for future research in this field.
    \item  \textbf{Classical wave propagation and light-matter interactions.} Classical electrodynamic bounds are still highly relevant for quantum optics and photonics, where systems typically rely on a predominantly classical ``infrastructure'', such as cavities and waveguides, to control interactions with optically addressable quantum elements. We do not, however, discuss bounds on genuine quantum-optical properties and phenomena, such as entanglement. Causality constraints on classical scattering and conservation-law-based bounds on wave control may be generalized, respectively, to quantum scattering and quantum control problems, but with some crucial differences (see, e.g., \cite{Nussenzveig1972,zhang2021conservation}, and footnote \ref{footnote_quantum_scat}).
    \item \textbf{Optical material responses and basic electromagnetic phenomena.} As illustrated in Fig. 1, this Tutorial focuses more on the response of materials and structures than on devices and system-level applications. Specifically, limits on optoelectronic devices such as modulators, amplifiers, detectors, and photovoltaic cells are beyond the scope of this work (see, e.g., \cite{Shockley1961,liu2016fundamental,xu2015generalized}). In this context, we also note that identifying limits on hybrid systems that involve both optical, electronic, and algorithmic design, such as end-to-end optimized systems for imaging and inference, is an important research frontier in this field.
    \item \textbf{Deterministic electrodynamics.} This Tutorial does not cover statistical or fluctuation-based phenomena (thermal or quantum, such as radiative heat transfer or Casimir effects) or problems related to the propagation of wave signals in noisy channels. By extension, we do not focus on thermodynamics-based or information-based bounds (see, e.g., \cite{landsberg1980thermodynamic,fan2022photonics,franceschetti2017wave}). However, as we briefly discuss in Section~\ref{sec:channels}, bounds and scaling laws on the number of available channels or degrees of freedom in a photonic structure may serve as a bridge between wave-based and information-theoretic-based quantities and constraints, an area that we believe could provide very fertile ground for future research.
\end{itemize}
Every bound discussed in this Tutorial rests on specific assumptions, such as linearity, passivity, time invariance, reciprocity, or spatial locality, and relaxing any one of them may allow one to surpass the bound. Knowing which assumptions a bound relies on therefore indicates not only what cannot be achieved, but also under which conditions a limitation may be avoided. Table~\ref{tab:assumptions} collects, for the main classes of bounds, the assumptions they rest on and the routes around them discussed in the text.

\begin{table*}[tbp]
\centering
\caption{\textbf{Main classes of bounds discussed in this Tutorial, the assumptions they rest on, and alternatives.} Relaxing any listed assumption may allow one to surpass the corresponding bound. The last column lists, with section numbers in parentheses, the routes around each bound discussed in the text.}
\label{tab:assumptions}
\vspace{8pt}
\footnotesize
\renewcommand{\arraystretch}{1.0}\setlength{\tabcolsep}{4pt}
\begin{tabular}{p{0.2\linewidth}p{0.365\linewidth}p{0.365\linewidth}}
\hline
\textbf{Bounds (Sec.)} & \textbf{Key assumptions} & \textbf{Routes around the bound, and alternatives (Sec.)} \\
\hline
KK relations, Herglotz representations, and sum rules (\ref{sec:causality}, \ref{sec:passivity}) & Causality; linearity; time invariance; square integrability or known asymptotic behavior; passivity, for the Herglotz property and the positivity used in sum-rule bounds & Gain removes the Herglotz property (\ref{sec:how_large_imag}); conductors require modified KK formulas (\ref{sec:causality}); time-varying and nonlinear responses depend on several frequencies (\ref{sec:symmetry_time}, \ref{sec:nonlinear}) \\
Material response: refractive index and the real and imaginary parts of $\chi$ (\ref{sec:large_susceptibility}--\ref{sec:small_imaginary}) & Linear, passive, causal, bulk media; total oscillator strength fixed by the electron density ($f$-sum rule); for the index bounds, transparency and a specified dispersion or bandwidth constraint & Gain (\ref{sec:how_small_real}, \ref{sec:how_large_imag}); a relaxed dispersion constraint, i.e., operation closer to resonance, at the cost of loss (\ref{sec:large_susceptibility}); potential loopholes for gyrotropic plasmonic media (\ref{sec:large_susceptibility}) \\
Focusing, wavefront shaping, channels, and thickness (\ref{sec:FSFocus}--\ref{sec:thickness}) & Linear medium; sources confined to a given aperture or region, with fixed total power & Larger apertures or solid angles (\ref{sec:FSFocus}); sub-diffraction spots via superoscillations, within the bounds and at the cost of power in sidelobes (\ref{sec:FSFocus}); in non-reciprocal media the optimal-focusing recipe requires transposed material tensors (\ref{sec:CMNonreciprocal}) \\
Reciprocity constraints, e.g., extinction symmetries (\ref{sec:GlobalSym}) & Linear media with $\varepsilon=\varepsilon^T$, $\mu=\mu^T$, $\xi=-\zeta^T$, Eq.~(\ref{eq:reciprocal}): no bias that is odd under time reversal & Bias by magnetic field, velocity, or angular momentum; nonlinear systems, with strict limitations (\ref{sec:tr_symmetry}) \\
Time-reversal-symmetry constraints (\ref{sec:GlobalSym}) & Time-reversal-symmetric systems, e.g., lossless and reciprocal ones, Eq.~(\ref{eq:S_conditions}) & Loss or non-reciprocity, in general (\ref{sec:tr_symmetry}) \\
Brightness theorem: no perfect incoherent beam combining (\ref{sec:Brightness}) & Linear, passive, time-invariant systems at a single frequency (reciprocal or not) & Coherent inputs, incomplete combining, gain; time-varying systems combine directions only by spreading in frequency; nonlinearity (\ref{sec:Brightness}) \\
Single-frequency scattering bounds from conservation laws (\ref{sec:SingleFreq}, \ref{sec:comput_bounds}--\ref{sec:mode_volume}) & Linear, passive, local constitutive relations; given material and bounding volume; single frequency; positive material loss, for bounds set by finite loss; for mode-volume bounds, geometric constraints such as a minimum feature size & Gain, which removes the positive-semidefinite absorption term (\ref{sec:SingleFreq}); nonlocal constitutive relations require different constraints (\ref{sec:comput_bounds}) \\
All-frequency scattering bounds from sum rules (\ref{sec:CausalResp}) & Causality, passivity, linearity, and time invariance; known low- and high-frequency limits & Gain, which removes the positivity needed to turn the sum rules into inequalities; time-varying and nonlinear systems (cf.\ \ref{sec:cloaking}) \\
Cloaking, scattering reduction, and impedance matching (\ref{sec:cloaking}) & Linear, time-invariant, causal, stable systems; passivity for the bandwidth bounds; finite objects in free space, for the invisibility argument; non-diamagnetic materials, for the global cloaking bound & Active scatterers with prior knowledge of the incident field; time-varying or nonlinear systems (\ref{sec:cloaking}) \\
Delay--bandwidth product (\ref{sec:delay-bandwidth}) & Linear, time-invariant systems (reciprocal or not); a single resonance, or, for the general bounds, a given length and index contrast in one-dimensional structures & Nonlinear or time-varying systems (\ref{sec:delay-bandwidth}) \\
\hline
\end{tabular}
\end{table*}

\begin{figure*}[tbp]
\centering
\includegraphics[width=0.97\linewidth]{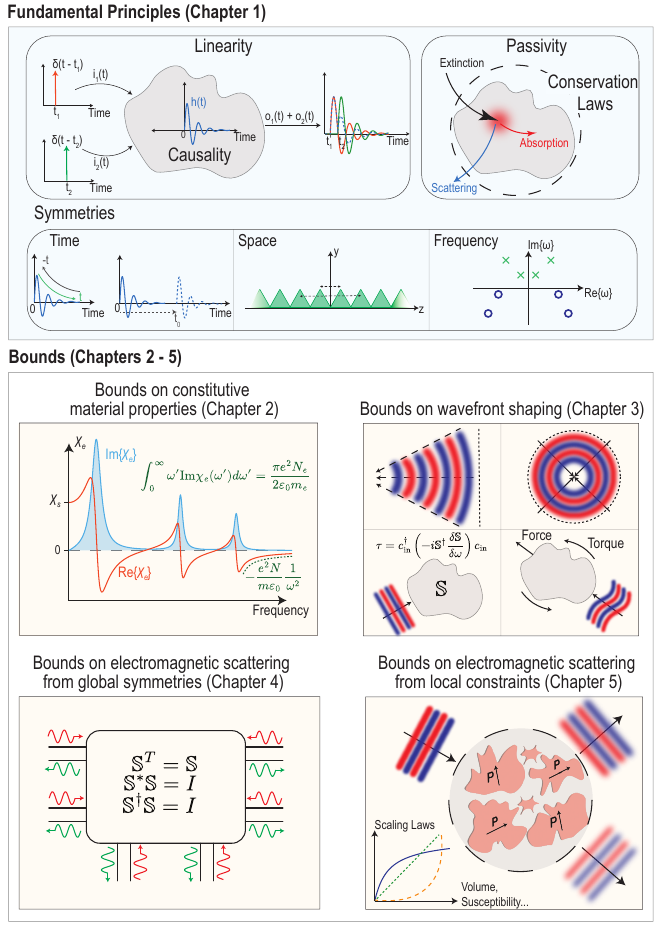}
\caption{\textbf{Overview of the manuscript}, illustrating the fundamental principles and broad classes of bounds discussed in this Tutorial. A brief sixth chapter looks forward.}
\label{fig:Overview}
\end{figure*}

Within this scope, and given the ubiquity of classical photonic and electromagnetic technologies, the range of applications influenced by the bounds discussed in this Tutorial is enormous, spanning areas from imaging and sensing to communications and computing. We predict that a bounds-first approach to photonic design will accelerate innovation: global bounds identify applications for which current devices are already operating close to their limits, and applications for which there is opportunity for orders-of-magnitude improvement over the current state of the art. While computational resources are expected to continue to grow, understanding fundamental limits can guide where and how these resources should be invested, while also revealing interpretable insights that would otherwise remain hidden within the complexity of a computational-design problem.

\section{Fundamental principles}
\subsection{Maxwell's equations}
For linear time-invariant media and in the frequency domain\footnote{Throughout this work, we assume the time-harmonic convention $e^{-i \omega t}$, which is common in physics and optics/photonics. The corresponding expressions under the $e^{j \omega t}$ convention, common in electrical engineering and applied electromagnetism (with $j=\sqrt{-1}$), are obtained by the substitution $i \to -j$ (equivalently, for real time-domain fields, by complex-conjugating the frequency-domain quantities and writing $j$ for $i$).}, electric and magnetic sources $\Jv_E$ and $\Jv_M$, respectively, give rise to electric and magnetic fields, $\Ev$ and $\Hv$, according to Maxwell's equations,
    \begin{align}
        \left[
        \begin{pmatrix}
             & \nabla \times \\
             -\nabla \times & 
        \end{pmatrix}
        +
        i\omega 
        \begin{pmatrix}
            \varepsilon & \xi/c \\
            \zeta/c & \mu
        \end{pmatrix}
        \right]
        \begin{pmatrix}
            \Ev \\
            \Hv
        \end{pmatrix}
        = 
        \begin{pmatrix}
            \Jv_{E} \\
            \Jv_{M}
        \end{pmatrix}
        \label{eq:MaxwellDiffEq}
    \end{align}
with appropriate boundary conditions. Radiating outgoing fields can be absorbed by perfectly matched layers~\cite{Berenger1994-lx,Sacks1995-la}, such that trivial zero-field boundary conditions are suitable. We have ignored the divergence equations, which are implied as long as $\omega$, $\varepsilon$, and $\mu$ are all nonzero. (These special scenarios can often be treated in static, quasi-static, or other specialized formulations.)
It is particularly useful to include independent magnetic current densities in \eqref{eq:MaxwellDiffEq}; in part for symmetry purposes, but especially for their utility as \emph{equivalent currents}. A common scattering-problem scenario has a medium excited by incoming fields $\Ev_{\rm inc}$ and $\Hv_{\rm inc}$. A simple version of the \textbf{surface equivalence principle}~\cite{Schelkunoff1936,Harrington1961,Oskooi2013} can be stated as follows: A scenario with fields $\Ev$ and $\Hv$ continuous across a surface $S$ with outward normal $\nhat$ is equivalent to a scenario with modified fields $\Ev'$ and $\Hv'$ to the interior of the surface as long as surface electric and magnetic currents 
\begin{align}
    \begin{pmatrix}
        \Jv_e \\
        \Jv_m
    \end{pmatrix}
    =
    \begin{pmatrix}
        & -\nhat \times \\
        \nhat \times 
    \end{pmatrix}
    \left[
    \begin{pmatrix}
        \Ev' \\
        \Hv'
    \end{pmatrix}
    -
    \begin{pmatrix}
        \Ev \\
        \Hv
    \end{pmatrix}
    \right]\delta(\xv-\xv_S)
    \label{eq:SEP}
\end{align}
are imposed, with $\delta(\xv-\xv_S)$ encoding a delta function specific to surface $S$ within the volume (domain). This result, \eqref{eq:SEP}, has two immediate uses.

First, one can specify currents $\Jv_e$, $\Jv_m$ that generate everywhere within $S$ an incoming field $\Einc$, $\Hinc$ while producing zero field to the exterior of $S$, by choosing $\Ev'=\Einc$, $\Hv' = \Hinc$, $\Ev=0$, and $\Hv=0$. If the surface $S$ contains all scatterers, then the corresponding sources
\begin{align}
    \begin{pmatrix}
        \Jv_e \\
        \Jv_m
    \end{pmatrix}
    =
    \begin{pmatrix}
        & -\nhat \times \\
        \nhat \times 
    \end{pmatrix}
    \begin{pmatrix}
        \Einc \\
        \Hinc
    \end{pmatrix}
    \delta(\xv-\xv_S)
    \label{eq:ECinc}
\end{align}
correctly produce the incident fields everywhere within $S$. Mathematically, this encodes the excitation in the right-hand side of the linear equation, which is often easier to work with than a specialized boundary condition, and arises repeatedly throughout this tutorial, in the conservation laws below and in the chapters that follow.

Second, one can conversely ``null'' the interior field by setting $\Ev'=\Hv'=0$, while producing the correct outgoing scattered fields, $\Ev=\Ev_{\rm scat}$ and $\Hv=\Hv_{\rm scat}$, in which case our currents at the surface $S$ are given by 
\begin{align}
    \begin{pmatrix}
        \Jv_e \\
        \Jv_m
    \end{pmatrix}
    =
    \begin{pmatrix}
        & \nhat \times \\
        -\nhat \times 
    \end{pmatrix}
    \begin{pmatrix}
        \Ev_{\rm scat} \\
        \Hv_{\rm scat}
    \end{pmatrix}
    \delta(\xv-\xv_S).
    \label{eq:ECscat}
\end{align}
These currents are effectively radiating in free space, and hence propagating according to the analytically known free-space Green's function. Convolutions against this Green's function can often leverage highly optimized discrete Fourier transforms. This insight underpins calculations of the ``far field'' from measurements close to a scatterer (i.e., near-to-far-field transformations). Although we do not use the near-to-far-field relation elsewhere in this tutorial, we will use a related trick to simplify and generalize key results in time-reversal symmetry in Sec.~\ref{sec:CMFocus}.

The separation of incident and scattered fields also lends naturally to an altogether different formulation of Maxwell's equations, via integral equations. If we encapsulate the material parameters in a 6$\times$6 tensor $\nu$, denote the free-space material tensor $\nu_0$, and denote their difference as the susceptibility $\chi$, the differential Maxwell equations can be written
\begin{align}
    \left(M_0 + i\omega \chi\right) \psi = s
\end{align}
If one rewrites the total field as the sum of its incident and scattered components, $\psi = \psi_{\rm inc} + \psi_{\rm scat}$, and uses the fact that the incident field is the solution in the absence of the scatterer, i.e., $M_0 \psi_{\rm inc} = s$, then one finds
\begin{align}
    M_0 \psi_{\rm scat} = -i\omega \chi \psi.
\end{align}
Physically, this means that the scattered field is the field produced in free space (because the Maxwell operator on the left-hand side is $M_0$) by a polarization density (both electric and magnetic, generally) given by $\chi \psi$. Because the scattered field is effectively produced by polarization fields in free space, it can be written as the convolution
\begin{align}
    \psi_{\rm scat}(\xv) = \int_V \Gamma_0(\xv,\xv') \chi(\xv') \psi(\xv') \,{\rm d}\xv',
\end{align}
where $\Gamma_0(\xv,\xv')$ denotes the free-space Green's function from a unit-amplitude dipole at $\xv'$ to a field at $\xv$. It is common (and quite useful for our purposes) to treat the polarization fields as the single unknown, replacing $\phi = \chi \psi$ and writing the total field $\psi = \psi_{\rm inc} + \psi_{\rm scat} = -\xi \phi$, with $\xi=-\chi^{-1}$ taking the role of the material parameter~\footnote{Note that $\phi$ corresponds to $(\Pv \quad \vect{M})^T$, for electric polarization $\Pv$ and magnetization $\vect{M}$, and that $\chi$ is nonzero everywhere that the polarization field is nonzero, which is our domain of interest when solving for $\phi$, so that $\xi$ is well-defined. If the medium is electric-only, with zero magnetic susceptibility, or magnetic-only, with zero electric susceptibility, then one can simply write an analogous integral equation only for $\Pv$ or $\Mv$, respectively.}. We then have a \textbf{volume integral equation} for the polarization fields:
\begin{align}
    \xi(\xv) \phi(\xv) + \int_V \Gamma_0(\xv,\xv') \phi(\xv') \,{\rm d}\xv'  = -\psi_{\rm inc}(\xv).
\end{align}
In the same spirit as writing Maxwell's equations $M\psi = s$, we can notationally simplify the volume integral equation to
\begin{align}
    \left(\Gamma_0+\xi\right) \phi = -\psi_{\rm inc}.
    \label{eq:IE}
\end{align}
This can be understood either as an operator equation, with the overloaded term $\Gamma_0$ now a ``Green's function operator'' that acts on a six-vector polarization density $\phi$ and produces the corresponding field $\psi$, or as a discretized matrix equation with sufficiently high numerical resolution to approximate the original integral equation to arbitrary accuracy (and preserving symmetries). As an alternative to \eqref{eq:IE}, it is quite popular to use \emph{surface integral equation} formulations of Maxwell's equations, with effective surface currents at interfaces between homogeneous materials, for efficient numerical solvers with the fewest possible degrees of freedom~\cite{Chew1995}. Yet surface integral equations have (thus far) offered little benefit for fundamental limits, and hence we do not discuss them further. The volume integral equation of \eqref{eq:IE}, by contrast, sometimes referred to as the Lippmann--Schwinger equation, forms the basis for a wide range of techniques in scattering theory~\cite{Chew1995,Newton2013,Nussenzveig1972} and features prominently in our discussion of conservation laws in Sec.~\ref{sec:conslaws} and the corresponding fundamental limits in Chapter~\ref{sec:Scatt_Local}.
 
\subsection{Linearity} \label{sec:linearity}
A core concept that pervades our entire tutorial is \emph{linearity}. By assuming linear materials, we have either the differential- or integral-equation formulations,
\begin{align}
    M\psi &= s &\qquad\textrm{ (differential), } \nonumber \\
    \left(\Gamma_0+\xi\right) \phi &=-\psi_{\rm inc} &\qquad\textrm{ (integral) }, \nonumber
\end{align}
which are linear ``input-output'' relations. In the differential version, the inputs are currents densities and the outputs are total fields, whereas in the integral version, the inputs are incident fields and the outputs are the induced polarization densities. Linearity implies superposition: additive inputs produce additive outputs.

It is often useful to directly write $\psi$ as a linear\footnote{Linear as long as no source produces no field; otherwise the map is affine.} function of $s$, which defines the \textbf{Green's function} $\Gamma(\xv,\xv')$:
\begin{align}
    \psi(\xv) = \int_V \Gamma(\xv,\xv') s(\xv') \,{\rm d}\xv', \quad\textrm{ or }\quad \psi = \Gamma s,
\end{align}
where this Green's function is now specific to a given scattering configuration and not necessarily the response in free space. By direct comparison, one can see that the Green's function operator is formally the inverse of the differential operator, i.e., $\Gamma = M^{-1}$. Similarly, one can write the induced polarization fields $\phi$ generally as a convolution of an operator with the ``source'' $\psi_{\rm inc}$. This operator can be denoted the $\TT$ operator, giving
\begin{align}
    \phi(\xv) = \int_V \TT(\xv,\xv') \psi_{\rm inc}(\xv') \,{\rm d}\xv', \quad\textrm{ or }\quad \phi = \TT \psi_{\rm inc}.
\end{align}
The operator/matrix on the right-hand side is the ``volume T matrix,'' and it maps incident fields on a scatterer's domain to the polarization fields they produce, on the same domain. Formally, the $\TT$ matrix is the negative inverse of the relevant integral-equation operator, $\TT = -\left(\Gamma_0 + \xi\right)^{-1}$~\footnote{Throughout this Tutorial, we reserve double-struck symbols (such as $\SS$, $\TT$, $\GG$, $\DD$, $\II$, $\mathbb{A}$, $\mathbb{E}$, $\XX$, $\mathbb{Y}$, $\mathbb{Z}$, $\mathbb{V}$) for named physical operators with fixed roles across the manuscript --- scattering matrices, $T$-matrices, Green's-function operators, conservation-law operators, identity, absorptivity/emissivity, immittance/admittance/impedance, vector spherical waves. Plain capital letters are used for generic matrix variables in proofs and intermediate constructions (such as the $C$ and $D$ in the Herglotz representation, Eq.~(\ref{eq:GenHN})).}.

Finally, another canonical linear object in photonics and electromagnetism is the \textbf{scattering matrix}, or $\SS$ matrix. The conventional definition of the $\SS$ matrix decomposes incoming fields on a bounding surface $S$ into power-carrying ``channels'' (power orthonormalized modes on the surface, such as plane waves or vector spherical waves), which for a given excitation are excited in proportion to coefficients in a vector $\cin$. Outgoing fields are decomposed on outgoing power-carrying channels, which are almost always the time-reversal partners of the incoming waves\footnote{Cf.~\cite{Zhou2016} for an interesting case in which a different choice is useful.}, according to coefficients $\cout$, with the $\SS$ matrix connecting the two:
\begin{align}
    \cout = \SS \cin.
    \label{eq:SMatrix}
\end{align}
There are many other linear matrices/operators that one can define for different choices of inputs and outputs, and in Sec.~\ref{sec:CausalResp} we discuss a few other possibilities in the context of causal response matrices. But the two linear Maxwell equations, alongside Green's functions, $\TT$ matrices, and $\SS$ matrices, cover almost the entirety of this tutorial.

There is a key sense in which the ``linear'' Maxwell equations are, in fact, nonlinear. They are nonlinear with respect to geometric degrees of freedom, in the susceptibility patterns $\chi(\xv)$. For a \emph{fixed} structure, the fields $\psi$ are linear superpositions for source current or source fields. But when the structure can vary, as in design problems, the resulting optimization problems are nonlinear, because the relevant term in the Maxwell's equation is quadratic, as the product $\chi \psi$~\footnote{Indeed, this distinction has recently been exploited to create nonlinear neural networks in ``linear'' electromagnetic systems~\cite{Yildirim2023-hq,Xia2024-nx}.}. The nonlinearity with respect to geometry is crucial to the richness of electromagnetic scattering phenomena\footnote{If the scattering from two nearby spheres were simply the additive sum of the scattering from each individual sphere, and so on for any collection of scatterers, there would be far less variety ultimately possible.}, but it also tremendously increases the complexity of the design process. 

\subsection{Symmetry}
Symmetry is the property that a physical system or a mathematical object remains invariant under a class of transformations.
Mathematically, a transformation operator $  T$ (e.g., a translation operator) represents a symmetry of the system if it \emph{commutes} with the operator $  L$ describing the system dynamics (e.g., the Hamiltonian in quantum mechanics or the differential operator governing the electromagnetic wave equation\footnote{In frequency domain, for systems made of non-magnetic, isotropic media, this operator can be written as $  L=\nabla \times \frac{1}{\varepsilon(\xv)}\nabla \times $, which enters the wave equation for the magnetic field $\Hv$ as $  L \Hv=\left(\frac{\omega}{c}\right)^2\Hv$, defining a standard Hermitian eigenvalue problem in the dissipationless case \cite{Joannopoulos2011}.}. To see what this means, note that if $  T$ represents a symmetry of the system, it should be equivalent whether we apply the governing operator $ L$ directly, or we first apply the transformation (e.g., translating a field in space), then apply $ L$, and finally transform the system back: $ L= T^{-1} L  T$ \cite{Joannopoulos2011}\footnote{Note that, instead, the equality $ L= L  T$ is generally false, even if $ T$ represents a symmetry of the system. For example, applying the differential operator governing the electromagnetic wave equation to a generic field, i.e. $ L \psi$, yields a different result if the field is first translated in space, i.e., $ L  T \psi$ (the resulting function is shifted in space). The result of operating with $ L$ after a translation $ T$ must be translated back, i.e., $ T^{-1} L  T \psi$, to make a meaningful comparison between the two results.}. This implies that $ T L- L  T=0$, which can also be written as $\left[ T, L\right]=0$, where the left-hand side is the commutator of the two operators, defined as in quantum mechanics. An important consequence is that, if two operators commute, they share a common set of eigenfunctions. This property allows one to construct the eigenfunctions (e.g., the electromagnetic eigenmodes) of the governing operator $ L$ using the eigenfunctions of the transformation operator $ T$, which are typically much easier to determine, while also providing insight into the classes of eigenfunctions supported by the system and the associated conserved quantities (see also Section \ref{sec:conslaws}).

As these considerations suggest, symmetry and related concepts play a central role in all branches of physics, and electromagnetics and photonics are no exception. In this section, we briefly review some of the key symmetries relevant to the topics covered in this Tutorial.

\subsubsection{Time-translation symmetry} \label{sec:symmetry_time}
    Time-translation symmetry means that a physical system, and its response to an input, are invariant under a time shift. If the time shift can be arbitrary, then the system has the property of \emph{continuous} time-translation symmetry or time-invariance. In this case, the absolute time instant is irrelevant, and only time differences matter. In other words, if the input to the system is shifted by a certain time interval, the output is simply shifted by the same time interval, with no other change. As a result, the output of an LTI system, more generally expressed as a linear functional of the input with a two-time response (kernel) function $h(t,\tau)$, becomes a standard convolution integral,
    \begin{equation}
    o(t) = \int_{ - \infty }^{ + \infty } h(t,\tau )i(\tau )d\tau =\int_{ - \infty }^{ + \infty } h(t-\tau )i(\tau )d\tau , 
    \label{eq:LTI_response}
    \end{equation}
    since the response of a time-translation-invariant system can only depend on the difference between the arguments of the kernel function. This temporal convolution captures the non-instantaneous (delayed) response of virtually all electromagnetic systems, a property often referred to as \textbf{temporal nonlocality or time- (frequency-) dispersion}. Vacuum is the special case in which the temporal response kernel reduces to a Dirac delta function, yielding a frequency-independent permittivity. The convolution theorem then implies that the Fourier transform of the output is simply given by the product of the Fourier transforms of the input and the kernel function, all evaluated at the same frequency variable $\omega$, which also implies that frequency is conserved in an LTI system. We also note that, aside from requiring the Fourier transform of $h(t)$, i.e., $h(\omega)$, to be a single-variable (single-frequency) function, time-translation symmetry itself does not impose constraints on the possible form of this function. The most important restrictions on its frequency dependence (i.e., its dispersion) arise instead from causality and passivity, as discussed in Sections \ref{sec:causality}, \ref{sec:passivity}.
    
    The observation above regarding the conservation of frequency in LTI systems is consistent with the general symmetry arguments mentioned in the introduction of this section (see also \cite{Joannopoulos2011}, which our discussion here closely follows). Defining a time-translation operator $ T_{t_0}$, if the system is time-translation invariant then $ T_{t_0} \varepsilon(\xv,t-t_0)=\varepsilon(\xv,t)$ (and similarly for the other constitutive parameters), and the operator $ T_{t_0}$ commutes with the differential operator governing the electromagnetic wave equation, i.e., $\left[ T_{t_0}, L_t\right]=0$. The eigenmodes of the system can then be chosen to be eigenfunctions of $ T_{t_0}$ for any value of $t_0$. Specifically, one can easily see that a function of the form $e^{a t}$ is an eigenfunction of any $ T_{t_0}$, where $a$ is a complex constant, since $ T_{t_0}e^{a t}=e^{a (t-t_0)}=e^{-a t_0}e^{a t}$, and $e^{-a t_0}$ is the corresponding eigenvalue. Without loss of generality, and consistent with the $e^{-i\omega t}$ convention used throughout, we can choose $a=-i\omega$, where $\omega$ is the complex frequency. Thus, all eigenmodes of our time-invariant system must have a separable $t$-dependence of the form $e^{-i \omega t}$, and they can be classified by their frequency $\omega$, which is conserved, i.e., it is not modified by the LTI electromagnetic interactions described by $ L_t$. If the LTI system is composed of lossless materials and is therefore dissipationless (equivalently, if the number of photons is conserved), it follows that electromagnetic field energy must also be conserved. This is consistent with Noether’s theorem: in a dissipationless system with \emph{continuous} time-translation symmetry, energy is a conserved quantity. Additional considerations related to conservation laws are discussed in Section \ref{sec:conslaws}. 
    
    Finally, we note that many of the properties discussed above are no longer preserved in the time-varying case, even if linearity still holds. In a linear time-varying system, frequency is not conserved, and the frequency-domain response is described by a convolution involving a function of two frequency variables, as discussed, for example, in \cite{koutserimpas2024time}. Equivalently, if the modulation is itself regarded as an input (which may be optical, electrical, acoustic, mechanical, etc.), a linear time-varying system is nonlinear with respect to the two inputs taken together, since its output depends on products of the signal and the modulation; the resulting generation of sum and difference frequencies is analogous to parametric frequency mixing in nonlinear optics. Energy and photon number are also generally not conserved in time-dependent media because energy can be exchanged between the signal and the modulation, even if the system is otherwise lossless. Interesting special cases can arise in which energy and photon-number conservation become fully decoupled, such as systems with a fixed number of photons that gain energy by climbing a ``frequency ladder'' \cite{pendry2022photon}. The study of time-varying systems is currently one of the most active areas of research in electromagnetics and photonics, motivated in part by the possibility of overcoming theoretical limitations and performance bounds that apply to LTI systems \cite{hayran2023using,ciabattoni2025observation,asgari2024photonic}.

\subsubsection{Space-translation symmetry}
    Continuous space-translation symmetry in a given spatial direction means that the system is spatially invariant (homogeneous) along that direction. Although any real system lacks perfect spatial homogeneity, it is still instructive to consider this idealized case, at least along one direction, for which the spatial analogue of Eq. (\ref{eq:LTI_response}) holds. In this case, the convolution integral captures the \textbf{spatially nonlocal} response of the system, a property also known as spatial dispersion. Taking a spatial Fourier transform then shows that the system response depends on the wavevector (and therefore on the linear momentum of the wave). Nonlocality in the intrinsic response of a material itself (e.g., in the polarization response induced by an applied electric field) is typically weak, and often neglected, but plays a crucial role in regularizing extreme physical behaviors \cite{monticone2020truly,hassani2024dynamical} and in establishing fundamental limits to the response of plasmonic materials and nanostructures, as discussed, for example, in Section \ref{sec:small_imaginary}. It is useful to distinguish nonlocality associated with the motion of mobile excitations---free carriers (as described, e.g., by hydrodynamic models of the electron gas \cite{wubs2016nonlocal}) or excitons (whose center-of-mass motion makes the exciton resonance wavevector dependent)---whose range is set by how far these excitations travel during the response, from nonlocality that involves no such transport, e.g., due to the finite size of bound-charge distributions (molecules, crystal unit cells), which underlies natural optical activity, or to the structure of metamaterials \cite{monticone2025nonlocality}. The study of nonlocality in natural materials, as well as engineered effective nonlocality in metamaterials and metasurfaces, is currently a very active area of research. We refer the reader to \cite{monticone2025nonlocality} for a comprehensive review of this field.

    Analogous to the discussion of continuous time-translation symmetry, if a system has continuous space-translation symmetry in a given direction $z$, the corresponding translation operator $ T_\textbf{d}$ commutes with the governing differential operator, i.e., $\left[ T_\textbf{d}, L\right]=0$ for any choice of $\textbf{d}=d\hat \zv$. The eigenmodes of the system can then be chosen to be eigenfunctions of all the operators $ T_\textbf{d}$. Following the same argument used for the time-invariant case, all eigenmodes of our $z$-invariant system must have a separable $z$-dependence of the form $e^{i k_z z}$, and they can therefore be classified by their wavevector $k_z$. The wavevector in this direction, and thus the corresponding linear momentum, is conserved, i.e., it is not modified by the electromagnetic interactions described by $ L$, consistent with Noether's theorem for systems with continuous translational invariance. We refer the reader to \cite{Joannopoulos2011} for further details, particularly for systems with discrete space-translational symmetry, i.e., periodic structures (such as photonic crystals).
    
    Finally, we note that many common configurations in optics and electromagnetics (such as free-space lens-based systems) are modeled as systems that lack both space-translation symmetry and spatial locality. Even when nonlocality at the material level is neglected, spatial nonlocality is ubiquitous in the input-output response of general optical and electromagnetic systems due to the physics of wave propagation: the output at a given point typically depends on the input over an extended spatial region.
    In such cases, if the input and output are defined on two surfaces or volumes separated by a certain distance (with coordinates $\xv_T$ on the input space and $\xv_R$ on the output space), the input-output response of the system (assumed LTI) at a given frequency can be written as
    \begin{equation}
    o(\xv_R) = \int D(\xv_R; \xv_T)i(\xv_T)d\xv_T , 
    \label{eq:system_operator}
    \end{equation}
    where the integral is evaluated over the input space and $D$ is the spatial kernel of the integral operator (device operator) defined by the equation above (this operator corresponds to the electromagnetic Green’s function if the input function is defined as a current source and the output as the resulting field). The additional assumption that the kernel is square integrable (a reasonable assumption for many wave-based systems of interest\footnote{This can be verified directly for the electromagnetic Green's function provided the input and output spaces are disjoint, avoiding the singularity of the Green’s function at the origin.} \cite{Miller2019-hl}) implies that it is a \textbf{Hilbert-Schmidt integral operator}.
    As a result, it is compact and, in any orthonormal basis, it can be approximated arbitrarily well by a sufficiently large finite matrix  \cite{Miller2019-hl,miller2019introduction}. The singular value decomposition (SVD) of this matrix yields the orthogonal communication ``channels'' or ``modes'' of the system between its input and output surfaces~\footnote{These are not the conventional electromagnetic modes of the system (i.e., the eigenfunctions of the homogeneous wave equation), but rather pairs of functions in the input and output spaces. In this sense, a useful perspective is to view any linear optical system as a device (a ``mode converter'' \cite{miller2012all}) that converts orthogonal functions in the input space to orthogonal functions in the output space,
    with the corresponding pairs determined by the SVD of the device operator/matrix.}. These concepts provide a powerful framework for analyzing and optimizing the spatial response of a wide range of electromagnetic and optical systems \cite{Miller2019-hl,miller2012all}, as well as for establishing fundamental limits on the minimum thickness required for such systems to perform a desired spatial operation (e.g., imaging or signal processing) \cite{miller2023optics,monticone2023toward}. We will further expand on these concepts in Sections \ref{sec:channels}, \ref{sec:thickness}.

\subsubsection{Time-reversal symmetry and reciprocity} \label{sec:tr_symmetry}
Time-reversal symmetry means that a physical system, and its response to an input, are invariant under time reversal, namely, under the transformation $t \to t'=-t$. In electromagnetics, time-reversal symmetry implies that if a given initial field configuration evolves according to Maxwell’s equations into a final configuration, then the time-reversed final configuration (obtained by reversing the magnetic field, wavevectors, and Poynting vector, but not the electric field\footnote{\label{footnote_odd_t}While the classical laws of physics are \emph{form-invariant} under time reversal (they do not acquire new terms, different signs, etc.), the individual physical quantities are time-reversal symmetric or antisymmetric (even or odd). The charge density is even under time reversal, and so are the polarization density, displacement field (via the form invariance of Gauss’s law), and electric field. The wavevector, magnetic field/induction (via Faraday's Law), magnetization density, current density, and Poynting vector are instead odd under time reversal; see \cite[Sect.~6.10]{Jackson1999}.}) will evolve back into the time-reversed initial configuration, with exactly the same field level.
Importantly, in this process, one must also convert outgoing-wave boundary conditions to appropriate incoming-wave conditions, and vice versa. 

In the frequency domain, a time-reversal transformation corresponds to a frequency-reversal transformation (according to the properties of the Fourier transform), which in turn corresponds to complex conjugation due to the reality of classical time-domain fields: a real function $h(t)$ implies $h(-\omega)=h(\omega)^*$. It then follows that the frequency-domain Maxwell's equations are form-invariant under this transformation (conjugation), with  $\textbf{E}^*$ and $-\textbf{H}^*$ as the new fields\footnote{If the frequency is complex, as in the case of complex frequency excitations \cite{kim2025complex}, $\omega$ is also conjugated, corresponding to transforming exponentially growing fields into exponentially decaying ones, and vice versa.}, provided that the constitutive parameters satisfy specific conditions \cite{Haus1984}. For the most general linear material [bianisotropic, including anisotropy and magneto-electric coupling; see Eq.~(\ref{eq:const_bian})], these conditions are
        \begin{equation}
        \varepsilon = \varepsilon^*, ~~ 
        \mu = \mu^*, ~~
        \xi=-\xi^*, ~~
        \zeta=-\zeta^* \quad \textrm{ (time-reversal invariant)}
        \label{eq:t-reversal}
        \end{equation}
where $\varepsilon$ is the permittivity, $\mu$ is the permeability, and 
$\xi$ and $\zeta$ are the magneto-electric coupling tensors~\footnote{For simplicity of notation, the same symbols are used for constitutive parameters whether they are scalar or tensorial.}. Interestingly, these conditions are generally different from those on the constitutive parameters of a \emph{lossless} medium, obtained by requiring the absorbed power to vanish [see Eq.~(\ref{eq:absorbed})], which impose the permittivity and permeability be Hermitian, and the magneto-electric tensors to be related:
        \begin{equation}
        \varepsilon = \varepsilon^\dagger, ~~ 
        \mu = \mu^\dagger, ~~
        \xi=\zeta^\dagger \quad \textrm{ (lossless)}
        \label{eq:lossless}
        \end{equation}
Here, $\dagger$ denotes the Hermitian adjoint (conjugate transpose), $A^\dagger = (A^*)^T = (A^T)^*$. The two sets of conditions coincide for isotropic materials with no magneto-electric coupling; in this case, time-reversal symmetry implies that the material is lossless. (At the level of macroscopic electrodynamics, a system composed of lossy isotropic media (i.e., with a complex scalar permittivity) is therefore not time-reversal symmetric from the standpoint of electromagnetic field evolution, since full time reversal symmetry would also require converting dissipation into gain, and vice versa.) A simple time-domain picture of this result: in a lossless Lorentz medium, $\partial_t^2\Pv+\omega_0^2\Pv=\varepsilon_0\omega_p^2\Ev$, the material and wave equations contain only second-order time derivatives and are therefore invariant under $t\to-t$, whereas a damping term $\gamma\,\partial_t\Pv$, being first order in time, changes sign under this transformation and breaks time-reversal symmetry.

Time-reversal symmetry also imposes conditions on the scattering matrix $\SS$ of the system: $\SS^*=\SS^{-1}$, as shown in \cite[Sec. 3.3]{Haus1984}. This is again different from the lossless condition, which requires the scattering matrix to be unitary, $\SS^\dagger=\SS^{-1}$ (Refs.~\cite{Haus1984,pozar2011microwave}), in order to satisfy power conservation with no dissipation.

A concept related to, but distinct from, time-reversal symmetry is electromagnetic reciprocity. Quoting Ref. \cite{caloz2018electromagnetic}, ``a reciprocal system is defined as a system that exhibits the same received-transmitted field \emph{ratios} when its source(s) and detector(s) are exchanged.'' From this definition, it follows that a system lacking time-reversal symmetry can still be reciprocal, since the received-transmitted field \emph{ratios} may remain the same upon exchanging source and detector even if the absolute field level obtained by time-reversing the received field and ``sending it back'' to the source differs from the initial one. This is the case, for example, for simple lossy systems, which break time-reversal symmetry but are perfectly reciprocal. Indeed, the standard derivation of the Lorentz reciprocity theorem \cite{caloz2018electromagnetic,pozar2011microwave} shows that a system is reciprocal if it is composed of materials whose constitutive parameters satisfy
        \begin{equation}
        \varepsilon = \varepsilon^T, ~~ 
        \mu = \mu^T, ~~
        \xi=-\zeta^T \quad \textrm{ (reciprocal)}
        \label{eq:reciprocal}
        \end{equation}
        which are clearly different from the time-reversal symmetry conditions in Eq. (\ref{eq:t-reversal}). Breaking reciprocity enables a wide range of useful devices and systems, including isolators and circulators, asymmetric emission and absorption, and truly unidirectional waveguides with strong (Chern-type) topological protection, among many others. These opportunities have motivated intense research efforts over the past several years. What is required to break reciprocity? First, as noted above, linear loss and/or gain cannot induce nonreciprocity: for example, a true isolator cannot be realized using any combination of linear loss and/or gain. Similarly, reciprocity cannot be broken through geometrical structuring of linear isotropic materials \cite{jalas2013and}. Instead, in linear systems, reciprocity can only be broken by externally biasing the system with a physical quantity that is odd under time reversal (e.g., velocity, linear or angular momentum, magnetic field; see also footnote \ref{footnote_odd_t}), consistent with the Onsager--Casimir principle \cite{caloz2018electromagnetic}. (In the Lorentz model above, a static magnetic bias $\mathbf{B}_0$ adds a Lorentz-force term proportional to $\partial_t\Pv\times\mathbf{B}_0$ to the equation for $\Pv$; like damping, this term is first order in time, but, unlike damping, it gives the permittivity tensor antisymmetric off-diagonal components, $\varepsilon\neq\varepsilon^T$, and hence makes the medium nonreciprocal.) Reciprocity can also be broken in nonlinear systems through a combination of self-biasing and structural asymmetry, albeit with strict limitations \cite{shi2015limitations}. We refer the reader to \cite{caloz2018electromagnetic} for a detailed discussion of nonreciprocal systems. We also note here that such systems still obey \textbf{generalized reciprocity}: if, when transposing sources and receivers, one \emph{also} transposes the material susceptibilities, then the corresponding reciprocity relations are recovered.

Reciprocity also imposes conditions on the scattering matrix of the system, requiring it to be symmetric, $\SS=\SS^T$ \cite{Haus1984,pozar2011microwave}. For convenience, we summarize here the various relevant conditions on the scattering matrix:
\begin{subequations}
    \begin{align}
    &\SS^{*} \SS = \mathbb{I} \quad \textrm{ (time-reversal invariant)} \\
    &\SS^{\dagger} \SS = \mathbb{I} \quad \textrm{ (lossless)}\\
    &\SS^{T} = \SS \quad \textrm{ (reciprocal)}.
    \end{align}
    \label{eq:S_conditions}
\end{subequations}
From Eqs. (\ref{eq:t-reversal})--(\ref{eq:S_conditions}), one can see that time-reversal symmetry and power conservation (losslessness) together imply reciprocity. In other words, for lossless systems, the concepts of time-reversal symmetry and reciprocity coincide. More generally, any two of these constraints imply the third. It also follows that, for LTI materials, three cases are possible \cite{Guo2022}: (i) materials that satisfy all three constraints (e.g., a lossless isotropic dielectric), (ii) materials that violate all three constraints (e.g., a lossy magnetized plasma), and (iii) materials that satisfy only one of them (e.g., a lossless magnetized plasma or a lossy isotropic dielectric)~\footnote{A particularly unusual case is that of materials that are nonreciprocal and non-Hermitian (i.e., with loss and gain), yet satisfy time-reversal symmetry \cite{Buddhiraju2020}, for example with a real but asymmetric permittivity tensor. Ref.~\cite{Buddhiraju2020} also proposed a possible implementation based on van der Waals heterostructures composed of transition-metal dichalcogenide monolayers pumped by a circularly polarized laser.}.

As seen in this section, time-reversal symmetry and reciprocity, when present, already impose significant constraints on the response of materials and systems. Moreover, these global symmetries underpin several important results on the fundamental limits of wave propagation and scattering, as discussed in Chapters \ref{sec:WaveShaping} and \ref{sec:GlobalSym}.

\subsection{Causality} \label{sec:causality}
\begin{figure*}[tb]
\centering
\includegraphics[width=0.6\linewidth]{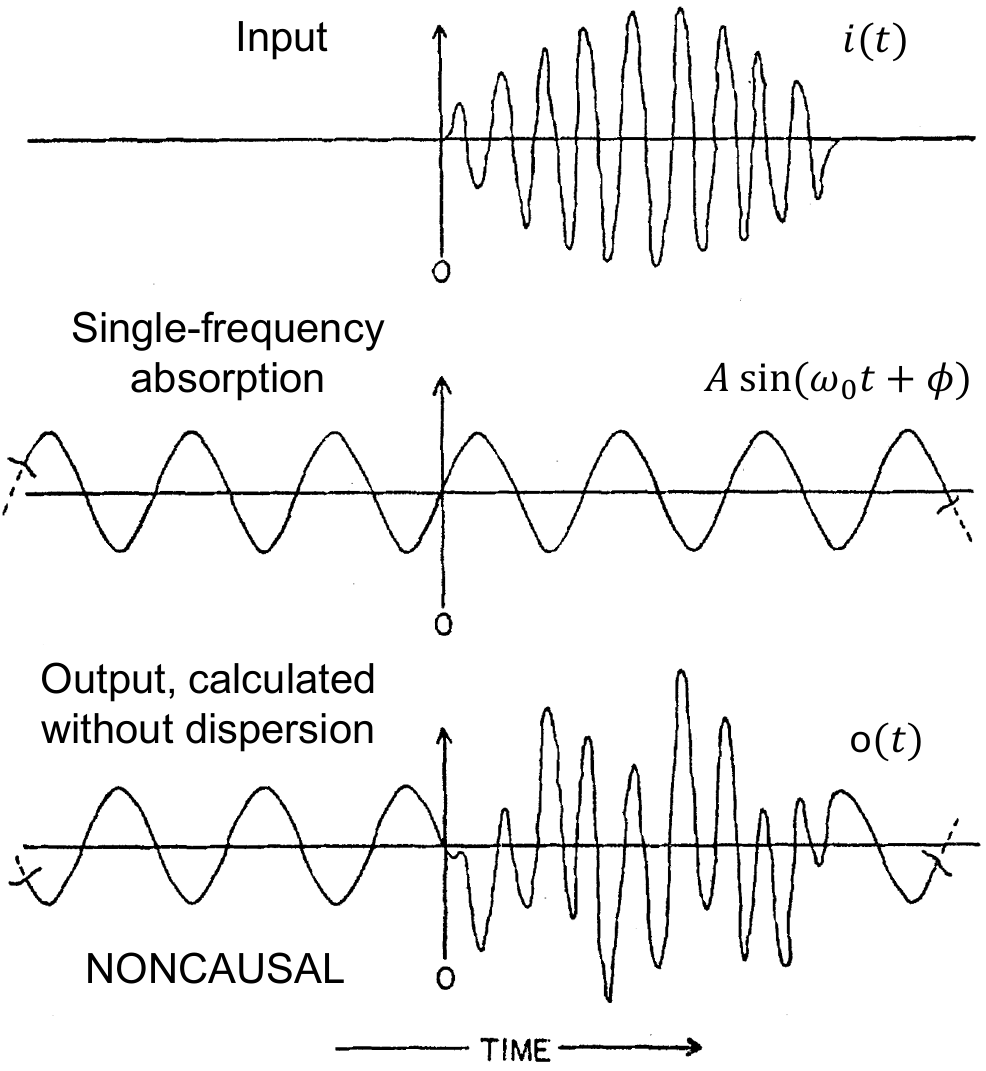}
\caption{\textbf{Causality, absorption, and frequency dispersion}. Thought experiment illustrating the fundamental connection between causality and dispersion. An input signal that is zero for $t<0$ can be expressed as a superposition of harmonic components, each of which individually extends from $t=-\infty$ to $t=+\infty$. A system that absorbs only a single frequency component while leaving all others unaffected would necessarily be noncausal. Causality therefore requires that absorption at one frequency be accompanied by phase shifts at other frequencies, enabling destructive interference that cancels the output for $t<0$. (Figure adapted from Ref.~\cite{toll1956causality}.)}
\label{fig:causality}
\end{figure*}

Causality is the condition that the output of a system cannot temporally precede the input.
This requirement is technically referred to as the \emph{primitive} causality condition and implies that the response function of the system must vanish, $h(t)=0$, for $t<0$ (Fig. \ref{fig:Overview}). If a spatial separation $d$ exists between input and output, one can also define a \emph{relativistic} causality condition---no signal can propagate with velocity greater than the speed of light in vacuum $c$---which implies that $h(d,t)=0$, for $t<d/c$. In both cases, it follows, as formally stated by the remarkable \textbf{Titchmarsh's theorem} \cite{Nussenzveig1972}, that the Fourier transform $h(\omega)$ of a sufficiently well-behaved, i.e., \textbf{square integrable}\footnote{A function is said to be \emph{square integrable} if $ \int_{ - \infty }^{ + \infty } |h(x)|^2 dx < A $    where $A$ is a finite real constant. Instead, a function is said to be \emph{bounded} if $|h(x)| \leq B, \forall x$, where $B$ is a finite real constant. Boundedness does not imply square integrability.}, causal response function is holomorphic on the upper half of the complex frequency plane (UHP; assuming the $e^{-i \omega t}$ convention for time-harmonic fields). From standard complex analysis arguments (contour integration over this half-plane), one can show that the real and imaginary parts of $h(\omega)$ are \textbf{Hilbert transforms} of each other. Depending on the context, these relations are also known as the \textbf{Kramers-Kronig (KK) relations} (the term we will use throughout this tutorial), or as \textbf{dispersion relations} or \textbf{Plemelj formulas}, and are commonly written as
\begin{equation}
 \mathop{{\rm Re}} h(\omega) = \frac{2}{\pi } ~ \Pvl\!\int_0^\infty  \frac{\Omega \mathop{{\rm Im}} h(\Omega) } {{\Omega^2} - {\omega ^2}}d\Omega,
 \label{eq:KKRe}
\end{equation}

\begin{equation}
 \mathop{{\rm Im}} h(\omega) = -\frac{2 \omega}{\pi } ~ \Pvl\!\int_0^\infty  \frac{\mathop{{\rm Re}} h(\Omega) } {{\Omega^2} - {\omega ^2}}d\Omega,
 \label{eq:KKIm}
\end{equation}

\noindent where $\Pvl\!\int$ indicates the Cauchy principal-value integral, and we have used the symmetry relation, $h(-\omega)^*=h(\omega)$, which holds for response functions that are real in the time domain. The KK relations strongly constrain the admissible frequency dependence of $h(\omega)$. When the response function represents the causal response of a material to an electromagnetic field, these relations imply that absorption is intrinsically linked to frequency dispersion, and vice versa. A simple argument to see why this must be true is given in \cite{toll1956causality} (Fig. \ref{fig:causality}): Consider a hypothetical medium that absorbs only a single frequency $\omega_0$, while leaving all other frequencies unaffected. Such a medium would have the effect of removing a single harmonic component from the input signal $i(t)$. If $i(t)$ starts at a specific time, it is then evident that the resulting output of this system would be noncausal, $o(t)=i(t)-A\sin(\omega_0 t+\phi)$ (bottom panel of Fig. \ref{fig:causality}). To preserve causality, the phases of the remaining frequency components must be readjusted so that their interference cancels the noncausal portion of the output. This frequency-dependent phase shift corresponds to dispersion, which therefore must accompany any absorption process.

Notice the \emph{spectrally nonlocal} nature of the KK relations: absorption anywhere in the electromagnetic spectrum affects frequency dispersion everywhere, even in spectral regions where absorption is negligible or zero. To see this, consider a frequency range in which $\mathop{{\rm Im}} h(\Omega )$ can be neglected (i.e., a lossless region). In this range, there is no need to take the principal value in Eqs. (\ref{eq:KKRe},\ref{eq:KKIm}), as the point $\Omega=\omega$ (the pole of the integrand) no longer lies within the region of integration \cite{landau2013electrodynamics}. One can then take the frequency derivative of the integral in Eq. (\ref{eq:KKRe}) with respect to $\omega$, obtaining
\begin{equation}
 \frac{d \mathop{{\rm Re}} h(\omega)}{d \omega} = \frac{4 \omega}{\pi } ~ \int_0^\infty  \frac{\Omega \mathop{{\rm Im}} h(\Omega) } {({\Omega^2} - {\omega ^2})^2}d\Omega,
 \label{eq:dKKRe}
\end{equation}
which explicitly shows that the frequency derivative of $\mathop{{\rm Re}} h(\omega )$ at a given frequency depends on $\mathop{{\rm Im}} h(\omega )$ over the entire spectrum, including frequencies far from $\omega$. Moreover, if $h(\omega)$ represents the response of a \emph{passive} material, this derivative is always positive \footnote{\label{footnote_biased_plasma}An interesting exception occurs in conducting materials under a magnetic bias, for which an additional term appears in \emph{both} KK relations \cite{abdelrahman2020broadband}, in contrast to unbiased conducting materials, for which an additional term appears only in Eq. (\ref{eq:KKIm}), as discussed later in this section.} because $\mathop{{\rm Im}} h(\Omega ) \ge 0$ (see Chapter \ref{sec:bounds_materials}). The resulting monotonically increasing dispersion in low-loss regions is commonly referred to as \emph{normal dispersion}, in contrast to the \emph{anomalous dispersion} (negative derivative) observed near lossy resonances~\footnote{An analogous result applies to the impedance of passive causal circuits (i.e., circuits containing only positive resistances). In this context, the statement that the reactance of a one-port network must increase monotonically with frequency in lossless ranges is known in applied electromagnetics as Foster’s reactance theorem \cite{foster1924reactance}.}.

The requirement of square integrability in the derivation of the KK relations suggests interesting connections between causality and the concepts of boundedness, stability, and passivity. A common definition of stability is that the output of a system remains bounded for all bounded inputs \cite{Triverio2007,oppenheim1997signals}. Clearly, stability does not imply causality or passivity: a bounded output could still precede the bounded input, and the output energy could exceed the input energy at any time instant. Moreover, stability does not imply square integrability. As seen from Eq. (\ref{eq:LTI_response}), bounded-output bounded-input stability only requires that $\int_{ - \infty }^{ + \infty } |h(\tau )|d\tau \le C $, where $C$ is a finite real constant, a condition weaker than square integrability. Interestingly, even the stronger requirement that a square-integrable input (finite total input energy) produces a square-integrable output (finite total output energy), and $\int_{-\infty}^{\infty} |o(t)|^2 \, dt
\le A \int_{-\infty}^{\infty} |i(t)|^2 \, dt$, for some finite constant $A$, only implies boundedness of $|h(\omega)|$, namely, $|h(\omega)|^2\le A$, again a condition weaker than square integrability, as discussed in \cite[Section 1.7]{Nussenzveig1972}. For response functions that are causal but not square integrable in this sense, i.e., with $|h(\omega)|$ bounded by a constant, one can still derive KK relations after suitably subtracting a constant from the original function; more specifically, by defining a new square-integrable function, $D(\omega)=[h(\omega)-h(\omega_0)]/(\omega-\omega_0)$, where $\omega_0$ is an arbitrarily chosen real frequency \cite{Nussenzveig1972,srivastava2021causality}, yielding so-called KK relations with one subtraction. More generally, a similar approach applies to non-square-integrable functions with $|h(\omega)|=O(\omega^n)$, leading to KK relations with $n+1$ subtractions \cite{Nussenzveig1972,srivastava2021causality}. 

While stability does not imply causality, the condition of passivity---as defined in Eq. (\ref{eq:passivity}), often referred to as \emph{strong} passivity---directly implies a causal response \cite{Nussenzveig1972,Youla1959,Triverio2007,yaghjian2021simplified}, as we will see in Section \ref{sec:passivity}. Conversely, causality does not imply passivity, and the standard KK relations also apply to non-passive but stable systems \cite{nistad2008causality} (see also Section \ref{sec:how_large_imag}).
Moreover, passivity does not imply square integrability. Thus, analogous to the discussion in the previous paragraph, one can have passive and causal linear systems for which KK relations in the form of Eqs. (\ref{eq:KKRe},\ref{eq:KKIm}) cannot be derived. A simple example, outside the cases discussed above, is provided by passive conducting materials. In such materials, the susceptibility is not square integrable because of a pole at $\omega=0$ (as can be seen from a lossy Drude model), arising from their nonzero real conductivity $\sigma_0$ for static electric fields. This issue can, however, be resolved by slightly modifying the integration path around this pole, which yields a modified version of the KK relations containing an additional term, $+\sigma_0/(\varepsilon_0\omega)$ for $h=\chi$, on the right-hand side of Eq. (\ref{eq:KKIm}) \cite{Lucarini2005} (see also footnote \ref{footnote_biased_plasma}).

A necessary condition for square integrability is that the function $h(\omega)$ vanishes as $|\omega| \to \infty$. The asymptotic behavior of this function depends on the specific physical problem under consideration. For example, in a physical material the electric susceptibility $\chi$ must approach zero at very large frequencies because the material cannot respond to (i.e, polarization cannot be induced by) an applied electric field that varies sufficiently rapidly (think of a harmonic oscillator driven by a force oscillating at a frequency far above its resonance; any nonzero mass would require infinite energy to oscillate at infinite frequency)~\footnote{Importantly, this behavior of $\chi$ is not a consequence of primitive causality. Moreover, relativistic causality only requires the high-frequency limit of the refractive index $n=\pm \sqrt{\chi+1}$ not be smaller than unity, in order to prevent the front of a signal pulse, which contains the highest frequency components, from propagating superluminally without dispersion, corresponding to superluminal information velocity \cite{abdelrahman2021physical}. More broadly, the asymptotic form of the response function must be determined from additional physical considerations.}. Moreover, in this high-frequency regime, \emph{any} material behaves universally as an isotropic gas of free electrons \cite[Sec. 78]{landau2013electrodynamics},\cite{Lucarini2005}. The asymptotic susceptibility tensor of any material, whether metal or dielectric, isotropic or anisotropic, is therefore diagonal and given by
\begin{equation}
 \chi_{i,j} (\omega) = - \frac{e^2 N_e} {m_e \varepsilon_0} \frac{1} {\omega^2} \delta_{i,j} +o(\omega^{-2}) ,
 \label{eq:asymp}
\end{equation}

\noindent where $e$ and $m_e$ are the electron charge and rest mass, respectively, $\varepsilon_0$ is the free-space permittivity, $N_e$ is the total electron density, and $\delta_{i,j}$ is the Kronecker delta, equal to 1 if $i = j$ and 0 otherwise.
This universal asymptotic behavior plays a key role in deriving fundamental limits to the linear optical response of materials, as we will see in Section \ref{sec:large_susceptibility}. More broadly, evaluating the KK relations in the limits of zero or infinite frequency, together with the physically determined asymptotic forms of the function of interest, leads to integral \textbf{sum rules} of great importance across many areas of physics, from optical materials research to classical and quantum scattering theory. Throughout this tutorial, we will see several examples of such sum rules, both for optical material properties (Chapter \ref{sec:bounds_materials}) and for scattering parameters and operators (Chapter \ref{sec:Scatt_Local}), as well as the fundamental limits that can be derived from them.

Finally, we note that causality, i.e., a response that is one-sided in time, is only one of four possible one-sided responses that can be defined in time or space, and in the physical or Fourier domain \cite{salehi2026unified}. According to Titchmarsh's theorem, each of them is equivalent, for square-integrable responses, to KK relations in the conjugate variable. Unlike temporal causality, however, the other three are not imposed by nature: for instance, the response of a material at a point generally depends on the fields on all sides of it, so there is no general KK relation with respect to the wavevector. One-sided frequency responses (\emph{spectral causality}) and one-sided wavevector responses can be engineered in time-varying and inhomogeneous media, respectively, to achieve one-sided frequency or momentum conversion~\cite{hayran2021spectral,horsley2015spatial}, whereas whether a response that is one-sided in space (a one-sided spatially nonlocal response) can be realized at all remains open~\cite{salehi2026unified}.

\subsection{Passivity} \label{sec:passivity}
There is a long history of identifying constraints associated with passive linear systems~\cite{Raisbeck1954,Youla1959,Beltrami1967,Wohlers1969,Willems1972,Boyd1982,Mahanta2018,srivastava2021causality}, with applications in fields ranging from circuit theory and control~\cite{Kalman2010} to electrical interconnects~\cite{Triverio2007} to elastic materials~\cite{Khodavirdi2022} to quantum field theory~\cite{Hartman2016}. In this section, we review some of the classic results of what is known in passive linear systems. We will specialize to $N$-port systems, in which there is a finite number $N$ of orthogonal input/output channels that are normalized to each carry unit power into or out of the system. Particularly useful pedagogical introductions include Refs.~\cite{Wohlers1969,Boyd1982,Triverio2007,srivastava2021causality}.

We can start with the ``scattering'' picture of a linear system, with input amplitudes collated into an $N \times 1$ vector $\av$ scattered into an $N \times 1$ vector of output amplitudes $\bv$ via the scattering matrix $\SS$ of \eqref{eq:SMatrix}:
\begin{align}
    \bv = \SS \av.
    \label{eq:SMat}
\end{align}
\emph{Passivity} is the condition that the total power outflow be smaller than the total power inflow, up to (any) time $t$:
\begin{align}
    \int_{-\infty}^{t} \left[\av^T(\tau) \av(\tau) - \bv^T(\tau) \bv(\tau) \right] \,{\rm d}\tau \geq 0.
    \label{eq:passivity}
\end{align}
This condition is sometimes referred to as \emph{strong passivity}, with \emph{weak passivity} defined \emph{only} for the limit $t \rightarrow \infty$. Strong passivity \emph{implies} causality, whereas weak passivity does not~\cite{Boyd1982}. Causality can be defined as the requirement that a zero input signal, $\av(\tau) = 0$, up to some time $t$, implies zero output, $\bv(\tau) = 0$ up to the same time; this condition is an immediate consequence of \eqref{eq:passivity}.

We will use the frequency domain (whereas some engineering literature prefers the Laplace domain); hence, the key right-half plane of the Laplace domain rotates to the upper-half plane for the frequency domain, with key conditions relating to positive imaginary parts (frequency) instead of positive real parts (Laplace). 

The frequency-domain scattering matrix $\SS(\omega)$ represents a passive linear system if and only if:
\begin{enumerate}
    \item $\SS(\omega)$ is analytic for $\Im \omega > 0$,
    \item $\II - \SS^\dagger(\omega) \SS(\omega)$ is positive semidefinite for $\Im \omega > 0$, \textrm{ and}
    \item $\SS^*(\omega) = \SS(-\omega^*)$,
\end{enumerate}
where $\II$ is the $N\times N$ identity matrix. These conditions define \emph{bounded-real} functions~\cite{Youla1959,Triverio2007}. A simple physical interpretation of the conditions is that the first is a consequence of causality, the second a consequence of passivity (for inputs with both oscillating and growth/decay terms), and the third is a consequence of real-valued time-domain signals, though such an interpretation only implies that they are necessary, and not their sufficiency. (One can alternatively characterize these functions entirely by their real-line values~\cite{Wohlers1969,Triverio2007}, which is useful for applications such as real-time passivity verification~\cite{Mahanta2018}.) The scattering matrix of a passive linear system everywhere in the UHP can be written:
\begin{align}
   \SS(\omega = \omega_0 + i\gamma) = \frac{\gamma}{\pi} \int_{-\infty}^{\infty} \frac{\SS(\omega')}{(\omega' - \omega_0)^2 + \gamma^2} \,{\rm d}\omega'.
   \label{eq:SMatRep}
\end{align}
More rigorous derivations of \eqref{eq:SMatRep} start with the function $\SS(\omega)$ defined only in the UHP, then prove \eqref{eq:SMatRep} with $\SS(\omega')$ as boundary values of the function, suitably defined~\cite{Boyd1982}. An alternative, slightly less rigorous approach, is to use the analyticity of $\SS(\omega)$ and take a contour integral of $\SS(\omega) / \left[(\omega-\omega_0)^2 + \gamma^2\right]$. There is an important subtlety that requires caution. Typical photonic basis functions (e.g. vector spherical waves) are spatially distributed, and scatterers of nonzero size can excite outgoing-wave scattering coefficients \emph{before} they would have been excited in the absence of the scatterer, violating strong passivity. Recovering strong passivity requires the introduction of size-dependent phase shifts that must be included in the definition of the scattering matrix, $\SS(\omega)$ (cf. Section \ref{sec:cloaking}).

The complementary \emph{immittance} formalism uses port ``currents'' $\iv$ and ``voltages'' $\vv$. (``Immittance'' encapsulates both impedance and admittance.) Abstractly, immittance variables can be derived from scattering variables via $\vv = 2\left(\av - \bv\right)$ and $\iv = 2\left(\av + \bv\right)$, or vice versa. The immittance strong passivity condition is
\begin{align}
    \int_{-\infty}^t \vv^T(\tau) \iv(\tau) \,{\rm d}\tau \geq 0.
    \label{eq:passiveim}
\end{align}
An immittance matrix $\XX$ (impedance $\mathbb{Z}$ or admittance $\mathbb{Y}$) represents a passive linear system if and only if:
\begin{enumerate}
    \item $\XX(\omega)$ is analytic for $\Im \omega > 0$,
    \item $\Re \XX(\omega) = \frac{1}{2}\left[\XX(\omega) + \XX^{\dagger}(\omega)\right]$ is positive semidefinite for $\Im \omega > 0$, \textrm{ and}
    \item $\XX^*(\omega) = \XX(-\omega^*)$.
\end{enumerate}
These conditions define \emph{positive-real} matrices. (Sometimes, though not always~\cite{Youla1959}, they are only defined as such in the Laplace domain.) There is a quite useful representation theorem for immittance matrices, although for compatibility with our $\TT$-matrix discussions, we will first make a small pivot. A $\TT$ matrix relates a field to a dipole density, rather than a current, and hence a $\TT$ matrix is analogous to an immittance matrix divided by $-i\omega$, for frequency $\omega$. Hence, a $\TT(\omega)$ matrix represents a passive $N$-port linear system if and only if:
\begin{enumerate}
    \item $\omega\TT(\omega)$ is analytic for $\Im \omega > 0$,
    \item $\Im \left[\omega \TT(\omega)\right]$ is positive semidefinite for $\Im \omega > 0$, \textrm{ and}
    \item $\omega^*\TT^*(\omega) = -\left[\omega\TT(\omega)\right]_{\omega = -\omega^*}$,
\end{enumerate}
where $\Im \left[\omega \TT(\omega)\right]$ is the anti-Hermitian part of $\omega\TT(\omega)$. The first two of these conditions define a matrix-valued \emph{Herglotz--Nevanlinna} function~\cite{Gesztesy2000,Fritzsche2012,Luger2022}, which have a well-known representation theorem:
\begin{align}
    \omega\TT(\omega) = \CC + D \omega + \int_{-\infty}^{\infty} \left[\frac{1}{\lambda - \omega} - \frac{\lambda}{1+\lambda^2}\right] \,{\rm d}\Omega(\lambda),
    \label{eq:GenHN}
\end{align}
where $\omega$ is in the UHP, $\CC$ is Hermitian, $D$ is Hermitian positive semidefinite, and ${\rm d}\Omega(\lambda)$ is a matrix-valued measure satisfying certain integrability conditions. (An analogous representation in the Laplace domain was recognized by Youla~\cite{Youla1959}; Beltrami connected this work to earlier results by Herglotz and Cauer~\cite{Beltrami1967,Cauer1932}.) The values of $\CC$ and $D$ are specified by $\TT$: $\CC = \Re \left[i\TT(i)\right]$ and $D = \lim_{y\rightarrow\infty}\left[\TT(iy)\right]$. The $\TT$ matrix decays as $1/\omega^2$, which enables significant simplification of the representation. From Remark 2.8.3 and Theorem 2.4.2 of \citeasnoun{Luger2022}, the growth condition $\int_0^{\infty} \xv^\dagger \TT(iw) \xv \,{\rm d}w < \infty$ simplifies the representation. This growth condition is satisfied by the $\TT(\omega)$ matrix thanks to its quadratic decay at high frequencies. Then, the representation is~\cite{Luger2022}
\begin{align}
    \omega\TT(\omega) = \int_{-\infty}^{\infty} \frac{{\rm d}\Omega(\lambda)}{\lambda - \omega}.
\end{align}
In the abstract, this is an ``oscillator'' representation of the $\TT$ matrix. We make this more concrete in Sec.~\ref{sec:T-matrix-response}.

\subsubsection{Herglotz functions: Sum rules and bounds} \label{sec:herglotz}
Under certain conditions, Herglotz functions yield integral identities (sum rules), associated with their low- or high-frequency asymptotic expansions, with important implications for the derivation of physical bounds for passive linear systems. More broadly, the approach outlined below can often be used as a useful ``shortcut'' to derive sum rules for passive systems directly, without the need to first derive causality-based integral relations (such as KK relations). This is ultimately possible because strong passivity implies causality, as discussed above.

For simplicity, here we consider scalar-valued Herglotz functions $h(z)$ of a complex variable $z$. From the general Herglotz representation in Eq. (\ref{eq:GenHN}), specialized to the scalar case, it follows that $h(z)/z$ and $h(z)z$ are bounded as $z \to \infty$ and $z \to 0$, respectively (details on how these limits are taken are discussed in \cite{bernland2011sum}). It is therefore natural to assume that $h(z)$ admits the following low- and high-frequency asymptotic expansion \cite{bernland2011sum,nedic2018herglotz}
\begin{equation}
\left\{
\begin{aligned}
h(z) &= b_{1} z + b_0+ b_{-1} z^{-1} + \dots = \sum_{n=1-2N_{\infty}}^{1} b_{n} z^{n} + o\!\left( z^{\,1-2N_{\infty}} \right)
\qquad \text{as } z \,\to\, \infty, \\[6pt]
h(z) &= a_{-1} z^{-1} + a_0+ a_{1} z + \dots =  \sum_{n=-1}^{2N_{0}-1} a_{n} z^{n} + o\!\left( z^{\,2N_{0}-1} \right)
\qquad \text{as } z \,\to\, 0.
\end{aligned}
\right.
\end{equation}
Here the expansions on the right-hand-side have been written up to orders $2N_{0}-1$ and $1-2N_{\infty}$ with $N_{0}\ge 0, N_{\infty}\ge 0$; the reason for this will become clearer below. If we also assume that $h(z)$ satisfies the symmetry  $h(z) = -h^{*}(-z^{*})$, which for $z=\omega$ implies a real-valued time-domain response, then the odd-order coefficients in these asymptotic expansions are real-valued, whereas the even-order coefficients are purely imaginary. Then, we choose $N_{0}$ and $N_{\infty}$ such that the expansions stop at the appearance of the first nonzero imaginary term. This is done because the sum rules below are only associated with the real coefficients in these expansions (see \cite{bernland2011sum,nedic2018herglotz} for technical details). Importantly, one can show that the general Herglotz representation (\ref{eq:GenHN}) also implies that $-a_{-1}$ and $b_{1}$ are non-negative.

One can then prove, either by Cauchy integration of $h(\omega)$ in $\mathbb{C}^+$~\cite{sohl2008general} or through more general arguments~\cite{bernland2011sum}, that the following family of sum rules holds for Herglotz functions~\footnote{More precisely, the left-hand side should be written as $\lim_{\varepsilon \to 0^{+}} \, \lim_{y \to 0^{+}} 
\frac{2}{\pi} \int_{\varepsilon}^{1/\varepsilon} 
\frac{\operatorname{Im} [h(x + i y)]}{x^{2p}} \, dx$, but we have omitted these limits to simplify the notation.}:
\begin{equation}
\frac{2}{\pi} \int_{0}^{\infty} \frac{\operatorname{Im}[h(x)]}{x^{2p}}\, dx 
= a_{2p-1} - b_{2p-1}
=
\begin{cases}
- b_{2p-1}, & p < 0,\\
a_{-1} - b_{-1}, & p = 0,\\
a_{1} - b_{1}, & p = 1,\\
a_{2p-1}, & p > 1,
\end{cases}
\label{eq:h_sum_rules}
\end{equation}
where $x$ is a real variable and $1 - N_{\infty} \le p \le N_{0}$.

The two most common cases, which we will encounter in Sections \ref{sec:how_small_real},\ref{sec:cloaking}, are those with $p=0$ and $p=1$. If the low-frequency expansion is known and $a_1\ge0$, the $p=1$ sum rule gives
\begin{equation}
\frac{2}{\pi} \int_{0}^{\infty} \frac{\operatorname{Im}[h(x)]}{x^{2}}\, dx 
= a_{1} - b_{1}\le a_1
\label{eq:h_sum_rule_p1}
\end{equation}
where we used the fact that $b_1\ge0$. Instead, if the high-frequency expansion is known and $b_{-1}\le0$, the $p=0$ sum rule gives 
\begin{equation}
\frac{2}{\pi} \int_{0}^{\infty} \operatorname{Im}[h(x)]\, dx 
= a_{-1} - b_{-1}\le - b_{-1}
\label{eq:h_sum_rule_p0}
\end{equation}
where we used the fact that $a_{-1}\le0$. 

Since the integrand in these expressions is non-negative by definition for Herglotz functions, identities and inequalities of this type impose strong positivity constraints on the response of linear passive systems over finite bandwidths, as we will see later in this Tutorial.

\subsection{Conservation laws} \label{sec:conslaws}
In many dynamical systems, there are ``conserved quantities'' that are invariant in the system evolution. Noether's theorem, a foundational physical principle, relates continuous symmetries to conserved quantities. Of particular interest in photonics and electromagnetism are \textbf{quadratic invariants}, of the form $\psi^{\dagger} Q \psi$, whose value remains fixed in the absence of sources (e.g., currents) or sinks (e.g., absorbing boundaries). It is known that there are fifteen (inequivalent) quadratic conservation laws for Maxwell's equations, which correspond to quantities including energy, linear momentum, angular momentum, and ``zilch'' (related to helicity)~\cite{Lipkin1964,Fushchych1992}. To identify these quantities, one can start with the time-dependent, source-free Maxwell equations in free space,
\begin{align}
    \frac{\partial}{\partial t} \begin{pmatrix}
        \Ev \\
        \Hv
    \end{pmatrix} = 
    \begin{pmatrix}
        & \varepsilon_0^{-1} \nabla \times \\
        -\mu_0^{-1} \nabla \times &
    \end{pmatrix}
    \begin{pmatrix}
        \Ev \\
        \Hv
    \end{pmatrix} \rightarrow
    \frac{\partial \psi}{\partial t} = M\psi.
    \label{eq:TDME}
\end{align}
We can search for a quadratic invariant $I(\psi)$, generally written as
\begin{align}
    I(\psi) = \int_V \psi^{\dagger}(\xv,t) Q \psi(\xv,t) \,{\rm d}\xv = \psi^{\dagger} Q \psi
\end{align}
by enforcing the condition that its time derivative be zero,
\begin{align}
    \frac{dI}{dt} = \dot{\psi}^{\dagger} Q \psi + \psi^{\dagger} Q \dot{\psi} = \psi^{\dagger} M^{\dagger} Q \psi + \psi^{\dagger} Q M \psi = \psi^{\dagger} \left( M^{\dagger} Q + Q M \right) \psi.
\end{align}
This quantity will be zero for all $\psi$ for all $Q$ such that $M^{\dagger} Q = -QM$. Note that \emph{without} the factors of $\varepsilon_0$ and $\mu_0$ in \eqref{eq:TDME}, the matrix $M$ would be anti-Hermitian, i.e. $M^{\dagger} = -M$, for fields propagating in finite wavepackets (with the field decaying to zero sufficiently quickly at infinity). Hence the choice
\begin{align}
    Q = 
    \frac{1}{2}\begin{pmatrix}
        \varepsilon_0 & \\
        & \mu_0
    \end{pmatrix}
\end{align}
creates the desired anti-Hermiticity relation, $M^{\dagger} Q = -QM$, as well as providing a useful prefactor, and we have our first invariant,
\begin{align}
    I = \int_V \psi^{\dagger}(\xv,t) Q \psi(\xv,t) \,{\rm d}\xv = \frac{1}{2}\int_V \left[ \varepsilon_0 |\Ev(\xv,t)|^2 + \mu_0 |\Hv(\xv,t)|^2 \right] \,{\rm d}\xv,
    \label{eq:TEInv}
\end{align}
which is the total field energy. The other fourteen conserved quantities proceed from similar logic but different $Q$ matrices. (Often, they are easier to derive from symmetry principles applied to the scalar and vector potential, instead of the fields.) 

In the presence of a non-dispersive scattering medium, one can modify $\varepsilon_0 \rightarrow \varepsilon$ and $\mu_0 \rightarrow \mu$ and total field energy is still conserved. In the presence of electric and magnetic current sources $\Jv_e$ and $\Jv_m$ (varying in space and time), the time-dependent Maxwell equations can be written
\begin{align}
    \frac{\partial \psi}{\partial t} = M\psi - \begin{pmatrix}
        \varepsilon_0^{-1} \Jv_e \\
        \mu_0^{-1} \Jv_m
    \end{pmatrix}
    = M\psi - s.
\end{align}
Now a quadratic quantity such as the energy of \eqref{eq:TEInv} is no longer invariant, because the currents can produce or absorb energy, but a simple calculation gives
\begin{align}
    \frac{dI}{dt} &= \frac{d}{dt} \left[ \frac{1}{2} \int_V \left[ \varepsilon_0 |\Ev(\xv,t)|^2 + \mu_0 |\Hv(\xv,t)|^2 \right] \,{\rm d}\xv \right] \nonumber \\
                  &= -\int_V \left[ \Jv_e(\xv,t) \cdot \Ev(\xv,t) + \Jv_m(\xv,t) \cdot \Hv(\xv,t)\right] \,{\rm d}\xv,
    \label{eq:TETD}
\end{align}
with the term on the right-hand side being the work done by a current distribution on an electromagnetic field. An analogous procedure yields the linear momentum, angular momentum, zilch, etc., produced by a set of currents.

Fundamental limits are typically developed for the frequency domain, where a harmonic excitation produces steady-state fields propagating from or to infinity. In this case, finite wavepackets cannot be assumed and the harmonic version of \eqref{eq:TETD}, for example, is no longer a complete statement of energy conservation: the Maxwell operator $M$ is no longer anti-Hermitian when multiplied by $Q$, due to the boundary fluxes. An additional term corresponding to the Poynting flux can be added (giving the usual statement of energy conservation, cf. Sec.~6.7 of Ref.~\cite{Jackson1999}), but such boundary flux terms can impede fundamental limits as they introduce indefinite quadratic forms. 

A simple(r) approach to conservation laws in the frequency domain arises through the integral-equation formulation of Maxwell's equations, \eqref{eq:IE}, which encode open boundaries in the Green's function and do not require field integrals along bounding surfaces to measure outgoing power flows. If we start anew with the frequency-domain integral equation,
\begin{align}
    \left(\Gamma_0+\xi\right) \phi = -\psi_{\rm inc},
    \label{eq:IE2}
\end{align}
we can form a quadratic conservation law by integrating both sides against the conjugated polarization fields (multiplying from the left by $\phi^{\dagger}$, in the simplified notation), and taking the imaginary part. It is also useful for interpretation (and discussions of causality) to multiply by the prefactor $\omega/2$, ultimately yielding
\begin{align}
    \frac{\omega}{2}\phi^{\dagger}\left(\Im \Gamma_0+ \Im \xi\right) \phi = \frac{\omega}{2} \Im \left(\psi_{\rm inc}^{\dagger} \phi\right),
    \label{eq:OpticalThm}
\end{align}
where we incorporated the negative sign on the right-hand side into the argument of the imaginary part. \eqref{eq:OpticalThm} is a (particularly useful) version of the \textbf{optical theorem}. Each term has a simple interpretation. The first term is the power scattered from the incident field by the scattering body (with susceptibility $\chi$), i.e.,
\begin{align}
    P_{\rm scat} = \frac{\omega}{2} \phi^{\dagger} \left(\Im \Gamma_0\right) \phi = \frac{\omega}{2} \Im \int_V \int_V \phi^{\dagger}(\xv) \Gamma_0(\xv,\xv') \phi(\xv') \,{\rm d}\xv {\rm d}\xv' = \frac{1}{2} \Re \int_S \Ev_{\rm scat} \times \Hv_{\rm scat}^* \cdot \nhat,
\end{align}
where the $S$ in the last term on the right-hand side is a bounding surface, connecting the less common volume-integral expression at the left to the conventional (but less useful for fundamental limits) surface Poynting flux term on the right. The second term in the optical theorem of \eqref{eq:OpticalThm} is the power absorbed in the scatterer itself, which can be seen through the equivalent expressions,
\begin{align}
    P_{\rm abs} &= \frac{\omega}{2} \phi^{\dagger} \left(\Im \xi\right) \phi = \frac{\omega}{2} \int_V \phi^{\dagger}(\xv) \left[ \Im \xi(\xv) \right] \phi(\xv) \,{\rm d}\xv \nonumber \\
    &= \frac{\omega}{2} \int_V \phi^{\dagger}(\xv) \left[ \frac{\Im \chi(\xv)}{|\chi(\xv)|^2} \right] \phi(\xv) \,{\rm d}\xv = \frac{\omega}{2} \int_V \psi^{\dagger}(\xv) \left[ \Im \chi(\xv) \right] \psi(\xv)\,{\rm d}\xv,
\end{align}
with the last expression being the usual (generalized) expression with the squares of the electric and magnetic fields being multiplied by the imaginary part of the material susceptibility. 

The final term in the optical theorem, on the right-hand side of \eqref{eq:OpticalThm}, is the \emph{extinction}. Extinction is the sum of absorption and scattering; correspondingly, it is the total power ``removed'' from the incident wave. It has a simple physical justification. Rewriting the polarization fields $\phi$ as induced polarization currents, $-i\omega\phi = \left( \Jv_{e,\textrm{ind}} \quad \Jv_{m,\textrm{ind}} \right)^T$, we can rewrite the extinction term as
\begin{align}
    P_{\rm ext} &= \frac{\omega}{2} \Im \left(\psi_{\rm inc}^{\dagger} \phi\right) = \frac{\omega}{2} \Im \int_V \psi_{\rm inc}^{\dagger}(\xv) \phi(\xv) \,{\rm d}\xv \nonumber \\
                                                                                  &= \frac{1}{2} \Re \int_V \left[ \Einc^*(\xv) \cdot \Jv_{e,\textrm{ind}}(\xv) + \Hinc^*(\xv) \cdot \Jv_{m,\textrm{ind}}(\xv) \right],
\end{align}
which is analogous to the negative of the harmonic version of the right-hand side of \eqref{eq:TETD}, as expected~\footnote{The negative sign is missing here because the incident field is driving the induced currents, whereas in \eqref{eq:TETD}, the free currents are driving the fields that they produce.}. For the purposes of fundamental limits, two important properties of the optical theorem should be mentioned here. First, both terms on the left-hand side of \eqref{eq:OpticalThm} are \textbf{positive semidefinite quadratic forms}. The operators/matrices $\Im \Gamma_0$ and $\Im \xi$ (where ``$\Im$'' denotes anti-Hermitian part) are both positive semi-definite\footnote{A matrix $A$ is positive semidefinite if $\xv^{\dagger} A \xv$ is nonnegative for all $\xv$.}, as expected from the fact that absorbed and scattered powers must be positive in passive systems~\footnote{Without passivity, $\Im \xi$ can be negative, or take negative eigenvalues, but $\Im \Gamma_0$ remains positive semidefinite.}. Second, the extinction on the right-hand side within the imaginary part is \emph{linear} in the fields (polarization or electromagnetic). It is a quadratic overlap, but the incident field is fixed for a given excitation, and hence the term within the imaginary part is linear in the response. This has important ramifications for causality, as discussed in Sec.~\ref{sec:CausalResp}, as well as conservation-law-based bounds, as discussed in Sec.~\ref{sec:SingleFreq}.

Finally, we can note that instead of taking the imaginary part in forming the optical theorem of \eqref{eq:OpticalThm}, we can instead take the real part, giving:
\begin{align}
    \frac{\omega}{2}\phi^{\dagger}\left(\Re \Gamma_0+ \Re \xi\right) \phi = -\frac{\omega}{2} \Re \left(\psi_{\rm inc}^{\dagger} \phi\right).
    \label{eq:ReacOpticalThm}
\end{align}
This counterpart to the optical theorem is a statement of \textbf{``reactive power'' conservation}, with the term in quotes because it does not correspond to real power flow, but instead to conservation laws around stored energies~\cite{Jackson1999}. The terms on the left-hand side are quadratic forms, but they need not be positive, even in passive systems. Hence the polarization fields that satisfy \eqref{eq:ReacOpticalThm} are not bounded in norm, and that equation cannot alone be used for fundamental limits. However, it can be used in tandem with the optical theorem, as we discuss in Sec.~\ref{sec:SingleFreq}.

One can generate additional quadratic constraints, many of which are ``conservation laws,'' by modifying the steps from the integral equation of \eqref{eq:IE2} to the optical theorem of \eqref{eq:OpticalThm}. Instead of multiplying just by $\phi^{\dagger}$ on the left, if we multiply by $\phi^{\dagger} \DD$, where $\DD$ is an arbitrary 6$\times$6 tensor operator, then we arrive at
\begin{align}
    \phi^{\dagger}\DD\left(\Gamma_0+\xi\right)\phi &= -\phi^{\dagger}\DD\psi_{\rm inc}, \nonumber \\
    &\int_V\!\!\int_V \phi^{\dagger}(\xv)\,\DD(\xv,\xv')\bigl[(\Gamma_0\phi)(\xv') + \xi(\xv')\phi(\xv')\bigr]\,{\rm d}\xv\,{\rm d}\xv' \nonumber \\
    &\qquad = -\int_V\!\!\int_V \phi^{\dagger}(\xv)\,\DD(\xv,\xv')\,\psi_{\rm inc}(\xv')\,{\rm d}\xv\,{\rm d}\xv',
    \label{eq:ConsDMatrix}
\end{align}
where $(\Gamma_0\phi)(\xv') \equiv \int_V \Gamma_0(\xv',\xv'')\,\phi(\xv'')\,{\rm d}\xv''$; no $\omega/2$ prefactor is needed here. Presumably, the full set of inequivalent, spatially local and constant $\DD$ tensors produces the 15 possible conservation laws, though we do not know of any work showing this. What has become clear, however, is that among spatially local $\DD$ tensors, the full set of constraints implied by \eqref{eq:ConsDMatrix} can be the foundation of a general computational approach to fundamental limits, as discussed in Sec.~\ref{sec:comput_bounds}.

\subsection{Rayleigh quotients}
\label{sec:RQs}
Energy, momentum, and related quantities in electromagnetism are \textbf{quadratic forms}. Fundamental limits thereby revolve around maximizing or minimizing quadratic forms, so here we review a set of basic results with regards to optimizing quadratic forms. We consider finite-dimensional vectors $\xv$ (size $M\times 1$, for arbitrary $M$), and corresponding quadratic forms given by vector-matrix-vector products such as $\xv^{\dagger} A\xv$, but all of the results we describe can be seamlessly generalized to infinite-dimensional vectors and quadratic forms that might be written in inner product notation as $\left\langle x,Ax\right\rangle$.

The Rayleigh quotient is a real-valued quantity of the form 
\begin{align}
    \frac{\xv^{\dagger} A \xv}{\xv^{\dagger} \xv},
\end{align}
where $A$ is Hermitian~\footnote{Any matrix can be decomposed into its Hermitian and anti-Hermitian parts, and an anti-Hermitian part of $A$ could only contribute an imaginary value to the quadratic form $\xv^{\dagger} A\xv$.}. Intuitively, this quantity arises when one wants to maximize a quadratic quantity (the numerator) subject to some normalization constraint on the source/field/excitation $\xv$; the presence of the denominator prohibits unbounded trivial solutions. The maximum of the Rayleigh quotient over all possible $\xv$ vectors (excluding the zero vector) is the largest eigenvalue of $A$,
\begin{align}
    \max_{\xv} \frac{\xv^{\dagger} A \xv}{\xv^{\dagger} \xv} = \lambda_{\rm max}(A),
\end{align}
and the $\xv$ for which this value is achieved is the eigenvector corresponding to the largest eigenvalue. Both statements are easily proven by writing $A$ in its eigenbasis, $A = U\Lambda U^{\dagger}$ for unitary $U$, defining a new variable $y=U^{\dagger}x$, and then optimizing over all possible $y$. Similarly, the \emph{minimizer} of the Rayleigh quotient is the smallest eigenvalue of $A$, with this value achieved when $\xv$ is the corresponding eigenvector. These \textbf{extremal eigenvalue} statements are used repeatedly in the chapter on wavefront shaping, Chapter~\ref{sec:WaveShaping}, with the vector $\xv$ typically a set of excitation coefficients and the quadratic form arising from a focusing metric or related objective. Linearity (for a fixed geometry) plays a key role in connecting quadratic forms of outputs to quadratic forms of inputs.

There is a simple extension to the above results that is useful for ``communication channels,'' discussed in Sec.~\ref{sec:channels}. Suppose one knows the largest eigenvalue $\lambda_1$ and corresponding eigenvector $\xv_1$, and asks: among all possible $\xv$ \emph{orthogonal} to $\xv_1$, what is the maximum value of the Rayleigh quotient (and which $\xv$ achieves it)? Perhaps not surprisingly, the largest value is the \emph{second largest} eigenvalue, $\lambda_2$, achieved\footnote{This can be proven by the same eigendecomposition, with the orthogonality condition now removing the largest eigenvalue from ``access'' by $\xv$.} with corresponding eigenvector $\xv_2$:
\begin{align}
\max_{\xv,\xv_1^{\dagger}\xv=0} \frac{\xv^{\dagger} A \xv}{\xv^{\dagger} \xv} = \lambda_{\rm 2}(A). 
\end{align}
Clearly, this can continue for the $N$ largest (or smallest) values of the Rayleigh quotient achieved by orthogonal vectors. (In Sec.~\ref{sec:channels}, the orthogonal vectors represent sources that can serve to excite independent communication channels.)
There is a compact mathematical representation of the more general fact. Instead of an $M\times1$ vector $\xv$, consider the $M\times N$ \emph{matrix} $X$. A matrix generalization of the numerator of the Rayleigh quotient is $\Tr\left(X^{\dagger} A X\right)$. Instead of the vector normalization in the denominator, we can enforce orthogonal unit vectors by enforcing $X^{\dagger} X = I_{N\times N}$, where $I_{N\times N}$ is the $N \times N$ identity matrix~\footnote{$X$ may not be, and likely is not, square in these scenarios.}. Hence, the maximization statement for the matrix generalization is
\begin{align}
\max_{X^{\dagger}X=I_{N\times N}}  \Tr\left(X^{\dagger} A X\right) = \sum_{i=1}^N \lambda_{i}(A). 
\label{eq:FanThm}
\end{align}
where $\Tr$ denotes the trace, with the right-hand side being the sum of the first (largest) $N$ eigenvalues, with the columns of the optimal $X$ given by the $N$ corresponding eigenvectors. This is known as Fan's theorem~\cite{Overton1992-mb,Fan1949-lu}. If one considers this result for a sequence of $N$ values, the statements above about each individual eigenvector optimizer and corresponding eigenvalue optimum is recovered.

A final related mathematical result that we will state, but not prove, relates to the above matrix product of \eqref{eq:FanThm}, now with an additional matrix $B$ in the product: $\Tr\left(X^{\dagger}AXB\right)$, with $X$ still constrained to have orthonormal columns ($X^{\dagger}X = I_{N\times N}$). If we consider how large this product can be, it should make intuitive sense that the optimal $X$ might simultaneously diagonalize both $A$ and $B$, in an ordering such that the largest eigenvalue of $A$ multiplies the largest of $B$, the second largest eigenvalues multiply, and so on. Indeed, for Hermitian $A$ and $B$, and provided that $B$ is positive semidefinite (or that $X$ is square), one can prove that this product is a bound on the matrix product:
\begin{align}
\max_{X^{\dagger}X=I_{N\times N}}  \Tr\left(X^{\dagger} A X B\right) \leq \sum_{i=1}^N \lambda^{\downarrow}_{i}(A)\lambda_i^{\downarrow}(B),
\label{eq:HW1}
\end{align}
where the superscripts denote that the eigenvalues are in descending order. Note that for this case of maximization, if $B$ is positive semidefinite and the $N$ largest eigenvalues of $A$ are nonnegative (even for square $X$), one can relax the orthogonality constraint to $X^{\dagger} X \leq I_{N\times N}$ and need not require equality. Alternatively, assuming $B$ is positive semidefinite, the \emph{minimum} of the trace product occurs when the $N$ smallest eigenvalues of $A$ are ``anti-ordered'' against the eigenvalues of $B$ (smallest eigenvalue of $A$ multiplying the largest of $B$, and so on), giving the lower bound:
\begin{align}
\min_{X^{\dagger}X=I_{N\times N}}  \Tr\left(X^{\dagger} A X B\right) \geq \sum_{i=1}^N \lambda^{\uparrow}_{i}(A)\lambda_i^{\downarrow}(B),
\label{eq:HW2}
\end{align}
where $\lambda_i^{\uparrow}(A)$ denotes the $i$-th smallest eigenvalue of $A$. In this case it is important to keep the equality in the constraint $X^{\dagger}X=I_{N \times N}$; otherwise, $X$ could simply be the zero matrix. The two results of \eqref{eq:HW1} and \eqref{eq:HW2} are versions of the Hoffman--Wielandt theorem~\cite{Hoffman1953-yr,Laurent2005-tc}, and they are useful in understanding the ``incompressibility'' of passive linear photonic systems and corresponding ``brightness'' theorems, as discussed in Sec.~\ref{sec:Brightness}.

\section{Bounds on constitutive material properties}
\label{sec:bounds_materials}
In this chapter, we discuss fundamental limits on the constitutive properties of natural and artificial media, which underlie the electromagnetic response of any material body under arbitrary illumination. Our discussion focuses primarily on LTI media. Under this assumption, the response $\eta$ of a material (the induced polarization or magnetization density) to an input field $\psi$ (electric or magnetic) can be expressed by the linear term in a Volterra series, which generalizes the concept of Taylor series by its ability to capture spatio-temporal ``memory'' effects, i.e., temporal and spatial dispersion \cite{monticone2025nonlocality,svirko2000polarization,wubs2016nonlocal},
\begin{equation}
    \eta(t,\boldsymbol{r}) \propto \int_{-\infty}^t dt'\,\int d\boldsymbol{r}'\, \chi(t-t'; \boldsymbol{r},\boldsymbol{r}') \psi(t',\boldsymbol{r}'),
    \label{eq:volterra}
\end{equation}
where $\chi(t-t'; \boldsymbol{r},\boldsymbol{r}')$ is a spatio-temporally dispersive response function (corresponding to the electric, magnetic or magneto-electric susceptibility; written here as a scalar for simplicity), with the spatial and temporal integrals extending over all space and time, subject to the restriction of causality. Extending this to nonlinear media requires considering higher-order terms in the series, expressed as multiple convolutions involving multiple time variables\footnote{Note, however, that there are circumstances under which such a series representation does not converge (see, e.g., \cite[Ch. 6]{boyd2020nonlinear}).} \cite{Lucarini2005,boyd2020nonlinear}. Instead, extending (\ref{eq:volterra}) to linear time-varying media simply requires replacing $\chi(t-t'; \boldsymbol{r},\boldsymbol{r}')$ with a temporally dispersive response function that changes with time, i.e., $\chi(t,t'; \boldsymbol{r},\boldsymbol{r}')$ (the response is no longer time-shift invariant) \cite{koutserimpas2024time}.
In the following, we further simplify our analysis by assuming that the medium is homogeneous and not spatially dispersive, i.e., $\chi(t-t'; \boldsymbol{r},\boldsymbol{r}')=\chi(t-t')\delta( \boldsymbol{r}-\boldsymbol{r}')$. We note, however, that the concept of spatial nonlocality will play an important role, in an indirect way, for some of the bounds discussed below.

Under these assumptions, the most general form of the constitutive relations for a \textbf{linear bianisotropic medium}, incorporating both anisotropy and magneto-electric coupling (chirality), can be written in matrix form as
\begin{equation}
\begin{pmatrix}
\mathbf{D} \\
\mathbf{B}
\end{pmatrix}
=
\begin{pmatrix}
\varepsilon & \dfrac{1}{c}\xi \\
\dfrac{1}{c}\zeta & \mu
\end{pmatrix}
\begin{pmatrix}
\mathbf{E} \\
\mathbf{H}
\end{pmatrix},
\label{eq:const_bian}
\end{equation}
where $\varepsilon$, $\mu$, $\xi$, and $\zeta$ are the permittivity, permeability, and magneto-electric coupling tensors introduced in Sec.~\ref{sec:tr_symmetry}, and $c$ is the speed of light in vacuum. If the medium is isotropic yet retains magneto-electric coupling, it is referred to as a bi-isotropic medium. In the reciprocal case, the constitutive relations are then typically written as, $\mathbf{D} = \varepsilon \mathbf{E} + \frac{i}{c}\,\kappa \mathbf{H}$ and $\mathbf{B} = -\frac{i}{c}\,\kappa \mathbf{E} + \mu \mathbf{H}$, where $\kappa$ is the scalar chirality parameter.

The key concepts typically used to derive bounds on constitutive material properties involve a combination of single-frequency and dispersion-related constraints, along with other relevant physical considerations, as summarized below.

    \paragraph{1. Single-frequency constraints} There are several important restrictions on the electromagnetic response of a material at any given frequency, which however impose no constraints on how the constitutive parameters must vary with frequency. An important example is the set of restrictions imposed by passivity. In the bulk of a homogeneous material, the only ``escape channel'' for the input electromagnetic energy is absorption\footnote{\label{footnote_Ewald_Oseen}According to the Ewald-Oseen theorem, the scattered fields (radiation damping) at the level of the individual elements of a crystalline (meta)material sum up perfectly to produce a single wave propagating according to the constitutive relations of the macroscopic effective medium. This is only strictly true in the periodic case: if the (meta)material constituent elements are not arranged in a perfectly periodic array, spurious waves are present, which act as additional escape channels for the input energy, and result in a new contribution to the imaginary part of the effective, macroscopic constitutive parameters \cite{tretyakov2003analytical}. This contribution, however, is always strictly non-negative and, therefore, cannot change the passive character of the medium, as expected.}. From Poynting's theorem \cite{Jackson1999}, the time-averaged power absorbed by a material at a given frequency is
    \begin{align}
    P_a=\frac{1}{2} \Re[-i\omega(\textbf{E}^* \cdot \textbf{D}+\textbf{H}^* \cdot \textbf{B})],
    \label{eq:absorbed}
    \end{align}
    which must be non-negative in a passive medium. Using the constitutive relations, one can then derive passivity conditions on the constitutive parameters of the material. In the simplest case, for an isotropic, non-chiral material, this implies $\Im{\varepsilon}\ge0$ and $\Im{\mu}\ge0$ at any given (positive) frequency. Instead, passivity, per se, imposes no constraints on the sign of the real part of the permittivity and permeability. More broadly, passivity considerations typically imply positivity constraints that play a key role in deriving fundamental limits from sum rules, as discussed in the following sections. (Active media are briefly discussed in Section \ref{sec:how_large_imag}.) Other important single-frequency physical constraints on the constitutive parameters are the reciprocity conditions in Eq. (\ref{eq:reciprocal}) and the so-called reality condition, namely, the fact the response functions must be real in the time-domain, which implies the symmetry $\varepsilon(-\omega)^*=\varepsilon(\omega)$, in the frequency domain.

    \paragraph{2. Dispersion or bandwidth constraints} In addition to single-frequency constraints, there are also restrictions on how constitutive parameters may depend on frequency, namely, on the functional form of this dependence or its derivatives. These are typically related to the principle of causality, KK relations, and all-frequency sum rules, as discussed in Section \ref{sec:causality}.
    An important example is the fact, already mentioned above, that the derivative of the susceptibility or permittivity of a passive causal medium must be strictly positive in transparency ranges (frequency windows where the absorption is negligible). A more stringent restriction can be derived by considering the time-averaged stored electromagnetic energy density in a transparent dispersive medium\footnote{\label{footnote_stored_energy}These expressions are invalid in highly lossy and dispersive frequency ranges (typically, near a material resonance). More broadly, the problem of determining the stored electromagnetic energy in lossy dispersive media from $\varepsilon(\omega)$, $\mu(\omega)$ is known to have no satisfactory solution, which represents an intriguing example of the shortcomings of macroscopic electrodynamics. One can even construct examples of systems with identical macroscopic parameters, at all frequencies, that store different amounts of energy \cite{askne1970energy,ramakrishna2008physics}. Knowledge of the microscopic structure and degrees of freedom, or additional information about the functional form of the constitutive parameters with respect to other variables (collision frequencies, etc.), is necessary to determine the stored energy in these cases.}, which, in the isotropic non-chiral case, is given by \cite[p. 275]{landau2013electrodynamics}
    \begin{align}
    W=\frac{1}{4} \left[\frac{d\left(\varepsilon(\omega) \omega\right)}{d \omega}\left|\textbf{E}\right|^2+\frac{d\left(\mu(\omega) \omega\right)}{d \omega}\left|\textbf{H}\right|^2\right].
    \label{eq:stored}
    \end{align}
    The stored energy in a passive medium should be positive, such that if the system is closed, absorption will ultimately convert this energy into heat. One can also argue that a material should be able to store more electromagnetic energy than free space, leading to the inequalities: 
    \begin{align}
    \frac{d\left(\varepsilon(\omega) \omega\right)}{d \omega}\ge \varepsilon_0, ~\frac{d\left(\mu(\omega) \omega\right)}{d \omega}\ge \mu_0
    \label{eq:derivative_bounds}
    \end{align}
    which imply a lower bound on the slope of the permittivity and permeability functions: $\frac{d\varepsilon(\omega)}{d \omega}\ge \frac{\varepsilon_0-\varepsilon(\omega)}{\omega}$. For $\varepsilon>\varepsilon_0$, the condition $d\varepsilon/d\omega\ge0$ is the stronger one. A sharper bound can also be derived from Eq.~(\ref{eq:KKRe}): in a transparency window, $\varepsilon(\omega)-\varepsilon_0 = (2/\pi)\int_0^\infty \Omega \Im\varepsilon(\Omega)\,(\Omega^2-\omega^2)^{-1}{\rm d}\Omega$ with $\Omega\Im\varepsilon(\Omega)\ge 0$, so that
    \begin{align}
    \omega\frac{d\varepsilon}{d\omega} - 2\left[\varepsilon_0-\varepsilon(\omega)\right] = \frac{2}{\pi}\int_0^\infty \frac{2\Omega^3 \Im\varepsilon(\Omega)}{(\Omega^2-\omega^2)^2}\,{\rm d}\Omega \ge 0, \quad\text{i.e.,}\quad \frac{d\varepsilon(\omega)}{d \omega}\ge \frac{2\left[\varepsilon_0-\varepsilon(\omega)\right]}{\omega},
    \label{eq:derivative_bounds_2}
    \end{align}
    with equality for a lossless Drude permittivity (similarly for the susceptibility, permeability, and refractive index; for $\varepsilon>\varepsilon_0$, the condition $d\varepsilon/d\omega\ge0$ is again the stronger one).
    
    \paragraph{3. Transformations of the macroscopic constitutive relations} The derivation of the equations of macroscopic electrodynamics from the microscopic Maxwell’s equations is not unique. Instead of splitting electric and magnetic effects into distinct electric and magnetic constitutive parameters, as usually done \cite{Jackson1999}, one may include all the terms resulting from the multipole expansion of the averaged microscopic currents in a new definition of \textbf{D}, dependent on \textbf{E} \emph{and} its spatial derivatives, while setting $\textbf{B}=\mu_0 \textbf{H}$ \cite{landau2013electrodynamics,agranovich2004linear,silveirinha2007metamaterial}. 
    This results in a spatially dispersive (nonlocal) equivalent permittivity, which can be related to the standard constitutive parameters of a bianisotropic medium [Eq. (\ref{eq:const_bian})] by the formula:
    \begin{align}
    \varepsilon_{nl}(\omega, \mathbf{k}) 
    & = \varepsilon 
    - \frac{1}{c^2}\xi ~ \mu^{-1} ~ \zeta 
    + \frac{1}{c^2}\left( \xi ~ \mu^{-1} \times \frac{\mathbf{k}}{k_0} 
    - \frac{\mathbf{k}}{k_0} \times \mu^{-1} ~ \zeta \right) 
    \nonumber \\
    &+ \frac{1}{c^2}\frac{\mathbf{k}}{k_0} \times \left( \mu^{-1} - \mu_0^{-1}\II \right) \times \frac{\mathbf{k}}{k_0}.
    \label{eq:eff_nonlocal}
    \end{align}
    In other words, a local bianisotropic material can be equivalently described as a nonlocal, anisotropic, nonmagnetic medium with the equivalent permittivity given above. In this sense, magnetic and magneto-electric (that is, chiral) effects can be equivalently interpreted as (weak) forms of spatial dispersion of the permittivity \cite{landau2013electrodynamics,agranovich2004linear,silveirinha2007metamaterial}. 
    For complex media and metamaterials with a chiral or bianisotropic response, this description can be advantageous when deriving limits and bounds, particularly because KK relations, and several related sum rules, can be applied to a spatially dispersive permittivity $\varepsilon_{nl}(\omega,\textbf{k})$ for each individual wavevector, since causality constrains its frequency dependence at fixed $\textbf{k}$ (as, e.g., for the Lindhard permittivity of a free-electron gas) \cite{landau2013electrodynamics,Lucarini2005}. 

    \paragraph{4. Conditions of physical realizability} Key properties of plane-wave propagation in a homogeneous material---such as the wavevector, defined as $\kv=\frac{\omega n}{c}~\hat\kv$, the refractive index $n=\pm\sqrt{\varepsilon\mu/\varepsilon_0\mu_0}$, and the wave impedance $Z=E/H=\pm\sqrt{\mu/\varepsilon}$ (all written here in the isotropic case for simplicity)---depend on the constitutive parameters of that material and are therefore constrained by the restrictions outlined above. They are also subject, however, to independent physical conditions, which, in turn, can often be used to identify additional constraints on the constitutive parameters themselves. Notably, these include so-called ``conditions of physical realizability'': (a) Energy should flow \emph{away} from a source (Sommerfeld radiation condition). Using the time-averaged Poynting vector, this implies that the real part of the wave impedance for a plane wave in a homogeneous material must be positive (whereas the sign of the imaginary part is unconstrained). (b) In a passive homogeneous system, the amplitude of a wave cannot diverge as it propagates away from the source, which directly implies that the imaginary part of the wavenumber (the magnitude of the wavevector) and of the refractive index must be non-negative. Note that this is independent of the passivity conditions on the imaginary part of the permittivity and permeability. Having independent conditions is necessary due to the multi-valued square-root nature of the relation between constitutive parameters and wave-propagation properties. These conditions can be used, among other things, to prove that a negative index of refraction is possible or, to be more precise, that the correct sign of the real part of the refractive index for a passive medium with negative real parts of the permittivity and permeability is indeed negative \cite[Sec. 11.2.1]{Novotny2012-py}.

    \paragraph{5. Order-of-magnitude physical arguments} When a problem does not appear amenable to the derivation of rigorous fundamental limits, or when a fundamental limit depends on specific material parameters, one can often resort to physical arguments based on the fact that many key material properties, such as the electron density, mass density, binding energy, vary by only one or two orders of magnitude across a broad range of solid-state materials of practical interest. We will see a few examples below.

\subsection{How \textbf{large} can the \textbf{real} part of the linear susceptibility be?} \label{sec:large_susceptibility}

The real part of the electric susceptibility $\chi$, permittivity $\varepsilon$, or refractive index $n$, of a transparent medium can, in principle, take any value, from negative to positive infinity, around a ``lossless'' Lorentz resonance, i.e., with absorption ideally localized as a delta function at the resonant frequency (see also Section \ref{sec:small_imaginary}). Why, then, do we not observe materials with refractive indices of 10 or even 100 at optical frequencies?

Since the single-frequency value of $\Re \chi$ can, in principle, be arbitrarily large, any meaningful bound must incorporate some additional constraints, such as dispersion or bandwidth constraints, and depend on the ``strength'' of the resonance, as well as on the number of material resonances. Following Ref. \cite{shim2021fundamental}, a rigorous bound of this type can be derived from just three ingredients: causality, passivity, and a sum rule. This approach also provides an instructive example and a general template for establishing bounds on other physical properties subject to similar constraints (such as for scattering problems \cite{zhang2023all}; see Section \ref{sec:T-matrix-response}).

The critical constraint needed to bound the susceptibility is the so-called \textbf{\textit{f}-sum rule}, which can be derived from the KK relation for $\Re \chi$, Eq. (\ref{eq:KKRe}), in the high frequency limit $\omega \to \infty$ (more precisely, by invoking the superconvergence theorem \cite{Lucarini2005}), together with the universal asymptotic form of the susceptibility in this regime, Eq. (\ref{eq:asymp}), yielding\footnote{A corresponding sum rule for the real part of the susceptibility can also be derived in the same way, yielding $\int_{0}^{\infty} \Re \chi(\omega') d\omega' =0$ for non-conducting media (a conductor contributes $-\pi\sigma_0/(2\varepsilon_0)$), which implies that the average value of the real part of the susceptibility (relative permittivity) over the entire spectrum is zero (unity).}
\begin{equation}
\int_{0}^{\infty} \omega' \, \text{Im}\,\chi(\omega') \, d\omega' 
= \frac{\pi e^2 N_e}{2 \varepsilon_0 m_e} 
= \frac{\pi \omega_p^2}{2},
\label{eq:f-sum rule}
\end{equation}
with \(e\), \(\varepsilon_0\), \(m_e\), and \(N_e\) as defined after Eq.~(\ref{eq:asymp}). A plasma frequency \(\omega_p\) is also often defined to collect these parameters, although the material is not a plasma. Since the integrand is related to the electromagnetic absorption of the material, this sum rule implies that the total absorption over the entire spectrum must be proportional to the total electron density~\footnote{A different form of this sum rule can also be given, which admits another intriguing physical interpretation. Considering the symmetry condition $\chi(-\omega)=\chi(\omega)^*$ ($\Re\chi(-\omega)=\Re\chi(\omega)$ and $\Im\chi(-\omega)=-\Im\chi(\omega)$) required by a real time-domain response, we can write $S=\int_{0}^{\infty} \omega' \Im \chi(\omega') d\omega' 
=\frac{1}{2}\int_{-\infty}^{\infty} \omega' \Im \chi(\omega') d\omega'=\frac{1}{2}\int_{-\infty}^{\infty} \omega' \Im \chi(\omega') d\omega'-\frac{i}{2}\int_{-\infty}^{\infty} \omega' \Re \chi(\omega') d\omega'$, where we used the fact that the second term in this last expression is zero due to symmetry. We can further simplify this expression by collecting everything under a single integral,
which then allows us to write the $f$-sum rule as: 
\begin{equation*}
\int_{-\infty}^{\infty} -i\omega' \,\chi(\omega') \, d\omega' 
= \pi \omega_p^2,
\end{equation*}
which is the form given by Schwinger in \cite[p. 51]{schwinger1998classical}. Note that the integrand is the frequency-domain polarization current density for an electric field $E=1$ V/m. Furthermore, one can recognize that the left-hand-side of this expression is simply an inverse Fourier transform evaluated at $t=0$, i.e., $\int_{-\infty}^{\infty} -i\omega' \chi(\omega') e^{-i\omega't} d\omega'|_{t=0}$. This, finally, implies that the left-hand-side of the $f$-sum rule can be interpreted as the time-domain polarization current density induced in the medium, at $t=0^+$, in response to an applied impulse of electric field, $ E(t)\propto\delta(t)$. This interpretation also provides a physical explanation for the fact that the $f$-sum rule only depends on the electron charge and mass, but not the resonant frequencies or collision frequencies of the medium \cite{Caloz2019_private}. Quoting Schwinger \cite{schwinger1998classical}, ``\emph{in the response to \emph{[a spatially uniform]} electric pulse localized at time zero, only the inertia \emph{[and charge]} of the electrons matter---frictional and binding forces have not time in which to act.}'' This explanation is also consistent with the fact that the response of a system to very high frequencies (in the limit, for $\omega\to \infty$) is associated with its response at very short time scales (nearly instantaneous response, for $t=0^+$), which is indeed constrained by the inertia of the electrons.
}. In different terms, this sum rule constrains the distribution and strength of oscillators in a material. KK relations and the \textit{f}-sum rule can then be discretized using localized basis functions, such as delta functions, yielding 
\begin{equation}
\operatorname{Re}\chi(\omega) = \sum_{i=1}^{N} \frac{c_i \, \omega_p^2}{\omega_i^2 - \omega^2}
\label{eq:chi_causal_repres}
\end{equation}

\begin{equation}
\sum_i c_i = 1
\label{eq:normalized_f_sum}
\end{equation}

The first equation can be interpreted as a general, first-principle representation of the susceptibility, valid for any material, in terms of a set of ``lossless'' Drude-Lorentz oscillators with transition frequencies $\omega_i$ and normalized oscillator strengths $c_i$, the latter being constrained by a normalized version of the \textit{f}-sum rule, Eq. (\ref{eq:normalized_f_sum}). In addition, passivity provides another critically important restriction: $c_i\ge0$ (the integrand of (\ref{eq:f-sum rule}) must be nonnegative for passive media). Microscopically, the $c_i$ are (normalized) quantum-mechanical oscillator strengths, $f_{n0}=(2m_e\omega_{n0}/\hbar)|\langle n|\hat{x}|0\rangle|^2$, set by the transition dipole matrix elements between the initial (here, ground) state $|0\rangle$ and excited states $|n\rangle$, and Eq.~(\ref{eq:normalized_f_sum}) is the macroscopic counterpart of the Thomas--Reiche--Kuhn sum rule, $\sum_n f_{n0}=1$ per electron (Section~\ref{sec:nonlinear}). Oscillator strengths of downward transitions (from an excited initial state $|n\rangle$ to a lower state, $\omega_{0n}<0$) are negative, so a medium with population inversion can have a net negative oscillator strength (gain) at the corresponding frequency (see footnote \ref{footnote_active_media}). In crystals, the intraband (free-carrier) contribution of a partially filled band is conventionally expressed through the effective mass, as an oscillator strength $m_e/m^*$ per carrier.

\begin{figure*}[tb]
\centering
\includegraphics[width=0.99\linewidth]{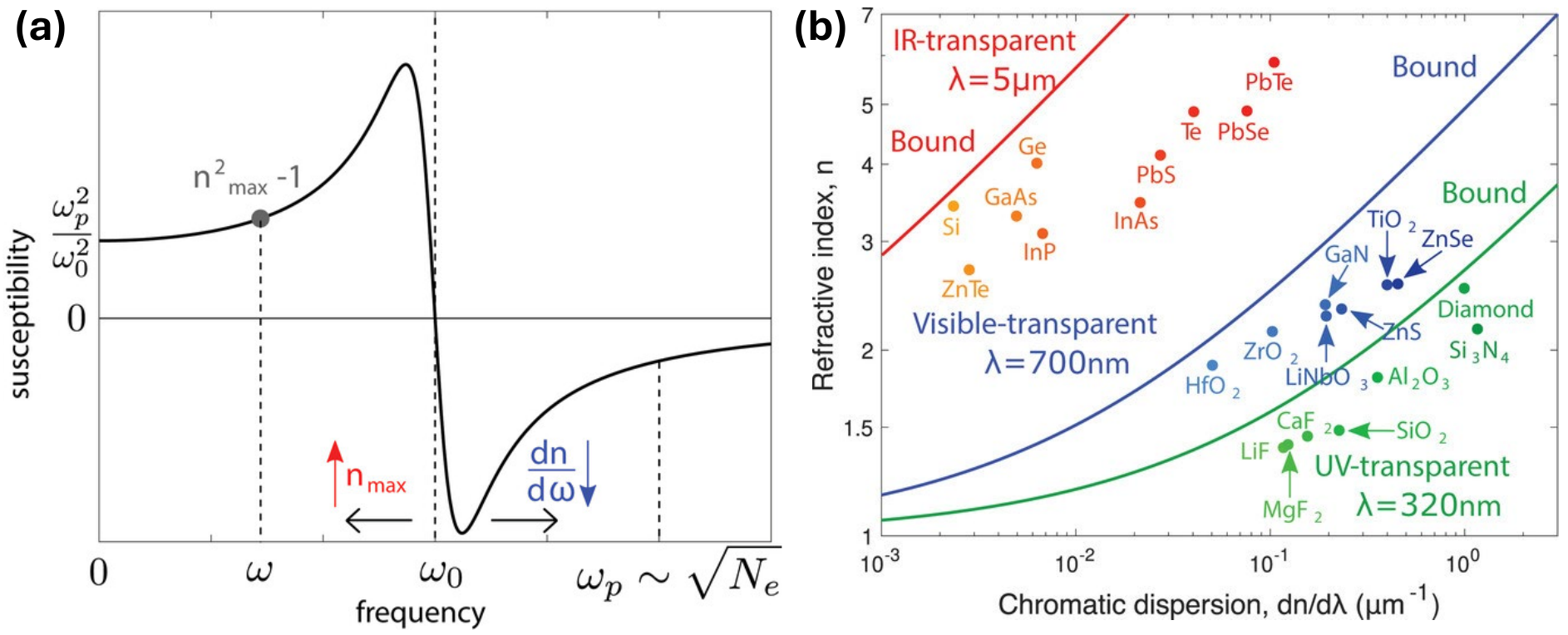}
\caption{\textbf{Maximum refractive index}. (a) Schematic of a single Drude--Lorentz oscillator illustrating the tradeoff between refractive index and dispersion. Lowering the resonance frequency $\omega_0$ increases the ratio $\omega_p^2/\omega_0^2$, thereby enhancing the maximum refractive index $n_{\max}$ at frequency $\omega$, but at the expense of increased dispersion $\mathrm{d}n/\mathrm{d}\omega$. The plasma frequency $\omega_p = \sqrt{Ne^2/(\varepsilon_0 m_e)}$ and, hence, the oscillator strength, are set by the material's electron density. (b) Refractive index versus chromatic dispersion for various optical materials, evaluated at three representative wavelengths within their respective transparency windows. The solid curves denote the corresponding physical bounds, computed at each wavelength using the average electron density of the materials considered. The results clearly illustrate the increase of refractive index with increasing dispersion.  (Figure adapted from Ref.~\cite{shim2021fundamental}.)}
\label{fig:refr_index}
\end{figure*}

With this representation and constraints, and adding the requirement that the dispersion of $\chi$ be smaller than a given constant $\chi'$, one can then set up an optimization problem to maximize $\Re \chi$:
\begin{align}
\max_{c_i} \quad & \operatorname{Re}\chi(\omega) 
= \sum_{i=1}^{N} c_i f_i(\omega) \\
\text{subject to} \quad 
& \frac{d\,\operatorname{Re}\chi(\omega)}{d\omega} 
= \sum_{i=1}^{N} c_i f_i'(\omega) \leq \chi' \\
& \sum_{i=1}^{N} c_i = 1, \quad c_i \geq 0
\end{align}
Note that the optimization is performed over the oscillator strengths of a very large number of oscillators with fixed transition frequencies, rather than over the oscillator frequencies themselves. This is advantageous because the susceptibility depends linearly on the coefficients $c_i$ (whereas it depends nonlinearly on the frequencies $\omega_i$) resulting in a linear program that, although large, can be rigorously solved to global optimality using well-established tools. As shown in \cite{shim2021fundamental}, one can rigorously prove that the optimal solution is always realized by a single nonzero oscillator that exhausts the total oscillator strength allowed by the \textit{f}-sum rule, i.e., with strength $c_0=1$ and oscillator frequency $\omega_0^2 = \omega^2 \left(1 + \sqrt{2\omega_p^2/(\omega^3 \chi')}\right)
$. The intuition behind the tradeoffs associated with this optimal solution is illustrated in Fig. \ref{fig:refr_index}(a).

After some simple algebra, one can finally derive an analytical bound on the largest achievable refractive index for any passive, linear, isotropic material:
\begin{equation}
\frac{\left(n^2 - 1\right)^{2}}{n} \leq \frac{\omega_p^2 n'}{\omega},
\label{eq:max_n}
\end{equation}
which can be further simplified if the index is moderately large:
\begin{equation}
n \lesssim \left( \frac{\omega_p^2 n'}{\omega} \right)^{1/3},
\label{eq:max_n_large}
\end{equation}
where $n'=\chi'/2n$ is the maximum allowed dispersion of the refractive index. From these analytical bounds---which depend only on three parameters: frequency of interest, dispersion constraint, and electron density---the reasons why it is fundamentally difficult to achieve a large refractive index at optical frequencies become apparent. First, as the frequency increases, the maximum achievable refractive index decreases, consistent with the typical refractive-index values observed in the visible range, which are substantially lower than those attainable, for example, in the microwave regime. More importantly, it is clear from Eq. (\ref{eq:max_n_large}), that a ten-fold increase in refractive index, for fixed $\omega$ and $n'$, would require a $10^3$ increase in electron density, which is arguably unfeasible given the limited range of electron density in solid-state materials (on the order of $10^{22}-10^{23}~\mathrm{cm}^{-3}$). Since the frequency of interest is fixed and a massive increase in electron density is highly unlikely, the only remaining option is to tolerate larger dispersion, namely, to operate closer to resonance. In optical materials, however, resonances are broadened by various loss mechanisms. As a result, operating very close to resonance does not typically yield very large increases in refractive index and implies significantly higher dissipation. Interestingly, Ref.~\cite{shim2021fundamental} showed that many naturally occurring transparent materials already operate close to the upper bound [Fig.~\ref{fig:refr_index}(b)], given their electron density, in the regime of small to moderate dispersion, whereas significant room for improvement remains in the high-dispersion region.

These results imply that a transparent material with the highest possible refractive index would ideally have an electronic band structure that is as close as possible to two flat bands separated by an energy gap $\hslash \omega_0$ (a single dipole transition using up the \textit{f}-sum rule), as if the material were a collection of ideal two-level systems with no coupling (other than via light scattering) and no broadening. This picture is closely related to the Penn model of covalent and partially ionic solids~\cite{penn1962wave}, which represents the valence electrons by a single effective transition at an average (Penn) gap $E_g$ between bonding and antibonding states, giving an electronic (clamped-ion) permittivity $\varepsilon_\infty/\varepsilon_0\approx1+(\hbar\omega_{p,v}/E_g)^2$, with $\omega_{p,v}$ the plasma frequency of the valence electrons. Greater bond polarity (ionicity) widens this average gap, which helps explain why wide-bandgap materials tend to have relatively small electronic permittivities~\cite{phillips1970ionicity,khurgin2026are}. Clearly, realizing an idealized two-level-like response would be more feasible in the form of a dilute atomic gas than a solid-state material. Interestingly, these observations are qualitatively consistent with the results of Refs. \cite{andreoli2021maximum,andreoli2023maximum}, where it was found that in an atomic crystal with low density (quantum optics regime), light-matter interactions can have a single-mode character, allowing a purely real refractive index that grows with the cube root of the electron density. At larger atomic densities, however, the electronic orbitals begin to overlap (quantum chemistry regime), breaking the single-mode regime. Other important considerations regarding the requirements for realizing high-index optical materials---and the fundamental challenges associated with this goal---are discussed from a solid-state physics perspective in \cite{khurgin2022expanding}. Excitonic and other many-body effects fit within the same framework: electron--hole Coulomb interactions redistribute oscillator strength, e.g., transferring part of it from the interband continuum to sharp exciton resonances below the single-particle gap, which can locally enhance both $\Re\chi$ and $\Im\chi$, but they cannot increase the total, since the \textit{f}-sum rule (\ref{eq:f-sum rule}) depends only on the total electron density, irrespective of the interactions among electrons.

As an alternative route toward achieving high-index materials---without relying on quantum chemistry or solid-state material discovery/design---we note that, if metals with significantly reduced losses could be realized (see Section \ref{sec:small_imaginary}), it would be possible to engineer metallo-dielectric composites and metamaterials with refractive indices exceeding 100, although accompanied by unavoidably large dispersion according to Eq. (\ref{eq:max_n_large}) (Ref.~\cite{shim2021fundamental}). Regarding the broader role of metamaterials in this context, we also note that the same bounds derived above for isotropic media also apply to the refractive index of (nearly) all linear, passive bianisotropic media, including anisotropic, magnetic, and chiral metamaterials~\footnote{There are some intriguing potential loopholes for gyrotropic plasmonic media
(which have a modified KK relation \cite{abdelrahman2020broadband}) and
hyperbolic metamaterials, which may be difficult, however, to exploit in practice \cite{shim2021fundamental}.}. While the full derivation is somewhat involved, the key idea relies on the equivalence, mentioned above, between local bianisotropic constitutive relations and a nonlocal, purely electric, alternative representation, and the fact that the same constraints used above---KK relations, passivity, and the \textit{f}-sum rule---apply to the diagonal components of the equivalent nonlocal permittivity, Eq. (\ref{eq:eff_nonlocal}), diagonalized with a suitable choice of polarization basis. This result implies that, if one tries to increase the refractive index of a (meta)material by maximizing both its electric and magnetic susceptibilities, the resulting refractive index would not exceed the one predicted by the bound above. This makes intuitive sense: trying to maximize the effective magnetic response of a material requires shaping the microscopic currents to induce strong magnetic dipoles, using the same underlying matter and available electrons that could be used, instead, to increase the electric response. The overall response of the material, described by the nonlocal permittivity (\ref{eq:eff_nonlocal}), is still governed by the available number of electrons and is, therefore, limited by the same upper bound.

\subsection{How \textbf{small} can the \textbf{real} part of the linear susceptibility be?} \label{sec:how_small_real}
No fundamental principle prohibits the real part of the linear susceptibility, permittivity, permeability, or refractive index from being zero or negative at any given (nonzero and finite) frequency, even in the absence of losses \cite{milton2020further_comments}. As already mentioned, ``lossless'' Drude or Lorentz models (with the imaginary part localized as delta functions at zero or resonant frequencies, respectively) are perfectly consistent with causality (KK relations) and passivity.

A more interesting question is whether the constitutive parameters can be equal to zero or a specified negative value over a nonzero bandwidth. In the introduction of this chapter we clarified that causality and passivity set a lower bound on their frequency dispersion. In the lossless case, \eqref{eq:derivative_bounds_2} shows that $\frac{d\varepsilon(\omega)}{d \omega}\ge \frac{2\left[\varepsilon_0-\varepsilon(\omega)\right]}{\omega}$ (similarly for the susceptibility, permeability, and refractive index; for $\varepsilon>\varepsilon_0$, the condition $d\varepsilon/d\omega\ge0$ is stronger). For instance, if the permittivity vanishes at a frequency $\omega_0$, as in epsilon-near-zero media, the derivative cannot be smaller than $2\varepsilon_0/\omega_0$, regardless of the type of material under consideration, which is only required to be LTI, causal, and passive (lossless). While a permittivity lower than the free-space value cannot remain perfectly constant over a nonzero-measure bandwidth, the next natural question to ask is whether it can remain within certain limits around the desired value $\varepsilon(\omega_0)=\varepsilon_m$, over a specified bandwidth. This is easy to answer, approximately, for very small frequency ranges $\mathcal{B}=\left[\omega_1,\omega_2\right]$, where one can assume the minimum derivative above remains constant, yielding a maximum deviation from the desired value:  $\left| \varepsilon(\omega_{1,2}) - \varepsilon_{\mathrm{m}} \right|\ge  (\varepsilon_{0} - \varepsilon_{\mathrm{m}})B $, where $B=(\omega_2-\omega_1)/\omega_c$ is the fractional bandwidth. For broader bandwidths, this argument is clearly invalid, as the derivative does not need to remain constant (but this result can still be generalized; see \cite{gustafsson2010sum_metamaterials}). In addition, the derivative bound is strictly valid only if losses are zero. These problems can be overcome, and the question posed above answered in full generality, by using the theory of Herglotz functions \cite{bernland2011sum}, introduced in Section \ref{sec:herglotz}. Since the response of a material is ``immittance passive'' according to Eq. (\ref{eq:passiveim}), with the input identified as the electric field $v(t)=E(t)$ and the output as the displacement current $u(t)=\partial D(t)/\partial t$, one can define a Herglotz function as $h(\omega)=i w(\omega)$, where  $w(\omega)$ is the Fourier transform of the impulse response of the system, which is, in this case, $w(t)=\partial \varepsilon(t)/\partial t$. Hence, $h(\omega)= \omega \varepsilon(\omega)$, and one can derive sum rules and bounds on $h$ (or suitable compositions of $h$ with other Herglotz functions) as in Eq. (\ref{eq:h_sum_rules}). We refer the reader to \cite{gustafsson2010sum_metamaterials,bernland2011sum} for additional details and only give here the final result. For an arbitrary, passive, lossless material, one finds that the maximum deviation from the desired value, $\varepsilon_m<\varepsilon_0$, over a frequency range $\mathcal{B}$ with fractional bandwidth $B$, is bounded from below as \cite{skaar2006bounds,cassier2017bounds}:
\begin{equation}
\max_{\omega \in \mathcal{B}} \left| \varepsilon(\omega) - \varepsilon_{\mathrm{m}} \right|
\ge
\frac{B}{1 + B^2/4} \left( \varepsilon_{0} - \varepsilon_{\mathrm{m}} \right),
\label{eq:low_epsilon_bound}
\end{equation}
which further confirms that media with permittivity smaller than $\varepsilon_0$ must be frequency dispersive. This bound is sharp and it is attained by a single lossless Drude term (additional resonances cannot help). The bound derived in Ref.~\cite{gustafsson2010sum_metamaterials}, $B/(1+B/2)$, agrees with it to first order in $B$, and both are consistent with the previous approximate result, derived by assuming constant minimum derivative, for small fractional bandwidths $B/2\ll1$. If losses are allowed, the smallest achievable deviation is exactly $(B/2)(\varepsilon_0-\varepsilon_m)$ for every $B$~\cite{lindjohansen2009perfect}, smaller than the lossless value by the factor $2/(1+B^2/4)$. Note, however, that this bound concerns the magnitude of the complex difference, $|\varepsilon-\varepsilon_m|$, which does not distinguish between a deviation of the real part and a nonzero imaginary part, i.e., absorption. Allowing more loss, the real part can even be kept exactly constant over the band, and the minimum loss required for this is known~\cite{seip2005extremal}. However, many applications, such as epsilon-near-zero media or a perfect lens, require both a flat real part and low loss, which raises the question of the optimal trade-off between the two. 
\begin{openq}
How much can a frequency-shaped in-band loss improve the flatness of the real part of the permittivity? Specifically, for a given upper bound on the in-band absorption, what is the minimum achievable maximum deviation of $\Re\,\varepsilon$ from a target value over a band, and how should the absorption spectrum be shaped to attain it?
\end{openq}

Finally, we note that all these limitations break down if the medium is not passive, that is, in the presence of gain. Specifically, as discussed in \cite{nistad2008causality,milton2020further_comments}, causality does not preclude a medium from exhibiting a dispersionless near-zero or negative permittivity (or permeability, or refractive index). In other words, in active media it is, in principle, possible to achieve $\max_{\omega \in \mathcal{B}} \left| \varepsilon(\omega) - \varepsilon_{\mathrm{m}} \right| \approx 0$ over a finite nonzero-measure bandwidth. However, attention should be paid to the possible onset of instabilities \cite{nistad2008causality} (see also the next section and footnote \ref{footnote_active_media}). More generally, in the presence of loss and/or gain, meaningful and application-relevant trade-offs may arise between the attainable bandwidth, the maximum deviation from the target constitutive parameter value, and the permissible level of absorption or gain compensation.

\subsection{How \textbf{large} can the \textbf{imaginary} part of the linear susceptibility be?} \label{sec:how_large_imag}
The imaginary part of the electric susceptibility is constrained by the \textit{f}-sum rule, Eq. (\ref{eq:f-sum rule}). For a passive medium, in which $\Im \chi \ge 0$ at all frequencies, this sum rule directly implies an upper bound on how large the imaginary part of the susceptibility can be---that is, on how lossy the material can be--- over any finite bandwidth. Assuming constant $\Im \chi$ over a bandwidth $\Delta \omega$ around a central frequency $\omega_c$, one finds
\begin{equation}
\Im \chi \le \frac{\pi \omega_p^2}{2 \Delta \omega ~ \omega_c}.
\end{equation}

Clearly, if the medium was active, with $\Im \chi < 0$ at some frequencies, this bound would no longer apply, and arbitrarily large values of loss and gain would, in principle, be possible, provided that the \textit{f}-sum rule, Eq. (\ref{eq:f-sum rule}), remains satisfied. An interesting question, then, is whether such a medium would be intrinsically unstable, and whether upper bounds exist on the achievable gain in a medium free of intrinsic \emph{absolute} instabilities (i.e., fields diverging with time at any fixed point in space)~\footnote{\label{footnote_active_media}Some clarifications about active media and stability are warranted. The susceptibility or permittivity of a medium with gain does not need to have poles in the upper half of the complex frequency plane (but might have zeros there). For example, the susceptibility of an inverted Lorentzian medium, a simple model for a linear gain medium, has a negative oscillator strength but all the poles are in the lower half-plane. This represents a stable \emph{polarization} response to an input field, in the sense that the polarization response at a fixed point in space does not diverge with time. However, from the standpoint of wave propagation, the electromagnetic field will still exponentially grow as it propagates, since the refractive index $n$ and wavenumber $k$ have a negative imaginary part, leading to the so-called \emph{convective} instabilities of standard gain media, using the terminology used in \cite{nistad2008causality}. Systems composed of media with convective instabilities are truly unstable (fields diverging with time) only if they are spatially infinite or placed in an appropriate resonator configuration. Moreover, a stronger form of intrinsic instability occurs if the susceptibility has zeros in the upper half-plane, yielding branch points for $n$ and $k$ there ($n$ and $k$ lose meaning at real frequencies). This produces so-called \emph{absolute} instabilities, with fields diverging with time at any fixed point in space. We refer the readers to \cite{nistad2008causality} for additional details, including about the causality properties of active media.}. Intriguingly, the answer to this latter question is negative. As shown in \cite{nistad2008causality}, one can construct examples of physical, causal, material models that exhibit no absolute instabilities regardless of how large the gain $(-\Im \chi )$ is. In other words, causality alone does not prohibit absolutely stable media with arbitrarily large gain.

\begin{figure*}[tb]
\centering
\includegraphics[width=0.75\linewidth]{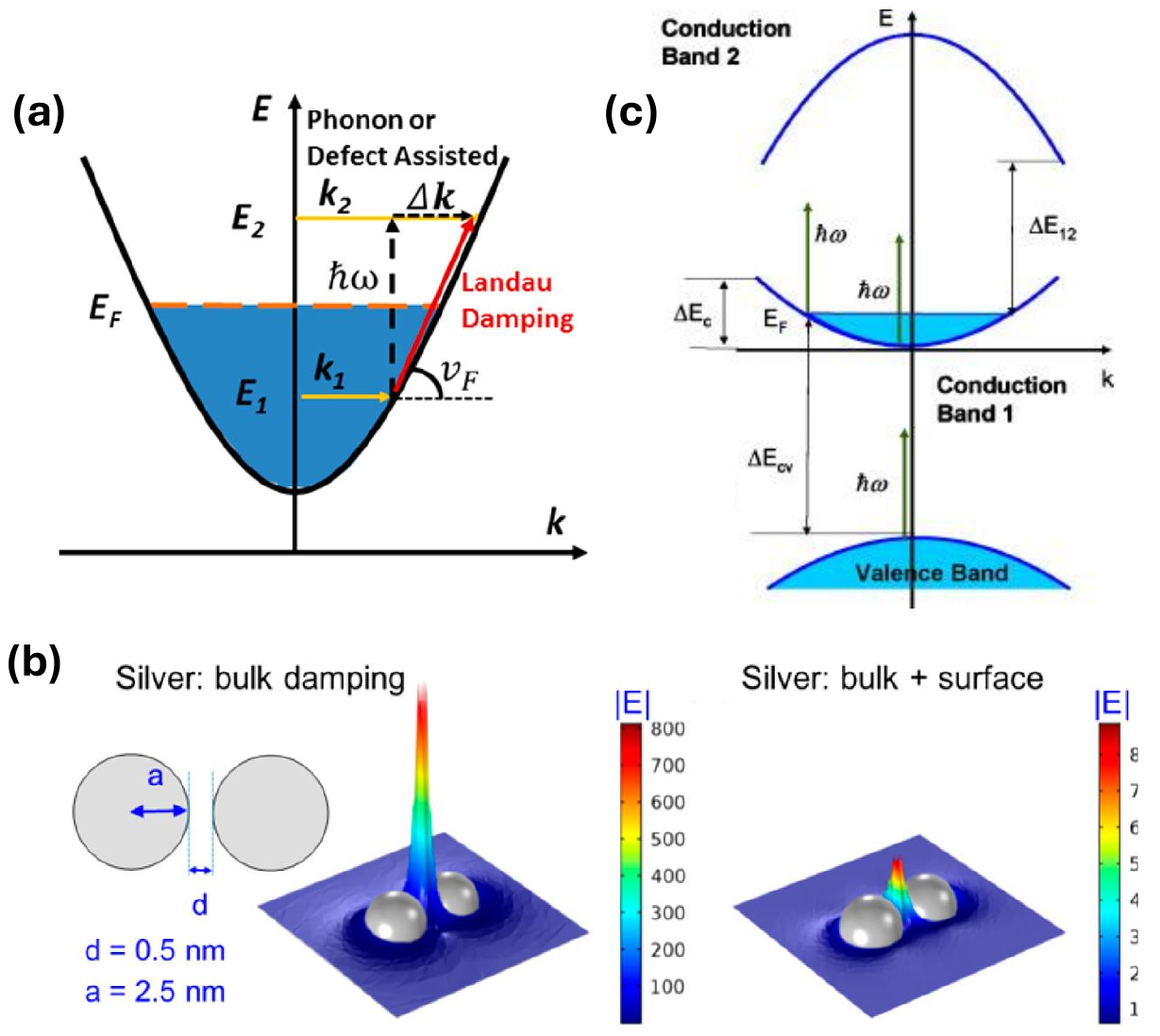}
\caption{\textbf{Lossy and lossless metals.} (a) Absorption in metals: Absorption of a photon with small momentum/wavevector assisted by a phonon or defect (dashed arrows), or direct absorption of a confined surface plasmon-polariton (SPP) with large momentum/wavevector (solid arrow), corresponding to Landau damping (surface-collision assisted damping). (b) Example illustrating the impact of Landau damping: SPP field distribution in a silver dimer, calculated by including only bulk damping (phonon/defect scattering; left) and by including both bulk and surface damping (Landau damping; right). (c) Band diagram of a hypothetical lossless metal exhibiting zero absorption at optical frequencies. If the photon energy falls into the gap between two conduction bands, absorption vanishes while the permittivity can remain negative. (Panels (a,b) adapted from Ref.~\cite{khurgin2017landau}; panel (c) adapted from Ref.~\cite{khurgin2010search}.)}
\label{fig:loss_metals}
\end{figure*}

\subsection{How \textbf{small} can the \textbf{imaginary} part of the linear susceptibility be?} \label{sec:small_imaginary}
Can a material be perfectly lossless? The KK relations, Eqs. (\ref{eq:KKRe}),(\ref{eq:KKIm}), indicate that a material with an identically zero imaginary part of the permittivity (or susceptibility) at all frequencies would necessarily have a real part equal to the permittivity of free space. In other words, the only perfectly lossless material over the entire spectrum is vacuum. A stronger constraint follows from the oscillator-strength sum rule, Eq. (\ref{eq:f-sum rule}), which implies that for a material with a given total electron density $N$, the frequency integral of the imaginary part of the susceptibility must be proportional to $N$. However, this imaginary part could, in principle, be concentrated at isolated frequencies in the form of delta functions, as discussed in previous sections. Thus, causality does not forbid materials that are perfectly lossless over all but a discrete set of frequencies~\footnote{Physically, it is well established that any non-steady process in a material body is accompanied by some degree of thermodynamic irreversibility \cite[p. 274]{landau2013electrodynamics}. Electromagnetic losses (the imaginary part of the permittivity and permeability) in a variable electromagnetic field are therefore never exactly zero, for any nonzero frequency. In transparency ranges, however, losses can be so small that they can be safely neglected, except in extreme cases. Technically, in fact, the frequency-domain uniqueness theorem for the solution of Maxwell's equation is only valid if losses are not exactly zero, $\Im \varepsilon,\Im \mu \ne 0$ \cite{balanis2012advanced,mystilidis2024uniqueness}. In the perfectly lossless case, thermodynamic paradoxes may arise, especially in  scenarios involving nonreciprocity and unidirectional waves \cite{ishimaru2017electromagnetic,mann2019nonreciprocal,fernandes2019topological,monticone2020truly}. Moreover, poles of a transfer function may move onto the real axis, yielding undamped resonances at real frequencies, which implies that the transfer function is undefined at these frequencies and the solution is not unique (any input leads to the same infinite output).
nonzero losses or nonlocal effects regularize these issues and guarantee a unique solution to a boundary value or initial value problem.}. Indeed, dielectrics can often be accurately approximated as nearly lossless across broad frequency windows, sufficiently above or below resonances. Interestingly, the reason why optical glass is so transparent is nontrivial and cannot be fully explained within a purely classical treatment of light-matter interactions. In particular, modeling a transparent dielectric or semiconductor below its energy bandgap using a simple Lorentz oscillator model (where the bandgap sets the resonance frequency) dramatically overestimates absorption at photon energies well below the bandgap. Within the Lorentz model, the imaginary part of the wavenumber, $\Im k$, scales as $\omega^2$ at frequencies much lower than resonance, whereas in reality it decays exponentially with increasing detuning from the band edge (a behavior known as Urbach tail \cite{dow1972toward,khurgin2010search}). Using the scattering rate of valence electrons for optical glass, Lorentz formula would predict $\Im k$ on the order of 10 cm$^{-1}$ \cite{khurgin2010search}, which would correspond to a power attenuation on the order of 90\% for a 1 mm propagation length, which is clearly off by orders of magnitude. The reason for this discrepancy is that \emph{light absorption requires a transition between two real electronic states}. If there are no states within a given energy range, absorption cannot occur. (More precisely, energy must be conserved for the system as a whole: phonon-assisted, including multi-phonon, transitions allow weak absorption below the gap and contribute to the exponential Urbach tail even in pristine crystals \cite{khurgin2014multi,dow1972toward}, and, in the presence of free carriers, Auger-like processes can also contribute; these higher-order processes are much weaker than direct interband absorption above the gap.) 

This insight also suggests a possible route toward realizing a long-sought-after lossless metal~\footnote{A lossless Drude model, with the imaginary part of the permittivity perfectly localized at zero frequency, is not prohibited by the KK relations. A physical realization approximating such behavior is provided by superconductors below their superconducting energy gap $\Delta$ (where the electromagnetic excitation energy becomes sufficient to break Cooper pairs), which depends on their critical temperature as $\Delta (T=0) \approx 1.764 k_b T_c$, where $k_b$ is Boltzmann constant, according to BCS theory \cite{tinkham2004introduction}. Not even room temperature superconductors would behave as perfect conductors at optical frequencies.}. Absorption in metals is associated primarily with transitions within the conduction band, enabled by scattering processes that guarantee momentum matching [Fig. \ref{fig:loss_metals}(a)]: scattering of conduction electrons by defects, lattice vibrations, and spatial inhomogeneities such as interfaces. Even if one could fabricate a perfectly defect-free metal and lower the temperature to minimize lattice vibrations---that is, even if the material had zero intrinsic bulk losses---dissipation would still occur due to collisions of electrons with the boundaries of a finite structure, leading to surface-collision induced damping or \textbf{Landau damping} \cite{khurgin2015deal,hassani2020physical} [Fig. \ref{fig:loss_metals}(a)]. This effect becomes particularly important in nanostructures with highly confined fields (such as in plasmonics), as the spatial confinement $\Delta x$ implies a wider wavevector bandwidth (according to the standard Fourier uncertainty principle, $\Delta x \Delta k \ge 1/2$ for root-mean-square widths) which can compensate for the momentum mismatch between the initial and final states in a transition. Surface-induced damping is one of the dominant loss mechanism in high-quality nano-plasmonic systems and represents the minimum achievable loss with existing metallic materials (more on this below). This effect, which is fundamentally \emph{nonlocal} since it depends on the field wavevector, establishes a physical upper bound on the maximum possible field confinement and enhancement in plasmonic systems [Fig. \ref{fig:loss_metals}(b)]. We refer the reader to \cite{khurgin2017landau} for a detailed discussion (we only note here that the loss rate associated with Landau damping must be calculated in a self-consistent iterative manner, as the initially calculated damping affects the field confinement, which in turn modifies the damping rate, and the process must be iterated until convergence is reached \cite{khurgin2017landau,hassani2020physical}). Finally, we stress that even if scattering and surface collisions are present (and arbitrarily strong), absorption is only possible if a final unoccupied state exists, as mentioned above. Following \cite{khurgin2010search}, a metal that is virtually lossless within a given frequency window could, in principle, be realized if a material were discovered or engineered with an energy gap between its first and second conduction bands, as illustrated in Fig. \ref{fig:loss_metals}(c), subject to some additional conditions (for example, absorption assisted by electron--electron scattering, in which a single photon excites two conduction electrons and its energy is shared between them, makes the requirements on the band structure even more stringent wherever umklapp processes or band nonparabolicity allow such scattering to relax the current). Importantly, the permittivity of such a material would need to remain negative within this window to retain its metallic characteristic. Such a material platform has, so far, remained elusive.

\begin{openq}
        Are there other fundamental reasons that prevent a lossless metal from existing? Can emerging computational solid-state approaches, as for example in \cite{hu2022high,hu2025hypergap} (see also the recent experimental demonstration of a hyper-gap transparent conductor \cite{wu2025hypergap}), accelerate the search for materials with band structures consistent with lossless metallic behavior at optical frequencies?
\end{openq}

\subsection{Magneto-electric coupling and chirality}
Physically, magneto-electric coupling arises when the currents induced in a structure by an electric or magnetic field produce both electric and magnetic dipole moments. A simple model of a chiral molecule or metamaterial inclusion is a conducting helix \cite{lindell1994electromagnetic}: an electric field along its axis drives a current along the wire that both separates charge along the axis (an axial electric dipole) and circulates around it (an axial magnetic dipole); conversely, a time-varying axial magnetic field induces a circulating current and hence an axial magnetic and electric dipole. The sign of this magneto-electric coupling is set by the handedness of the helix, so mirror-image structures have chirality parameters $\kappa$ of opposite sign. How strong can the chiral response of a material be? A first insight into this question can be obtained by examining the implications of passivity. Starting from Eq. (\ref{eq:absorbed}) for the power absorbed in a medium and substituting the constitutive relations for an isotropic, reciprocal, chiral medium (a bi-isotropic medium), one finds \cite{lindell1994electromagnetic,silveirinha2010comment}
\begin{equation}
P_a = \frac{\omega}{2} 
\begin{pmatrix} \mathbf{E}^* & \mathbf{H}^* \end{pmatrix}
\begin{pmatrix}
\Im \varepsilon & i\Im \frac{\kappa}{c} \\
-i\Im \frac{\kappa}{c} & \Im \mu
\end{pmatrix}
\begin{pmatrix} \mathbf{E} \\ \mathbf{H} \end{pmatrix}.
\label{eq:chiral_absorbed}
\end{equation}

To satisfy passivity, i.e., $P_a\ge0$, the matrix appearing in this expression must be positive semidefinite, which implies the usual conditions on the imaginary part of the permittivity and permeability, namely, $\Im{\varepsilon}\ge0$ and $\Im{\mu}\ge0$, but also the additional constraint that the determinant of the matrix must be positive \cite{lindell1994electromagnetic,silveirinha2010comment}. This latter condition imposes a limit on the strength of the imaginary part of the chirality coefficient: 
\begin{equation}
(\Im \kappa/c)^2\le\Im \varepsilon \Im \mu
\label{eq:chiral_inequality}
\end{equation}
This inequality implies that, if the permittivity and permeability are real, a passive chiral medium cannot dissipate energy. While Eq. (\ref{eq:chiral_inequality}) may indirectly impose restrictions on the real part of the chirality coefficient $\Re \kappa/c$, if a dispersion model is assumed for the constitutive parameters, this quantity is not, in general, bounded by an analogous inequality. In fact, metamaterials can exhibit strong chirality, where $(\Re \kappa/c)^2 \gg \Re \varepsilon \Re \mu $, leading to intriguing effects such as negative refraction without negative permittivity and permeability \cite{zhou2009negative}. One can, however, use different arguments to derive constraints on $\Re \kappa/c$. Using the transformation of constitutive relations discussed in the introduction of this chapter, we can use Eq. (\ref{eq:eff_nonlocal}) to represent the local, chiral, constitutive parameters with an equivalent nonlocal permittivity, which, for the case $\mathbf{k}=0$, yields \cite{silveirinha2010comment}
\begin{equation}
    \varepsilon_{nl}(\omega, \mathbf{k}=0)   = \varepsilon(\omega)-\frac{[\kappa(\omega)/c]^2}{\mu(\omega)}.
    \label{eq:eff_nonlocal_chiral}
\end{equation}
Since, as noted above, a nonlocal permittivity satisfies causality for each fixed wavevector \cite{Lucarini2005}, it obeys KK relations. If, in addition, the medium is passive---more precisely, if the passivity condition above holds throughout the upper half of the complex-frequency plane, and $\mu$ has no zeros there---then $\omega\varepsilon_{nl}(\omega,0)$ is a Herglotz function, just like $\omega\varepsilon(\omega)$ for a local medium. Consequently, within transparency windows $\Re[\varepsilon-(\kappa/c)^2/\mu]$ must increase with frequency \cite{silveirinha2010comment}, and the sharper inequality, Eq.~(\ref{eq:derivative_bounds_2}), applies to the equivalent permittivity (\ref{eq:eff_nonlocal_chiral}), yielding a lower bound on the frequency dispersion of the medium:
\begin{equation}
    \frac{d\varepsilon_{nl}(\omega)}{d \omega}\ge 2\frac{[\kappa(\omega)/c]^2/\mu(\omega)-(\varepsilon(\omega)-\varepsilon_0)}{\omega},
    \label{eq:derivative_bound_nonlocal}
\end{equation}
which is a more stringent condition than the simple requirement of a positive derivative if $(\kappa/c)^2>\mu(\varepsilon-\varepsilon_0)$ (for $\mu>0$), i.e., in the strong chirality regime. (Other bandwidth-related bounds analogous to Eq. (\ref{eq:low_epsilon_bound}) can also be derived.) These results indicate that a transparent medium can exhibit strong chirality only if it is sufficiently frequency dispersive, with a lower bound on the frequency derivative directly proportional to $(\kappa/c)^2$. Since strong chirality can give rise to negative refraction, as discussed above, this provides yet another indication that nature does not permit dispersionless negative-refraction effects in linear time-invariant passive systems.

\subsection{Nonlinear susceptibilities} \label{sec:nonlinear}
In the context of perturbative nonlinear optical phenomena, the higher-order terms of the Volterra series in Eq. (\ref{eq:volterra}) define causal and square-integrable nonlinear relations between the applied fields and the resulting polarization response. In the frequency domain, the Fourier transform of the response functions correspond to the nonlinear susceptibilities, which depend on multiple frequency variables, such as $\chi^{(3)}(\omega_\sigma;\omega_1,\omega_2,\omega_3)$, where $\omega_\sigma=\omega_1+\omega_2+\omega_3$. Due to causality and square integrability (see Section \ref{sec:causality}), nonlinear susceptibilities always admit KK relations with respect to an individual frequency variable, if \emph{all others} can be kept fixed. This is, however, not always possible or meaningful for relevant nonlinear processes, where a pair or more of the frequency variables are mutually dependent and are varied at the same time, as in harmonic generation processes, for instance, where $\chi^{(3)}(3\omega;\omega,\omega,\omega)$. In such cases, additional conditions must be satisfied in order to derive KK relations with respect to $\omega$, as formalized by the so-called Scandolo's theorem \cite{Lucarini2005}. It turns out that KK relations do exist for susceptibilities describing harmonic generation and many other nonlinear processes, but notably not for the self-induced change in refractive index $\chi^{(3)}(\omega;\omega,\omega,-\omega)$; in this latter case, the susceptibility (and corresponding reflectivity \cite{peiponen2002dispersion}) is meromorphic in the complex frequency plane. We refer the interested readers to \cite{Lucarini2005,boyd2020nonlinear,koutserimpas2024time} for additional details. 

For all cases in which KK relations hold, one can also derive corresponding integral sum rules, analogous to the \emph{f}-sum rule (\ref{eq:f-sum rule}), encoding global constraints linking all-frequency integrals of the susceptibility to fundamental material properties, such as the electron density (as in the linear case) and the spatial derivatives of the electric potential experienced by the electrons \cite{Lucarini2005}. Given the existence of KK relations and sum rules, it is natural to ask whether similar bounding strategies can be applied to nonlinear susceptibilities using the approach described in Section \ref{sec:large_susceptibility} for the linear susceptibility. Nonlinear KK relations and sum rules have indeed been employed to derive useful trade-offs and relations, for example, in the rigorous derivation (under specific assumptions) of Miller’s Empirical Rule \cite{scandolo1995miller,koutserimpas2025exploring}, an important relationship indicating that the nonlinear susceptibility is proportional to the product of linear susceptibilities \cite{boyd2020nonlinear}.\footnote{While the off-resonant, electronic $\chi^{(3)}$ of transparent materials follows relatively simple scaling rules (wider bandgaps generally implying smaller nonlinear susceptibilities), the situation for $\chi^{(2)}$ and other even-order susceptibilities is considerably more complex, since in the bulk, local (electric-dipole) approximation they vanish in centrosymmetric media, and their magnitude therefore also depends on the degree of structural asymmetry \cite{khurgin2026are}.} However, these results fall short of establishing general fundamental limits, and the approach used to bound the linear susceptibility cannot be directly extended to the nonlinear case. The underlying reason for this issue lies in the difficulty of identifying suitable positivity constraints, even for passive media. Unlike the linear case, the imaginary part of nonlinear susceptibilities is not always directly related to absorption or gain. In some cases, such as for harmonic generation susceptibilities, this imaginary part simply represents a phase relationship between nonlinear polarization and applied field \cite{bloembergen1996nonlinear}. In the absence of a clear positivity constraint, one cannot derive bounds from sum rules whose integrands can assume both positive and negative values.

\begin{figure*}[tb]
\centering
\includegraphics[width=0.75\linewidth]{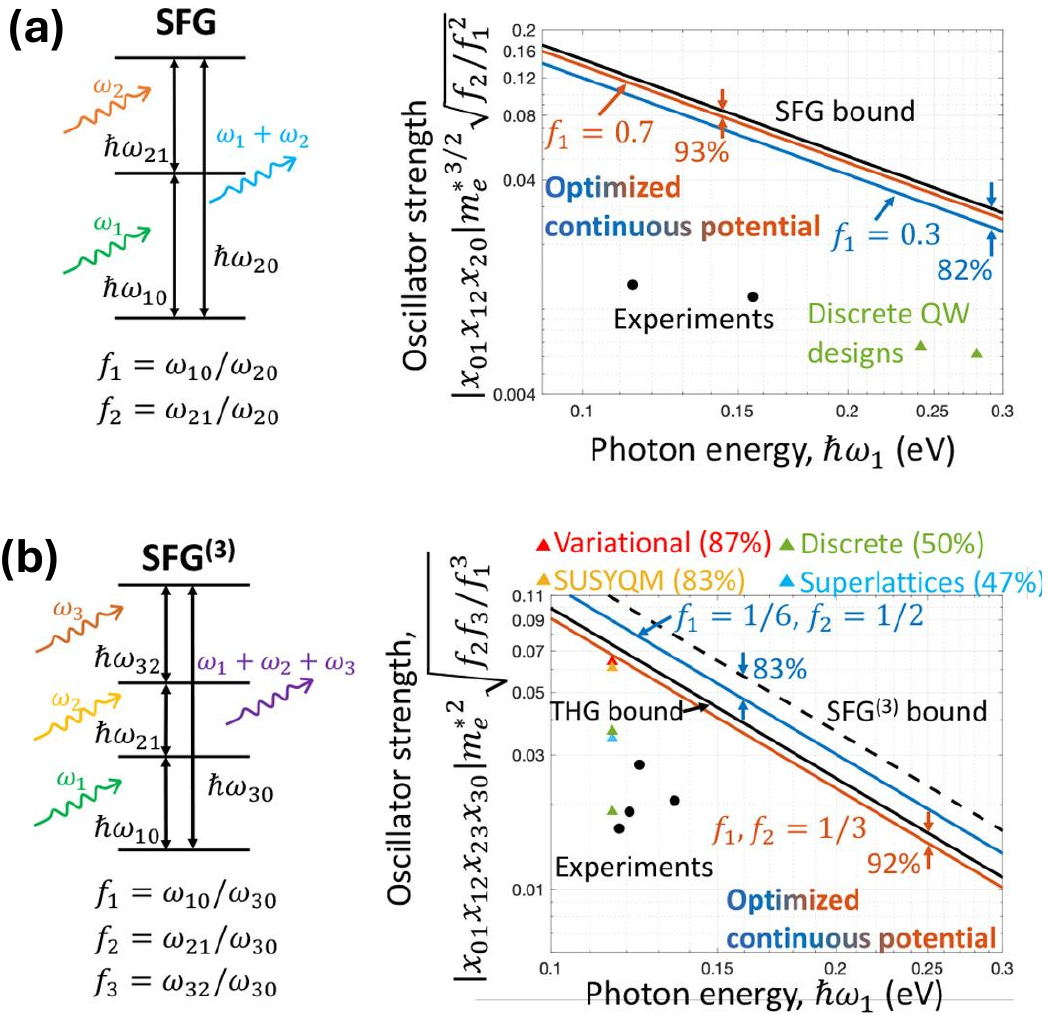}
\caption{\textbf{Bounds on resonant nonlinear susceptibilities.} (a) Schematic of a three-level system used for a second-order sum-frequency generation (SFG) process (left), together with the corresponding physical bound on the normalized oscillator strength of the doubly resonant second-order susceptibility, $\chi^{(2)}$, as a function of photon energy. The plot compares the physical bound with experimental measurements (black dots) and theoretical predictions (colored markers) from the literature, as well as near-optimal quantum-well structures with an inverse-designed continuous potential. (b) Similar to (a) but for a four-level system involved in a third-order SFG process, with susceptibility $\chi^{(3)}$. (Figure adapted from Ref.~\cite{li2025approaching}.)}
\label{fig:nonlinear}
\end{figure*}

This difficulty can be partially circumvented, as demonstrated in \cite{kuzyk2000physical,kuzyk2003erratum,li2025approaching}, by using standard quantum-mechanics-based models for the nonlinear susceptibilities, rather than a causality-based representation as in Eq. (\ref{eq:chi_causal_repres}), together with discrete sum rules on the dipole transition matrix elements, instead of integral sum rules on the imaginary part of the susceptibilities as in Eq. (\ref{eq:f-sum rule}) for the linear case\footnote{Rather surprisingly, TRK sum rules can be derived independently either from quantum-mechanics arguments, or from the $f$-sum rule (\ref{eq:f-sum rule}) by discretizing the integral through a dipole-moment-based oscillator representation of the linear susceptibility (see, e.g., \cite{li2025approaching,koutserimpas2025exploring}).}. These  discrete sum rules are known as the \textbf{generalized Thomas--Reiche--Kuhn (TRK) sum rules}, which encode constraints between transition dipole moments $x_{ij}$ and the energy/frequency spectra: 
\begin{equation}
\sum_{n=0}^{\infty}
\left[
\omega_n - \frac{1}{2}(\omega_p + \omega_q)
\right]
x_{pn} x_{nq}
=
\frac{\hbar}{2 m_e}\,\delta_{pq}
\end{equation}
where $\omega_i$ is the frequency of state $i$, $m_e$ is the bare electron mass (or conduction band effective mass in the context of quantum wells \cite{li2025approaching}), and, for each pair $(p,q)$ the sum is over all states indexed by $n$. By considering specific index pairs, one can derive bounds on the oscillator strengths (products of transition dipole moments) for the dominant terms of the susceptibility, leading to bounds on nonlinear susceptibilities for off-resonant \cite{kuzyk2000physical,kuzyk2003erratum} and doubly-resonant second-/third-order nonlinear responses \cite{li2025approaching} (Fig. \ref{fig:nonlinear}; see also Ref.~\cite{khurgin1989second} for early estimates of resonant intersubband $\chi^{(2)}$ in asymmetric quantum wells). This bounding procedure works because it automatically encodes positivity restrictions through the TRK sum rules (for example, considering the $p=q=0$ case, the resulting TRK relation yields a constraint with strictly positive coefficients, precluding unbounded transition dipole moments). However, unlike the causality-based bounding approach employed for the linear susceptibility in Section \ref{sec:large_susceptibility}, this framework does not inherently encode bandwidth or dispersion constraints. It would therefore be interesting in the future to explore whether these different approaches can be unified into a general framework to answer the following, and other, open questions:
\begin{openq}
        Can general bounds on nonlinear susceptibilities be established that incorporate dispersion and bandwidth constraints, without relying on specific quantum-mechanical assumptions, as in the linear case?
\end{openq}
However, it remains unclear whether one can derive a fully general, causality-based representation of nonlinear susceptibilities directly from KK relations, without any assumptions about the underlying microscopic interaction physics, as in Eq. (\ref{eq:chi_causal_repres})---and with oscillator-strength coefficients that can be constrained through positivity conditions, as in Eq. (\ref{eq:normalized_f_sum}).

Finally, we note that, for many practical problems, one is actually more interested in the strength and efficiency of a desired nonlinear effect, such as frequency conversion, rather than the strength of the nonlinear susceptibility itself. Even if the intrinsic nonlinearity of a material is weak, the overall nonlinear response can be enhanced by increasing the propagation length or by exploiting resonances and slow-light effects, which, however, are ultimately constrained by \textbf{delay-bandwidth limitations}, as discussed in Section \ref{sec:delay-bandwidth}. More broadly, we echo the argument made by Khurgin in \cite{khurgin2023nonlinear} that, in practice, given the fundamental difficulty of engineering nonlinear optical materials with orders-of-magnitude improvements in oscillator strengths (as implied by the considerations above), the most viable route to enhance nonlinear optical responses is to increase the interaction time between light and nonlinear material. This can be achieved either intrinsically, by operating near absorption regions and exciting real carriers with long characteristic lifetimes (introducing tradeoffs with dissipation and speed) or externally, through appropriate geometrical designs such as resonant or slow-light structures, as mentioned above (introducing tradeoffs with device length and bandwidth).

\subsection{Refractive-index modulation}
The refractive index of a material can be changed not only via all-optical nonlinearities, but also through many other interactions such as electro-optical, acousto-optical, magneto-optical, and thermal effects. Regardless of the underlying physical mechanism, KK relations can be formulated not only for the complex refractive index itself, but also for a \emph{change} in refractive index $\Delta n$ (due to the linearity of these relations), which explains how a change in absorption, $\Im[\Delta n]$, must be accompanied by a corresponding change in $\Re[\Delta n]$, and vice versa. One might therefore hope to combine this property with appropriate sum rules to derive fundamental limits on $\Re[\Delta n]$; however, as in the previous section, this approach is hindered by the difficulty of establishing suitable positivity constraints, since a change in absorption, $\Im[\Delta n]$, can assume either positive or negative values. In fact, if the modulation does not change the electron density, the sum rule $\int_0^\infty \omega\, \Im[n(\omega)]\, d\omega = \pi\omega_p^2/4$ (the refractive-index counterpart of the \emph{f}-sum rule, Eq.~(\ref{eq:f-sum rule})) implies $\int_0^\infty \omega\, \Im[\Delta n(\omega)]\, d\omega = 0$, so that $\Im[\Delta n]$, unless it vanishes identically, \emph{must} change sign.

\begin{figure*}[tb]
\centering
\includegraphics[width=0.8\linewidth]{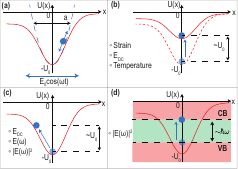}
\caption{\textbf{Mechanisms and energy requirements for large refractive-index modulation.} (a) Schematic of a typical anharmonic potential $U(x)$ of depth $U_0$. The refractive index depends on its local curvature, averaged over all electrons, $\langle d^2U(x)/dx^2 \rangle_N$, and on the electron density $N$ \cite{khurgin2024energy}. (b) The overall curvature can be changed through lattice distortion induced by temperature, strain, or an applied electric field (or through a phase transition). (c) The curvature experienced by the electrons can be altered by moving them upward within the potential, through an applied electric field, while remaining in the same energy band (virtual transition). (d) Electrons can be promoted to a higher energy band (from valence band to conduction band), saturating absorption through band filling and inducing a refractive-index change via KK relations. In all cases, achieving a large refractive-index modulation (50\% or more) requires an energy comparable to $U_0$ (not necessarily dissipated). (Figure adapted from Ref.~\cite{khurgin2024energy}.)}
\label{fig:energy_n}
\end{figure*}

Other important questions in this context concern the energy or power required to impart a desired modulation, as well as the ultimate speed at which such modulation can occur. Quantitative answers to these questions depend strongly on the specific class of materials, structures, and modulation mechanisms employed. One of the few general, albeit qualitative, arguments that can be made is about the minimum energy required to induce a large change in refractive index. While rigorous bounds are not known, physically intuitive orders-of-magnitude considerations can still be useful. Following Khurgin \cite{khurgin2024energy}, one can argue that a large change in refractive index (50\% or more) always requires an energy comparable to the depth $U_0$ of the anharmonic electronic potential experienced by the electrons (Fig. \ref{fig:energy_n}).
$U_0$ is comparable to the binding energy of each bond (but choosing the cohesive energy, or the fundamental bandgap energy, or the Penn bandgap energy (Sec.~\ref{sec:large_susceptibility}), would yield similar orders-of-magnitude estimates \cite{khurgin2024energy}). This energy varies by less than an order of magnitude, within the range 1-10eV, across a wide range of solid-state materials of interest. Then, the total energy density needed to impart a large change of refractive index is on the order of $W\sim N U_0$, where $N$ is the density of valence electrons, which also does not vary significantly across materials and falls in the range $10^{22}-10^{23}~\mathrm{cm}^{-3}$. Therefore, one can estimate that, regardless of the material or modulation scheme used, the energy density required to significantly change the refractive index is on the order of: $W\sim 10^{3}-10^{5}~\mathrm{J/cm}^{3}$. Note that, depending on the considered mechanism, this energy does not necessarily have to be dissipated within the material to induce the desired refractive-index change (dissipation is unavoidable, however, when the modulation relies on real electronic transitions, i.e., light absorption). We refer the reader to \cite{khurgin2024energy} for a detailed discussion supporting the validity of this qualitative argument for different modulation schemes and materials. The same energy scale also sets a characteristic time scale for modulation: through the energy--time relation $\Delta E\,\Delta\tau \gtrsim \hbar$, it corresponds to $\hbar/U_0 \approx 0.07$--$0.7$~fs for $U_0$ in the range 1--10~eV, the same order as the sub-femtosecond interband dynamics observed in silicon~\cite{schultze2014attosecond}. This is a dimensional estimate rather than a bound, since $U_0$ has not been identified with the energy spread that enters a rigorous energy--time (quantum speed limit) relation for the modulation dynamics. 

An interesting question the reader might ask is the following: Given a material body occupying a fixed volume, is more energy required to change the refractive index by altering the material properties themselves, as described above, or by physically displacing the material, through mechanical translation or rotation, so that the same volume is now occupied by air, with a much lower refractive index? An order-of-magnitude estimate can again be made. The energy (work) needed to move an initially-at-rest material body of mass $m$ by a distance $d$ in a time $t$ at constant acceleration is: $U_m=2md^2/t^2$ (twice if the object also needs to be stopped at the new location). Since the room-temperature mass density of most solid-state materials is approximately in the range $\sim 10^{-3}-10^{-2}~\mathrm{kg/cm}^{3}$, the energy density required to move a material body over a distance of, for example, tens of micrometers (the typical size of micromirrors \cite{ren2015tailoring}) can be below $1~\mathrm{J/cm}^{3}$ for timescales $t$ longer than a few microseconds (corresponding to modulation rates of hundreds of kHz), and below the previously estimated energy density $W$ for timescales longer than tens of nanoseconds (tens of MHz modulation rates). In liquid-crystal systems, the required energy densities are even lower, since the moving elements are much smaller. These orders-of-magnitude estimates show that, as argued in \cite{khurgin2024energy}, for relatively slow modulation speeds, mechanically displacing a material body to modify the optical properties within a fixed volume can fundamentally require far less energy than any scheme based on directly altering the material’s intrinsic properties.

\section{Bounds on wavefront shaping}
\label{sec:WaveShaping}
In this chapter, we consider limits to control over electromagnetic-wave propagation in scenarios where the wave sources can be freely varied but the scatterers themselves are fixed. Such scenarios comprise \textbf{wavefront shaping}. ``Wavefront'' should not be interpreted literally, instead referring generally to wave fields, to be shaped for objectives ranging from free-space focusing to trajectory control in complex media.

We describe bounds for free-space focusing and superoscillations (Sec.~\ref{sec:FSFocus}), focusing in/through complex media (Sec.~\ref{sec:CMFocus}), and various other wavefront shaping metrics (Sec.~\ref{sec:OtherObjectives}). For each, bounds typically utilize the same four steps:
\begin{enumerate}
    \item A linear matrix/operator relation between the source degrees of freedom, $s$, and the output field/ports of interest. Generically, the ``output'' $\psi = \AA s$ for controllable sources $s$ and a matrix/operator $\AA$, which might be a (known) Green's function or scattering matrix.
    \item An objective $f$ that is a quadratic function of the output field,
    \begin{align}
        f(\psi) = \psi^{\dagger} \QQ \psi = s^{\dagger} \AA^{\dagger} \QQ \AA s,
        \label{eq:WPObj}
    \end{align}
    where $\QQ$ may encode field intensity at point, momentum transfer through a surface, time delay in a scattering process, etc. 
    \item A source-normalization constraint, often based on fixing supplied energy of power, such as a ``unit-norm'' constraint:
    \begin{align}
        s^{\dagger} s = 1.
    \end{align}
    \item The Rayleigh-quotient result that the maximum or minimum of a quadratic objective normalized by a unit-norm constraint is given by the largest or smallest eigenvalue of the relevant operator. For example,
    \begin{align}
        \operatorname{max}_{s} \frac{s^{\dagger} \AA^{\dagger} \QQ \AA s}{s^{\dagger} s} = \lambda_{\rm max}(\AA^{\dagger} \QQ \AA).
        \label{eq:OptSrc}
    \end{align}
\end{enumerate}
This Rayleigh-quotient-based approach (\secref{RQs}) to quadratic wavefront shaping objectives was developed in Ref.~\cite{Mazilu2011-fg} under the name ``optical eigenmodes,'' though we advise against this terminology given the confusion and overloading of ``modes'' in the photonics literature~\cite{Miller2019-hl}. We propose calling this approach the \textbf{extremal-eigenvalue approach}.

Hence, bounds for many wavefront shaping problems reduce to computations of extremal eigenvalues and their corresponding eigenvectors. In the many cases we consider below, we will see that they differ primarily by the relevant forms of the operators $\AA$ and $\QQ$, as well as the relevant basis for the source coefficient vector $s$. In some cases (e.g. superresolution) there are additional steps required to enable the four steps above. A useful simplification occurs if the operator $\AA^{\dagger} \QQ \AA$ is \emph{rank one}, with only one nonzero eigenvalue, $\lambda$ and corresponding eigenvector $v$. Then it can be represented in outer-product form:
\begin{align}
    \AA^{\dagger} \QQ \AA = \lambda v v^{\dagger},
\end{align}
in which case the optimal wavefront shaping objective is proportional to $\lambda$, and the optimal source distribution is proportional to $v$. The individual elements of extremal eigenvectors can take continuous values constrained only by a total power/energy normalization; in some cases, experimental conditions might be more constrained (e.g., binary amplitude variables in a spatial light modulator). Then, one typically uses local-optimization techniques to find high-performance source distributions~\cite{Bender2019-tg,Cao2023-bh}, and the extremal-eigenvalue approach can offer complementary global bounds.
  
\subsection{Free-space focusing}
\label{sec:FSFocus}
Free-space focusing is the canonical objective for control of electromagnetic wave propagation. We consider four variants of the question: to what extent can electromagnetic waves be focused? A long mathematical literature characterizes this through uncertainty principles relating a wave's spatial extent to the support of its plane-wave (Fourier) spectrum, beginning with the Heisenberg--Pauli--Weyl inequality and refined into many forms (see~\cite{Folland1997} for a survey). A related framework introduced by Slepian, Pollak, and Landau in a foundational series of papers~\cite{Slepian1961,Landau1961,Landau1962,Slepian1964} identifies the prolate spheroidal wavefunctions (PSWFs) as the eigenfunctions of the joint spatial-and-spectral concentration problem. The near-bimodal distribution of these eigenvalues was subsequently used to count the ``degrees of freedom'' of a band-limited image~\cite{ToraldoDiFrancia1969,Miller2019-hl}. This thread provides perhaps the most rigorous mathematical foundation for the ``diffraction limit,'' though it has been developed primarily for scalar signals in one dimension. Our complementary perspective targets specific operational objectives---focal-point intensity or beam width---for \emph{vector} wave fields generated by sources within a constrained domain. The maximum-intensity case underlies the eigenvalue framework of the first three subsections; the width case, treated in the fourth subsection, leads to the notion of super-oscillations.
\subsubsection{From any aperture (near or far field)}
Consider any surface $S$, the ``aperture,'' on which sources $s$ can be placed to create the largest field intensity at a point $\xv_0$. Then the field at $\xv_0$ produced by the sources $s$ are given by a convolution of the free-space Green's function $\Gamma_0$ with the sources: $\psi(\xv_0) = \int_S \Gamma_0(\xv_0,\xv_S) s(\xv_S) \,{\rm d}\xv_S$, which we can write in more compact notation as
\begin{align}
    \psi_0 = \Gamma_{0S} s.
\end{align}
Hence the linear operator $\AA$ of the extremal-eigenvalue discussion above is identically $\Gamma_{0S}$. The objective is the field intensity at $\xv_0$, 
\begin{align}
    f(\psi) = \psi_0^{\dagger} \psi_0,
\end{align}
which implies that the corresponding quadratic operator $\QQ$ is simply the identity operator $\II$ (at $\xv_0$). If the source vector $s$ is normalized such that $s^{\dagger} s = 1$ (subject to subtle physical considerations, discussed below), then the maximum intensity at $\xv_0$ is given by the largest eigenvalue of $\AA^{\dagger} \QQ \AA$, i.e.,
\begin{align}
    \operatorname{max}\left(\|\psi(\xv_0)\|^2\right) = \lambda_{\rm max}\left(\Gamma_{0S}^{\dagger} \Gamma_{0S}\right),
    \label{eq:MaxPsi0}
\end{align}
and the optimal source distribution is given by the corresponding eigenvector. This is a simple computational bound, and we provide reciprocity / time-reversal intuition for it in Sec.~\ref{sec:CMFocus} below. Beyond the largest eigenvalue, the full spectrum of $\Gamma_{0S}^{\dagger}\Gamma_{0S}$ characterizes the joint spatial--spectral concentration of the aperture-to-target map; reading off the knee of this spectrum gives the ``degrees of freedom'' of the map, generalizing Slepian's 1D scalar result to vector waves~\cite{Miller2019-hl}.

\subsubsection{From $\leq 2\pi$ solid angle in the far field}
Next, we specialize \eqref{eq:MaxPsi0} to the scenario in which one can excite plane waves propagating with wavevector $\kv$ coming from within a $2\pi$ solid angle, i.e., one side of the $x$-$y$ plane: $\hat{\zv}\cdot \kv = k_z\geq 0$. The plane waves can propagate within a solid angle defined by a numerical aperture $\sin\theta$, which restricts the parallel wavenumbers ($\kp = k_x \hat{\xv} + k_y \hat{\yv}$) of the plane waves, $|\kp| \leq k_0\sin\theta$, where $k_0=\omega/c$. For a given $\kp$, the two transverse polarizations are denoted $\ep{\kp}$ for $p=1,2$. Then, the total electric field is the linear combination of each plane wave's electric field multiplied by coefficients $c_p(\kp)$,
\begin{align}
    \Ev = \sum_{p} \int \dkk c_p(\kp) \ep{\kp} e^{i\kv\cdot\xv},
    \label{eq:PWEv}
\end{align}
where the range of the integral corresponds to the allowed solid angle. What coefficients produce the largest field intensity at $\xv_0$ (taken as the origin)? We employ an approach similar to that of \citeasnoun{Dias2020} to identify how the maximum intensity and optimal field profiles vary as a function of allowed solid angle.

How to normalize the plane-wave amplitudes? One possibility is an ``amplitude normalization,'' in which the sums/integrals of the squared amplitudes are normalized to 1, treating each plane-wave excitation as equally ``costly:''
\begin{align}
    \sum_p \int \dkk |c_p(\kp)|^2 &= 1 && \text{[amplitude normalization]},
\end{align}
which we can write in simplified vector notation $\vect{c}^{\dagger} \vect{c} = 1$. Another possibility is ``power normalization,'' in which the power flowing through the $x$-$y$ plane is the metric cost. For a superposition of plane waves as in \eqref{eq:PWEv}, the $z$-directed power $P$ is
\begin{align}
    P = \frac{2\pi^2}{Z_0} \sum_p \int \dkk \frac{k_z}{k} |c_p(\kp)|^2.
\end{align}
One can define new coefficients $\alpha_p(\kp) = \left[ \sqrt{2\pi^2 k_z / \left(Z_0 k\right)} \right] c_p(\kp)$, with which the power constraint simplifies to 
\begin{align}
    \sum_p \int \dkk |\alpha_p(\kp)|^2 &= 1. && \text{[power normalization]}
\end{align}
The two normalization constraints are identical in the paraxial limit, but otherwise yield different maximum field intensities, with the largest gap for $2\pi$ excitation solid angle.

For a target polarization $\qhat$, the focal-point intensity $|\qhat^* \cdot \Ev(0)|^2 = |\vect{v}_1^\dagger \vect{c}|^2$ is a rank-one quadratic form in $\vect{c}$, with $\vect{v}_1$ the projection of the plane-wave polarization basis onto $\qhat$. The optimal coefficients for the two cases follow immediately:
\begin{align}
    c_p^{\rm (opt)}(\kp) &= \left[\qhat^{*} \cdot \ep{\kp} \right]^{*}, \label{eq:cOpt} \\
    \alpha_p^{\rm (opt)}(\kp) &= \sqrt{\frac{k}{k_z}} \left[\qhat^{*} \cdot \ep{\kp} \right]^{*}, \label{eq:alphaOpt}
\end{align}
for amplitude and power normalization, respectively.

Given the optimal coefficients of \eqrefs{eq:cOpt}{eq:alphaOpt}, the optimal focal-point intensities are:
\begin{align}
    \frac{\varepsilon_0 |\Ev(0)|^2}{P_{\rm in}} &\leq \frac{k^2}{2\pi c} \left[\frac{4}{3} - \cos\theta - \frac{\cos^3\theta}{3} \right] && \text{[power normalization]} \label{eq:OptPowerNorm} \\
    \frac{\varepsilon_0 |\Ev(0)|^2}{P_{\rm amp}} &\leq \frac{k^2}{2\pi c}\sin^2\theta \left[1 - \frac{\sin^2\theta}{4}\right] && \text{[amplitude normalization]} \label{eq:OptAmpNorm}
\end{align}
where we included factors of $\varepsilon_0$ to simplify the units. (For power normalization at $2\pi$ solid angle, all possible polarizations at the origin have the same optimal intensity; with amplitude normalization the $z$-polarized maximum is $2/3$ of the in-plane one.) In the paraxial limit, both expressions reduce to $k^2 \theta^2/(2\pi c)$. At $2\pi$ solid angle ($\theta \to \pi/2$), the bracket in \eqref{eq:OptPowerNorm} reduces to $4/3$, giving the energy-density bound
\begin{align}
    \frac{U(\xv_0)}{P_{\rm in}} \leq \frac{k^2}{6\pi c} \qquad (\Omega = 2\pi),
    \label{eq:TwoPiBound}
\end{align}
in terms of the energy density $U(\xv_0) = \tfrac{1}{4}\varepsilon_0|\Ev(\xv_0)|^2$. The power-normalized maximum is $16/9\approx1.78$ times the amplitude-normalized one; because the latter is normalized by $P_{\rm amp}\equiv\frac{2\pi^2}{Z_0}\sum_p\int|c_p|^2$ rather than by the physical power $P_{\rm in}$, the amplitude-optimal beam measured per unit physical power delivers $\tfrac{135}{128}\,k^2/(2\pi c)$, 1.26X less than the optimum. The optimum exceeds the diffraction-limited Gaussian reference beam of \citeasnoun{Dias2020} by 1.43X. For intuition, a naive diffraction-limited estimate, in which $P_{\rm in}$ is concentrated into a spot of area $(\lambda/2)^2$ with plane-wave-like fields (time-averaged electric energy density $I/2c$ at intensity $I$), gives $U\approx 2P_{\rm in}/(\lambda^2 c)=k^2P_{\rm in}/(2\pi^2 c)$, within a factor of $\pi/3\approx1.05$ of \eqref{eq:TwoPiBound}. The near-agreement depends on the assumed spot area (a disk of diameter $\lambda/2$ would instead exceed the bound by about 20\%), whereas \eqref{eq:TwoPiBound} fixes the prefactor rigorously. 

\begin{figure*}[tb]
\centering
\includegraphics[width=1\linewidth]{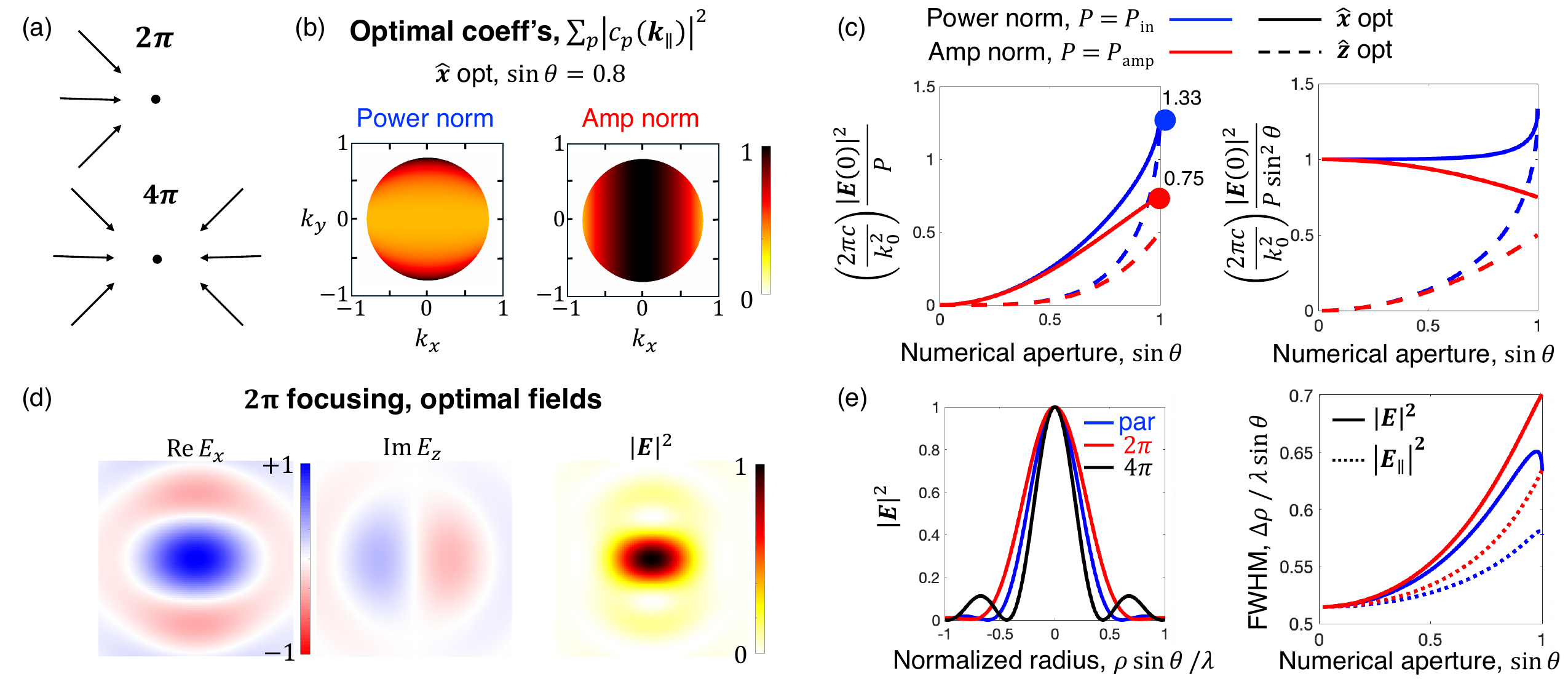}
\caption{\textbf{Plane-wave focusing.} (a) Schematic of the $2\pi$ and $4\pi$ aperture limits. (b) Optimal focusing coefficients, as a function of wavenumbers $k_x$ and $k_y$, for $\sin\theta = 0.8$. When normalizing by amplitude, the smaller $k_x$ values have the largest relative contribution to $|E_x|^2$; when normalizing by power, the large-$|\kp|$ waves with polarization aligned to $\hat{\xv}$ (large $|k_y|$) carry little $z$-directed power per unit field and hence are the coefficients to maximize. (c) Maximum value of field intensity at the origin, for maximum in-plane intensity $|\hat{\xv}\cdot \Ev|^2$ (solid) versus maximum out-of-plane intensity, $|\hat{\zv}\cdot\Ev|^2$ (dashed); on the right, normalized by $\sin^2\theta$. Intensities are per $P=P_{\rm in}$ for power normalization (blue) and per $P=P_{\rm amp}$ of \eqref{eq:OptAmpNorm} for amplitude normalization (red). (d) Optimal focal-plane fields ($|x|,|y|\le\lambda$, $\hat{\xv}$ horizontal) for $2\pi$ focusing, for maximum $|\hat{\xv}\cdot\Ev|^2$; $\Re E_x$ and $\Im E_z$ share one color scale. There is an unavoidable out-of-plane $E_z\propto\cos\phi$ field that pushes the first minimum along the polarization axis to a larger radius. (e) A comparison of the intensity distributions as a function of radius for paraxial (blue), $2\pi$ aperture (red, along the polarization axis), and $4\pi$ aperture (black, perpendicular to the polarization). Along the polarization axis the relative beam size increases from paraxial to $2\pi$ (right: full width at half maximum, FWHM, along the polarization axis for power (blue) and amplitude (red) normalization, for $|\Ev|^2$ (solid) and $|\Ev_\parallel|^2$ (dotted)); perpendicular to the polarization it instead decreases, and at $2\pi$ coincides with the $4\pi$ profile.}
\label{fig:PWFocusing}
\end{figure*}

The $\sqrt{k/k_z}$ factor in \eqref{eq:alphaOpt} concentrates the power-normalized coefficients near the largest $|\kp|$ with aligned polarizations [\figref{PWFocusing}(b)]. \Figref{PWFocusing}(c) compares the resulting intensities versus solid angle: globally optimal in-plane polarizations (solid lines) versus $z$-polarized intensities (dashed), the latter falling as $\theta^4$ in the paraxial limit and, for power normalization, reaching the in-plane optimum only at full $2\pi$. Along the polarization axis, the focal-spot full widths at half maximum (FWHMs) [\figref{PWFocusing}(e)] increase with aperture in normalized-radius units (decrease in absolute units); power-normalized fields are slightly tighter than amplitude-normalized, and their in-plane components $|\Ev_\parallel|^2$ (dotted) tighter still. Perpendicular to the polarization the trend reverses: the normalized FWHM of the power-normalized spot falls from 0.51 to 0.40 at $2\pi$, so high-NA optimal spots are elliptical, elongated along the polarization.

Computing the optimal fields across the plane of interest (beyond just the focal point) leads to complicated integrals that may not have analytical solutions for all numerical apertures. Closed-form expressions in the paraxial and $2\pi$-solid-angle limits (and for amplitude normalization at any NA) are derived, and verified numerically, in the companion repository~\cite{CompanionRepo}.
In the paraxial limit, the power- and amplitude-normalized optimal coefficients are identical, and the resulting field can be entirely polarized in-plane (say, the $\hat{\xv}$ direction), leading to the well-known Airy pattern:
\begin{align}
    \left|\Ev(\xp)\right|^2_{\rm paraxial} \sim \left[ \frac{J_1(k_0 \rho \sin\theta)}{k_0 \rho \sin\theta} \right]^2.
    \label{eq:E2Paraxial}
\end{align}
where $J_n$ is the (cylindrical) Bessel function of the first kind of order $n$.

Strikingly, there is an equally simple expression for the field intensity in the $2\pi$ solid angle limit. Extensive cancellations in products of fractional Bessel functions lead to a field intensity proportional to
\begin{align}
    \left|\Ev(\rho\hat{\xv})\right|^2_{\Omega = 2\pi} \sim 4 \left( \frac{j_1(k_0 \rho)}{k_0 \rho} \right)^2 + \left( \frac{J_2(k_0 \rho)}{k_0 \rho} \right)^2,
    \label{eq:E2TwoPi}
\end{align}
along the polarization ($x$) axis, where $j_n$ is the spherical Bessel function of the first kind of order $n$, the first term comes from the in-plane ($x$-polarized) field, while the second comes from an unavoidable $z$-directed component $E_z\propto\cos\phi\,J_2(k_0\rho)/(k_0\rho)$ that is zero at the origin and along the $y$ axis, but nonzero elsewhere. The spot is elliptical: perpendicular to the polarization, $|\Ev(\rho\hat{\yv})|^2 \sim \bigl[j_0(k_0\rho)-j_1(k_0\rho)/(k_0\rho)\bigr]^2$, the same profile as the $4\pi$ optimum derived below. (For equal plane-wave amplitudes, the in-plane focal-plane field of the $2\pi$ optimum is exactly half that of the $4\pi$ optimum.) This appears to be the first derivation of the field distribution of an optimally focused field excited from a $2\pi$ solid angle. This spherical Bessel function does not lead to a tighter beam, though it may seem somewhat surprising that a \emph{spherical} Bessel function arises for the (\emph{two-dimensional}) in-plane field intensity distribution. But it ideally foreshadows our analysis of the $4\pi$ solid angle case, in which spherical waves are the natural basis, which we tackle next.

\subsubsection{From $4\pi$ solid angle in the far field}
\begin{figure*}[tb]
\centering
\includegraphics[width=1\linewidth]{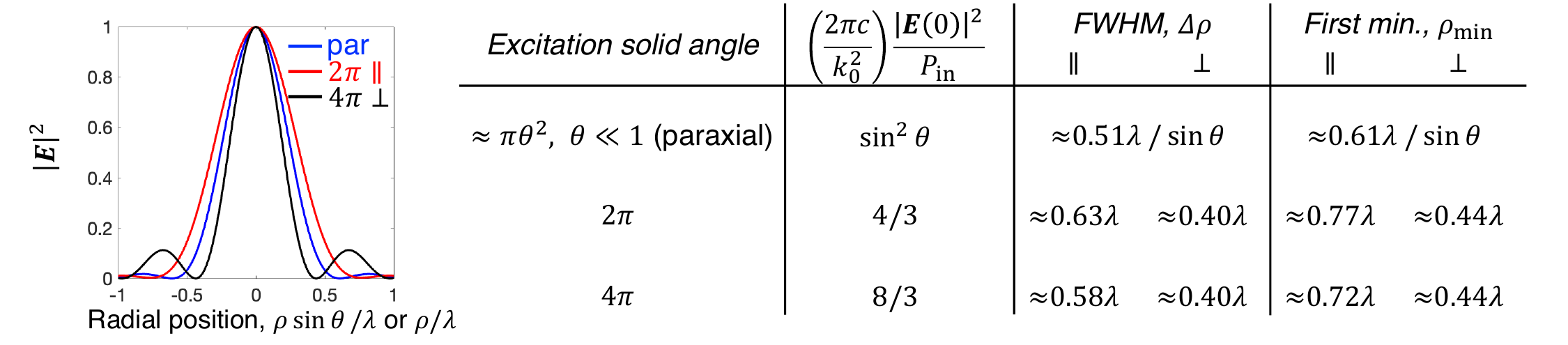}
\caption{\textbf{Comparison of optimal focusing fields}. Radial fields in the paraxial, $2\pi$, and $4\pi$ apertures. Along the polarization axis, maximally intense fields broaden a bit, in normalized-radius units, as their numerical aperture increases from paraxial to $2\pi$; the $4\pi$ profile shown (perpendicular to the polarization) is tighter, but the $2\pi$ profile perpendicular to its polarization is identical to it. The $4\pi$ limit doubles the intensity and removes the axial elongation of the $2\pi$ spot. In the table, $\parallel$ and $\perp$ denote directions along and perpendicular to the focal-point polarization; along the polarization the $2\pi$ intensity has a first minimum ($I/I_0\approx3\times10^{-3}$) at $0.77\lambda$ rather than a zero.}
\label{fig:PWFocusingFourPi}
\end{figure*}

For $4\pi$ (``full-sphere'') excitation, the maximum focal-point intensity doubles from the $2\pi$ maximum: combining optimal single-sided excitations from each hemisphere doubles the incident power, doubles the field amplitude, quadruples the intensity, and therefore doubles the relative intensity~\cite{Dias2020}. This leads to the doubling (from $4/3$ to $8/3$) of the normalized intensity maximum in Fig.~\ref{fig:PWFocusingFourPi}. This result was first derived by Bassett~\cite{Bassett1986} in a spherical-wave basis, which is sufficiently natural and insightful that we include its key points here as well. Finally, we connect to the time-reversal discussion of Sec.~\ref{sec:CMFocus}.

The fields within any sphere that has sources only to its exterior can be decomposed into the ``regular'' (non-diverging) vector spherical waves, centered at $\xv_0$:
\begin{align}
    \psi(\xv) = \sum_{\ell m\sigma} c_{\ell m\sigma} \mathbb{V}^{\rm (reg)}_{\ell m\sigma}(\xv),
\end{align}
where $\ell$ and $m$ are the spherical-harmonic integers and $\sigma$ denotes TE- or TM-polarized waves. The $\mathbb{V}^{\rm (reg)}_{\ell m \sigma}(\xv)$ comprise both the electric and magnetic vector waves~\cite{Jackson1999,Tsang2000}, while the coefficients $c_{\ell m \sigma}$ are the degrees of freedom to be optimized. One can follow the extremal-eigenvalue prescription from above: compute the quadratic objective (intensity at the origin) relative to a norm constraint (unity incoming power) and find the largest eigenvalue and its corresponding eigenvector. The spherical-wave description dramatically simplifies this analysis. Of the full set of vector spherical waves, only the electric-dipole waves with
\begin{align}
    \Ev(\xv) \sim \Nv_{1,0}(\xv), \Nv_{1,-1}(\xv), \textrm{ or } \Nv_{1,1}(\xv)
\end{align}
have nonzero amplitude at the origin, where the normalized vector spherical waves are $\Nv_{\ell,m}(\xv) = \frac{1}{k\sqrt{\ell(\ell+1)}}\nabla \times \bigl[\nabla \times \bigl( \xv z_{\ell}^{\textrm{reg}/+/-}(kr)Y_{\ell m}(\theta,\phi)\bigr)\bigr]$, $Y_{\ell m}$ are the spherical harmonics, and $z_{\ell}^{\textrm{reg}/+/-}$ corresponds to the three spherical Bessel functions $j_{\ell}, h_{\ell}^{(1)}, h_{\ell}^{(2)}$, respectively. These three channels are mutually orthogonal in both field amplitude and power flow (a standard property of multipole expansions), so the bound on intensity-per-power follows from any single channel; by symmetry all three give the same value. Take $\Nv_{1,0}$, whose amplitude at the origin is $(2/3)\sqrt{3/(8\pi)}$ along $\hat{\zv}$.

If one normalizes the intensity to the incoming part of the regular vector spherical wave (corresponding to the second Hankel function), then $P_{\rm in} = \frac{1}{8Z_0 k^2}$, where $Z_0=\sqrt{\mu_0/\varepsilon_0}$ is the impedance of free space. For the energy density $U(\xv) = \frac{1}{4}\varepsilon_0 |\Ev(\xv_0)|^2$ at a single point, relative to ``incoming power'' $P_{\rm in}$, the bound for energy density relative to incoming power is
\begin{align}
    \frac{U(\xv_0)}{P_{\rm in}} \leq \frac{k^2}{3\pi c} \qquad (\Omega = 4\pi),
    \label{eq:FourPiBound}
\end{align}
exactly double the $2\pi$-excitation optimum of \eqref{eq:TwoPiBound}~\footnote{\citeasnoun{Dias2020} reports half this value and argues that Bassett's bound is too large by a factor of two. The discrepancy is an artifact of an inconsistent prefactor in their energy-density formula rather than a physics correction. They define complex amplitudes via $\Ev_{\rm phys}(\xv,t) = 2\Re[\Ev(\xv) e^{-i\omega t}]$, so that $\langle \Ev_{\rm phys}^2\rangle = 2|\Ev|^2$ and the Gaussian time-averaged energy density is $U = (\epsilon_h|\Ev|^2 + |\Hv|^2)/(4\pi)$. The expression they use, $(\epsilon_h|\Ev|^2 + |\Hv|^2)/(8\pi)$, has a prefactor smaller by a factor of two and therefore equals $U/2$. (The factor of $8\pi$ appears in the \emph{instantaneous} energy density, in Gaussian units.) Their plane-wave incident-power expression, by contrast, does correctly evaluate to the physical time-averaged Poynting flux under the same convention (as can be checked by direct substitution of a single plane wave). Multiplying their reported bound by two recovers \eqref{eq:FourPiBound}. This also explains why the expression in \citeasnoun{Dias2020} for $2\pi$ excitation is smaller than our \eqref{eq:TwoPiBound}, by the same factor of two.}.\footnote{Although nothing in the derivation assumes that sources lie in the far field of $\xv_0$, \eqref{eq:FourPiBound} relies on the spherical-wave power normalization and does not apply to near-field sources, whose evanescent components can contribute energy density without carrying power; near-field bounds require a plane-wave or dipole-Green's-function representation instead.} This is \emph{not} a bound for the energy density with respect to \emph{work done by the source currents}, which can be exactly zero: One can set up currents that produce self-sustaining fields without any outgoing energy flow (``nonradiating currents''~\cite{Devaney1973}), in which case the energy-density-to-work-done ratio would diverge. But this requires driving currents that absorb and reuse power.

The scaling of this bound with $k^2/c$ can be understood intuitively: energy density is in the numerator, while the incoming power will be proportional to energy density multiplied by cross-section (a regular electric dipole wave occupies an effective area $\sim \lambda^2$ at its focus) and the speed of light $c$, giving $\sim 1/(\lambda^2 c) \sim k^2/c$.

One can additionally identify the spot size of the focused field that achieves the intensity bound of \eqref{eq:FourPiBound}. The vector spherical wave in the $x$-$y$ plane is given by
\begin{align}
    \Nv_{1,0}(r,\phi,\theta=\pi/2) = \sqrt{\frac{3}{8\pi}} \left( \frac{j_1(kr) - \sin(kr)}{kr}\right) \hat{\vect{e}}_{\theta} = -\sqrt{\frac{3}{8\pi}} \left( \frac{j_1(kr) - \sin(kr)}{kr}\right) \hat{\zv},
    \label{eq:N10xy}
\end{align}
where the second form uses $\hat{\vect{e}}_\theta = -\hat\zv$ in the equatorial plane. The $4\pi$-optimal field is purely $z$-polarized throughout the $x$-$y$ plane, which is impossible for $\leq 2\pi$ excitations: a one-sided aperture cannot constructively enhance the out-of-plane polarization beyond the in-plane optimum, whereas counter-propagating $4\pi$ waves can cancel all in-plane components. The resulting profile is significantly tighter than \eqrefs{eq:E2Paraxial}{eq:E2TwoPi}, although it coincides with the $2\pi$ profile perpendicular to the polarization. For an $x$-polarized focus with optical axis $z$, the benefit of $4\pi$ excitation is along the optical axis (FWHM $0.89\lambda\to0.40\lambda$) and, by cancelling $E_z$, along the polarization axis ($0.63\lambda\to0.58\lambda$). By symmetry the maximum intensities along $\hat{\xv}$, $\hat{\yv}$, and $\hat{\zv}$ are equal, but the in-plane spatial profile is asymmetric: a $z$-optimized field is tightest in the $x$-$y$ plane, an $x$-optimized field tightest in the $y$-$z$ plane, etc. The intensity of the $4\pi$-optimal excitation, in the plane perpendicular to the focal-point polarization, is given by
\begin{align}
    \left| \Ev(\xp) \right|^2_{\Omega=4\pi} \sim \left( \frac{j_1(k\rho) - \sin(k\rho)}{k\rho}\right)^2.
    \label{eq:E2optFourPi}
\end{align}
A comparison with the paraxial \eqref{eq:E2Paraxial} and $2\pi$ \eqref{eq:E2TwoPi} distributions is shown in \figref{PWFocusingFourPi}, alongside the FWHMs and first-minimum radii of each.

The field $\Nv_{1,m}^{\rm (reg)}(\xv)$ is, up to appropriate choice of polarization, proportional to the imaginary part of the free-space dyadic Green's function with source at the focal point, which is exactly the time-reversal-optimal field (place a source at the origin, measure its outgoing radiation, and time-reverse it), as discussed in Sec.~\ref{sec:CMFocus}.

A note of caution is warranted. Despite our discussions of the widths of the various optimized beams in this subsection and the last, \emph{the objectives for which these beams are optimal are focal-point intensities, not measures of the beam width}. Hence, although these and related wave fields are often referred to as ``diffraction-limited'' beams, they are \emph{not} beams operating at some fundamental width lower limit, or lower bound. They are beams that maximize intensity at a single focal point. If one does optimize for minimum ``width,'' it is possible to achieve widths significantly smaller than the diffraction limit. The realization of this ``super-oscillation'' possibility, even for waves generated by far-field sources, led to significant excitement, albeit tempered by theoretical analyses of requisite tradeoffs. These are the topics of our next section.

\subsubsection{To a sub-diffraction-limited spot (``superoscillations'')}
What if, instead of maximizing focal-point intensity, one directly minimizes beam width in a plane of interest? In many imaging setups, one might be willing to pay an energy cost for reducing the width of a point-spread function. There are a number of ways to achieve superresolution, including computational techniques (based on \emph{a priori} information~\cite{Hell1994,Betzig2006} or large signal-to-noise ratios~\cite{Shahram2006-nk,Narimanov2019-ml}) or near-field optical scattering (to interact with large-wavenumber waves)~\cite{Hillenbrand2025,Synge1928}. In this section, we focus exclusively on efforts to generate, \emph{from the far field}, free-space waves with ``faster-than-Nyquist'' oscillations.

What is a suitable width objective? There are a variety of possible options, but probably the simplest and most-studied asks: \textbf{for a wave field with unity amplitude at the origin, what is the smallest in-plane radius $\rho_{\rm zero}$ at which the field can be forced to zero?} The answer, strikingly, is that there is no lower bound, and one can make arbitrarily small zeros, despite controlling waves with a finite spatial bandwidth. This is the origin of the term ``super-oscillations,'' referring to the notion of forcing oscillations at a scale $\rho_{\rm zero} \ll 1/k_{\rm max}$, contrary to Fourier intuition. Such oscillations come at an extreme energy cost if one wants to create beams orders-of-magnitude smaller than the diffraction limit. However, the energy cost for factor-of-two reductions is less extreme, and may be beneficial for certain applications.

Sub-diffraction-limited focal spots have been investigated since at least 1952, when Toraldo di Francia~\cite{ToraldoDiFrancia1952} constructed beams with successively narrower main lobes (at the cost of increasingly large side-lobe energy) in a scalar, weak-scattering limit. The phenomenon was reframed in the 1990s as a special case of \emph{superoscillation}---band-limited functions whose local rate of oscillation exceeds their highest Fourier component---an idea pointed to in the context of quantum weak values by Aharonov and colleagues~\cite{Aharonov1988}, then made explicit for waves by Berry~\cite{Berry1994}. A signal-processing reanalysis~\cite{Ferreira2006} placed the energy-versus-spot-size tradeoff on quantitative footing. Our presentation here is one-dimensional, mirroring most of this literature: the beam propagates in two dimensions, say $x$ and $z$, with focusing intended on the line $z=0$, and we ask how narrow the field can be in the $x$ direction. The electric field can be decomposed into plane waves with $\kv = k_x \hat{\xv} + k_z \hat{\zv}$, giving the Fourier series
\begin{align}
    E(x,z) = \int_{-k_m}^{k_m} c(k_x) e^{i(k_x x + k_z z)} \,{\rm d}k_x,
    \label{eq:EPW2D}
\end{align}
where $k_m$ is the largest transverse wavenumber (e.g., set by an aperture). There are a few simple arguments that showcase the tradeoff between wave fields that oscillate rapidly and their relative energy costs~\cite{Ferreira2006}. First, a signal oscillating as $a \cos(k_x x)$ can have a derivative much larger than $k_x$, and hence oscillate from 1 to 0 over a distance much smaller than the wavelength $\lambda$, if $a$ is large, suggesting an energy-derivative tradeoff. Second, consider the Fourier series of \eqref{eq:EPW2D}, along the $z=0$ line, at two points $x_1$ and $x_2$ separated by $x_2-x_1 = \delta \ll \lambda, 2\pi/k_m$. The fields have a difference of unity, i.e., $E(x_2,0)-E(x_1,0)=1$, yet the arguments of their Fourier series are nearly identical. Taking the differences of the Fourier series:
\begin{align}
    E(x_2,0) - E(x_1,0) &= \int c(k_x) \left[ e^{ik_x x_2} - e^{ik_x x_1} \right] \,{\rm d}k_x &\simeq i\delta \int c(k_x) k_x \,{\rm d}k_x.
\end{align}
The left-hand side equals 1. The right-hand side has $\delta$, the small spacing, which implies the Fourier integral must be large. This implies a large energy requirement, which can be quantified via the Cauchy-Schwarz inequality: $\int c(k_x) k_x \,{\rm d}k_x \leq \left[ \int |c(k_x)|^2 \,{\rm d}k_x \int |k_x|^2 \,{\rm d}k_x \right]^{1/2}$, where the integral of the squared Fourier coefficients is proportional to the real-space energy of the field, $E$, by Parseval's theorem. Then one finds that the energy must scale at least as fast as inverse quadratically with the separation width, $E \gtrsim 1/\delta^2$. This result, with precise constants, is given in \cite{Ferreira2006}.

Yet for proper beam formation, one would want the signal to go from 0 to 1 \emph{then back to} 0, over a total distance $2\delta$. (It is not useful simply for the field to quickly go from 0 to 1, if it then remains of order unity.) An expression for fixing the field at an arbitrary number of points $N$ is given in \cite{Ferreira2006}, but it is intended for large-$N$ asymptotics and does not appear to give the optimal scaling with respect to $\delta$. Instead, we can generalize the argument above to three points, setting $E(x_1)=E(x_3)=0$, $E(x_2)=1$, and $x_3 = x_2 + \delta = x_1 + 2\delta$. We take the differences between each pair of neighboring points, add them together, and Taylor expand, with first-order terms cancelling, leaving only second-order terms:
\begin{align}
    \left[E(x_2,0) - E(x_1,0)\right] + \left[E(x_2,0) - E(x_3,0)\right] &= \int c(k_x) \left[ 2e^{ik_x x_2} - e^{ik_x x_1} - e^{ik_x x_3} \right]. \nonumber \\
                                                                        &\simeq \delta^2 \int c(k_x) k_x^2.
\end{align}
Again, the left-hand side is order-one (it is exactly two), while now the right-hand side has a coefficient that scales \emph{quadratically} with the ``beam width'' $\delta$. Hence, the energy of the field must now scale with the inverse of $\delta$ raised to the \emph{fourth} power:
\begin{align}
    E \gtrsim \frac{1}{\delta^4}.
\end{align}
This inverse-fourth-power scaling of energy with separation is arguably the most important metric for understanding what is possible when designing super-oscillation point-spread functions. The derivations above were for the case of two-dimensional propagation and focusing on a one-dimensional line, but the same scaling arises if one analyzes three-dimensional propagation and focusing on a two-dimensional plane. In that case, the analogous scenario is a ring of radius $\delta$ at which the field should equal zero, relative to a unit amplitude at the origin, as considered in \cite{Shim2020}. In that case, the mathematical approach to derive bounds (with correct constants) is more involved, but the key scaling result is the same. This polynomial---rather than exponential---scaling is what makes superoscillation a practically viable route to sub-diffraction focusing: a superoscillatory beam achieves rapid local variation by interfering high-eigenvalue PSWF modes, avoiding the exponential cost that would arise from operating in the (suppressed) below-the-knee subspace of Slepian's band-concentration framework~\cite{Kuang2025,DMiller2025}.

\begin{figure*}[tb]
\centering
\includegraphics[width=1\linewidth]{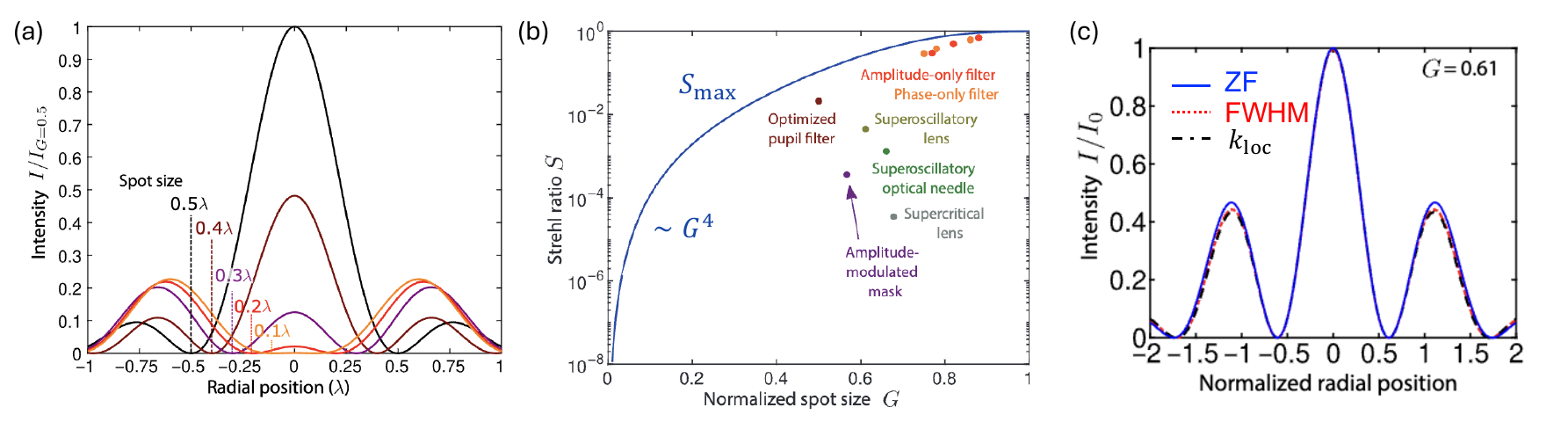}
\caption{\textbf{Bounds on focal-point intensity subject to a spot-size constraint.} (a) Optimal radial intensity profiles, in the paraxial limit, for normalized zero-field spot sizes ranging from $G = 0.5\lambda$ down to $0.1\lambda$. (b) Maximal Strehl ratio $S_{\rm max}$ as a function of $G$, exhibiting the universal $G^4$ scaling at small $G$. Markers indicate previously proposed designs, which fall well below the bound at moderate $G$ (see Ref.~\cite{Shim2020} for individual citations). (c) Optimal beam intensity profiles for three spot-size definitions: zero-field (ZF), full-width at half maximum (FWHM), and local wavenumber ($k_{\rm loc}$). The nearly indistinguishable profiles show that the bound is robust to spot-size definition. (Figure adapted from Ref.~\cite{Shim2020}.)}
\label{fig:Superres}
\end{figure*}
Fig.~\ref{fig:Superres}(a) depicts field intensity profiles for the fields that minimize total intensity while achieving a zero-field ring at a prescribed ``spot size'' $G$ (equivalent to $\delta$), for a paraxial excitation. The black curve, optimized for $G=0.5$, nearly coincides with the ``diffraction-limited'' focal-point-intensity optimal field of the previous sub-section. As the spot size is reduced, the energy in the side lobes rapidly increases relative to the intensity at the peak. In Fig.~\ref{fig:Superres}(a), the fields are normalized to have total intensity of 1, in which case the intensity at the origin decreases with the fourth power of the spot size. This scaling is more readily apparent in Fig.~\ref{fig:Superres}(b), which plots the Strehl ratio as a function of normalized spot size, alongside many theoretical proposals from the literature~\cite{de-Juana2003-oq,Gundu2005-kc,Kosmeier2011-fj,Rogers2012-ve,Rogers2013-df,Qin2017-pc}. One can see that for nearly diffraction-limited spots (normalized $G$ close to 1), there are designs that closely approach the bounds. But for smaller spot sizes, there is ample room for improvement. Ref.~\cite{Shim2020} demonstrates the possibility for inverse design~\cite{Jensen2011,Miller2012,Bendsoe2013-ed,Molesky2018-xc} to discover metasurface patterns that can approach the bounds.

Perhaps surprisingly, the bounds on focal-point intensity subject to a fixed spot size are quite robust to the \emph{definition of spot size}. One can imagine that prescribing exactly zero field might be a stringent constraint, and that smoother definitions might ameliorate the intensity penalty. However, this is not the case. In Ref.~\cite{Shim2020}, in addition to the zero-field (``ZF'') constraint, two alternative constraints were considered. First, one which specifies a full-width-at-half-maximum (``FWHM'') radius. Second, a metric that measures a ``local wavenumber'' via the Laplacian of the fields at the origin. Analytical expressions for optimal fields for the alternative metrics were not found, but global optima are efficiently computable and the results show nearly identical field patterns, as seen in Fig.~\ref{fig:Superres}(c), and virtually identical numerical values for each of the possible objectives of interest (cf. Fig.~6 of Ref.~\cite{Shim2020}.)

Hence, it is possible to arbitrarily reduce the ``size'' of a focused electromagnetic field, but one must pay a quartic energy penalty. In practice, this likely prohibits factor-of-ten reductions in beam size (corresponding to 10,000X energy increases). However, a factor of 2 reduction and a corresponding $\approx$10X increase in required energy (even better than the anticipated 16X, which is required in the small-$\delta$ asymptotic limit) may be feasible, and designs developed to approach the above bounds may be desirable.

\begin{openq}
The Slepian--Pollak--Landau diffraction-limit framework---the eigenvalue distribution of the joint spatial--spectral concentration problem, the bimodal degrees-of-freedom counting, and the associated capacity bounds---was developed for scalar signals in one dimension. Can it be rigorously extended to vector electromagnetic fields, and reconciled with the bounds on intensity and beam width discussed above?
\end{openq}

\subsubsection{The space-time diffraction limit}
\label{sec:spacetime_focusing}

The bounds derived above all consider focusing in space at a single frequency. A natural generalization is to ask: what is the limit on focusing in both space \emph{and} time? Schr{\"o}dinger-equation analogs were studied in the context of quantum matter waves by Brukner and Zeilinger~\cite{BruknerZeilinger1997}, an exactly solvable diffraction-in-time model was developed by Goussev~\cite{Goussev2013}, and the spacetime focusing concept was experimentally realized for water waves by Bacot et al.~\cite{Bacot2016}. Intuitively, one might think the answer is simply a four-dimensional ``diffraction limit,'' but the space and time dimensions are not independent. Within the temporal bandwidth that is available, and presumably the primary determinant of temporal focusing, for each frequency that is available, there is a different range of spatial frequencies available, since propagating waves lie within the light cone, $|\kv| \le \omega/c$. 

Writing the field as a generalized Fourier integral over wavevector \emph{and} frequency,
\begin{align}
    \psi(\xv,t) = \int \dkk \int d\omega \, c(\kv,\omega) \, e^{i(\kv\cdot\xv - \omega t)},
    \label{eq:STField}
\end{align}
where $\psi$ denotes a scalar wave amplitude---e.g., the Schr\"odinger amplitude in the matter-wave analog, or a chosen polarization component of the electric field in the electromagnetic case (specifically, in the 3D vector treatment below, the radial amplitude of the $\Nv_{1,0}$ electric-dipole channel of \eqref{eq:N10xy}). By the same extremal-eigenvalue argument used above, with amplitude normalization $\int\!\!\int|c|^2=1$, the optimal $c(\kv,\omega)$ for maximizing intensity at the spacetime origin, $(\xv=0,t=0)$, is uniform across the set of possible plane-wave spatial and temporal frequencies:
\begin{align}
    c_{\rm opt}(\kv,\omega) =
        \begin{cases}
            1, & |\kv| \le |\omega|/c \ \text{and}\  |\omega| \le \omega_{\max} \\
            0, & \text{otherwise,}
        \end{cases}
\end{align}
where the symmetric extension to negative $\omega$ ensures that $\psi$ is real-valued. For two-dimensional propagation observed along a line ($k$ is the wavevector component along the line, so propagating waves fill $|k| \le |\omega|/c$), plus time, the field profile that maximizes intensity at $(x=0,t=0)$ is
\begin{align}
    \psi_{\rm 1D}(x,t) &\propto \int_{-\omega_{\max}}^{\omega_{\max}}\! d\omega \int_{-|\omega|/c}^{|\omega|/c}\! dk\, e^{i(kx - \omega t)} \nonumber \\
    &\propto \frac{1}{x}\left[\frac{\sin^2\!\bigl(\omega_{\max}(x-ct)/(2c)\bigr)}{x-ct} + \frac{\sin^2\!\bigl(\omega_{\max}(x+ct)/(2c)\bigr)}{x+ct}\right].
    \label{eq:ST1d}
\end{align}
At $t=0$ this reduces to $\psi(x,0) \propto \sin^2(\omega_{\max} x/(2c))/x^2$, to be contrasted with the single-frequency 1D-band-limited optimum $\psi(x) \propto \sin(\omega_{\max} x/c)/x$ that emerges from the analyses of the previous subsections.

\begin{figure*}[tb]
\centering
\includegraphics[width=0.9\linewidth]{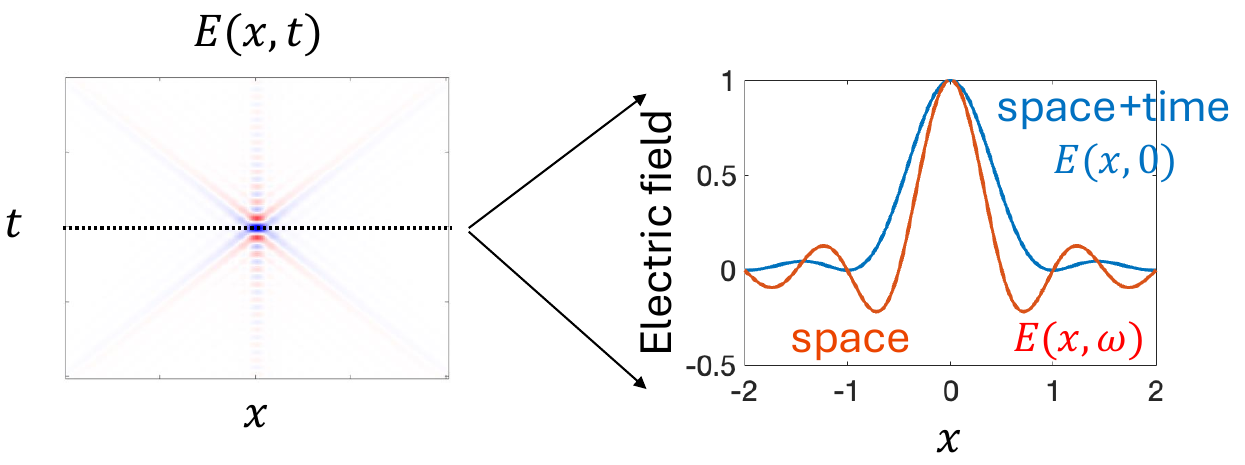}
\caption{\textbf{Space-time-optimal focusing (focal line, 2D propagation).} Left: The full electric field $E(x,t)$, with two counter-propagating wave packets along $x=\pm ct$ coherently summing to a sharp focus at the spacetime origin; the dotted line marks the $t=0$ cross-section. Right: The spacetime-focused field $E(x,0)$ (blue) is broader than its single-frequency counterpart (orange), due to the collective contributions of smaller frequencies which localize in time but delocalize in space.}
\label{fig:SpacetimeFocus}
\end{figure*}

The 3D vector case is analogous. With the same symmetric extension in $\omega$ and uniform weight per unit solid angle and frequency (i.e., normalizing $\int\!\!\int\sum_p|a_p(\Omega,\omega)|^2\,{\rm d}\Omega\,{\rm d}\omega = 1$ for the transverse plane-wave amplitudes $a_p$), each frequency contributes the radial $\Nv_{1,0}$-like profile of \eqref{eq:N10xy}, modulated in time by $\cos(\omega t)$, and integrating over $\omega \in [0,\omega_{\max}]$ yields a space-time-optimal radial field
\begin{align}
    \psi_{\rm 3D}(\rho,t) \propto \int_0^{\omega_{\max}} \cos(\omega t) \left[ \frac{\sin(\omega\rho/c)}{\omega\rho/c} - \frac{j_1(\omega\rho/c)}{\omega\rho/c} \right] d\omega,
\end{align}
where the bracket is precisely the $\Nv_{1,0}$ profile evaluated at frequency $\omega$. At $t=0$ this admits a closed-form expression in terms of the sine integral, $\psi_{\rm 3D}(\rho,0)\propto[\mathrm{Si}(U)+j_1(U)]/U$ with $U=\omega_{\max}\rho/c$, a profile with no zeros, decaying as $\pi/(2U)$ with only shallow side lobes. For the 1D profile there are two interesting features, especially if one compares the spatial localization at $t=0$ to the spatial localization of a single-frequency excitation. First, spatial side lobes are suppressed faster with spacetime focusing, explained mathematically by the $1/x^2$ scaling of \eqref{eq:ST1d} versus the conventional $1/x$. Second, and perhaps more interesting, is that one \emph{loses} spatial localization upon adding temporal localization. There are extra factors of $1/2$ in the $\sin^2$ terms of \eqref{eq:ST1d} that lead to a \emph{doubling} of the distance to the first zero for the spacetime-optimized wave versus the space-optimized wave. (For a strictly one-dimensional wave equation, $k=\pm\omega/c$ and the $t=0$ optimum is $\sin(\omega_{\max}x/c)/x$, the same as the single-frequency band-limited profile: this spatial cost of temporal localization arises from propagation in two or more dimensions.) As shown in Fig.~\ref{fig:SpacetimeFocus}, this is the price of localizing in time: a pulse sharply concentrated at $t=0$ requires energy spread across a range of frequencies, and the resulting spatial profile is an average of single-frequency profiles spanning $\omega \in [0,\omega_{\max}]$. Smaller $\omega$ contributes broader spatial main peaks, and their inclusion in the average widens the net profile relative to the single-frequency optimum at $\omega_{\max}$ (in 3D, the half-maximum width grows by a factor of 1.8). 

Hence focusing in both space and time is fundamentally distinct from focusing in space alone: temporal localization extracts an irreducible spatial cost, set by the spread of frequencies needed to form a short pulse. This tradeoff must occur in applications such as pump-probe spectroscopy, ultrafast metrology, or time-gated imaging, where temporal sharpness is the primary objective.

\subsection{Focusing in complex media, and time reversal}
\label{sec:CMFocus}
The bounds of the previous subsection are specific to free space. In complex media---where scatterers are present and the Green's function is not analytically known---we cannot derive bounds in closed form, yet a striking result holds: the maximum focal-point intensity from currents supplied on a bounding aperture has a clean expression in terms of the field that a dipole at the focal point would produce on that aperture. The recipe for achieving the bound is often loosely called ``time reversal,'' but its mathematical foundation is in fact \emph{reciprocity}; the two coincide only in special cases. The bound parallels its free-space counterpart, \eqref{eq:MaxPsi0}, and the framework extends naturally to broadband spacetime focusing.

\subsubsection{The intensity bound}
\label{sec:CMIntensityBound}
Consider currents $s_{\rm supp}$ supplied on a bounding aperture $A$, generating a field at the focal point $\xv_0$ via the (generally unknown) Green's function $\Gamma$ of the medium:
\begin{align}
    \psi(\xv_0) = \int_A \Gamma(\xv_0, \xv') s_{\rm supp}(\xv') \,{\rm d}\xv' \equiv \Gamma_{0,A}\, s_{\rm supp}.
\end{align}
We seek the currents that maximize the focal-point intensity, subject to a unit-norm constraint on $A$. As in \secref{FSFocus}, we measure the vector field through a polarization $\phi$ (separately optimized), reducing the objective to a rank-one quadratic form $|\phi^{\dagger}\Gamma_{0,A} s_{\rm supp}|^2 = s_{\rm supp}^{\dagger}(\Gamma_{0,A}^{\dagger}\phi\phi^{\dagger}\Gamma_{0,A}) s_{\rm supp}$, with optimal current $s_{\rm supp,opt} \propto \Gamma_{0,A}^{\dagger}\phi$. The maximum focal-point intensity is then
\begin{align}
    \bigl|\phi^{\dagger}\psi(\xv_0)\bigr|^2_{\rm opt} = \phi^{\dagger} \Gamma_{0,A} \Gamma_{0,A}^{\dagger} \phi = \int_A \bigl|\Gamma_{0,A}^{\dagger}\phi\bigr|^2(\xv') \,{\rm d}\xv'.
    \label{eq:CMFocusBound0}
\end{align}
For a \emph{reciprocal} medium, $\Gamma_{0,A}^T = \Gamma_{A,0}$, so $(\Gamma_{0,A}^{\dagger}\phi)(\xv') = (\Gamma_{A,0}(\xv',\xv_0)\phi^*)^*$ is the complex conjugate of the field at $\xv'$ produced by a dipole at $\xv_0$ with polarization $\phi^*$. Calling that dipole field $\psi_0(\xv')$,
\begin{align}
    \bigl|\phi^{\dagger}\psi(\xv_0)\bigr|^2_{\rm opt} = \int_A \bigl|\psi_0(\xv')\bigr|^2 \,{\rm d}\xv' = \int_A \bigl[|\Ev_0(\xv')|^2 + |\Hv_0(\xv')|^2\bigr] \,{\rm d}\xv'.
    \label{eq:CMFocusBound}
\end{align}
\Eqref{eq:CMFocusBound} is the central bound of this subsection. The maximum focal-point intensity is set entirely by the field on the bounding aperture produced by a dipole at the focal point. In the far zone this is proportional to the local density of states (LDOS); in the near field it can be orders of magnitude larger. Crucially, \eqref{eq:CMFocusBound} requires neither full $4\pi$ solid angle nor time-reversal invariance---only reciprocity. The aperture $A$ may be arbitrarily shaped, the measurement may be in the near field, and the medium may contain loss or gain.

\subsubsection{Reciprocity, not time reversal}
\label{sec:CMReciprocity}
Substituting $\Gamma_{0,A}^{\dagger}\phi = (\Gamma_{A,0}\phi^*)^*$ in the reciprocal medium, the optimal current for \eqref{eq:CMFocusBound0} acquires a physical interpretation:
\begin{align}
    s_{\rm supp,opt} \propto \bigl(\Gamma_{A,0}\phi^*\bigr)^*.
    \label{eq:CMTRRecipe}
\end{align}
In words: \emph{to focus from $A$ onto $\xv_0$, place a (conjugated) dipole at $\xv_0$, allow it to radiate to the aperture in the same medium, and drive currents proportional to the conjugates of the measured fields}. This is the recipe associated with ``time-reversal'' or ``phase-conjugate'' focusing in the wavefront-shaping literature.

But the operation we just used is not time reversal. It is \emph{source transposition}---placing the source where the detector was, and vice versa---which is a statement about \emph{reciprocity}. Reciprocity is a symmetry between source and detector; time reversal is a symmetry between forward and backward propagation. The two are distinct, and the optimality of \eqref{eq:CMFocusBound0} depends only on the former. In a reciprocal but non-TRI medium (e.g., one with loss or gain), the recipe \eqref{eq:CMTRRecipe} still gives optimal focusing, even though time-reversal symmetry is broken. Similarly, if the aperture does not capture all outgoing radiation, the currents prescribed by \eqref{eq:CMTRRecipe} do not produce the time-reversed fields, but they are still optimal. Conversely, in a TRI but non-reciprocal medium (rare in practice~\cite{Buddhiraju2020}), the optimization is \emph{not} achieved by time-reversing the dipole field---a TR-derived prescription would be suboptimal. In the common case where both symmetries hold, the two prescriptions coincide, but the optimality of \eqref{eq:CMTRRecipe} and the bound \eqref{eq:CMFocusBound} require only reciprocity.

\subsubsection{Field patterns and time reversal}
\label{sec:CMFieldPattern}
That said, scenarios \emph{with} time-reversal symmetry appear to be the only ones in which we can say more about the field patterns near the maximum-intensity point in a complex medium. (The analog of our analysis of the ``widths'' of the optimal beams in previous subsections.) We know the time-reversed fields when two conditions are met: (i) the medium is time-reversal invariant (TRI), and (ii) the fields are measured over a closed (e.g., full-sphere) aperture. These fields are optimal, producing \eqref{eq:CMFocusBound} for the target-point intensity, if two additional conditions are met: (iii) measurements are performed in the far zone of $\xv_0$ (and all scatterers of the complex medium), and (iv) the medium is reciprocal. Classical time-reversal focusing is a technique with a rich experimental history spanning ultrasound~\cite{Fink1992}, optical wavefront-shaping in biological tissue~\cite{Horstmeyer2015}, and ``instantaneous time mirrors'' in classical wave systems~\cite{Bacot2016}.

\begin{figure*}[tb]
\centering
\includegraphics[width=1\linewidth]{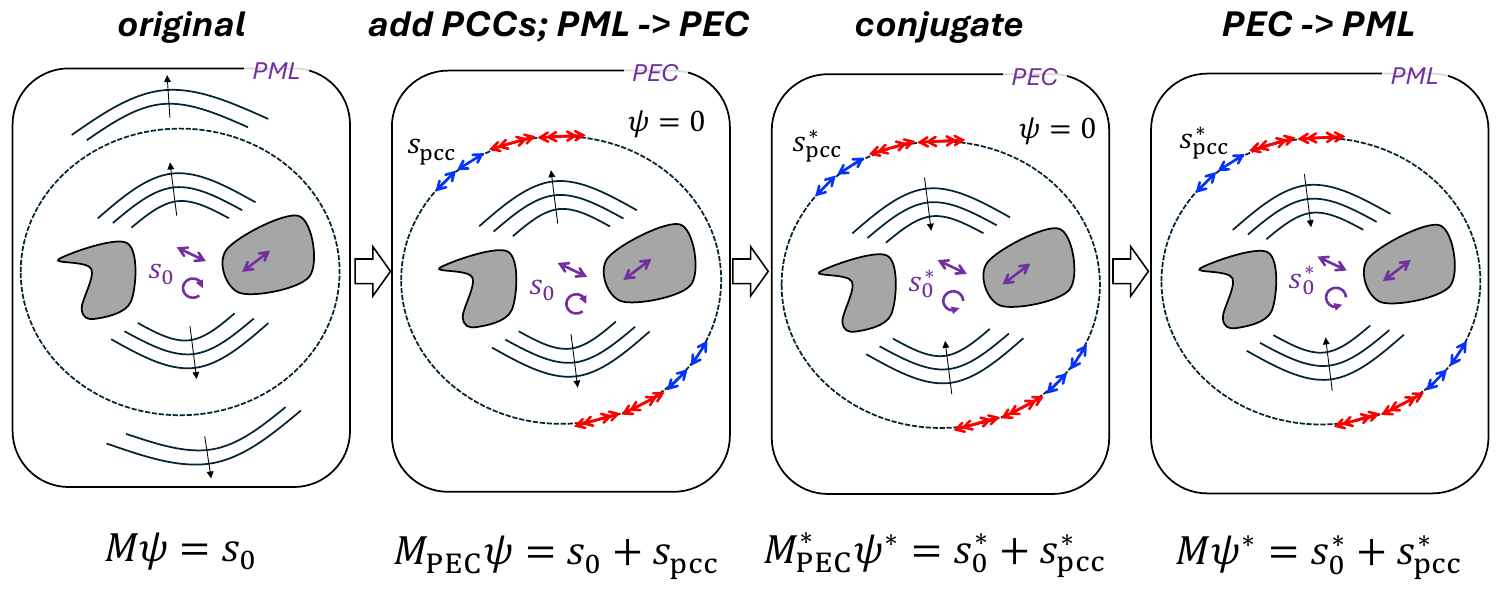}
\caption{\textbf{Partial time reversal at an outer boundary.} Four-step construction underlying \eqref{eq:ConjFields}. (1)~The interior sources $s_0$ (purple, possibly within scatterers shown in grey) radiate fields $\psi$ under outgoing boundary conditions (PML). (2)~``Perfectly cancelling currents'' $s_{\rm pcc}$ are added on an exterior surface (dashed) via the surface-equivalence principle, exactly nulling the field outside; the now-zero exterior allows the PML to be replaced by a PEC boundary. (3)~Because PEC is time-reversal invariant ($M_{\rm PEC}^* = M_{\rm PEC}$), conjugating the system gives $M_{\rm PEC}^*\psi^* = s_0^* + s_{\rm pcc}^*$. (4)~Since the exterior remains zero, we revert to outgoing BCs (PEC $\to$ PML) to arrive at $M\psi^* = s_0^* + s_{\rm pcc}^*$, i.e., \eqref{eq:Msintconj}.}
\label{fig:setup}
\end{figure*}

We write Maxwell's equation in a form amenable to time-reversal and reciprocity symmetries:
\begin{align}
    \begin{pmatrix}
         & \nabla \times \\
         \nabla \times &
    \end{pmatrix}
    \begin{pmatrix}
        \Ev \\
        i\Hv
    \end{pmatrix}
    - \omega
    \begin{pmatrix}
        \varepsilon & -i\alpha \\
        i\beta & \mu
    \end{pmatrix}
    \begin{pmatrix}
        \Ev \\
        i\Hv
    \end{pmatrix}
    =
    \begin{pmatrix}
        i \Jv_e \\
        -\Jv_m
    \end{pmatrix} , \quad \text{ or} \quad M \psi = s,
\end{align}
where $\Jv_e$ and $\Jv_m$ are electric and (fictitious) magnetic currents, and $\alpha = \xi/c$ and $\beta = \zeta/c$ in terms of the off-diagonal constitutive tensors of \eqref{eq:MaxwellDiffEq}. The factors of $i$ in the field and source six-vectors are chosen so that time-reversal corresponds precisely to conjugation~\cite{deRosny2010,Guo2022}. A naive application of TRI symmetry $M^* = M$ would suggest that conjugating the source ($s \to s^*$) conjugates the field, but this overlooks the boundary conditions: time-reversal flips outgoing to incoming. The fix is the following four-step construction, illustrated in Fig.~\ref{fig:setup}.

Suppose the original interior sources $s_0$ (e.g., a guide-star dipole at the focal point) produce a field $\psi$ with outgoing boundary conditions,
\begin{align}
    M \psi = s_0,
    \label{eq:Msint}
\end{align}
where $M$ without a subscript denotes outgoing BCs. By the surface-equivalence principle~\cite{Schelkunoff1936,Jin2011,Oskooi2013}, we can introduce ``perfectly cancelling currents'' $s_{\rm pcc}$ on the exterior bounding surface that radiate only outward and exactly cancel the originally outward propagating field there. With the exterior nulled, the outgoing BC can be replaced by a PEC condition, which \emph{is} time-reversal invariant: $M_{\rm PEC}^* = M_{\rm PEC}$. Conjugating the PEC system $M_{\rm PEC}\psi = s_0 + s_{\rm pcc}$ gives a TRI solution $M_{\rm PEC}\psi^* = s_0^* + s_{\rm pcc}^*$, and since the exterior remains zero we can revert to outgoing BCs:
\begin{align}
    M \psi^* = s_0^* + s_{\rm pcc}^*.
    \label{eq:Msintconj}
\end{align}
The conjugated surface currents $s_{\rm pcc}^*$ are the ``phase-conjugate'' currents implemented in TR-focusing experiments, where the interior sources $s_0^*$ are \emph{not} present. To isolate the field $\psi_{\rm TR}$ produced by $s_{\rm pcc}^*$ alone, we subtract $\psi_{\rm TR} = \psi^* - \psi_2$, where $\psi_2$ satisfies $M\psi_2 = s_0^*$. Writing both in terms of the $6\times 6$ Green's-function tensor $\Gamma$ gives the key result,
\begin{align}
    \psi_{\rm TR}(\xv) = -2i \int \Im\bigl[\Gamma(\xv,\xv')\bigr]\, s_0^*(\xv') \,{\rm d}\xv',
    \label{eq:ConjFields}
\end{align}
for $\xv$ inside the closed surface (outside it, $\psi_{\rm TR} = -\Gamma s_0^*$), where ``$\Im$'' acts entrywise on the GF tensor. \Eqref{eq:ConjFields} was first derived by~\citeasnoun{deRosny2010} via an alternative argument that assumed reciprocity in addition to TRI; the PEC-extension construction here requires only TRI of the medium and is, in our view, somewhat cleaner.

\Eqref{eq:ConjFields} has a nice physical interpretation: the field everywhere produced by phase-conjugated surface currents is the conjugated original sources convolved with the entrywise-imaginary part of the medium's Green's function. Taking $\Im\Gamma$ smooths the GF, removing its near-field singularities; the most rapid variations that can arise are wavelength-scale, unless there are near-field interactions with a scatterer that generate new near-field singularities. Note that the location at which the field is maximized need not be a local optimum in the field intensity---the scattering profile that creates the largest field at $\xv_0$ can certainly lead to even larger fields at neighboring points (e.g., through resonant enhancement). The derivation does not require far-field measurements or any minimum surface radius, contradicting a common assumption that perfectly time-reversed sources would create a divergent conjugate field at the original source location.

A simpler form emerges when the original sources are entirely real-valued, $s_0 = s_0^*$ (recall that $s$ is built from $i\Jv_e$ and $-\Jv_m$). Subtracting \eqref{eq:Msint} from \eqref{eq:Msintconj} gives $M(\psi^* - \psi) = s_{\rm pcc}^*$, so
\begin{align}
    \psi_{\rm TR} = -2 i \Im \psi.
\end{align}
In free space, $\Im\Gamma$ centered at $\xv_0$ is, up to a polarization factor, the regular vector spherical wave $\Nv_{1,m}^{\rm (reg)}$, recovering \eqref{eq:N10xy} of \secref{FSFocus}. For media with arbitrary scatterers, the focused field is the same smoothing of the original dipole field by $\Im\Gamma$ of the actual medium---the natural complex-medium analog of free-space diffraction-limited focusing.

\begin{openq}
The construction above should be readily extensible to nonlinear media. Can it help us understand the diffraction limit, or its circumvention, for nonlinear time reversal?
\end{openq}

\subsubsection{Spacetime focusing}
\label{sec:CMSpacetime}
The framework above extends naturally from monochromatic focusing in space to broadband focusing in spacetime. The relevant symmetry is time-domain reciprocity, established by~\citeasnoun{Welch1960} for impulse responses of LTI media and refined for time-translation-invariant media by~\citeasnoun{deHoop1987} and~\citeasnoun{Cheo2003}:
\begin{align}
    \Gamma_{\rm ret}(\xv, t; \xv', t') = \Gamma_{\rm ret}^T(\xv', t; \xv, t'),
\end{align}
where $\Gamma_{\rm ret}$ is the retarded time-domain Green's function. We now ask: among all space-and-time-dependent currents $s_{\rm supp}(\xv,t)$ supplied on $A$ over a time window $[0,T]$, which maximizes the focal-point intensity $|\psi(\xv_0, t_0)|^2$ at a target spacetime point, subject to the unit-norm constraint $\int_0^T\!\!\int_A |s_{\rm supp}|^2 \,{\rm d}\xv\,{\rm d}t = 1$? The derivation parallels \eqref{eq:CMFocusBound}: the polarization-projected objective is again rank-one, and the optimal currents are the complex conjugate of the field on $A$ of a dipole at $\xv_0$ (with polarization $\phi^*$, as in \eqref{eq:CMFocusBound}), run backward in time from $t_0$: $s_{\rm supp,opt}(\xv,t)\propto\psi_0^*(\xv,t_0-t)$. For $t_0 = T$, the maximum intensity is
\begin{align}
    \bigl|\phi^{\dagger}\psi(\xv_0, t_0)\bigr|^2_{\rm opt} = \int_0^T\!\!\int_A \bigl|\psi_0(\xv,t)\bigr|^2 \,{\rm d}\xv\,{\rm d}t = \int_0^T\!\!\int_A \bigl[|\Ev_0(\xv,t)|^2 + |\Hv_0(\xv,t)|^2\bigr] \,{\rm d}\xv\,{\rm d}t,
    \label{eq:CMSTBound}
\end{align}
where $\psi_0(\xv,t)$ is the spacetime field on $A$ produced by an impulsive dipole at $(\xv_0, 0)$ in the same medium (for general $t_0$, the time integral runs over $[t_0-T,t_0]$). \Eqref{eq:CMSTBound} parallels both \eqref{eq:CMFocusBound} of the monochromatic case and the free-space spacetime bound of \secref{spacetime_focusing}. As in the monochromatic case, the recipe is reciprocity: place a dipole at the desired spacetime focus, propagate to the aperture, and time-reverse and conjugate the result---all in the same medium. The recipe reduces to true time reversal only when the medium is additionally TRI and the conditions of \secref{CMFieldPattern} are met. A recent example deriving generalized time-reversal recipes for pulse control in diffusive media (where TRI is broken by loss) is~\citeasnoun{McIntosh2025}.

\subsubsection{Non-reciprocal media}
\label{sec:CMNonreciprocal}
If the medium is non-reciprocal, then $\Gamma_{0,A}^T \neq \Gamma_{A,0}$ and the recipe \eqref{eq:CMTRRecipe} no longer applies as written. The fix is \emph{generalized reciprocity} (discussed in \secref{tr_symmetry}), which still exchanges source and detector locations provided the medium's constitutive tensors are simultaneously transposed ($\nu \to \nu^T$, where $\nu$ denotes the full bianisotropic constitutive matrix of \eqref{eq:MaxwellDiffEq}):
\begin{align}
    \Gamma_{0,A}^T = \Gamma_{A,0,\,\nu^T}.
\end{align}
The optimal current can then be written as $\Gamma_{0,A}^{\dagger}\phi = (\Gamma_{A,0,\,\nu^T}\phi^*)^*$: place dipolar sources at $\xv_0$, \emph{transpose the material tensors of the scattering medium everywhere}, measure the resulting field on the aperture, and then conjugate. The transposition step is rarely achievable in practice, but the broader point reinforces the message of \secref{CMReciprocity}: the optimization rests on source transposition---a (possibly generalized) reciprocity statement---not on time reversal. Finally, the aperture $A$ need not be closed; on an open surface the recipe still gives the optimal focal-point intensity, but the focused field is no longer the time-reversed pattern of \eqref{eq:ConjFields}.

\subsection{More general objectives}
\label{sec:OtherObjectives}
Having extensively discussed extremal eigenvalues for the canonical problem of focusing, it is relatively straightforward to extend this approach to a wide range of objectives of interest. In this section we describe three classes of objectives that exemplify the more general theory.
\begin{figure*}[tb]
\centering
\includegraphics[width=0.9\linewidth]{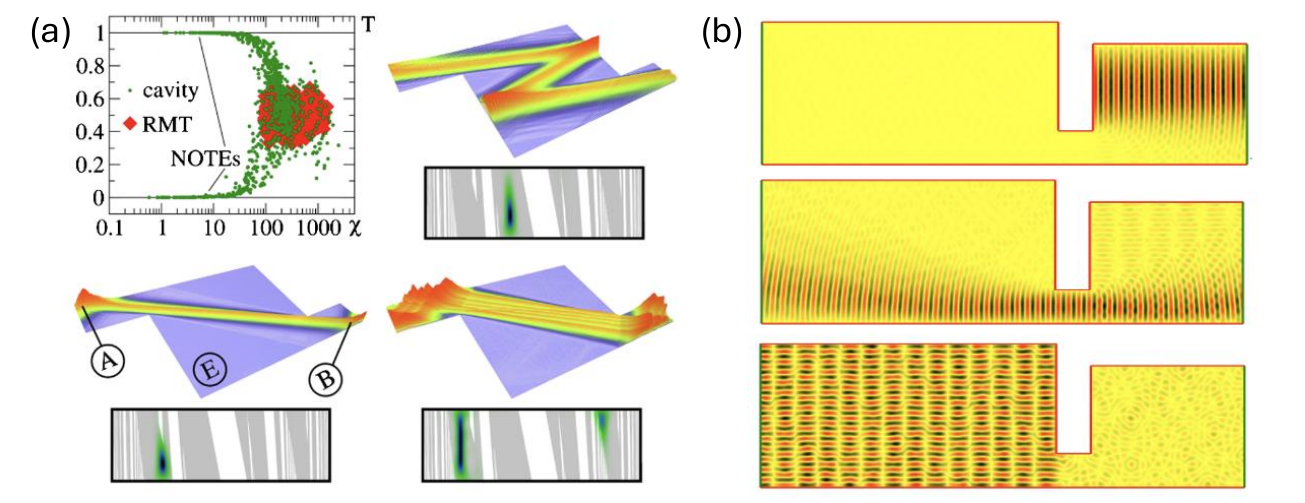}
\caption{\textbf{Wigner--Smith time-delay extremal modes.} (a) Examples of wave functions that minimize time delay between ports/leads for a strongly scattering rectangular cavity. These modes (termed ``NOTEs'') show quite different transmission statistics than those predicted by random matrix theory (RMT), and allow the possibility for a sender (``A'') to transmit to a receiver (``B'') while circumventing an eavesdropper (``E''). (b) Examples of modes of the time-delay operator in a notched waveguide with ports/leads along the left and right walls. Top: minimum-delay mode; Middle: a mode with time delay almost exactly matched to single-path transversal across the waveguide; Bottom: A highly reverberant mode with very long delay time. (Panel (a) adapted from Ref.~\cite{Rotter2011}; panel (b) adapted from Ref.~\cite{Patel2021}.)}
\label{fig:WSDelay}
\end{figure*}

\subsubsection{Wigner--Smith time delay}
Consider a wave packet of incoming-wave coefficients $\cin$ impinging upon a lossless scatterer with unitary scattering matrix $\SS$. A result first derived in the context of quantum scattering~\cite{Wigner1955,Smith1960}, rediscovered in the context of multimode fibers~\cite{Fan2005}, then translated to more general electromagnetic scattering~\cite{Rotter2011,Rotter2017,Patel2021,Patel2021b}, relates the \emph{time delay} $\tau$ experienced by the wave packet to the frequency derivative of the scattering matrix:
\begin{align}
    \tau = \cin^{\dagger} \left( -i \SS^{\dagger} \frac{\partial \SS}{\partial \omega}\right) \cin,
    \label{eq:TimeDelay}
\end{align}
where the operator in parentheses is Hermitian, since $\left[-i\SS^{\dagger} \partial \SS/\partial \omega\right]^{\dagger} = i (\partial \SS^{\dagger} / \partial \omega) \SS = i \partial (\SS^{\dagger} \SS)/\partial \omega - i \SS^{\dagger} (\partial \SS/\partial \omega) = - i \SS^{\dagger} (\partial \SS/\partial \omega)$. If one wants to minimize or maximize time delay for an incoming-wave normalization $\cin^{\dagger} \cin = 1$, the problem exactly follows the prescription at the beginning of the section, with the operator $\QQ$ given by $-i \SS^{\dagger} \partial \SS / \partial \omega$. The extremal values for the time delay, then, are
\begin{subequations}
\begin{align}
    \tau_{\rm min} &= \lambda_{\rm min} \left( -i \SS^{\dagger} \frac{\partial \SS}{\partial \omega}\right) \\
    \tau_{\rm max} &= \lambda_{\rm max} \left( -i \SS^{\dagger} \frac{\partial \SS}{\partial \omega}\right),
\end{align}
\end{subequations}
and the wave excitations that experience those delays, referred to as \textbf{principal modes}, are the corresponding extremal eigenvectors. Minimizing time delay ensures little interaction with strongly resonant modes that may be ``fragile'' or susceptible to small imperfections, and can generate ``particle-like'' wave patterns that appear nearly ballistic, as shown in Fig.~\ref{fig:WSDelay}(a) and the top panel of Fig.~\ref{fig:WSDelay}(b). Conversely, as exemplified in the bottom panel of Fig.~\ref{fig:WSDelay}(b), one can find modes with very long delay times that find a scattering path that maximally reverberates within the ``cavity'' formed in the waveguide. Note that the definition of time delay in \eqref{eq:TimeDelay} rigorously holds only in the lossless-material, unitary-$\SS$-matrix limit, although recent generalizations have been proposed for scenarios with losses and non-unitary $\SS$ matrices~\cite{Mao2023}.

\subsubsection{Force and torque generation}
Just as for power flow, forces and torques can be computed with appropriate quadratic expressions of the electromagnetic fields, suggesting the possibility for the extremal-eigenvalue approach. In free space, where vector spherical waves (VSWs) can serve as basis functions for incoming and outgoing coefficient vectors $\cin$ and $\cout$, respectively, straightforward algebra (cf. Ref.~\cite{Liu2019}) from the Maxwell stress tensor yields simple expressions for the $i$-directed force or torque on an arbitrary scatterer:
\begin{subequations}
\begin{align}
    F_i &= \frac{1}{c} \left[ \cin^{\dagger} \mathbb{P}_i \cin - \cout^{\dagger} \mathbb{P}_i \cout \right], \label{eq:ForceVSW}\\
    \tau_i &= \frac{1}{\omega} \left[ \cin^{\dagger} \mathbb{J}_i \cin - \cout^{\dagger} \mathbb{J}_i \cout \right].
    \label{eq:TorqueVSW}
\end{align}
\end{subequations}
The products $\vect{c}^{\dagger}\mathbb{P}\vect{c}$ and $\vect{c}^{\dagger}\mathbb{J}\vect{c}$ measure the linear and angular momenta carried in the propagating waves implied by $\vect{c}$, inward towards the scatterer for $\cin$ and outward, away from the scatterer, for $\cout$. The matrix elements of $\mathbb{P}_i$ and $\mathbb{J}_i$ are analytically given in the SM of Ref.~\cite{Liu2019}. In an optimal illumination problem, the scattering matrix $\SS$ is known, and one can substitute $\cout = \SS \cin$ in Eqs.~(\ref{eq:ForceVSW},\ref{eq:TorqueVSW}), yielding simple quadratic forms in terms of $\cin$ only. For example, the force equation simplifies to $F_i = (1/c) \cin^{\dagger} \left[ \mathbb{P}_i - \SS^{\dagger} \mathbb{P}_i \SS\right] \cin$. The optimal illumination fields are then given by the extremal eigenvectors of the matrix in square brackets. These are guaranteed to be globally optimal wavefronts for maximal force or torque generation in a desired direction, in the presence or absence of material loss, for free-space scattering.

Alternatively, in a lossless scattering process (free space or otherwise), it is possible to define a quadratic form for forces and torques via a generalization of the Wigner--Smith time-delay operator. By analogy of the time-delay operator being defined through frequency derivatives, it is possible to show that other Fourier pairs can be similarly related. For example, forces and torques on a specific particle can be computed or found through the quantity 
\begin{align}
    \cin^{\dagger} \left[ -i \SS^{\dagger} \frac{d\SS}{d\alpha} \right] \cin,
\end{align}
where $\alpha$ can be proportional to a center-of-mass coordinate for force or a rotation angle for torque. These generalized Wigner--Smith operators were proposed by Rotter et al. in Refs.~\cite{Ambichl2017,Horodynski2020}. Specific to forces and torques, a version of this matrix derivative had been discovered earlier by Rakich et al.~\cite{Rakich2009,Wang2011}. 

\begin{figure*}[tb]
\centering
\includegraphics[width=0.9\linewidth]{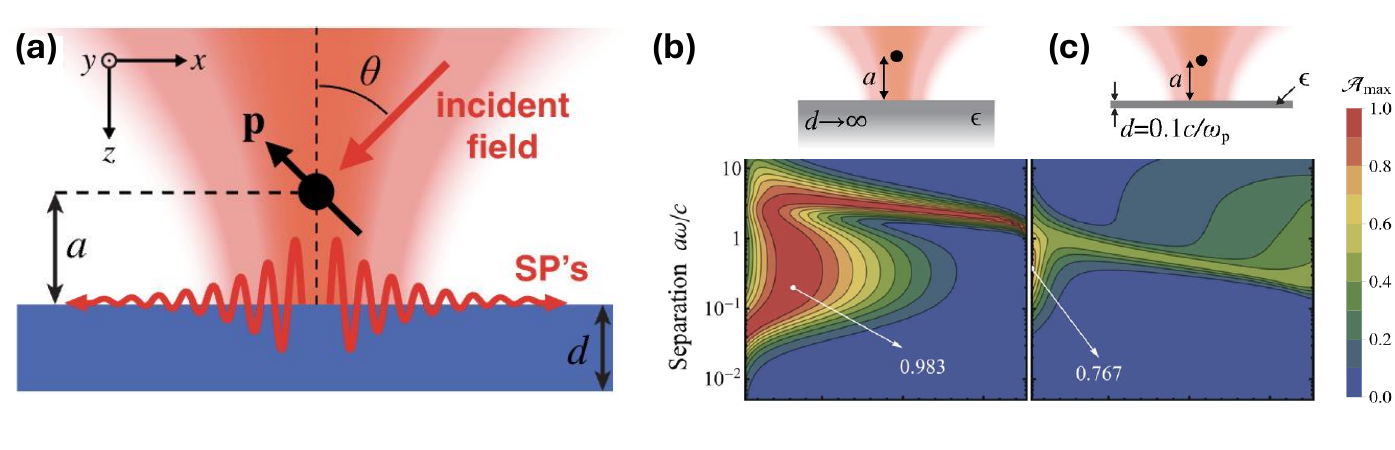}
\caption{\textbf{A powerful method for complete coupling of free-space waves to guided modes.} (a) Maximally focusing the incident field (including reflections from the surface) to the position of a point scatterer can generate large dipole moment, which then radiates in part to propagating modes (e.g., surface plasmons, ``SP's'') and in part to free space. (b,c) By adjusting the separation from the scatterer to the surface, one can adjust the coupling rates, until ``critical coupling'' is approached, offering near-unity maximum absorption percentages ($\mathcal{A}_{\rm max}$). (Figure adapted from Ref.~\cite{Dias2020}.)}
\label{fig:CompleteCoupling}
\end{figure*}

\subsubsection{Complete coupling to plasmons and guided modes}
For our final example, we consider a scenario in which one wants to efficiently and completely couple waves from free space to plasmons, or more generally any guided mode, propagating along a surface. Again we are confronted with what might appear to be a complex design problem, perhaps requiring tools such as ``inverse design''~\cite{Sigmund2003,Lu2011,Jensen2011,Miller2012,Lalau-Keraly2013,Lin2018}. There is an alternative, simple idea: place a dipole scatterer above the surface, which can independently couple both to free-space waves as well as to guided-mode waves, and can thereby serve to couple one to the other. The coupling can be treated in a sequential process: incident waves are focused to a point scatterer, generating a large dipole moment that then interacts with guided modes that are otherwise inaccessible directly from free space. Starting with the second half of the process: for a given dipole moment $\pv$ near a surface supporting a given set of guided modes, the power emitted by the dipole into those modes is determined simply by the local density of states~\cite{Joulain2003,Oskooi2013,Miller2023nfo}, which itself is directly proportional to the square of the original dipole moment, $|\pv|^2$. Maximizing the original induced dipole moment is exactly a focusing maximization: one should maximize the field intensity at the location of the particle. This parallels identically the derivations of Sec.~\ref{sec:FSFocus}, with the additional consideration that reflections from the planar surface back to the point scatterer should be taken into account in the optimization. This understanding is sufficient to maximize the power coupled to guided modes. But what fraction of the incoming power is converted to the guided modes? Here we can leverage intuition from coupled-mode theory, with the point-scatterer effectively having two ports (one to free space, and one to an appropriate superposition of guided modes). Optimal coupling occurs when the two rates are equal, which leads to near-perfect transmission approaching 100\% efficiency. Results in this direction are shown in Fig.~\ref{fig:CompleteCoupling}, from Ref.~\cite{Dias2020}. The coupling rates can be changed by changing the separation distance between the scatterer and the surface, and Fig.~\ref{fig:CompleteCoupling}(b,c) show coupling efficiencies well above 50\% for both a metallic half-space as well as a thin film. The ultimate limit of 100\% is nearly achievable with optimal focusing in tandem with a single parameter that controls coupling rates. The same approach should work equally well beyond plasmonics, for guided modes in dielectric or conducting systems.

\subsection{Communication channels} \label{sec:channels}
We end this chapter considering ``communication channels,'' a term pioneered by David Miller and colleagues~\cite{Miller2000-pr,Miller2019-hl,Miller2017}, which, in fact, constitute a beautiful extension of the approach used throughout the chapter. The central idea of the extremal-eigenvalue approach to wavefront shaping is that the single most effective source excitation for a given quadratic objective corresponds to the largest eigenvalue of the relevant matrix. Communication channels go beyond this and ask, if we can excite \emph{many} independent (or mutually incoherent) excitations, say $N$, then what excitations should we choose? The answer, roughly, corresponds to the $N$ extremal eigenvalues.
\begin{figure*}[tb]
\centering
\includegraphics[width=0.9\linewidth]{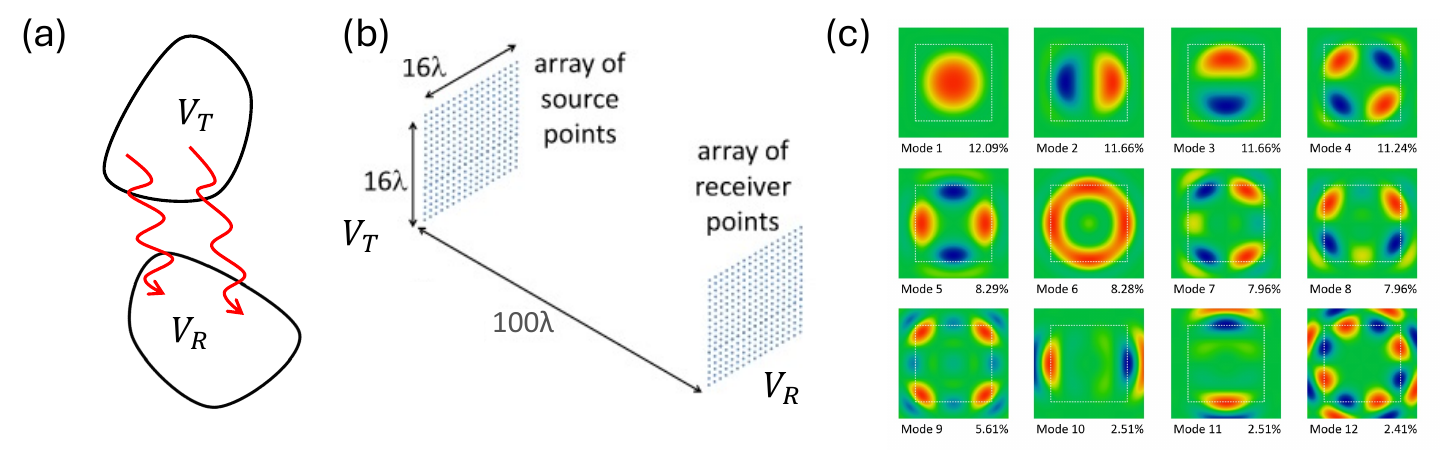}
\caption{\textbf{Communication channels.} (a) General setup: sources in a transmitting volume $V_T$ generate electromagnetic fields measured in a disjoint receiver volume $V_R$. (b) Specific configuration: two parallel $16\lambda \times 16\lambda$ arrays of $17\times17$ source/receiver points with spacing $\lambda$, separated by $100\lambda$. (c) The first 12 communication modes of this configuration, computed with the scalar Green's function: real part of the field that each source singular vector produces in the receiver plane, normalized to its maximum (dashed square: receiver array). Percentages are each mode's squared singular value as a percentage of the sum of all squared singular values ($=\sum_{ij}|T_{ij}|^2$). (Panel (a) adapted from Ref.~\cite{Miller2000-pr}; panel (b) adapted from Ref.~\cite{Miller2019-hl}, whose figure labels the separation $50\lambda$ although its coupling percentages appear to correspond to $100\lambda$; panel (c) is our computation following Ref.~\cite{Miller2019-hl}.)}
\label{fig:CommChannels}
\end{figure*}

More specifically, consider sources in a transmitting volume $V_T$ generating electromagnetic fields in a receiver volume $V_R$, as depicted in Fig.~\ref{fig:CommChannels}(a), and assume that the received fields can be coherently measured. This linear process, for any background environment, can be described by a convolution of the Green's function $\Gamma(\xv,\xv')$ with the sources $s(\xv)$: $\psi(\xv_R) = \int_{V_T} \Gamma(\xv_R,\xv_T) s(\xv_T)\,{\rm d}\xv_T$. As long as the two regions are disjoint, the collective Green's-function operator is ``compact'' (cf. Ref.~\cite{Miller2019-hl}) and can be described with arbitrary accuracy in a finite-dimensional discretization. Then the linear relationship from transmitting to receiver volume becomes a matrix equation,
\begin{align}
    c_{\rm receive} = T c_{\rm transmit},
    \label{eq:TransMatrix}
\end{align}
with $c_{\rm transmit}$ an $N_t \times 1$ vector, $c_{\rm receive}$ an $N_r \times 1$ vector, and $T$ an $N_r \times N_t$ ``transmission'' matrix. 

If one were to want to communicate information from $V_T$ to $V_R$, then a critical determinant of the information capacity would be the number of ``channels'' across which information can be simultaneously transmitted. To transmit in parallel, one needs orthogonal input vectors, and to receive information, one needs orthogonal measurement vectors~\footnote{Nonzero overlaps in input vectors reduce distinguishability and immediately lose information. Nonzero overlaps in output vectors mix inputs that were intended to be independent, again irrecoverably losing information.}. These are exactly the \emph{right} and \emph{left} singular vectors, respectively, of the matrix $T$. Moreover, the ``strength'' of the signal measured in a particular left singular vector for a unit-amplitude excitation of the corresponding right singular vector is the \emph{singular value} associated with the pair of vectors. Denoting the right and left singular vectors and corresponding singular values as $\sigma_{r,i}$, $\sigma_{l,i}$, and $\lambda_i$, respectively, the matrix $T$ can be written in its singular value decomposition:
\begin{align}
    T = \sum_i \lambda_i \sigma_{l,i} \sigma_{r,i}^{\dagger}
\end{align}
The tuples $\{\sigma_{r,i},\sigma_{l,i},\lambda_i\}$, for each $i$, are the \textbf{communication channels} of the system described by \eqref{eq:TransMatrix}, ordered from largest singular value to smallest. It should be clear from the preceding discussion that if one wants to communicate $N$ bits in parallel in such a system, they should select the first $N$ singular values of the $T$ matrix, excite the $N$ corresponding right singular vectors, and perform measurements on the $N$ corresponding left singular vectors. (Sometimes it is argued that orbital angular momentum and related wavefronts offer ``new'' degrees of freedom to enhance information capacities~\cite{Wang2012-OAM,Willner2015-OAM}. It is always beneficial to increase the total number of orthogonal functions that can be excited and measured at the source and measurement locations. Yet OAM offers no intrinsic advantage over other orthogonal profiles, and the global optimum will always be the fields comprising the communication channels.)

A more mathematical description of this intuition can be given. Suppose one can send any signal $c_{\rm transmit}$ with unit ``energy'' ($c_{\rm transmit}^{\dagger} c_{\rm transmit}$) and wants to maximize the total energy in the receiver volume. What signal should they send? The energy received is
\begin{align}
    c_{\rm receive}^{\dagger} c_{\rm receive} = c_{\rm transmit}^{\dagger} T^{\dagger} T c_{\rm transmit},
\end{align}
whose maximum is the largest eigenvalue of $T^{\dagger} T$, which is $\lambda_1^2$, with optimal excitation the first eigenvector of $T^{\dagger} T$, which is exactly the first right singular vector of $T$, $\sigma_{r,1}$. The receiver might then ask if they can measure the fields in a single coherent basis $\phi$ and capture all of the energy in that single basis? The optimal basis maximizes
\begin{align}
    \left| \phi^{\dagger} c_{\rm receive} \right|^2 = \left| \phi^{\dagger} T c_{\rm transmit} \right|^2 = \left| \phi^{\dagger} T \sigma_{r,1} \right|^2 = \lambda_1^2 \left| \phi^{\dagger} \sigma_{l,1} \right|^2.
\end{align}
If the receiver measures in precisely the basis of the first left singular vector, $\sigma_{l,1}$, then they will receive the full maximum of the energy transmission. The reasoning then proceeds in chain-like fashion: if one asks what \emph{two} independent (orthogonal) signals will maximize total energy delivery to the receiver, the first signal reasoning is identical, the second signal now must be orthogonal to the first, and the corresponding optimal excitation is exactly the second right singular vector, with optimal measurement in the second left singular vector. And so on. Hence, mathematically, communication channels seamlessly generalize the extremal-eigenvalue principle of wavefront shaping. Physically, they are the independent ``streams'' on which information can optimally be communicated.

In free space, the analytical and computational study of communication channels has historically focused on canonical configurations---most prominently paraxial rectangular/circular apertures, whose optimal modes are prolate spheroidal waves~\cite{Slepian1961,Landau1961,ToraldoDiFrancia1969}---which exhibit rapid, apparently exponential fall-off of coupling strengths past some number of well-coupled channels. A unified physical picture for this fall-off was recently provided in Ref.~\cite{DMiller2025}: spherical waves whose transverse variation on a bounding sphere is faster than a wavelength cannot directly propagate, and instead must tunnel through an effective angular-momentum barrier out to an ``escape radius'' beyond which they propagate; the channel-strength fall-off past some critical number is quasi-exponential precisely because of this tunneling. Complementary, shape-independent bounds on coupling strengths and information capacities between any two domains separable by a spherical surface were derived in Ref.~\cite{Kuang2025}, which showed that the optimal decay is in fact sub-exponential, and in three dimensions even slower (as the square root of the channel number), with domain size and configuration---rather than shape---controlling the achievable channel count. Closely related are fundamental bounds on MIMO antennas and the associated degrees of freedom, derived through current-based optimization and Q-factor constraints~\cite{ehrenborg2018fundamental_MIMO,ehrenborg2020physical_MIMO,ehrenborg2021capacity_MIMO_DOF}.

More broadly, the singular-value structure of $T$ provides a natural bridge to information-theoretic descriptions of imaging and sensing systems. Recent work by Waller and collaborators~\cite{pinkard2025information} develops a general framework for information-driven design of imaging systems, in which the number and quality of available communication channels effectively delimits the achievable mutual information between object and measurement. Sharpening this connection between the wave-physics-based channel counts of this section and information-theoretic figures of merit appears to us an especially promising direction.

\subsection{Optical thickness bounds} \label{sec:thickness}
\begin{figure*}[tb]
\centering
\includegraphics[width=0.99\linewidth]{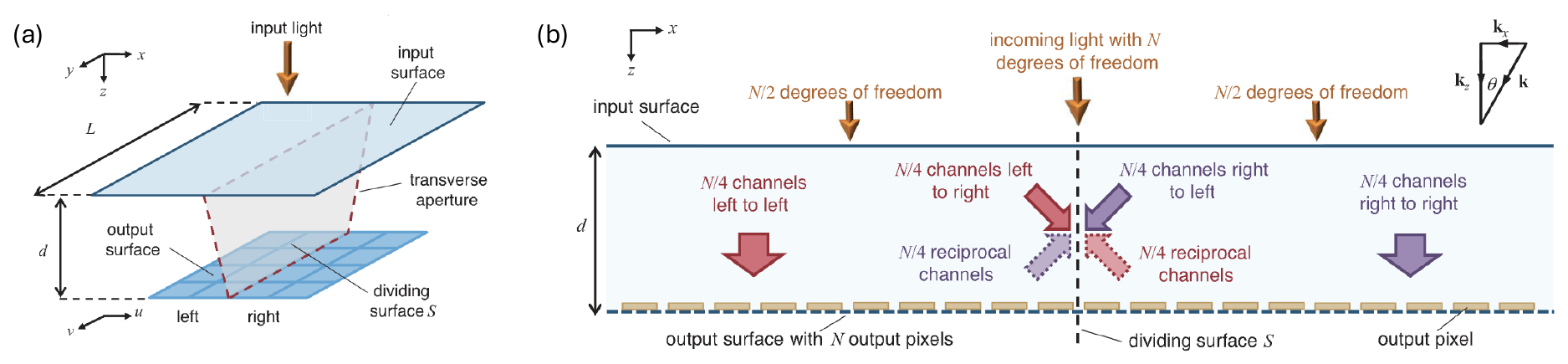}
\caption{\textbf{Overlapping nonlocality and thickness bounds.} (a) An optical system with input and output surfaces separated by distance $d$. A dividing surface $S$ cuts through both, defining a transverse aperture. (b) For an imager with $N$ output pixels split into left and right halves, the channels required decompose into four classes: $N/4$ left-to-left, $N/4$ right-to-right, and $N/4$ each crossing the dividing surface left-to-right and right-to-left. The $N/2$ cross-cutting channels are the overlapping nonlocality $C$. Diffraction then sets a minimum thickness $h \gtrsim C\lambda_0/(2\alpha n_{\max})$, with $\alpha$ the available $k$-space fraction and $n_{\max}$ the maximum available refractive index. (Figure adapted from Ref.~\cite{miller2023optics}.)}
\label{fig:OpticalThicknessBounds}
\end{figure*}

A novel twist on the idea of communication channels, which enabled resolution of a key outstanding problem in nanophotonics, was developed by David Miller in Ref.~\cite{miller2023optics}. Communication channels are essentially always defined at the input or output of an optical device. But in a linear wave system, if the wave evolution proceeds in ``sequential'' steps~\footnote{Allowing for possible ``reverberations,'' i.e., multiple scattering.}, interior boundaries can be understood as intermediate inputs/outputs, and can impose constraints in the same way that a pipe constricting at its center can impede water flow beyond the constraints of its input/output facets. For an optical device, this interior ``boundary'' can be an imaginary surface running perpendicular to the device's input/output planes, as depicted in Fig.~\ref{fig:OpticalThicknessBounds}(a). If inputs on the left half of the surface must become outputs on the right half (or vice versa), then independent waves must have passed through the dividing plane at some point in the process. The minimum number of such cross-cutting channels, Fig.~\ref{fig:OpticalThicknessBounds}(b), is termed the \emph{overlapping nonlocality} $C$, and is fixed by the input/output specification alone, before any design choice. Combining $C$ with a diffraction-limit heuristic yields a minimum thickness $h \gtrsim C\lambda_0/(2\alpha n_{\max})$ for one-dimensional nonlocality (e.g., in 2D devices), and a minimum transverse-aperture area $A \gtrsim C\left[\lambda_0/(2\alpha n_{\max})\right]^2$ for two-dimensional nonlocality (in 3D devices).

The $1/n_{\max}$ dependence is perfectly sensible: optical systems with access to higher refractive indices can pack more functionality per unit thickness. This dependence provides a direct quantitative connection between the material-bounds questions of Chapter~\ref{sec:bounds_materials} and the device-level constraints discussed here, and motivates the continued search for high-index dielectric platforms.

A related thickness bound was developed earlier in Ref.~\cite{Li2022}, specifically for wide-field-of-view (FOV) metalenses. Their approach quantifies the ``degree of nonlocality'' of a desired angle-dependent response through the lateral spreading $\Delta W$ of the transmission matrix in the spatial basis, and conjectures an empirical inequality $h \gtrsim \Delta W_{\max}$, which they observe to hold across a wide range of full-wave simulations of random multi-layer metasurfaces. For a wide-FOV lens this yields an explicit dependence on the lens diameter $D_{\rm out}$, numerical aperture $\rm NA$, and FOV. The underlying physics treats optical propagation as a \emph{diffusion process} --- a standard framing in disordered-media physics (e.g., the angular memory effect) but unconventional in the design of nanophotonic structures, where coherent wave dynamics are typically dominant. Like Miller's general framework, the Li \& Hsu bound is not derived rigorously, but inverse-designed multilayer metasurfaces have been observed to approach the predicted minimum thickness~\cite{Li2022}, providing compelling empirical evidence of tightness in those settings.

\begin{figure*}[tb]
\centering
\includegraphics[width=0.99\linewidth]{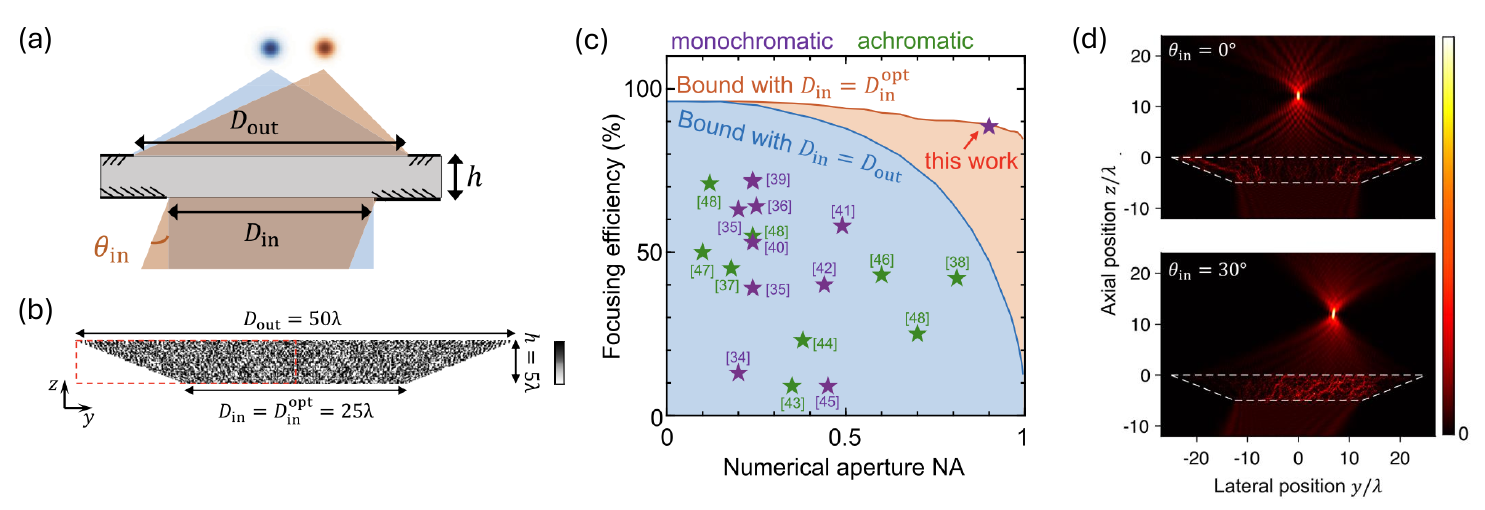}
\caption{\textbf{Active use of thickness and efficiency bounds in metalens design.} (a) A wide-FOV metalens whose input aperture $D_{\rm in}$ is intentionally smaller than its output aperture $D_{\rm out}$, mirroring conventional ray-optical imagers. (b) An inverse-designed monochromatic metalens with $D_{\rm out}=50\lambda$, $D_{\rm in}=D_{\rm in}^{\rm opt}=25\lambda$, thickness $h=5\lambda$ set at the Li \& Hsu thickness bound. (c) Focusing efficiency versus NA: the conventional bound with $D_{\rm in}=D_{\rm out}$ (lower, blue) is far below the bound with $D_{\rm in}$ optimized to maximize the efficiency limit (upper, orange); literature wide-FOV metalens designs (markers) lie below the lower bound, while the design here (red arrow) reaches the upper bound at NA $=0.9$. (d) Focal-plane intensity profiles at $\theta_{\rm in}=0^\circ$ and $30^\circ$ for the inverse-designed metalens, demonstrating diffraction-limited focusing across a $60^\circ$ FOV. (Figure adapted from Ref.~\cite{Li2024}.)}
\label{fig:LiHsuBoundsDesigns}
\end{figure*}

The bounds described above have largely been used to \emph{evaluate} existing designs. In Ref.~\cite{Li2024}, they are used more actively, woven into the design process itself: the Li \& Hsu thickness bound sets the device thickness, and the channel-averaged efficiency bound of Ref.~\cite{Li2023} determines the optimal aperture geometry. A central observation is that the conventional choice $D_{\rm in} = D_{\rm out}$, ubiquitous in the metasurface literature, is far from optimal --- the efficiency bound is maximized when the input aperture is significantly \emph{smaller} than the output aperture (Fig.~\ref{fig:LiHsuBoundsDesigns}(a)), mirroring the geometry of conventional ray-optical imagers and, to our knowledge, identified for the first time in the context of nanophotonic devices in Ref.~\cite{Li2024}. With this insight, a three-step recipe emerges: (i) choose $D_{\rm in}$ to maximize the efficiency bound, (ii) set $h$ at the thickness bound, and (iii) perform inverse design. Applied to a high-NA wide-FOV metalens (NA $= 0.9$, FOV $= 60^\circ$), the resulting design achieves a transmission efficiency of $98\%$ and a Strehl ratio of $92\%$ across the full FOV --- saturating the maximized bound and substantially exceeding all prior wide-FOV metalens designs (Fig.~\ref{fig:LiHsuBoundsDesigns}(c)).

The success of the active-design recipe above suggests that the bounds of Refs.~\cite{miller2023optics,Li2022} are tight enough in some regimes to function as design targets. But there is a substantial gap between the empirical evidence of tightness and a rigorous theory that would predict which device classes saturate the bounds and which do not.

\begin{openq}
    Can the heuristic thickness bounds above be derived rigorously? And which classes of devices realize or violate them --- particularly multilayered metasurfaces, where empirical tightness has been demonstrated in specific cases?
\end{openq}

More broadly, the channel-counting framework of Refs.~\cite{miller2023optics,Li2022} reframes a basic design question --- how thick must this device be? --- as something computable from the input/output specification alone, before any structure is drawn. We included these bounds here given their natural connection to the communication-channel ideas of Sec.~\ref{sec:channels}, but they would fit equally well in the next section, as bounds from global symmetries.

\section{Bounds on electromagnetic scattering from global symmetries} \label{sec:GlobalSym}
In this chapter, we consider scattering processes in \emph{designable} domains, expanding our purview beyond the fixed scatterers of the previous chapter. Here we focus on bounds that arise from ``global symmetries'' (e.g., reciprocity, time-reversal invariance) and their consequent restrictions on scattering matrices. These restrictions are generally independent of the volume and materials of the underlying design region, other than symmetries. Hence they cannot typically be used for scaling laws (which emerge in Chapter~\ref{sec:Scatt_Local}), but they form a useful starting point for delineating basic feasible/infeasible scattering response.

\subsection{Extinction symmetries in reciprocal systems}
\begin{figure*}[tb]
\centering
\includegraphics[width=1\linewidth]{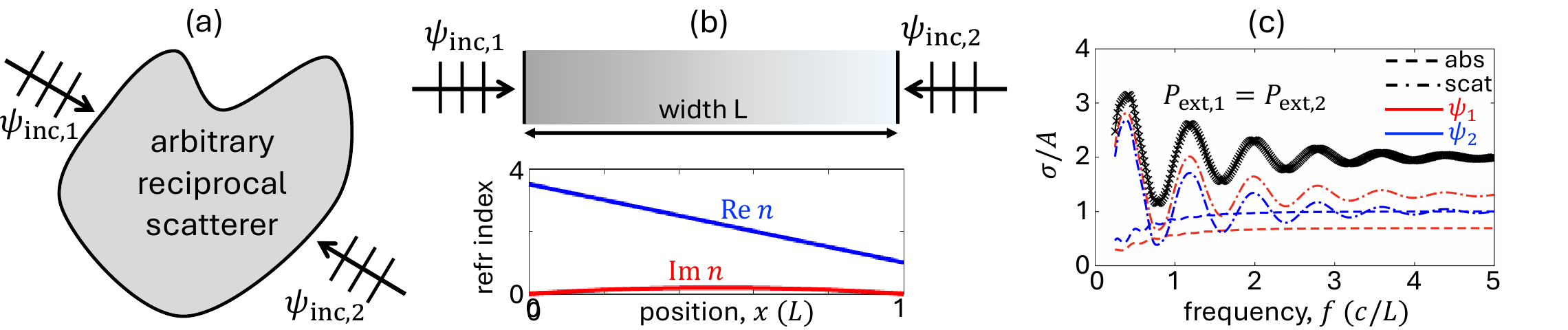}
\caption{\textbf{Reciprocity constraints on extinction.} (a) For any reciprocal scatterer (e.g. $\varepsilon = \varepsilon^T$ and $\mu = \mu^T$), extinction between two oppositely directed plane waves must be exactly equal, even if the absorption and scattering constituents take quite different values from the two directions. This result generalizes to any time-reversed incident waves such that $\psi_{\rm inc,2} = \psi_{\rm inc,1}^*$. (b) Example with a one-dimensional gradient-index scatterer. (c) Cross-sections per area, $\sigma/A$, for geometry in (b), exhibiting unequal absorption and scattering values but exactly equal extinction, at all frequencies.}
\label{fig:ExtSym}
\end{figure*}
We first discuss an elegant result from Sounas and Al\`{u}~\cite{Sounas2014}. The extinction of a plane wave with polarization $\hat{\sv}$ propagating in the $z$ direction by \emph{any} scatterer comprising reciprocal materials, is exactly equal to the extinction by that same scatterer of a time-reversed plane wave, with polarization $\hat{\sv}^*$, propagating in the $-z$ direction. This follows immediately from the optical theorem (Sec.~\ref{sec:conslaws}) and the $\TT$-matrix description of scattering (Sec.~\ref{sec:linearity}). The optical theorem dictates that the extinction in the two scenarios are
\begin{subequations}
\begin{align}
    P_{\textrm{ext},1} &= \frac{\omega}{2} \Im \int \psi_{\rm inc,1}^{\dagger} (\xv) \phi_{1}(\xv) \,{\rm d}\xv \\
    P_{\textrm{ext},2} &= \frac{\omega}{2} \Im \int \psi_{\rm inc,2}^{\dagger} (\xv) \phi_2(\xv) \,{\rm d}\xv,
\end{align}
\end{subequations}
where $\phi_{1,2}$ are the polarization fields induced by the two incident fields, $\psi_{\rm inc,1}$ and $\psi_{\rm inc,2} = \psi_{\rm inc,1}^*$ (by time reversal). The polarization fields are convolutions of the $\TT$ matrix with the incident fields; in compact vector notation, we have:
\begin{subequations}
\begin{align}
    P_{\textrm{ext},1} &= \frac{\omega}{2} \Im \left[ \psi_{\rm inc,1}^{\dagger} \TT \psi_{\rm inc,1} \right] \\
    P_{\textrm{ext},2} &= \frac{\omega}{2} \Im \left[ \psi_{\rm inc,2}^{\dagger} \TT \psi_{\rm inc,2} \right] = \frac{\omega}{2} \Im \left[ \psi_{\rm inc,1}^{T} \TT \psi_{\rm inc,1}^* \right] \\
    &= \frac{\omega}{2} \Im \left[ \psi_{\rm inc,1}^{\dagger} \TT^T \psi_{\rm inc,1} \right] = \frac{\omega}{2} \Im \left[ \psi_{\rm inc,1}^{\dagger} \TT \psi_{\rm inc,1} \right] = P_{\rm ext,1},
\end{align}
\end{subequations}
where the second equality uses the time-reversal symmetry of the incident waves (not necessarily the medium), the third takes the transpose of the entire expression, and the fourth uses the reciprocity of the $\TT$ matrix.

Intuitively: extinction is proportional to the imaginary part of the forward-scattering amplitude, and the forward-scattering amplitudes of time-reversed plane waves are reciprocal partners. It should be clear that the incident wave need not be a plane wave; the derivation above only requires that the second incident wave be the time reversal of the first. Nor is the derivation specific to electric fields or isotropic materials; any linear (magnetic, bi-anisotropic, etc.) system is subject to the above constraint. It is also important to remember that extinction is the sum of absorption and scattering, which need not individually be equal for the two incident waves. A Kindle e-reader displaying a blank page in a book may be dominated by scattering relative to absorption for light incident from above (e.g., the sun), whereas absorption may dominate relative to scattering at its black rear surface, producing a vivid visual asymmetry. But the extinction of the waves incident from the front and rear must be exactly equal.

This is illustrated in Fig.~\ref{fig:ExtSym}: a one-dimensional gradient-index scatterer (panel b) produces strongly asymmetric absorption and scattering cross-sections under front- vs. rear-incident plane waves, but its extinction is exactly equal in the two cases at every frequency (panel c). In this example, the slab's real refractive index decreases linearly from $n = 3.5$ at the front face to $n = 1$ at the rear (matched to free space), with a parabolic imaginary part peaking at $\Im n = 0.2$ in the slab interior; the strong front-side reflection vs.\ smooth rear-side coupling produces the front-dominated-scattering / rear-dominated-absorption asymmetry visible in panel~(c).

\begin{openq}
Reciprocity equates the extinctions of two time-reversed incident waves, but the individual absorption and scattering cross sections can differ markedly (cf.~Fig.~\ref{fig:ExtSym}). Are there fundamental bounds on this absorption--scattering asymmetry under time-reversed illumination?
\end{openq}

\subsection{Time-reversal symmetry constraints}
In a similar vein, Sounas and Al\`{u} derived bounds on field amplitude asymmetry ratios in linear time-reversal-symmetric scatterers. Consider a system with $N$ incoming/outgoing channels (``ports''). A natural question might be: if I excite the system from channel $A$ and generate field amplitude $\Ev_A$ at any point internal to the scatterer, are there any constraints on field amplitude $\Ev_B$ at the same point, but now excited from channel $B$? One might think not, but if the excitation from $A$ leads to nonzero outgoing-wave amplitude in channel $B$, then the asymmetry ratio is finite and bounded. 

Before developing the mathematical bounds, we can develop intuition around the question. Consider first the scenario in which $100\%$ of the energy excited in channel $A$ transmits to channel $B$. Then any unit-amplitude excitation from channel $B$ is in fact the time-reversed excitation of the original scenario (up to an unimportant global phase), and the scatterer's time-reversal symmetry requires $\Ev_B$ to have the same amplitude as $\Ev_A$. Going beyond this unit-transmission scenario to one in which that transmission to channel $B$ is large (say $90\%$) but not unity, the unit-amplitude excitation is no longer equivalent to the time-reversed of the original scenario. However, by linearity, it can be decomposed into the time-reversed scenario with amplitude $\sqrt{0.9}\approx 0.95$, producing $95\%$ of the original amplitude in $\Ev_B$, plus an orthogonal excitation carrying the remaining $10\%$ of the power. This reasoning holds at any fraction, not only near unity, and leads to asymmetry bounds that increase as the channel transmission from $A$ to $B$ decreases.

\begin{figure*}[tb]
\centering
\includegraphics[width=1\linewidth]{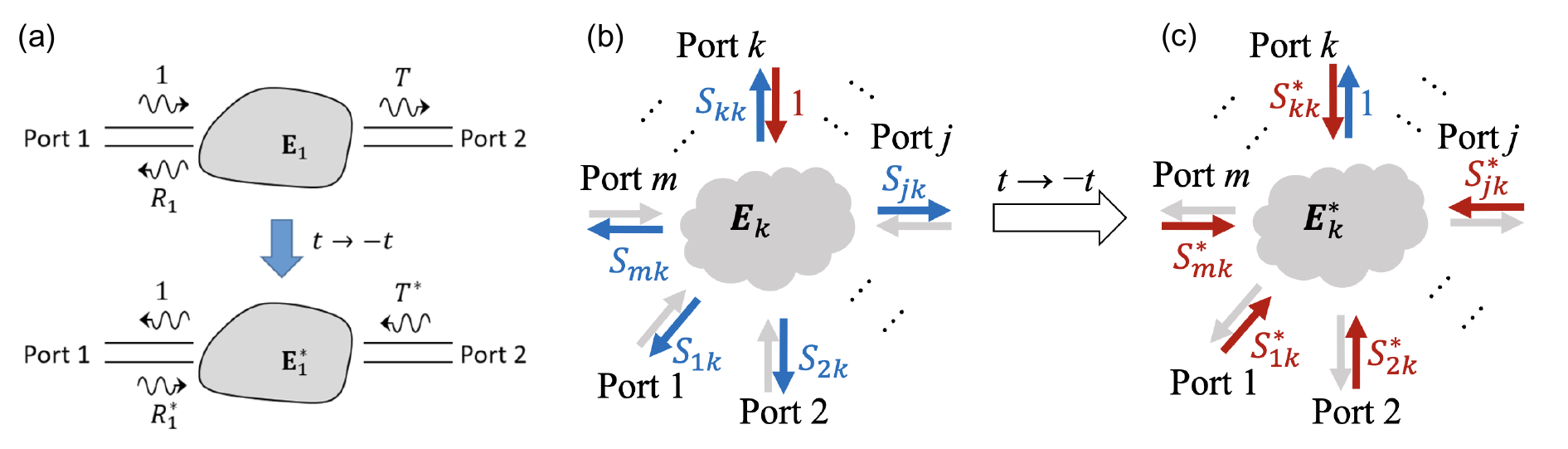}
\caption{\textbf{Time-reversal-symmetry field-ratio bounds.} (a-c) For a scatterer satisfying time-reversal symmetry, the magnitude of the internal field must be unchanged upon reversing all port power flows, both for two-port systems (panel a) and multi-port systems (panels b,c). (Panel (a) adapted from Ref.~\cite{Sounas2017}; panels (b,c) adapted from Ref.~\cite{Ma2025}.)}
\label{fig:TRRatioBounds}
\end{figure*}
We can now develop the mathematical reasoning that pairs with the intuitive explanation. Consider a lossless two-port system which, for simplicity, has reflection and transmission coefficients $R_1$ and $T$ for excitations from the first port, as depicted in Fig.~\ref{fig:TRRatioBounds}. Because the system is lossless, the reflection and transmission coefficients from the second port must be equal in amplitude to those from the first. The key observation is that if a unit-amplitude excitation from port 1 produces field $\Ev_{1}$ at any position, then the time-reversed excitation, given by a linear combination of $T^*$ at port 2 and $R_1^*$ at port 1, must produce everywhere within the scattering region exactly the time-reversed field $\Ev_1^*$. Hence,
\begin{align}
    \Ev_1^* = T^* \Ev_2 + R_1^* \Ev_1,
\end{align}
where $\Ev_2$ is the field produced by a unit-amplitude excitation at port 2. Clearly one can see if $T=1$ and $R_1=0$, then the amplitudes of $\Ev_1$ and $\Ev_2$ must be identical; and, more generally, their ratio can be bounded. A few algebraic steps (cf. SM of Ref.~\cite{Sounas2017}) leads to precisely such a bound:
\begin{align}
    \frac{1-R}{1+R} \leq \frac{|\Ev_2|^2}{|\Ev_1|^2} \leq \frac{1+R}{1-R},
    \label{eq:TRTwoPort}
\end{align}
where $R=|R_1|=|R_2|$. For reflection coefficients with near-zero amplitude, the field-intensity ratio is tightly bounded near one; as the reflection increases, so too can the field-intensity ratio.

This result can be generalized beyond two ports, as is done in Ref.~\cite{Sounas2017}. But it can also be generalized to a wide range of quadratic forms, beyond single-point field intensity ratios, as was developed recently by Ma, Pestourie, and Johnson~\cite{Ma2025}. We jump directly to their generalization. A unit-amplitude excitation from port $k$ will lead to amplitudes $S_{jk}$ in all ports $j$, and produce field $\Ev_k$ within the scatterer. The time-reversed excitations have amplitudes $S_{jk}^*$ in each port $j$, and produce internal field $\Ev_k^*$. These excitations are depicted in Fig.~\ref{fig:TRRatioBounds}(b). Hence the fields $\Ev_k$ must satisfy,
\begin{align}
    \Ev_k^* = \sum_{j} S_{jk}^* \Ev_j,
\end{align}
for each $k$. These separate constraints can be compiled into a larger constraint by forming larger vectors comprising each of the $\Ev_k$ in sequence. They can be generalized by allowing any real linear operation denoted by $L$ to act on the fields (e.g., selection of a specific point in space, or polarization). Forming the vector $\mathcal{E} = \begin{pmatrix}L \Ev_1 & L \Ev_2 & \ldots \end{pmatrix}^T$, the collective time-reversal symmetry constraint can be written
\begin{align}
    \mathcal{E}^* = S^{\dagger} \mathcal{E},
    \label{eq:ETRConstraint}
\end{align}
where $S$ is to be understood as a Kronecker product of the scattering matrix with the appropriate identity matrix~\cite{Ma2025}. Beyond local field-intensity ratios, one can bound any ratio of quadratic forms of $\mathcal{E}$ (with $W$ and $V$ Hermitian), provided the denominator is positive for every nonzero admissible $\mathcal{E}$:
\begin{align}
    f = \frac{\mathcal{E}^{\dagger}W\mathcal{E}}{\mathcal{E}^{\dagger}V\mathcal{E}}.
\end{align}
Treating $\mathcal{E}$ as an arbitrary complex vector, one can bound this ratio through extremal eigenvalues of the generalized eigenproblem $Wx=\lambda Vx$. Yet $\mathcal{E}$ is not an arbitrary complex vector; time-reversal symmetry requires that it must satisfy \eqref{eq:ETRConstraint}. 

\begin{figure*}[tb]
\centering
\includegraphics[width=1\linewidth]{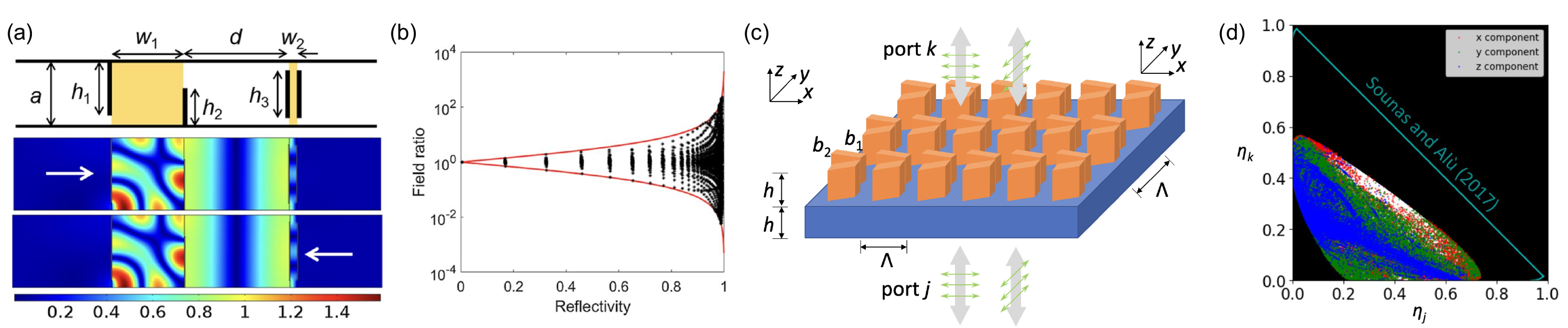}
\caption{\textbf{Numerical validation of the time-reversal-symmetry field-ratio bounds.} (a,b) Two-port system: an asymmetric multimodal cavity with an antireflection layer brings the input reflection close to zero, producing nearly identical field intensities for excitation from opposite ports (a); sweeping frequency and spatial location traces the full envelope $(1\mp R)/(1\pm R)$ predicted by \eqref{eq:TRTwoPort} (red curves in panel b). (c,d) Multi-port system: a square-lattice grating of silicon-nitride pillars on silica forms a four-port scatterer (c), whose simulated intensity fractions $\eta_k = \|\Ev_k\|^2/\sum_j \|\Ev_j\|^2$ for all three field components (red, green, blue) fill the feasible region predicted by \eqref{eq:fTRRatio} (white area in panel d), lying well inside the Sounas--Al\`{u} bound $\eta_j + \eta_k \le 1$ (cyan line). (Panels (a,b) adapted from Ref.~\cite{Sounas2017}; panels (c,d) adapted from Ref.~\cite{Ma2025}.)}
\label{fig:TRRatioExamples}
\end{figure*}

A particularly useful interpretation of \eqref{eq:ETRConstraint} is that the real and imaginary parts of $\mathcal{E}$ are not independent quantities. If we separate the real and imaginary parts of \eqref{eq:ETRConstraint} and solve for the imaginary part (assuming $\mathcal{I} + S_r^T$ is invertible), we find
\begin{align}
    \mathcal{E}_i = \left[\mathcal{I} + S_r^T \right]^{-1} S_i^T \mathcal{E}_r = F_S^T \mathcal{E}_r,   
\end{align}
where the subscripts denote entrywise real/imaginary parts. This leads to a new generalized Rayleigh quotient in terms of only the real part of $\mathcal{E}$,
\begin{align}
    f = \frac{\mathcal{E}_r^T \left[\mathcal{I} - iF_S^*\right] W \left[\mathcal{I} + iF_S^T\right]\mathcal{E}_r}{\mathcal{E}_r^T \left[\mathcal{I} - iF_S^*\right] V \left[\mathcal{I} + iF_S^T\right]\mathcal{E}_r} = \frac{\mathcal{E}_r^T W' \mathcal{E}_r}{\mathcal{E}_r^T V' \mathcal{E}_r}.
    \label{eq:fTRRatio}
\end{align}
Now, the time-reversal symmetry constraints on the fields are built directly into the quadratic forms themselves, and one can find bounds on the ratio of the two quadratic forms through the extremal eigenvalues of the real symmetric generalized eigenproblem $W'_r x = \lambda V'_r x$ over real $x$, where the subscript again denotes entrywise real parts: because $\mathcal{E}_r$ is real, the antisymmetric entrywise imaginary parts of $W'$ and $V'$ drop out of the quadratic forms, and solving the complex problem $W'x=\lambda V'x$ instead would return the unconstrained bound. Positivity of the denominator on the admissible fields means that $V'_r$ is positive definite; this is guaranteed if $V$ is positive definite, as for the total intensity that normalizes the fractions $\eta_k$ in Fig.~\ref{fig:TRRatioExamples}, and for the two-port ratio of \eqref{eq:TRTwoPort} it holds whenever $T \neq 0$. Without it the ratio can be unbounded, and the extremal eigenvalues are not bounds.

The bounds of Eqs.~(\ref{eq:TRTwoPort}) and (\ref{eq:fTRRatio}) have both been numerically validated. Sounas and Al\`{u}~\cite{Sounas2017} considered a multimodal cavity in a parallel-plate waveguide, fitted with an antireflection layer that brings the input reflection close to zero [Fig.~\ref{fig:TRRatioExamples}(a)]; the two field-intensity maps for excitation from opposite ports are essentially indistinguishable, the $R \to 0$ saturation of the two-port bound. Sweeping frequency and spatial location to span the full range $R \in [0,1]$ then traces out the envelope of the bound, as shown in panel (b). Ma, Pestourie, and Johnson~\cite{Ma2025} validated their generalized bound on a four-port system formed by a doubly-periodic grating of silicon-nitride pillars [panel (c)]; the simulated intensity fractions $(\eta_j, \eta_k)$, for all three field components and many sampling locations, fill the feasible region predicted by \eqref{eq:fTRRatio} (white area in panel (d)) but lie well inside the looser Sounas--Al\`{u} bound $\eta_j + \eta_k \le 1$ (cyan line), demonstrating that the new bound is meaningfully tighter. How these bounds evolve with the number of ports remains largely unexplored. Ports that are decoupled from the scatterer leave them unchanged, and once additional ports couple only to combinations of the same radiating fields of a finite scatterer (\secref{channels}), further ports should add little; where that crossover occurs likely depends on resonance quality factors, nearby scatterers, and the port geometry. Related questions about how design complexity scales with the number and function of ports, such as the minimal number of tunable parameters needed to route waves among many channels, are beginning to be addressed~\cite{alhulaymi2025coherent}.

\subsection{Scattering matrices: Changing the basis}
Conventionally, we associate a matrix change of basis with a unitary transform (e.g. $\mathbb{A} \rightarrow U^{\dagger} \mathbb{A} U$ for unitary $U$.) Yet this is not the correct transformation for a scattering matrix $\SS$. The rows and columns of a scattering matrix are defined by different basis functions---columns by incoming waves, rows by outgoing waves. To ensure simple symmetry conditions for time reversal and reciprocity, it is necessary for the outgoing-wave basis functions to be the time-reversed partners of the incoming waves. As a result, starting from a scattering matrix $\SS_1$ in a given set of incoming/outgoing waves, transforming to a new set of incoming/outgoing waves, defining $\SS_2$, by a change-of-basis matrix $U$, is in fact given by~\cite{Guo2023}
\begin{align}
    \SS_2 = U^T \SS_1 U,
    \label{eq:STransform}
\end{align}
with a transpose instead of a conjugate transpose on the first $U$. (To derive this, consider original input/output coefficients $\cout^{(1)}$ and $\cin^{(1)}$, such that $\cout^{(1)} = \SS_1 \cin^{(1)}$. We can write $\cin^{(1)} = U \cin^{(2)}$, in which case $\cout^{(1)} = U^{*} \cout^{(2)}$, by the time-reversal relation between incoming/outgoing-wave basis functions. Then, $\cout^{(2)} = U^T \SS_1 U \cin^{(2)}$, defining $\SS_2$.) Given the nonunitary nature of \eqref{eq:STransform}, a natural question is whether the choice of basis dictated by $U$ can have meaningful physical consequences? This question naturally fits into this broader section, as the basis for a scattering matrix represents the excitation ports of a (fixed) scattering system, as well as the additional choice of outgoing-wave measurement basis.

First, as expected, a change of basis \emph{cannot} change any of the fundamental global symmetries of the scattering matrix. If the original scattering matrix $\SS_1$ is lossless, reciprocal, and/or time-reversal invariant (TRI), the transformed scattering matrix $\SS_2$ will be as well:
\begin{subequations}
    \begin{align}
        \SS_1^{\dagger} \SS_1 = \mathbb{I} &\rightarrow \SS_2^{\dagger} \SS_2 = \mathbb{I} \quad \textrm{ (lossless)} \\
        \SS_1^{T} = \SS_1 &\rightarrow \SS_2^{T} = \SS_2 \quad \textrm{ (reciprocal)} \\
        \SS_1^{*} \SS_1 = \mathbb{I} &\rightarrow \SS_2^{*} \SS_{2} = \mathbb{I} \quad \textrm{ (TRI)},
    \end{align}
\end{subequations}
all of which can be verified by direct substitution of \eqref{eq:STransform}.

More generally, a change of basis cannot change any of the singular values of the scattering matrix. The singular values of $\SS_1$ can be gleaned from the eigendecomposition of $\SS_1^{\dagger} \SS_1$, which is given by $V \Sigma_1 \Sigma_1 V^{\dagger}$, where $\Sigma_1$ is the diagonal matrix comprising the singular values of $\SS_1$, and $V$ contains its right singular vectors. In the new basis, the singular values of the second scattering matrix $\SS_2$ can be found from
\begin{align}
    \SS_2^{\dagger} \SS_2 &= U^{\dagger} \SS_1^{\dagger} \SS_1 U \nonumber \\
    &= U^{\dagger} V \Sigma_1 \Sigma_1 V^{\dagger}  U \nonumber \\
    &= W \Sigma_1 \Sigma_1 W^{\dagger},
\end{align}
where $W$ is the \emph{unitary} matrix $U^{\dagger} V$. Hence the singular values of $\SS_2$ are identical to those of $\SS_1$. Physically, this makes sense: the singular values represent the number and strength of the input-output ``communication channels'' (discussed extensively in Sec.~\ref{sec:channels}), which cannot be modified from a change in basis.

\begin{figure*}[tb]
\centering
\includegraphics[width=0.6\linewidth]{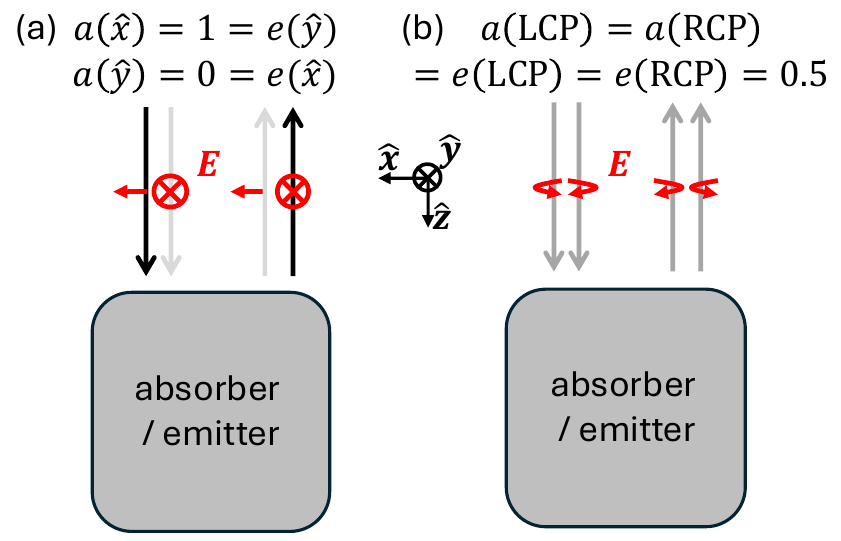}
\caption{\textbf{Kirchhoff's Law under a change of basis.} (a) Nonreciprocal absorber/emitter that perfectly absorbs $\hat{x}$-polarized waves and perfectly emits $\hat{y}$-polarized waves, demonstrating ideal violation of Kirchhoff's Law. (b) The same scatterer absorbs and emits equally into circularly polarized waves, now satisfying Kirchhoff's Law, simply by a change of basis.}
\label{fig:KirchhoffsLaw}
\end{figure*}
So what can a change of basis enable? Perhaps the key property that can change is the diagonal values of $\SS$, and corresponding matrices $\SS^{\dagger} \SS$ and $\SS \SS^{\dagger}$. These are the relevant quantities for thermal absorptivities and emissivities: defining absorptivity matrix $\mathbb{A} = \mathbb{I} - \SS^{\dagger}\SS$, the diagonal entries of $\mathbb{A}$ are the absorptivities for each scattering-matrix channel, and similarly the diagonal entries of $\mathbb{E} = \mathbb{I} - \SS \SS^{\dagger}$ are the thermal emissivities.

An interesting example in which absorptivity/emissivity non-trivially changes is an isolator-like nonreciprocal body as depicted in Fig.~\ref{fig:KirchhoffsLaw}(a) that perfectly absorbs $x$-polarized waves but emits only $y$-polarized waves. Such an emitter clearly violates Kirchhoff's Law (absorptivity equals emissivity in every channel). Right?

In fact, the violation of Kirchhoff's Law is basis-dependent. A simple change of basis creates a reciprocity-like scenario with Kirchhoff's Law satisfied. In the linear-polarization basis, the scattering, absorptivity, and emissivity matrices might be
\begin{align}
    \SS_1 = 
    \begin{pmatrix}
        0 & 1 \\
        0 & 0
    \end{pmatrix}, \quad
    \mathbb{A}_1 = \mathbb{I} - \SS_1^{\dagger} \SS_1 = 
    \begin{pmatrix}
        1 & 0 \\
        0 & 0
    \end{pmatrix}, \quad
    \mathbb{E}_1 = \mathbb{I} - \SS_1 \SS_1^{\dagger} = 
    \begin{pmatrix}
        0 & 0 \\
        0 & 1
    \end{pmatrix}.
    \label{eq:S1Matrices}
\end{align}
One can clearly see that the nonreciprocal scattering matrix leads to only $x$-polarized absorption and only $y$-polarization emission. We can transform to a circular-polarization basis by the unitary matrix
\begin{align}
    U = 
    \frac{1}{\sqrt{2}}
    \begin{pmatrix}
        1 & i \\
        i & 1
    \end{pmatrix},
\end{align}
which leads to scattering, absorptivity, and emissivity matrices given by
\begin{align}
    \SS_2 = 
    \frac{1}{2}
    \begin{pmatrix}
        i & 1 \\
        -1 & i
    \end{pmatrix}, \quad
    \mathbb{A}_2 = \mathbb{I} - \SS_2^{\dagger} \SS_2 = 
    \frac{1}{2}
    \begin{pmatrix}
        1 & i \\
        -i & 1
    \end{pmatrix}, \quad
    \mathbb{E}_2 = \mathbb{I} - \SS_2 \SS_2^{\dagger} = 
    \frac{1}{2}
    \begin{pmatrix}
        1 & i \\
        -i & 1
    \end{pmatrix}.
    \label{eq:S2Matrices}
\end{align}
Now, the emissivity and absorptivity matrices are equal, including along the diagonal, with equal values of 1/2. These diagonal values indicate equal 50\% absorptivities and emissivities from/to the circular-polarization basis functions, which makes sense: half of each incoming basis function is $\hat{x}$-polarized, and half of each outgoing basis function is $\hat{y}$-polarized, so each channel absorptivity and emissivity should be 1/2, as depicted in Fig.~\ref{fig:KirchhoffsLaw}(b).

Why does this nonreciprocal system violate Kirchhoff's Law? Whereas the original scattering matrix of \eqref{eq:S1Matrices} is nonreciprocal in \emph{amplitude}, the transformed scattering matrix of \eqref{eq:S2Matrices} is nonreciprocal in \emph{phase}, but ``reciprocal'' in amplitude: $\left|\SS^{(2)}_{21}\right| = \left|\SS^{(2)}_{12}\right|$. Kirchhoff's Law measures absorbed/emitted powers independent of phase, thus leading to the equality between absorptivity and emissivity matrices.

In a sequence of papers, Guo, Fan, and collaborators have derived extensive results identifying feasible and infeasible diagonal-matrix values through such change-of-basis transformations. Reflection coefficients are the diagonal entries of a scattering matrix and bounded in Ref.~\cite{Guo2022b}. Transmission values, defined as diagonal values of either a Green's function matrix $\mathbb{G}$ or its power analog, $\mathbb{G}^{\dagger} \mathbb{G}$, are bounded in Ref.~\cite{Guo2025}. Bounds to absorptivity, emissivity, and their contrast, generalizing the discussion above, are bounded in Ref.~\cite{Guo2023}. The reflection, transmission, and absorption metrics are extended to partially coherent systems in Refs.~\cite{Guo2024,Guo2024b}. In all cases, the key fact is that the magnitudes of the diagonal elements of a matrix are \emph{weakly majorized} by its singular-value vector (and, for Hermitian matrices such as $\mathbb{G}^{\dagger}\mathbb{G}$, majorized by its eigenvalues), leading to feasible sets for the relevant diagonal elements, existing in $n$-dimensional spaces for an $n$-port scattering matrix, defined by simple convex polytopes. In partially coherent systems, the relevant governing scattering equation is not $\cout = \SS \cin$ for coherent input/output vectors $\cout$ and $\cin$, but rather $\rho_{\rm out} = \SS \rho_{\rm in} \SS^{\dagger}$ (cf. Ref.~\cite{Zhang2019}), where $\rho_{\rm in,out}$ are density matrices (coherent-mode representations~\cite{Mandel1995}) describing the partial coherence.

\begin{openq}
The bounds above on absorptivity, emissivity, and their contrast follow from passivity alone (sub-unitarity of $\SS$). Can they be made tighter given material and volume constraints on the scatterer, perhaps in tandem with the methods of Chapter~\ref{sec:Scatt_Local}?
\end{openq}

\subsection{Brightness theorems for waves}
\label{sec:Brightness}
The \textbf{brightness theorem} in geometric optics is a statement about the ``incompressibility'' of a bundle of optical rays: {\'e}tendue, the differential spatial area and solid angle occupied by rays carrying a fixed amount of power, cannot be reduced in a passive linear system~\cite{Winston2018}. A simple consequence is that spatial concentration must be accompanied by angular spreading, and a corresponding limit to the angular range over which incoming rays can be spatially concentrated. The brightness theorem can be viewed more generally as a phase-space conservation law in linear systems, mathematically equivalent to Liouville's theorem~\cite{Goldstein1950}, which suggests the possibility for an analogous theorem in passive linear \emph{wave} systems as commonly used in photonics. 

In this section we describe such a ``brightness theorem'' for waves. Rays can be swapped for incoherent channels, {\'e}tendue can be defined by a channel count, while ``fixed power'' remains the same. We also describe simple generalizations to partially coherent cases and quantities of interest beyond power, such as focusing intensity (or spot size). Before describing the mathematical apparatus to quantify this behavior, we can describe some key intuitions.

\begin{figure*}[tb]
\centering
\includegraphics[width=0.95\linewidth]{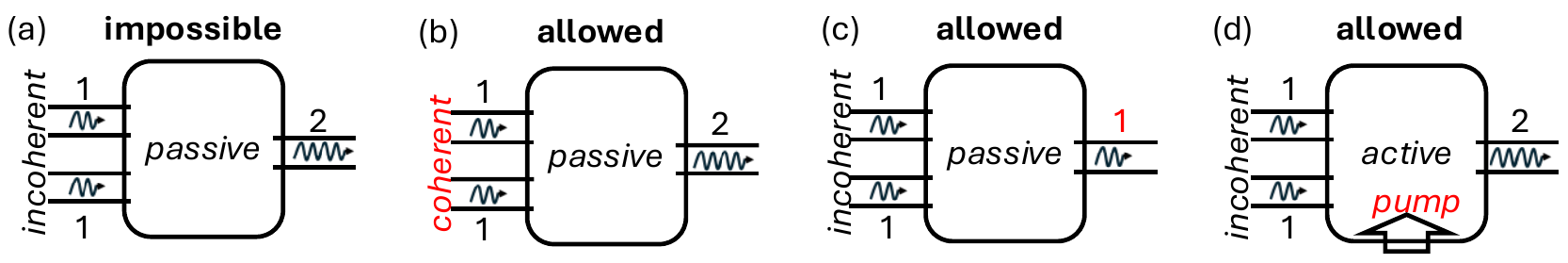}
\caption{\textbf{Beam combining: Allowed vs. impossible.} (a) Perfect incoherent beam combining in a passive linear system is impossible. Physically possible alternatives include: (b) coherent excitations, (c) imperfect beam combining, and (d) active media. In linear time-varying systems, inputs can share one output direction by occupying distinct frequency channels (see text)~\cite{wang2021spacetime}. Ports are power-carrying ``channels,'' and number labels refer to powers in each channel.}
\label{fig:BeamCombining}
\end{figure*}
Arguably, the core idea behind the brightness theorem is that ``beam combining'' is impossible in a passive linear system. One cannot excite two separate beams and combine them into a single one. Of course, one can excite distributed sources to create a strong focal point (as exemplified in the optimal-focusing discussion of Sec.~\ref{sec:FSFocus}), but they must be excited coherently to do so. Effectively, this is a single ``beam.'' A more precise statement is that two \emph{incoherent} beams cannot be combined into a single beam in a passive linear system. The term ``beam'' is also imprecise in a wave context: instead, one should refer to power-carrying \emph{channels} that can serve as basis functions for a scattering matrix. Plane waves, vector spherical waves, and ``communication channels'' are all examples.

Reciprocity offers a simple explanation of the prohibition of beam combining: suppose, in a reciprocal system, that one independently excites the two beams that combine into a single one, with 100\% fidelity. If one were to excite the single output in a reciprocal (or time-reversed) manner, then the two original input beams would have to be excited separately, both with 100\% efficiency, which is obviously impossible. Note the key role that perfect efficiency plays in this scenario; no law would be violated if half of the power of each beam were absorbed and the other half ``combined'' into one output beam. So a ``beam combiner'' that starts with 1 Watt of power in each of two beams and results in 1 Watt of power in a single beam is allowable, albeit of questionable utility. Here also we can appreciate the importance of the passivity condition: if gain is allowed, one can effectively absorb all of the power from one of the beams, while pumping new power into the desired output beam, in which case 2 Watts of incoming power on two beams could be converted to 2 Watts of outgoing power in a single beam, albeit at the cost of 1 Watt from the active system. These observations can help explain the possibilities, for example for loss-induced transparency~\cite{Ruter2010-oh}, greater-than-one reflectivity~\cite{Regensburger2012-fy}, and topological ``funneling''~\cite{Weidemann2020-yk} in gain-based PT-symmetric systems. 

The \textbf{prohibition on beam combining extends to non-reciprocal systems that are linear and passive}. Such systems must still obey ``generalized reciprocity,'' in which material susceptibilities are transposed with source and receiver definitions. This principle can be substituted into the above argument without modifying its conclusion. Fundamentally, it is the combination of linearity and passivity that prohibits efficient beam combining. Linear time-varying systems fit the same framework once frequency harmonics are treated as additional channels: a time modulation couples each incident frequency to other frequencies (Sec.~\ref{sec:symmetry_time}), so the scattering matrix connects (direction, frequency) channels. For example, Ref.~\cite{wang2021spacetime} designed a space--time-modulated metasurface that directs waves incident from two directions into a single outgoing direction, one at the incident frequency and the other shifted by the modulation frequency, with more than 99.9\% of the incident photon flux from each input reaching that direction. With amplitudes normalized to photon flux, its scattering matrix is unitary~\cite{wang2021spacetime}, and the two outputs occupy distinct frequency channels, one per incoherent input, exactly as the incompressibility statement developed below requires: concentration in angle is paid for with spread in frequency. In \emph{nonlinear} systems, by contrast, beam combining is of significant interest and very much allowed by fundamental principles~\cite{Pyrialakos2025-dh}.

A general mathematical formulation of ``incompressibility'' in passive linear systems can be described using the mathematics of \textbf{majorization}~\cite{Bhatia1997}. Consider a system with scattering matrix $\SS$ mapping input coefficients in a vector $\cin$ to output coefficients in $\cout$, $\cout = \SS \cin$. Incoherence and partial coherence in the input and output coefficients can all be modeled by forming density matrices (or coherency matrices~\cite{Goodman2000,Yamazoe2012,Okoro2017}) for each:
\begin{align}
    \rho_{\rm in} = \left\langle \cin \cin^{\dagger} \right\rangle, \qquad \rho_{\rm out} = \left\langle \cout \cout^{\dagger} \right\rangle.
\end{align}
Nonzero off-diagonal components indicate at least partial coherence. Eigenvectors of the density matrices are the ``coherent modes'' of Wolf~\cite{Wolf1982,Mandel1995}, and the eigenvalues correspondingly indicate the average power carried in each coherent mode over the ensemble. Direct substitution of the scattering matrix relationship between inputs and output coefficients yields a simple relationship between input and output density matrices:
\begin{align}
    \rho_{\rm out} = \SS \rho_{\rm in} \SS^{\dagger}.
\end{align}
For a unitary scattering matrix this corresponds simply to a unitary transformation, but one need not assume unitarity (no material loss). The general statement of incompressibility is as follows. If we arrange the eigenvalues of the respective density matrices in descending order, denoted $\boldsymbol\lambda^{\downarrow}(\rho)$ for density matrix $\rho$, then \textbf{the sum of the first $N$ eigenvalues of $\rho_{\rm out}$ is less than or equal to the sum of the first $N$ eigenvalues of $\rho_{\rm in}$, for any $N$}. This is exactly the definition by which one vector is weakly (sub)majorized by another, with the density-matrix relation given by
\begin{align}
    \boldsymbol\lambda^{\downarrow}\left(\rho_{\rm out}\right) \prec_w \boldsymbol\lambda^{\downarrow}\left(\rho_{\rm in}\right).
\end{align}
The sum-of-eigenvalues constraint, which is a definition of weak majorization (majorization proper also requires equal total power, i.e., no loss), is precisely an incompressibility condition: it prohibits any net power transfer from any one coherent mode to another subset whose modes initially have greater (or equal) power. To prove this impossibility of power ``concentration,'' we can compute the sums of the density-matrix eigenvalues. Both density matrices are diagonalizable (by Hermiticity): $\rho_{\rm out} = U\Lambda_{\rm out} U^{\dagger}$ and $\rho_{\rm in} = V \Lambda_{\rm in} V^{\dagger}$. Inserting these diagonalizations into $\rho_{\rm out} = \SS \rho_{\rm in} \SS^{\dagger}$ and defining $\SS' = U^{\dagger} \SS V$ (the still-subunitary $\SS$ matrix, with input/output bases transformed), we have
\begin{align}
    \Lambda_{\rm out} = \SS' \Lambda_{\rm in} \left(\SS'\right)^{\dagger}.
    \label{eq:LambdaOut}
\end{align}
The sum of the first $N$ eigenvalues of the outgoing-field density matrix on the left-hand side is given by $\sum_{i=1}^N \vect{e}_i^{\dagger} \Lambda_{\rm out} \vect{e}_i$. On the right-hand side, the same summation can be written $\sum_{i=1}^N \vect{s}_i^{\dagger} \Lambda_{\rm in} \vect{s}_i$, where $\vect{s}_i = \left(\SS'\right)^{\dagger} \ev_i$. This expression is a weighted average of the diagonal $\Lambda_{\rm in}$ matrix, with the vectors $\vect{s}_i$, which by passivity (and their definition via $\SS'$) have norms less than or equal to one. Intuitively, any $\vect{s}_i$ can at best pick out the largest diagonal (eigenvalue) of $\Lambda_{\rm in}$, any other $\vect{s}_i$ can at best pick out the second largest eigenvalue, and so on. Alternatively, as discussed in Sec.~\ref{sec:RQs}, the sum of the first $N$ eigenvalues of $\Lambda_{\rm out}$ is given by the maximum of $\Tr\left(I_N \Lambda_{\rm out}\right)$, where $I_N$ is all zeros except for the $N\times N$ identity matrix embedded to align with the largest diagonal entries of $\Lambda_{\rm out}$. This can be related to the incoming eigenvalues via $\Tr\left(I_N \Lambda_{\rm out}\right) =  \Tr\left(I_N \SS' \Lambda_{\rm in} (\SS')^{\dagger}\right)$. The maximum of this quantity over all subunitary matrices, per the Hoffman--Wielandt theorem as discussed at the end of Sec.~\ref{sec:RQs}, is bounded above by the $N$ largest eigenvalues of $\Lambda_{\rm in}$:
\begin{align}
    \sum_{i=1}^N \lambda_i^{\downarrow,\rm (out)} \leq \sum_{i=1}^N \lambda_i^{\downarrow,\textrm{in}},
\end{align}
for any $N$. This ``brightness theorem'' generalizes that of Ref.~\cite{Zhang2019}, which only included the restriction on the maximum-power channel (i.e., $N=1$). An equivalent formal characterization follows from the Hardy--Littlewood--P\'olya theorem~\cite{HardyLittlewoodPolya1934} and its extension to weak majorization~\cite{marshall2011inequalities}: for nonnegative spectra (zero-padded to equal length), $\lambda_{\rm out} \prec_w \lambda_{\rm in}$ if and only if $\lambda_{\rm out} = D \lambda_{\rm in}$ for some doubly-substochastic matrix $D$ (doubly stochastic when no power is lost). The brightness theorem therefore states that passive linear systems can only redistribute channel-power spectra in ways describable by doubly-(sub)stochastic mixing---a wave-physics analog of classical mixing. Entropy-like (Schur-concave) functions of the normalized spectrum are \emph{not} monotonic once loss is present: a polarizer acting on unpolarized light, for example, reduces the entropy of the normalized spectrum from $\log 2$ to zero.

\begin{figure*}[tb]
\centering
\includegraphics[width=0.95\linewidth]{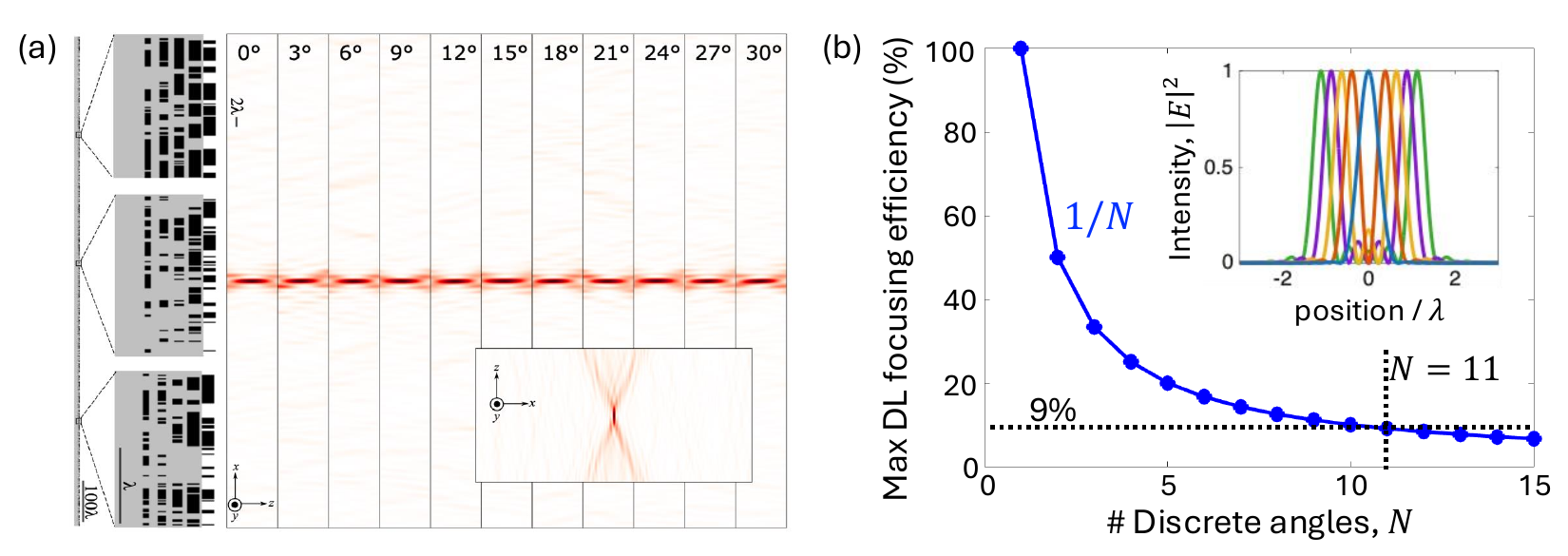}
\caption{\textbf{Metalens concentrator bounds.} (a) A metalens concentrator that focuses 11 incoming-wave angles (incoherently) to a diffraction-limited (DL) spot~\cite{Lin2019-ph}. (b) Bounds to maximum DL focusing efficiency for $N$ incoherent excitations; the design of (a) has at most 9\% DL focusing efficiency. (b, inset) The first five orthogonal-field-pattern intensities at the focusing plane, ordered from most to least concentrated. (Panel (a) adapted from Ref.~\cite{Lin2019-ph}.)}
\label{fig:AngleConcentrator}
\end{figure*}
A recently relevant example for which this ``brightness theorem'' can be applied is a ``metalens concentrator'' in which a discrete set of incoming angles are focused to \emph{the same diffraction-limited spot}~\cite{Lin2019-ph}, as depicted in Fig.~\ref{fig:AngleConcentrator}(a). Unlike an imaging lens, which has a continuously varying focal spot, here, the focal spot remains fixed. A diffraction-limited focal spot is effectively a single power-carrying channel: there is only one linear combination of plane waves that produces it (or two, for orthogonal polarizations with degenerate focal spots). Hence if 100\% of power from one incident angle is focused into that linear combination of plane waves, all other incoming-angle excitations must excite only plane-wave combinations that are orthogonal to the diffraction-limited one. The inset of Fig.~\ref{fig:AngleConcentrator}(b) shows five example orthogonal focal-plane fields (for 2D scalar-wave fields) that correspond to the five smallest eigenvalues of a minimum-spot-size objective. Five distinct incoming angles could at best achieve 100\% coupling to these five beams, with the amount of power coupled into the beam with the smallest spot, i.e., the ``diffraction-limited beam,'' averaging no more than 20\%. More generally, for $N$ independent (or incoherent) incident excitations, the maximum diffraction-limited (DL) focusing efficiency is $1/N$; for the design of Fig.~\ref{fig:AngleConcentrator}(a), with 11 angles, this implies a maximum average focusing efficiencies of $9\%$, as noted in Fig.~\ref{fig:AngleConcentrator}(b).

\begin{figure*}[tb]
\centering
\includegraphics[width=0.99\linewidth]{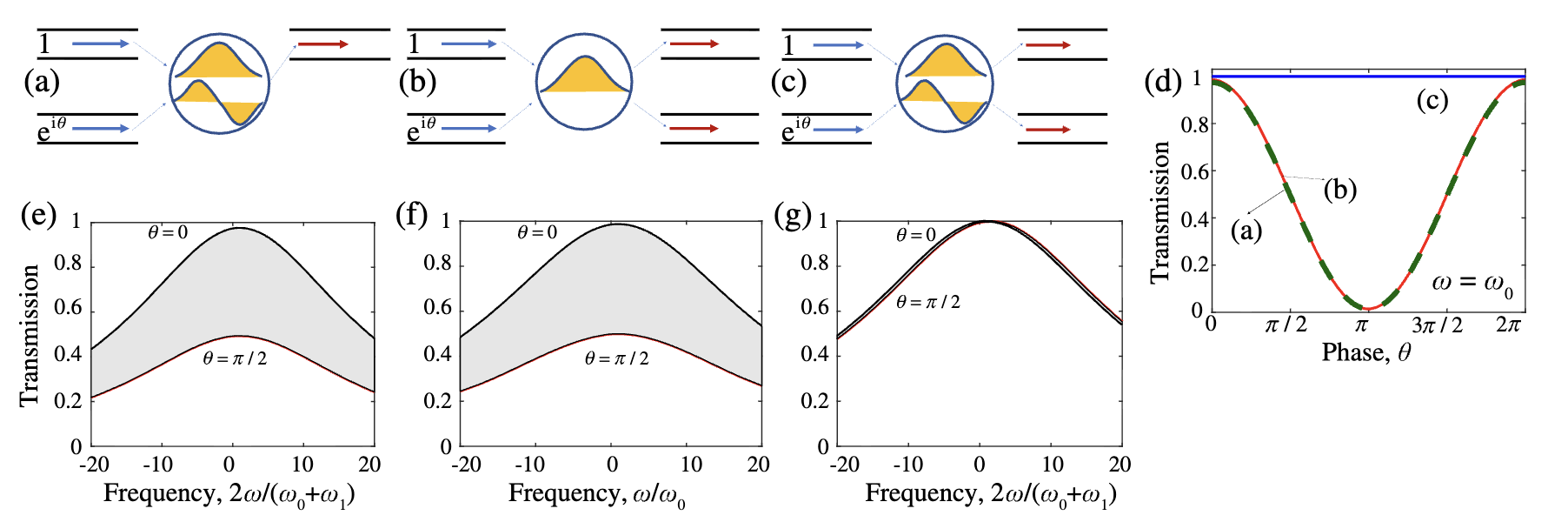}
\caption{\textbf{Brightness theorem constraints in resonant TCMT.} Incoherent excitations of a two-port waveguide (exemplified by a variable relative phase $\theta$), cannot lead to resonance-assisted unity transmission with fewer than two output ports (a,e) or fewer than two resonances (b,f). (c,g) A design via coupled mode theory that achieves unity transmission. (d) A comparison of the three transmissions, as a function of phase. (Figure adapted from Ref.~\cite{Zhang2019}.)}
\label{fig:WaveguideCombiner}
\end{figure*}
In many photonic scenarios it is of interest to use resonances to assist the scattering process, and often the resonant scattering processes can be modeled using temporal coupled mode theory (TCMT)~\cite{Suh2004-kz,Joannopoulos2011}. In TCMT, the ``incompressibility'' notion extends beyond the input/output channels to the resonances themselves. In particular, it is not possible for $N$ independent (incoherent) inputs to be directed to (at least) $N$ independent outputs through fewer than $N$ resonances. If there are $M<N$ resonances, then the transmission efficiency through the resonances is bounded above by $M/N$. An example of this is shown in Fig.~\ref{fig:WaveguideCombiner}, where incoherent waveguide routing from two inputs is only feasible with (at least) two resonances and (at least) two output ports, as in Fig.~\ref{fig:WaveguideCombiner}(c,g). A more detailed discussion and proofs in the resonance-assisted case are given in Ref.~\cite{Zhang2019}, which also delves into examples with partial coherence. The key mechanism for understanding partially coherent cases is to diagonalize the relevant matrices; in the diagonal basis, a partially coherent scenario has been transformed to fully incoherent excitations, albeit perhaps with unequal relative weights.

After this Tutorial was submitted and posted, D.~A.~B.~Miller~\cite{millerdab2026maximum} showed that for a fixed passive linear system at a single frequency, the brightness theorem above can be further strengthened, and in fact holds term by term: each ordered eigenvalue of the outgoing density matrix is bounded by the corresponding incoming one, $\lambda_i^{\downarrow,\rm (out)} \leq \lambda_i^{\downarrow,\rm in}$ for every $i$. (One way to see this must be true: the eigenvalues of $\SS' \Lambda_{\rm in} (\SS')^{\dagger}$ are the squared singular values of $\SS' \Lambda_{\rm in}^{1/2}$, none of which can exceed those of $\Lambda_{\rm in}^{1/2}$ when $\|\SS'\| \leq 1$.) The termwise bound is tight, since any output spectrum satisfying it can be reached by a passive filter acting in the eigenbasis of $\Lambda_{\rm in}$, and it is stronger than the weak majorization above: an incoming spectrum $(1, 0.3)$ weakly majorizes $(0.8, 0.4)$, yet no fixed passive system converts one into the other. Weak majorization remains the appropriate statement when the system itself fluctuates, as for ensemble or time averages over a randomly varying medium, or for quantities integrated over a bandwidth across which $\SS(\omega)$ changes appreciably, as near the resonances discussed above. For example, an equal mixture of the identity and a channel swap converts the spectrum $(1, 0)$ into $(1/2, 1/2)$, which violates the termwise bound but not weak majorization. (A deterministic, periodic modulation, by contrast, is a fixed system in the frequency-channel picture described earlier in this section.)

\section{Bounds on electromagnetic scattering from local constraints}\label{sec:Scatt_Local}
In this chapter, we consider bounds that depend on the constituent properties of a scatterer or its designable domain. The first three sections develop broadly applicable frameworks: single-frequency optical-theorem bounds (Sec.~\ref{sec:SingleFreq}), all-frequency sum rules built from causality (Sec.~\ref{sec:CausalResp}), and a general computational-bound framework based on local conservation laws (Sec.~\ref{sec:comput_bounds}). The remaining sections then dive into specific questions: how strong can the polarizability of a small scatterer be, how confined can a plasmonic field be, how small can a mode volume be, how invisible can an object be made, how sensitively can it sense its environment, and how long can light be delayed within it.

\subsection{Single-frequency bounds}
\label{sec:SingleFreq}
A particularly powerful method for identifying bounds to single-frequency scattering processes leverages the \textbf{optical theorem}, of \eqref{eq:OpticalThm} of the conservation-law section, which we copy here:
\begin{align}
    \frac{\omega}{2}\phi^{\dagger}\left(\Im \Gamma_0+ \Im \xi\right) \phi = \frac{\omega}{2} \Im \left(\psi_{\rm inc}^{\dagger} \phi\right),
    \label{eq:OpticalThm2}
\end{align}
for frequency $\omega$, polarization fields $\phi$, free-space (or background) Green's function operator $\Gamma_0$, and material parameter $\xi(\omega)=-1/\chi(\omega)$, for susceptibility $\chi(\omega)$. As discussed in the conservation-law section, the two terms on the left-hand side are the powers scattered and absorbed by the scatterer, respectively, while the right-hand side is the extinction. As also discussed, for passive scatterers, both left-hand-side terms are positive-semidefinite quadratic forms: $\Im \Gamma_0$ and $\Im\xi$ cannot have negative eigenvalues. By contrast, the right-hand side is linear in the polarization-field response, as the incident field is fixed for a given problem or application. 

The key idea for many fundamental limits is to drop Maxwell's equations themselves as a constraint and, instead, to use the optical theorem, on the designable volume of interest, as the only constraint to the objective of interest. This approach can get one surprisingly far. In the examples below, we assume nonmagnetic materials, in which case all tensors and vectors have 3 components in each dimension instead of 6. 

\textbf{Analytical bounds} can be derived for canonical objectives if the optical theorem is further simplified. The simplest example is for a lossy, scalar-susceptibility material, for which $\Im \xi = \Im \chi / |\chi|^2$ is a positive scalar. One can note that scattering must be positive, and reduce the optical theorem to the inequality that absorption be smaller than extinction, which can now be written
\begin{align}
    P_{\rm abs} \leq P_{\rm ext} \quad \Rightarrow \quad \phi^{\dagger} \phi \leq \frac{|\chi|^2}{\Im \chi}\Im \left(\psi_{\rm inc}^{\dagger}\phi\right).
\end{align}
This quadratic inequality clearly constrains how large of a polarization response can be excited in a scatterer---a quadratic function will always overtake a linear one. The polarization field with largest magnitude that can be excited across the scatterer is rigorously derived through variational calculus, but it can always be directly read off: $\phi_{\rm max} = i(|\chi|^2/\Im\chi)\psi_{\rm inc}$. This is exactly the optimal polarization field if the objective is extinction or absorption. This polarization field produces zero scattered power\footnote{As can be verified by computing $P_{\rm scat} = P_{\rm ext} - P_{\rm abs} = (\omega/2) \left[\Im\left(\psi_{\rm inc}^{\dagger}\phi\right)-\phi^{\dagger} (\Im\xi)\phi\right]$.}; the optimal polarization field for maximum scattered power is exactly half of $\phi_{\rm max}$, producing a quarter as much scattering as the maximum absorption or extinction\footnote{At the condition of maximum scattering, there is also an equal amount of absorption. One can maximize scattered power subject to more stringent constraints on absorption, for example, by specifying a maximum allowed radiative efficiency, ultimately yielding smaller total scattered powers in the trade-off for increased radiative efficiency~\cite{Miller2016,Yang2017,Miller2022}.}. Collectively, these bounds can be written~\cite{Miller2014,Miller2016}
\begin{align}
    P_{\rm abs,scat,ext} \leq \beta\frac{\omega}{2} \frac{|\chi|^2}{\Im \chi} \psi_{\rm inc}^{\dagger}\psi_{\rm inc},
\end{align}
where $\beta = 1$ for extinction and absorption and $1/4$ for scattering. For a plane wave incident upon a nonmagnetic medium, this bound can be simplified, especially written in terms of cross-sections $\sigma$ that are the respective powers divided by the incident plane-wave intensity, $I_{\rm inc} = \left\|\Ev_0\right\|^2/(2Z_0)$, where $Z_0 = \sqrt{\mu_0/\varepsilon_0}$ is the impedance of free space, and divided by the volume $V$ of the scatterer:
\begin{align}
   \frac{\sigma_{\rm abs,scat,ext}}{V} \leq \beta \frac{|\chi|^2}{\Im \chi} \frac{\omega}{c}.
   \label{eq:CSBoundML}
\end{align}
Lossy materials have cross-section bounds that depend only on a material loss metric $|\chi|^2/\Im \chi$, the wavelength, $\omega/c=2\pi/\lambda$, and the volume $V$.

What about a lossless or nearly lossless material? In this case, one can find recourse in the other term of the optical theorem, corresponding to scattered power. One can drop the absorption term, arriving at the inequality
\begin{align}
    P_{\rm scat} \leq P_{\rm ext} \quad \Rightarrow \quad \phi^{\dagger} \left(\Im \Gamma_0\right) \phi \leq \Im \left(\psi_{\rm inc}^{\dagger}\phi\right).
\end{align}
One can proceed with general bounds involving the $\Im \Gamma_0$ operator~\cite{Hugonin2015-eu}, but for simplicity we will specialize to the case of free space with compact (non-extended) scatterers. In this case, expanding the incident field and polarization response into vector spherical waves~\cite{Stratton2007-bm}, which have radial ``quantum'' number $n$ taking values $1,2,\ldots,\infty$, angular number $m=0,1,\ldots,n$ (with negative values accounted by symmetry), and polarization $p$ denoted ``TE'' or ``TM.''  One can show that the absorbed, scattered, and extinguished powers are given by~\cite{Kwon2009-ju}
\begin{subequations}
\begin{align}
    P_{\rm scat} &= \frac{|\Ev_0|^2}{2 Z_0 } \left(\frac{2\pi}{k^2}\right) \sum_{n=1}^{\infty} \sum_{m=0}^{n} \sum_{p=\textrm{TE,TM}} \gamma_{mn} \left|c^{\rm scat}_{mnp}\right|^2 \\
    P_{\rm ext} &= -\frac{|\Ev_0|^2}{2 Z_0 } \left(\frac{2\pi}{k^2}\right) \sum_{n=1}^{\infty} \sum_{m=0}^{n} \sum_{p=\textrm{TE,TM}} \gamma_{mn} \Re\left[\overline{c^{\rm inc}_{mnp}} c^{\rm scat}_{mnp}\right] \\
    P_{\rm abs} &= P_{\rm ext} - P_{\rm scat},
\end{align}
\label{eq:ASEvsw}
\end{subequations}
where $\gamma_{mn} = n(n+1)/(2n+1)\left((n+m)!/(n-m)!\right)$, and $c^{\rm inc}$ and $c^{\rm scat}$ are the coefficients of the incident and scattered fields into the appropriate regular and outgoing vector spherical waves~\footnote{The typical derivation of Eqs.~(\ref{eq:ASEvsw}) does \emph{not} use the volume Green's function operator. A much more direct route is through typical surface-flux expressions. Cf., for example, Ref.~\cite{Kwon2009-ju}. Yet the volume-integral approach is more useful for the later generalizations, and is itself a viable approach~\cite{Kuang2020-kl}.}. One can see in Eqs.~(\ref{eq:ASEvsw}) exactly the same ``quadratic-bounded-by-linear'' structure discussed above: scattered power is a quadratic function of the response coefficients, whereas extinction (which must be greater than scattered power) is linear in them. To maximize either extinction or scattering, it is readily apparent that the (globally) optimal scattering coefficients are given by $c^{\rm scat}_{mnp} = -c^{\rm inc}_{mnp}$. These coefficients imply no absorption. The maximal absorption occurs for the same coefficients but halved, i.e., $c^{\rm scat}_{mnp} = -(1/2)c^{\rm inc}_{mnp}$, leading to one-fourth as much absorbed power as that scattered in the maximum-scattering scenario. Together, the bounds are
\begin{align}
    P_{\rm abs,scat,ext} \leq \eta \frac{|\Ev_0|^2}{2 Z_0 } \left(\frac{2\pi}{k^2}\right) \sum_{n=1}^{\infty} \sum_{m=0}^{n} \sum_{p=\textrm{TE,TM}} \gamma_{mn} \left|c^{\rm inc}_{mnp}\right|^2,
    \label{eq:ASEvswBound}
\end{align}
where $\eta = 1$ for scattering and extinction, and $1/4$ for absorption. In the common scenario of a plane-wave excitation, only the $m=1$ waves are excited in the incoming field, with amplitudes given by $|c^{\rm inc}_{mnp}|^2=\left[(2n+1)/(n(n+1))\right]^2$ (Ref.~\cite{Kwon2009-ju}). The powers can be converted to cross-sections (dividing by the plane-wave intensity), and within each $m=1$ ``channel'' the cross-section bound is:
\begin{align}
    \left(\sigma_{\rm abs,scat,ext}\right)_{np} \leq \eta \left(\frac{\lambda^2}{2\pi}\right) \left(2n+1\right).
    \label{eq:CSvswChannelBound}
\end{align}
``Super-scatterers'' are objects whose scattering cross-section can exceed the single-channel bound of \eqref{eq:CSvswChannelBound}, which can be achieved by aligning resonances to activate multiple channels at or near their limits~\cite{Ruan2010-xx,Ruan2011-by,Wang2024-je}. To derive a bound over all channels, one could sum the terms of \eqref{eq:CSvswChannelBound}, but that summation diverges. For \emph{any} incoming wave carrying infinite power, the bound from \eqref{eq:ASEvswBound} will diverge. The bounds from the ``scattering-smaller-than-extinction'' approach simply produces unitary-like bounds in which all incoming power is scattered. A useful heuristic can regularize them, however: if one has confidence or knows \emph{a priori} that only a finite set of the vector spherical wave channels can be excited, then the summation over that finite set will converge. If we denote $N_{TE,TM}$ as the number of TE/TM channels, respectively (i.e., the largest index $n$ can take, for a given $p$), then the cross-section bounds are
\begin{align}
    \sigma_{\rm abs,scat,ext} \leq \eta \left(\frac{\lambda^2}{2\pi}\right)  \left(N_{\rm TE}^2 + 2N_{\rm TE} + N_{\rm TM}^2 + 2N_{\rm TM}\right).
    \label{eq:CSvswBoundFinite}
\end{align}
There is a simple physical intuition behind these bounds. Consider a scatterer in which only a single, $n=1$ channel can be excited. Then
\begin{align}
    \left(\sigma_{\rm scat,ext}\right)_{n=1,p} \leq \frac{3\lambda^2}{2\pi}, \quad \left(\sigma_{\rm abs}\right)_{n=1,p} \leq \frac{3\lambda^2}{8\pi}.
    \label{eq:CSvswChannelBound2}
\end{align}
The absorption bound in fact corresponds exactly to all incoming power being absorbed. This is achieved at a ``critical coupling'' condition, as generally described in temporal coupled mode theory~\cite{Joannopoulos2011}, in which the resonator comprising the scatterer couples to radiation at exactly the same rate that it couples to absorption. Then exciting the resonator from the time-reversed radiation channel will couple perfectly to the absorption ``port.'' How, then, can scattered power be four times as large? This is a related to the ``extinction paradox.'' An unperturbed incoming wave will pass through the origin and become an outgoing wave with amplitude, say, +1. The scatterer that creates maximum scattered power can produce a total outgoing wave in the same channel with amplitude -1. The ``scattered'' field is the difference between the total and the incident, which is now -2, and which has a square of 4. Hence the scattered power can be four times that counted in the incident wave. The absorption and scattering bound intuition can be generalized to channels beyond the lowest order simply by replacing the $3$ in \eqref{eq:CSvswChannelBound2} with $(2n+1)$.
\begin{figure*}[tb]
\centering
\includegraphics[width=0.95\linewidth]{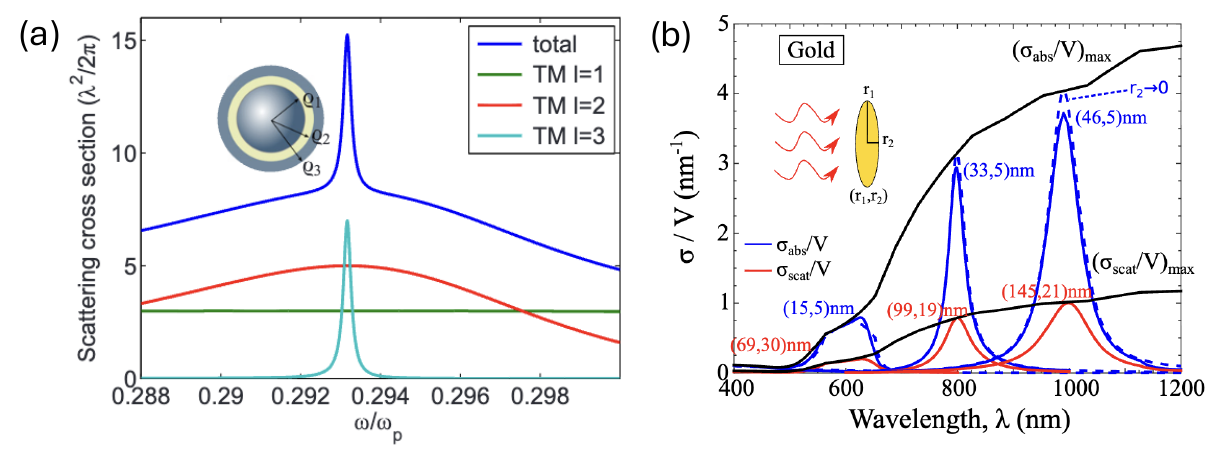}
\caption{\textbf{Saturating the optical-theorem bounds with plasmonic and superscattering designs.} (a) Plasmonic core--shell superscatterer~\cite{Ruan2011-by}: TM channels of total angular momentum $\ell = 1, 2, 3$ are tuned to a common frequency $\omega = 0.2932\,\omega_p$, so the total scattering cross-section (blue) approaches $15\,\lambda^2/2\pi$, well above any of the individual per-channel bounds $(2\ell+1)\lambda^2/2\pi$ of \eqref{eq:CSvswChannelBound}. (b) Per-volume absorption (blue) and scattering (red) cross-sections for several gold prolate nanorods of dimensions $(r_1, r_2)$; the resonance peaks track the material-loss bound $(\sigma_{\rm abs}/V)_{\max} \propto |\chi|^2/\Im\chi$ (black solid) from the visible into the near infrared, demonstrating that plasmonic resonators can saturate the bounds across a wide spectral range. (Panel (a) adapted from Ref.~\cite{Ruan2011-by}; panel (b) adapted from Ref.~\cite{Miller2016}.)}
\label{fig:SuperScat}
\end{figure*}

\begin{figure*}[tb]
\centering
\includegraphics[width=0.8\linewidth]{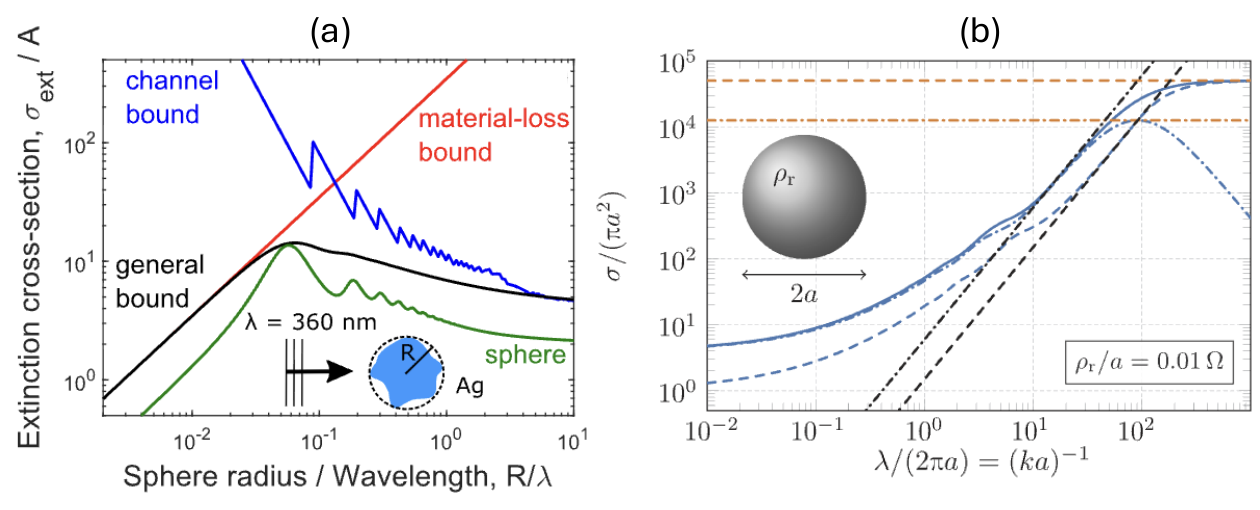}
\caption{\textbf{Optical-theorem bounds across all size regimes.} (a) Extinction cross-section bounds for an Ag scatterer enclosed in a spherical region of radius $R$, illuminated by a plane wave at $\lambda = 360$\,nm. The general bound \eqref{eq:CSOTBound} (black) tracks the realized Ag-sphere cross-section (green) within a small factor over four decades in size. (b) Absorption, scattering, and extinction bounds for an arbitrary obstacle of resistivity $\rho_r/a = 0.01\,\Omega$ confined within a sphere of radius $a$, plotted versus $(ka)^{-1}$. Together, the panels show these bounds remain quantitatively meaningful from the far-subwavelength (quasistatic, plasmonic) regime through resonance-based, wavelength-scale scattering to the geometric-optics limit. (Panel (a) adapted from Ref.~\cite{Kuang2020-kl}; panel (b) adapted from Ref.~\cite{Gustafsson2020}.)}
\label{fig:OptThmBndsFig}
\end{figure*}

The two sets of bounds above use complementary but distinct approaches: \emph{either} require absorption to be smaller than extinction \emph{or} require scattering to be smaller than extinction, leading to the different physical pictures shown in Fig.~\ref{fig:SuperScat}. It is possible to use the \textbf{full optical theorem to simultaneously enforce both constraints}, as recognized in Refs.~\cite{Gustafsson2020,Kuang2020-kl}. In this case, one rewrites the objective in terms of the polarization field $\phi$, imposes the full optical-theorem constraint, which can be written
\begin{align}
    \phi^{\dagger}\left(\Im \Gamma_0+ \Im \xi\right) \phi = \Im \left(\psi_{\rm inc}^{\dagger} \phi\right).
\end{align}
Identifying maximum exctinction, $(\omega/2)\Im \left(\psi_{\rm inc}^{\dagger} \phi\right)$ subject to this constraint is relatively straightforward. One can (with practice) read off the optimal solution: 
\begin{align}
    \phi_{\rm opt,ext} = i\left(\Im \Gamma_0+ \Im \xi\right)^{-1} \psi_{\rm inc},
\end{align}
with the corresponding optimal extinction given by $(\omega/2) \psi_{\rm inc}^{\dagger} \left(\Im \Gamma_0+ \Im \xi\right)^{-1} \psi_{\rm inc}$. Within a spherical bounding surface, one can diagonalize $\Im \Gamma_0$, with eigenvectors $\tau_i$ and eigenvalues $\rho_i$, so that the extinction bound becomes $\sum_i |\tau_i^{\dagger} \psi|^2 / (\rho_i + \Im \xi)$. For the prototypical scenario of a plane-wave incident upon a nonmagnetic scatterer, this bound can be simplified to a bound on the extinction cross-section:
\begin{align}
    \sigma_{\rm ext} \leq \frac{\lambda^2}{2\pi} \sum_{n,p} (2n+1) \frac{\rho_{n,1,p}}{\rho_{n,1,p}+\Im \xi},
    \label{eq:CSOTBound}
\end{align}
where the eigenvalues $\rho_{n,m,p}$ share an index with the vector spherical waves (which, normalized, are the eigenvectors of $\Im \Gamma_0$). \eqref{eq:CSOTBound} encapsulates the material-loss-dependent bound of \eqref{eq:CSBoundML} and the unitary-scattering bound of \eqref{eq:CSvswBoundFinite} into a single constraint. In the limit of small material loss, $\Im \xi = \Im \chi / |\chi|^2$ is very small, and the extinction bound simplifies to the bound over scattering channels. Conversely, in the limit of large material loss, or small volume (in which case the $\rho_i$ are small), only the first term in \eqref{eq:CSOTBound} is retained, yielding the material-loss-only bound. A key advantage of this bound is that it can ``interpolate'' between the two extremes. An example for a plasmonic scatterer is given in Fig.~\ref{fig:OptThmBndsFig}. 

Bounds for absorption, scattering, and other objectives, subject to the full optical theorem constraint, require a more sophisticated mathematical apparatus. In particular, the global optimum to a quadratic objective subject to a single quadratic constraint can be found by ``duality.'' The Lagrangian dual to the original optimization problem introduces Lagrange multiplier variables whose optima need to be found in tandem with the optima of the original variables. Further detail into this technique is beyond the scope of this paper, but is discussed in the context of electromagnetic bounds in, e.g., Refs.~\cite{Kuang2020-kl,Molesky2020,Gustafsson2020}, and in more general optimization theory in Ref.~\cite{boyd2004convex}. Why is the optima more challenging to find in these cases? Consider a scattered-power objective function, $(\omega/2) \phi^{\dagger} \left(\Im \Gamma_0\right) \phi$, subject to the optical-theorem constraint. One can imagine starting in the eigenbasis of the scattering operator, $\Im\Gamma_0$, but the optimal coefficients require a balancing act between the overlap of the eigenvectors with the incident field, as well as the relative amounts of radiation ($\rho_i$) versus absorption ($\Im \xi$). It turns out that one extra duality parameter $\nu$, properly tuned, can resolve this balancing question. The absorption and cross-section bounds, for plane-wave incidence on a nonmagnetic, compact scatterer, are
\begin{subequations}
\begin{align}
    \sigma_{\rm abs} &\leq \frac{\lambda^2}{2\pi} \frac{\nu^2}{4} \sum_{n,p} (2n+1) \frac{\rho_{n,1,p}}{\nu \rho_{n,1,p}+ (\nu-1)\Im \xi},\\
    \sigma_{\rm scat} &\leq \frac{\lambda^2}{2\pi} \frac{\nu^2}{4} \sum_{n,p} (2n+1) \frac{\rho_{n,1,p}}{(\nu-1)\rho_{n,1,p}+\nu\Im \xi}.
\end{align}
\label{eq:DualBounds}
\end{subequations}
The optimal $\nu$ is different in the absorption and scattering cases, and dependent on the material-loss parameters and scatterer size. It can be found computationally~\cite{Kuang2020-kl,Gustafsson2020}. The limits of Eqs.~(\ref{eq:DualBounds}) simplify to their material-only and scattering-channel-only counterparts in the appropriate asymptotic limits.

The canonical cross-section objectives discussed above can be generalized to a variety of scenarios. In Ref.~\cite{gustafsson2022upper_focusing}, bounds are found for the wave-focusing problems discussed in Sec.~\ref{sec:FSFocus}, now assuming fixed material-loss parameters that can be enforced by the optical theorem. In Ref.~\cite{Gustafsson2020}, extensions to \emph{directional} scattering bounds are given, as are results for the Purcell factor (LDOS enhancement). Optical-theorem-based bounds on absorption \emph{within} lossy media are derived in~\cite{ivanenko2019optical_theorems_lossy}. Closely related bounds---in some cases using $\TT$-operator generalizations of the optical-theorem constraint---extend the framework to thermal and fluctuational scenarios: radiative heat transfer between bodies~\cite{Miller2015,molesky2020RHT_theory,limited_role_structuring}, attractive and repulsive Casimir--Polder forces~\cite{venkataram2020CasimirPolder}, and angle-integrated absorption and thermal radiation for arbitrary objects~\cite{molesky2019Toperator}. Bounds on maximum antenna gain, effective areas, and related quantities, along with the optimal currents that generate them, have been developed in Refs.~\cite{gustafsson2012physical_optimal_currents,gustafsson2013optimal_currents,gustafsson2015physical_bounds_handbook,gustafsson2016antenna_current_matlab,gustafsson2019maximum_gain,gustafsson2022unified_characteristic,capek2023characteristic_scattering_dyadic}. In Ref.~\cite{Kuang2020-kl}, bounds to the minimum thickness of a perfectly absorbing patterned film are given, as are generalizations to any requisite absorption percentage. Another generalization combines the wavefront shaping bounds described in Chapter~\ref{sec:WaveShaping} with the optical-theorem-based bounds. Consider the extinction bound of \eqref{eq:CSOTBound}: it is a quadratic form of the incident field, like the objectives of Chapter~\ref{sec:WaveShaping}. Hence there is a globally optimal wavefront that maximizes the bound, which is given by the extremal eigenvector of the quadratic form. There is an example in Ref.~\cite{Kuang2020-kl} of a large separation opening in the extinction bounds for a plane wave versus a (properly normalized) optimal wavefront, at larger domain radii, and a numerical demonstration of an optimized wavefront scattering from a sphere at a level that surpasses even the plane-wave bounds, thereby guaranteeing that no structure of the same material, of any shape that fits in the same domain, can achieve the same extinction level. Such demonstrations showcase the utility of bounds in comparing architectural approaches (or architectural hyper-parameters) within an application area.

Finally, there is a single additional step in the optical theorem approach that is conceptually elegant and offers additional insights and restrictions on what is feasible. The step is simple to describe: when forming the optical theorem constraint of \eqref{eq:OpticalThm}, from the volume integral equation of \eqref{eq:IE2}, it is not necessary to take the imaginary part. One could equivalently take the real part. Or, do both! Taking both the imaginary and real parts yields two constraints (ignored the $(\omega/2)$ pre-factors):
\begin{subequations}
    \begin{align}
        \phi^{\dagger}\left(\Im \Gamma_0+ \Im \xi\right) \phi &= \Im \left(\psi_{\rm inc}^{\dagger} \phi\right), \\
        \phi^{\dagger}\left(\Re \Gamma_0+ \Re \xi\right) \phi &= -\Re \left(\psi_{\rm inc}^{\dagger} \phi\right),
    \end{align}
    \label{eq:TwoConstraints}
\end{subequations}
The first equation is the same optical theorem discussed above, which corresponds to (real) power conservation and is equivalent to a statement of conservation of Poynting flux. The second equation is akin to the imaginary part of the complex Poynting conservation law (cf. Sec.~6.9 of Ref.~\cite{Jackson1999}), and can be referred to as a \textbf{reactive power conservation law}. This conservation law does not comprise any positive semidefinite quadratic forms, and hence, cannot be used on its own to constrain the polarization response of a scattering medium. But it can be used in conjunction with the optical theorem, and offer an additional constraint that can further tighten scattering bounds~\cite{Gustafsson2020,Molesky2020-ig}.

\begin{figure*}[tb]
\centering
\includegraphics[width=0.5\linewidth]{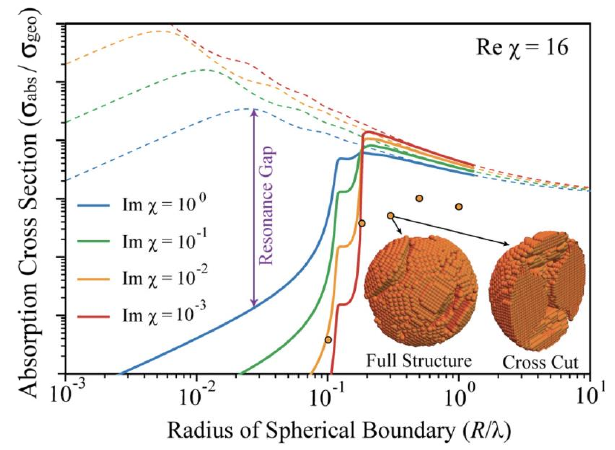}
\caption{\textbf{Tightened bounds via reactive-power constraint.} With the addition of a reactive power constraint, one can tighten bounds (solid lines) for subwavelength dielectric scatterers, relative to optical-theorem-only constraints (dashed lines). (Figure adapted from Ref.~\cite{Molesky2020-ig}.)}
\label{fig:DielScatBoundsFig}
\end{figure*}

Again, one can use the mathematics of duality to identify the global optimum of any quadratic objective subject to the two constraints of \eqref{eq:TwoConstraints}. In this case, there are two Lagrange-multiplier variables that enter the fray, and their optimal values can be systematically computed. Probably the most important new aspect that is captured when reactive power constraints are imposed is the inability of highly subwavelength dielectric ($\Re \varepsilon > 0$) scatterers to support strong resonances. As shown in Fig.~\ref{fig:DielScatBoundsFig}, power conservation only (dashed lines) leads to absorption cross-section bounds that scale with scatterer volume (thereby increasing relative to cross-section area as radius $R$ decreases). Enforcing the additional constraint of reactive power conservation dramatically reduces the bounds at small sizes ($R < 0.1\lambda$), accurately capturing the difference between highly subwavelength dielectric and metallic/polaritonic scatterers.

Are real and reactive power conservation, the foundation for this sub-section, the only useful constraints for single-frequency scattering bounds? It turns out that they are merely a \emph{global} starting point for an infinite set of \emph{local} power-conservation constraints, as developed in Sec.~\ref{sec:comput_bounds}. Before discussing those computation-heavy bounds, however, we discuss the key role of causality and sum rules, to widen our aperture for analytical bounds beyond a single frequency.

\subsection{Sum rules, for all-frequency bounds} \label{sec:CausalResp}
Frequency as a parameter occupies a special role in Maxwell's equations: it is the Fourier partner of time, and consequently encodes strong causality-based constraints. The essential characteristics of causal response functions, KK relations, and frequency dispersion are discussed in Sec.~\ref{sec:causality}. Here, we show how these ingredients can come together to yield fundamental bounds on scattering quantities. We will describe a variety of such bounds in this section, all of which satisfy four criteria:
\begin{enumerate}
    \item \textbf{An optical theorem}. Conventionally, the optical theorem is a relation relating the extinction (scattered plus absorbed powers) of a plane wave to the amplitude and phase of its far-zone scattered field (i.e., the object's shadow). More generally, for any incident generated outside a surface $S$ and any quadratic power/flow quantity $P(\omega)$, the sum of the total inward flow of $P$ through $S$ (e.g. absorption) and the outward flow in the waves (e.g. scattered power) can always be written as the imaginary part of an amplitude $s(\omega)$:
    \begin{align}
        P(\omega) = \Im s(\omega).
        \label{eq:optical_theorem}
    \end{align}
    This is the essence of the optical theorem: it links a power-carrying metric with an \emph{amplitude}, not the typical \emph{square} of any amplitude. Unlike squared amplitudes, amplitudes themselves can have strong dispersion constraints, from our next criteria.
    \item \textbf{A Kramers--Kronig (KK) relation}. Kramers--Kronig relations (or ``dispersion relations''~\cite{Nussenzveig1972} or Hilbert transforms~\cite{Nussenzveig1972}) allow us, when the conditions of Sec.~\ref{sec:causality} are satisfied, as detailed, below, to relate integrals of the imaginary part of the amplitude $s(\omega)$ to its real part (or vice versa). For example, a dispersion relation such as 
    \begin{align}
        \Re s(\omega) = \frac{2}{\pi} \,\Pvl\!\!\int_0^{\infty} \frac{\omega' \Im s(\omega')}{(\omega')^2 - \omega^2} \,{\rm d}\omega'.
        \label{eq:GenDisp}
    \end{align}
    \item \textbf{High- and/or low-frequency constants}. The next ingredient is to identify, or bound, the value of $\Re s(\omega)$, which is typically feasible at either zero frequency (statics) or in the infinite limit, $\omega \rightarrow \infty$. These limits also simplify the right-hand side of the dispersion relation (e.g. \eqref{eq:GenDisp}), yielding a ``sum rule'' along the lines of:
    \begin{align}
        \int_0^{\infty} \Im s(\omega)\,{\rm d}\omega = C,
        \label{eq:GenSumRule}
    \end{align}
    where $C$ is a ``constant'' that may depend on relevant low- or high-frequency parameters. Note also there there may be extra factors of $\omega$ (to positive or negative powers) multiplying $\Im s(\omega)$. Concrete examples are given below.
    \item \textbf{Passivity}. Finally, the sum rule of \eqref{eq:GenSumRule} is not useful as a bound unless the integrand is nonnegative at all (positive) frequencies. Otherwise, it is plausible for the integrand to be arbitrarily large in magnitude and negative outside a frequency band of interest and then take arbitrarily large and positive values within that band. This positivity constraint arises from passivity, which is the final ingredient for a causality-based sum rule bound.
\end{enumerate}
These four criteria have been identified and developed for three types of scattering response functions: plane-wave extinction, local density of states (dipole emission), and the volume-scattering $\TT$ matrix. We describe each, in turn.
\subsubsection{Plane-wave extinction} \label{sec:Plane_wave_extinction}
As discussed in Sec.~\ref{sec:conslaws}, the extinction (sum of absorption and scattered power) of an incoming plane wave by an arbitrary linear, nonmagnetic scattering body occupying volume $V$ is given by
\begin{align}
    P_{\rm ext} = \frac{\omega}{2} \Im \underbrace{\left[\int_V \Einc^* \cdot \Pv \,{\rm d}V\right]}_{s(\omega)}.
    \label{eq:PextOT}
\end{align}
\eqref{eq:PextOT} is a version of the \emph{optical theorem}, and indeed it relates a power quantity to the imaginary part of a scattering amplitude. The term in square brackets is referred to as the (normalized) \textbf{forward-scattering amplitude}---it measures the inner product between the induced polarization field and a forward-moving plane wave, and is equivalent to the far-zone scattered-field amplitude in the forward direction. Hence the first criteria is satisfied. For the next criteria, a KK relation, we need to satisfy the asymptotic criteria discussed in Sec.~\ref{sec:causality}. First, we modify the definition of $s(\omega)$ to $s(\omega) = \int_V \Einc(\xv,-\omega)\cdot \Pv(\xv,\omega) \,{\rm d}\xv$; at real frequencies, conjugating a field is equivalent to evaluating it at negative frequencies, but the latter is amenable to analytic continuation. In the high-frequency limit $|\omega|\rightarrow \infty$ in the upper-half plane, the material must become transparent (as discussed in Sec.~\ref{sec:causality}) and the susceptibility must decay to zero as $1/\omega^2$~\cite[Sec.~78]{landau2013electrodynamics}. Then the polarization field scales as $\Pv \sim \Einc/\omega^2$, decaying sufficiently quickly to enable a KK relation~\footnote{Here we can clarify why this section has been restricted to plane wave excitations, even though the optical theorem of \eqref{eq:PextOT} applies more generally. It appears that plane waves are the only propagating wave solutions for which $s(\omega)$ decays sufficiently fast in all directions in the complex plane. Consider two counter-propagating plane waves forming a standing wave, $\cos(\omega x/c)$. In the limit as $|\omega|\rightarrow \infty$ in the UHP, $s(\omega) \sim \cos(\omega x/c)\cos(-\omega x/c)/\omega^2$; both cosine terms diverge exponentially going up the imaginary axis, prohibiting closure of a KK integral.}. (Ensuring this decay is why the leading factor of $\omega$ in \eqref{eq:PextOT} is not included in the definition of $s(\omega)$.) Then a semicircular contour along the real line and closing in the UHP, with a semicircular ``hole'' at $\omega$, as seen in Fig.~\ref{fig:LDOSSumRule} (with $\omega = 0$ in that case), can be used to relate $s(\omega)$ at one frequency to an integral over all frequencies, giving a typical KK relation
\begin{align}
    \Re s(\omega) = \frac{2}{\pi} \,\Pvl\!\!\int_0^{\infty} \frac{\omega' \Im s(\omega')}{(\omega')^2 - \omega^2}\,{\rm d}\omega',
    \label{eq:KKExt}
\end{align}
a typical KK relation, now for the forward-scattering amplitude. Note that the optical theorem of \eqref{eq:PextOT} requires that the numerator of the integrand of the KK relation, $\omega' \Im s(\omega')$, be positive for all frequencies in passive scatterers. (Otherwise, extinction would be negative.) Finally, the value of $s(\omega)$ simplifies in two asymptotic limits: $\omega \rightarrow 0$ and $\omega \rightarrow \infty$. In the static case, the incident plane wave becomes a constant field with polarization $\hat{\vect{e}}$ and the integral simplifies to~\cite{Sohl2007-oo}
\begin{align}
    s(\omega\rightarrow0) \simeq \left\|\Ev_0\right\|^2 \hat{\vect{e}}^{\dagger} \bm{\alpha} \hat{\vect{e}},
\end{align}
where $\bm{\alpha}$ is a static polarizability tensor (for any shape, not necessarily with any symmetries), and $\Ev_0$ is the plane-wave amplitude. Inserting this expression into the KK relation of \eqref{eq:KKExt} yields a sum rule for extinction (using the connection from $s(\omega)$ to extinction of \eqref{eq:PextOT}), $\int_0^{\infty} P_{\rm ext}(\omega)/\omega^2 \,{\rm d}\omega = (\pi \left\|\Ev_0\right\|^2/4) \hat{\ev}^{\dagger} \bm{\alpha} \hat{\ev}$. For practical use, typically two modifications are made: rewriting the integral over wavelength (as ${\rm d}\omega / \omega^2 \sim {\rm d}\lambda$), and the extinction power is rewritten in terms of the extinction cross-section, $\sigma_{\rm ext}(\omega)$. The sum rule is then
\begin{align}
    \int_0^{\infty} \sigma_{\rm ext}(\lambda) \,{\rm d}\lambda = \pi^2 \left(\hat{\ev}^{\dagger} \bm{\alpha} \hat{\ev}\right).
    \label{eq:SRExt0}
\end{align}
A preliminary version of this sum rule, specific to ellipsoids, was derived by Purcell in 1969~\cite{Purcell1969}. The first version generalized to arbitrary shapes was derived by Sohl et al. in 2007~\cite{Sohl2007-oo}. Experimental verification was reported in~\cite{sohl2008scattering_absorption_identity}. This sum rule states that the extinction cross-section integrated over all wavelengths is determined by the static polarizability of the scatterer~\footnote{A material with magnetic static response simply has an extra term $\hat{\vect{h}}^{\dagger} \bm{\alpha}_m \hat{\vect{h}}$, for magnetic-field polarization $\hat{\vect{h}}$ and magnetic polarizability $\bm{\alpha}_m$.}. Static polarizabilities typically vary weakly with material permittivity~\cite{Sihvola2004-nx} but are directly proportional to scatterer volume; hence, one can interpret this sum rule as dictating that volume is the key determinant of all-wavelength scattering (see also Section \ref{sec:polarizabilities}). The wavelength variable in the integral is important: high frequencies (e.g., visible) tend to occupy small wavelength bands and hence can have relatively small contributions to this static-frequency-based sum rule.

Conversely, one can take the high-frequency asymptotic limit of the KK relation, \eqref{eq:KKExt}. As discussed in Sec.~\ref{sec:causality}, at high frequencies an electric susceptibility must go to zero, scaling as $-\omega_p^2/\omega^2$, where $\omega_p$ is a generalized plasma frequency given by $\omega_p^2 = q_e^2 N_e / (\varepsilon_0 m_e)$, for electron charge $q_e$, mass $m_e$ (the free-electron mass), density $N_e$, and vacuum permittivity $\varepsilon_0$. As the susceptibility goes to zero, the material becomes transparent, and one can write $\Pv(\omega\rightarrow\infty) \simeq \chi \Einc = -(\omega_p^2/\omega^2) \Einc$. Then the infinite-frequency limit of the KK relation of \eqref{eq:KKExt} is
\begin{align}
    -\frac{\omega_p^2}{\omega^2} \left\|\Ev_0\right\|^2 V = - \frac{2}{\pi \omega^2} \int_0^{\infty} \omega' \Im s(\omega') \,{\rm d}\omega'.
\end{align}
The integral on the right-hand side is proportional to the extinction, per \eqref{eq:PextOT}; rewriting this relation in terms of the extinction cross-section, we have
\begin{align}
    \int_0^{\infty} \sigma_{\rm ext}(\omega)\,{\rm d}\omega = \frac{\pi \omega_p^2}{2c} V.
    \label{eq:SRExtInf}
\end{align}
This sum rule, for extinction integrated over frequency, is directly proportional to the scatterer volume, just as for \eqref{eq:SRExt0}~\footnote{For a magnetic material, the high-frequency asymptote is much more subtle than the static limit, cf.~\cite{landau2013electrodynamics}.}. In this case, the material plays an important role as well: the coefficient $\omega_p^2$ is proportional to the (total) electron density. In some cases, the product $\omega_p^2 V$ is rewritten to express the integrated cross-section just in terms of the total number of electrons. This sum rule harkens to the origins of dispersion theory in the 1920's~\cite{Ladenburg1923-ge,Kramers1924-eh,Thomas1925-fd,Kuhn1925-vt}, was put on more rigorous footing in the 1960's~\cite{Gordon1963-dt}, and revisited in the last decade~\cite{Yang2015-wb}. Analogous sum rules have been developed for periodic structures, all-spectrum transmission through arrays, and perforated metal screens~\cite{gustafsson2009physical_allspectrum,gustafsson2012optical_theorem_periodic,ludvigosipov2020fundamental_perforated}.
\begin{figure*}[tb]
\centering
\includegraphics[width=0.99\linewidth]{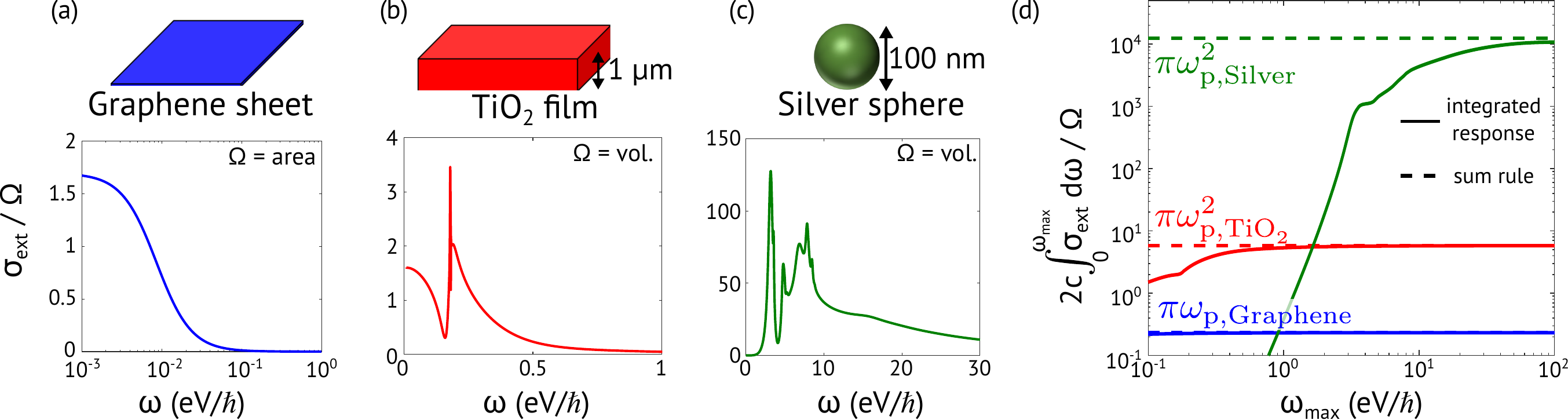}
\caption{\textbf{Spectral response and sum rules in three prototypical scatterers.} (a) A graphene sheet, (b) a TiO2 film, and (c) a silver sphere. (d) The extinction cross-section for a $z$-directed plane wave, normalized to the area/volume $\Omega$, displays wide-ranging spectral behavior, yet always converges to the sum rule of \eqref{eq:SRExtInf}.}
\label{fig:SumRuleThree}
\end{figure*}

Why are only electrons included in the sum rule, and not the nuclei? In fact, a most rigorous accounting for \eqref{eq:SRExtInf} includes them on equal footing~\cite{Franta2013-ml,Zhou2006-jt}. However, because the high-frequency (``plasma'') coefficients are proportional to the densities of each particle divided by their bare masses, the much heavier nuclei contribute negligibly towards the sum rule (neutrons, being uncharged, do not contribute at all). In fact, a reverse argument emerges: the heavy masses in the sum-rule constant explain why any optical susceptibilities mediated by nuclei can have significant contributions only at very long wavelengths, where their relative bandwidths must be small. This is a general argument for why phonon-mediated contributions to an electric susceptibility, which typically arise in the mid-infrared~\cite{Caldwell2015-fn}, in fact \emph{cannot} occur (with non-trivial strengths) at higher frequencies.

\begin{figure*}[tb]
\centering
\includegraphics[width=0.99\linewidth]{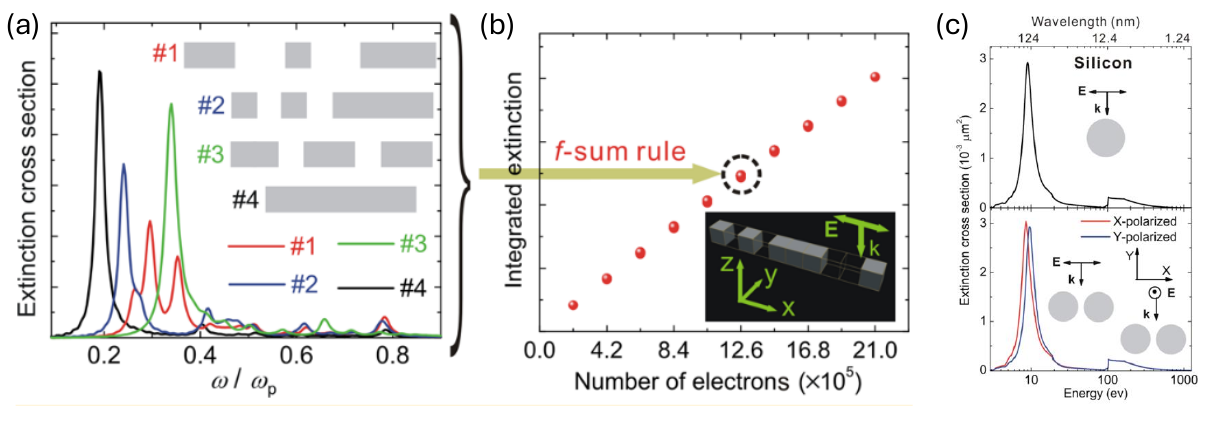}
\caption{\textbf{Confirmation of the plasma-frequency-based sum rule for a dielectric material.} (a) Simulated extinction cross-sections for a variety of antenna shapes and (b) their extinction as a function of total number of electrons. (c) Spectral contributions to extinction for a silicon sphere (top) and sphere dimer (bottom); notice the bump in extinction at energies above \SI{100}{eV}, from core electrons that are counted in the total electron density and which can contribute significantly to the sum-rule accounting. (Figure adapted from Ref.~\cite{Yang2015-wb}.)}
\label{fig:FSumRule}
\end{figure*}

To validate this sum rule, Fig.~\ref{fig:SumRuleThree} examines three prototypical scatterers spanning different material classes and dimensionalities: (a) a doped graphene sheet ($\mu = 0.5$\,eV, $T = 300$\,K), with the in-plane optical conductivity given by the standard intraband-plus-interband model~\cite{Stauber2008} and transmittance evaluated analytically for a sheet between vacuum half-spaces; (b) a $1$\,$\mu$m TiO$_2$ slab, with normal-incidence transmittance computed from a transfer-matrix calculation using a Drude--Lorentz permittivity fit~\cite{Siefke2016}; and (c) a silver sphere of radius $50$\,nm, with extinction obtained from a Mie T-matrix expansion truncated at multipole order $\ell = 10$, using a Drude--Lorentz fit for silver. Panel (d) shows the running integral $(2c/\Omega) \int_0^{\omega_{\rm max}} \sigma_{\rm ext}(\omega)\,{\rm d}\omega$, with $\Omega$ the relevant area (graphene) or volume (slab, sphere); in each case it converges to the corresponding $\pi \omega_p^2$ value (dashed lines) once $\omega_{\rm max}$ surpasses the dominant material resonances, confirming \eqref{eq:SRExtInf} across three decades in $\omega_p$. Figure~\ref{fig:FSumRule} confirms the same relation across a range of silicon antenna shapes, provided that core electrons are included in the total electron density---contributions that produce a distinct bump in the extinction spectrum above $\sim 100$\,eV.

\subsubsection{Local density of states} \label{sec:LDOS}
The \textbf{local density of states (LDOS)}, $\rho(\xv,\omega)$, proportional to the rate of energy loss by a radiating dipole in an arbitrary environment, is defined as
\begin{align}
    \rho(\xv,\omega) = \frac{1}{\pi\omega} \Im \Tr \Gamma(\xv,\xv,\omega),
    \label{eq:ldos}
\end{align}
where $\Gamma(\xv,\xv,\omega)$ is the Green's function, and the trace averages over orthogonal incoherent dipole orientations. The optical-theorem-like nature of LDOS, as the imaginary part of a scattering amplitude~\footnote{The optical theorem nature can be attributed to the fact that, for any environment, LDOS represents a sum of power lost to material absorption and power radiated to the far field.}, make it a plausible candidate for sum rules. At high frequencies, as material scatterers become transparent, the LDOS must asymptote to its free-space value, given by $\rho_0(\omega) = \omega^2 / 2\pi^2 c^3$~\footnote{This is the \emph{electric} LDOS, given by averaging over electric dipoles only. In some cases, the LDOS is defined as the sum of electric- and magnetic-dipole contributions (e.g., Ref.~\cite{Joulain2003}), in which case the free-space value doubles, $\rho_0(\omega) = \omega^2 / \pi^2 c^3$.}. The high-frequency decay is not sufficient to find a sum rule, with two possible resolutions.

One approach is to consider the quantity $(\rho - \rho_0)/\rho_0$, both subtracting and dividing by the free-space LDOS. In this case, the high-frequency decay is sufficiently fast that one need not even assume the material becomes transparent. Conversely, caution is needed near zero frequency, where the $\omega^2$ factor in $\rho_0$ can become a pole. To circumvent this singularity, one can use a modified Green's function with the longitudinal part removed. With this transverse-only Green's function, the zero-frequency singularity is removed, and one finds that the all-frequency integral is simply zero~\cite{Barnett1996}:
\begin{align}
    \int_0^{\infty} \frac{\rho(\xv,\omega)-\rho_0(\omega)}{\rho_0(\omega)}\,{\rm d}\omega = 0.
\end{align}
This sum rule states that on average (weighted by $1/\rho_0$), the difference $\rho-\rho_0$ is zero, meaning that any enhancements in LDOS at one frequency must be compensated by reductions at other frequencies. Away from free space, the longitudinal part must be removed on both indices of the Green's function, $P_T\Gamma P_T$, with $P_T$ the projector onto transverse (divergence-free) fields; this transverse spectral density obeys the zero sum rule for nonmagnetic bodies but generally differs from the physical decay rate of an emitter~\cite{Scheel2008}. This sum rule has been interpreted to give a wave-physics justification of the Yablonovitch $4n^2$ all-angle, broadband-frequency absorption enhancement limit~\cite{Buddhiraju2017}.

\begin{openq}
    Are there more rigorous physical interpretations or justifications of the transverse-only Green's function used to obtain the LDOS sum rule?
\end{openq}

A second approach is to simply subtract the free-space GF, without dividing by it, considering instead the quantity $\rho - \rho_0$. Now, the material must become transparent at high frequencies, but the low-frequency asymptote is well-behaved and contributes a nonzero total sum. Following the same process as for extinction, it is straightforward to find a KK relation and evaluate its asymptotic limit, giving the sum rule~\cite{Shim2019bandwidth}
\begin{align}
    \int_0^{\infty} \left[ \rho(\xv,\omega) - \rho_0(\omega)\right] \,{\rm d}\omega = \alpha_{\rm LDOS},
\end{align}
where the constant on the right-hand side is an LDOS version of a ``polarizability,'' defined as $\alpha_{\rm LDOS} = (1/2) \Re \Tr \Gamma(\xv,\xv,0)$. This can be bounded above by simple constants via monotonicity theorems. In the case of metallic ($\varepsilon\rightarrow \infty$ as $\omega\rightarrow 0$) half-space scattering, the sum rule simplifies to
\begin{align}
    \int_0^{\infty} \left[ \rho(\xv,\omega) - \rho_0(\omega)\right] \,{\rm d}\omega = \frac{1}{16\pi d^3},
    \label{eq:LDOSSumRuleHS}
\end{align}
where $d$ is the distance from the source position ($\xv$) to the scatterer.
\begin{figure*}[tb]
\centering
\includegraphics[width=0.8\linewidth]{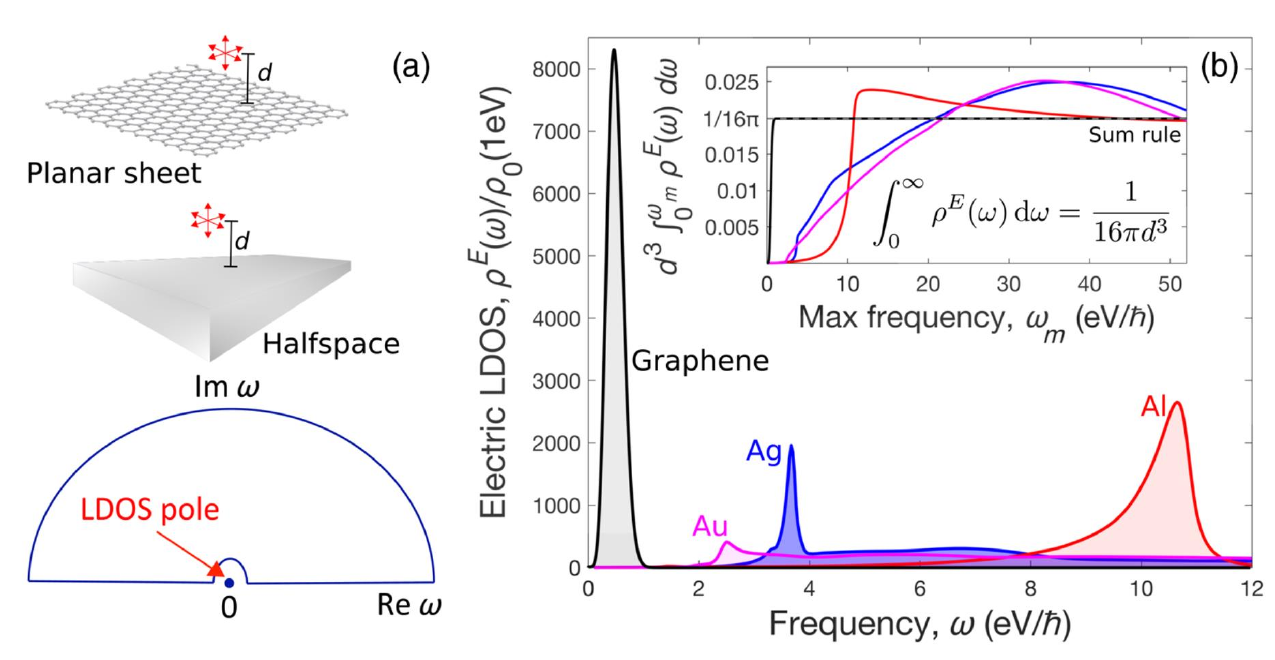}
\caption{\textbf{LDOS sum rules from contour integration.} (a) Schematic for LDOS calculation above a 2D sheet or 3D half-space, computed by a contour around the singularity at the origin (bottom). (b) Confirmation of the sum rule for 2D and 3D materials. The sum rules of these simply high-symmetry geometries also \emph{bound} the all-frequency integrals for any scatterer contained within them, via monotonicity theorems. (Figure adapted from Ref.~\cite{Shim2019bandwidth}.)}
\label{fig:LDOSSumRule}
\end{figure*}
The sum rule applies to bulk materials as well as to 2D materials, because both reduce to the same static limit: a perfectly conducting bulk half-space and a perfectly conducting infinitesimal sheet present identical zero-frequency responses to a nearby dipole, and the sum rule depends only on this static-limit polarizability $\alpha_{\rm LDOS}$. Both setups are illustrated schematically in Fig.~\ref{fig:LDOSSumRule}(a). Panel~(b) of the figure shows the electric-LDOS spectra for three metallic half-spaces (Au, Ag, Al) and a graphene sheet, exhibiting distinct resonance features at very different frequencies: graphene's plasmon near $0.5\,{\rm eV}$, surface-plasmon peaks for Au, Ag, and Al at progressively higher frequencies. All four spectra integrate to the same sum-rule value of $1/(16\pi d^3)$ (Eq.~\ref{eq:LDOSSumRuleHS}), as confirmed in the inset.

\subsubsection{$\TT$-matrix response}
\label{sec:T-matrix-response}
\begin{figure*}[tb]
\centering
\includegraphics[width=0.99\linewidth]{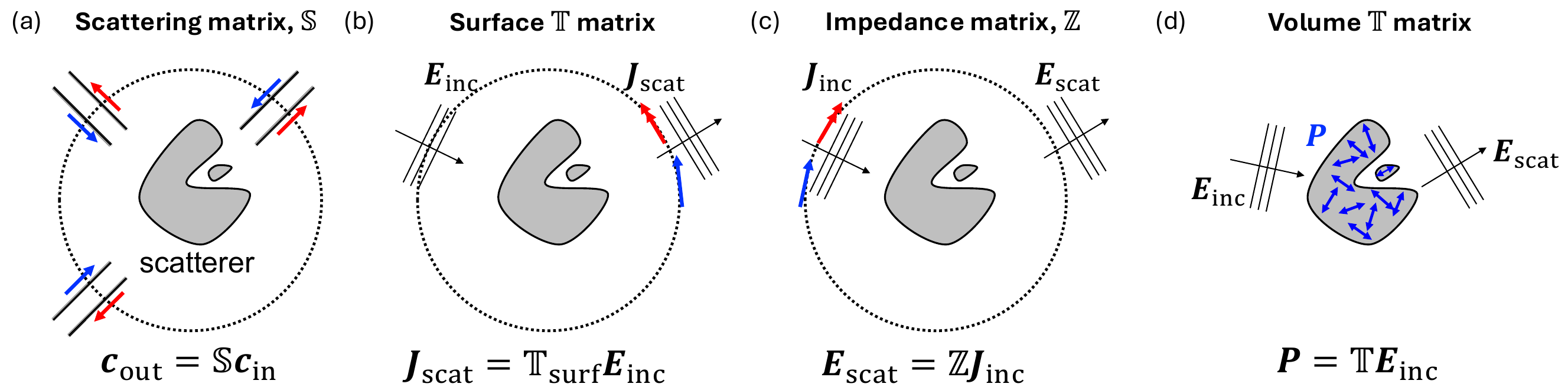}
\caption{\textbf{Common matrices constituting input--output relations in linear electromagnetism.} For the first three, (a)--(c), it is an open question whether a KK relation, a sum rule with material constants, and a simple passivity condition can be combined for meaningful bounds. (d) Volume $\TT$ matrices, which relate incident fields in a scatterer's volume to the polarization fields they induce, exhibit all three in tandem, yielding the simple causal relations of \eqref{eq:Tosc}.}
\label{fig:STZMatrices}
\end{figure*}

A new approach recently emerged that complements the optical-theorem nature of the previous subsections. Instead of starting with power objectives with optical theorems, one can target response functions themselves. Optical susceptibility and refractive-index sum rules~\cite{Nussenzveig1972,King1976-qu} are of this nature, and circuit theory, for example, has an extensive literature with causal representations of scattering and impedance matrices~\cite{Raisbeck1954,Wohlers1969,Boyd1982,bernland2011sum,srivastava2021causality}. Yet extending these ideas to electromagnetic waves requires care and caution. Fig.~\ref{fig:STZMatrices}(a--c) shows three input-output matrices that can be defined in electromagnetic scattering: the scattering matrix, $\SS$, the ``surface'' (or Waterman~\cite{Waterman1965-gp}) $\TT$ matrix, and an impedance matrix, $\mathbb{Z}$. In each case, it is not known how to achieve all four pre-requisites discussed above (an optical theorem, a KK relation, high/low-frequency constants, and a passivity constraint) in a form that carries material information. Consider the scattering matrix, $\SS$, for example, for the case of a scatterer in free space. If the $\SS$ matrix basis functions are the vector spherical waves, then a nonzero scatterer size implies that outgoing scattered waves can be generated \emph{earlier} than in the absence of the scatterer, modifying the causality condition which leads to additional phase shifts in the corresponding KK relation~\cite{Nussenzveig1972}. Once this geometric delay is removed, passivity ($\SS^{\dagger}\SS \leq \II$) and causality do combine: the delayed matrix $e^{2i\omega a/c}\SS$, where $a$ is the radius of the smallest sphere, centered on the origin of the expansion, that encloses the scatterer, is bounded-real~\cite{Youla1959}, which yields the spherical-wave sum rules of \eqref{eq:sum_rule_sn} and their recent extension to coherent channel superpositions~\cite{Han2026}. The only constant in these sum rules, however, is the geometric delay $2a/c$: they bound how strongly, and over what bandwidth, an object of a given size can reduce the outgoing spherical waves (through absorption or conversion into other channels), and even their refinements that include the static polarizability~\cite{bernland2011physical} bound only this reduction, not the magnitude of the response itself. Alternatively, pole--zero (Weierstrass) representations of the $\SS$ matrix have been proposed~\cite{Grigoriev2013-vw,Krasnok2019-ee}, but their generality beyond spherically symmetric scatterers is unclear~\cite{Benzaouia2021-fn}, and in any case they appear more useful for few-parameter scattering models rather than fundamental constraints.

\begin{openq}
    Is there a sum rule for conventional $\SS$ (or surface-$\TT$, or impedance) matrices whose constants carry material information, enabling bounds on the magnitude of the response rather than only its reduction?
\end{openq}

The ``volume'' $\TT$ matrix is a fourth linear scattering matrix, as depicted in Fig.~\ref{fig:STZMatrices}(d), and it resolves the causality/passivity issues that arise for the first three matrices. This $\TT$ matrix maps the incoming fields within the domain of a scatterer to the polarization fields that they induce; it can be defined from a volume-integral-equation formulation of electromagnetic scattering~\cite{Carminati2021-zb,zhang2023all}. The transparency of materials at high frequencies ensures sufficiently rapid decay of the elements of $\TT(\omega)$ at high frequencies, and the specific asymptotic scaling $\TT(\omega\rightarrow \infty) \sim -(\omega_p^2/\omega^2) \II_{V}$, where $\II_{V}$ denotes the identity matrix/operator on the scatterer domain $V$. One can similarly define a static $\TT$ matrix at zero frequency for a second sum-rule constant. The KK relation follows exactly the same procedure as a conventional material susceptibility, with one twist: the ``crossing symmetries'' around $\omega=0$ require careful analysis for matrices. The upshot is that the reciprocal (symmetric) and nonreciprocal (or ``anti-reciprocal,'' skew-symmetric) parts of the $\TT$ matrix have different symmetries and hence different functional KK relations. Together, the KK relation together with the asymptotic constants and a simple passivity condition converge in an oscillator representation for any linear scattering $\TT$ matrix~\cite{zhang2023all}, 
\begin{align}
    \TT(\omega) = \frac{2}{\pi} \lim_{\gamma\rightarrow 0} \int_0^{\infty} \frac{1}{\omega_i^2 - \omega^2 - i\gamma\omega} \left[ \mathbb{X}(\omega_i) + \frac{\omega_i}{\omega} \mathbb{Y}(\omega_i) \right] \,{\rm d}\omega_i.
    \label{eq:Tosc}
\end{align}
In this expression, the variables $\mathbb{X}(\omega_i)$ and $\mathbb{Y}(\omega_i)$ should be thought of as matrix-valued generalizations of oscillator strengths. The matrix $\mathbb{X}$ is symmetric ($\mathbb{X}^T = \mathbb{X}$) and the matrix $\mathbb{Y}$ is skew-symmetric ($\mathbb{Y}^T = -\mathbb{Y}$). In reciprocal scatterers, the skew-symmetric oscillator strength $\mathbb{Y}(\omega)$ is identically zero at all frequencies. \eqref{eq:Tosc} is an oscillator representation of the volume $\TT$ matrix, stating that every such matrix is a sum of Drude--Lorentz oscillator with matrix-valued coefficients. Its utility for bounds comes from the additional constraints on the oscillator-strength matrices, via passivity and the asymptotic constants, ultimately yielding~\cite{zhang2023all}
\begin{subequations}
\begin{align}
    &\mathbb{X}(\omega_i) \geq 0, \\
    -&\mathbb{X}(\omega_i) \leq \mathbb{Y}(\omega_i) \leq \mathbb{X}(\omega_i) \\
    \int_0^{\infty} &\mathbb{X}(\omega_i) \,{\rm d}\omega_i \leq \frac{\pi}{2}\omega_p^2 \II_V \\
    \int_0^{\infty} &\frac{\mathbb{X}(\omega_i)}{\omega_i^2} \,{\rm d}\omega_i \leq \frac{\pi}{2}\TT_0,
\end{align}
\end{subequations}
where the inequalities are understood in the sense of positive-definiteness. Hence the matrix coefficients $\mathbb{X}(\omega_i)$ are positive semidefinite, with a finite integral, offering the matrix analog of a positive oscillator strength with maximum allowable distribution. The matrix coefficients $\mathbb{Y}$ are not positive semidefinite, but are bounded at any frequency by the reciprocal coefficients $\mathbb{X}$. Collectively, these matrix coefficients parameterize every passive $\TT$ matrix, but the constraints are necessary rather than sufficient: passivity enters only as $\omega \Im \TT \geq 0$ (nonnegative extinction), not as the optical-theorem condition of nonnegative absorption. Adding that condition at any frequency can only tighten a bound, and doing so was part of identifying the ultimate limits to near-field radiative heat transfer~\cite{zhang2023all}.

\subsection{A general computational-bound framework} 
\label{sec:comput_bounds}
In this section, we describe a generalization of the optical-theorem-based approaches developed above. The approach described here is, in one sense, a seemingly complete theory of quadratic constraints for design problems with the linear Maxwell equations. There is still an interesting question at the heart of this claim, however, as described in the first open question below. 

As described in Sec.~\ref{sec:conslaws}, the real and reactive power conservation laws can be formed by starting with the volume integral equation and multiplying from the left with $\phi^{\dagger}$, in simplified vector notation, or in spatial notation by multiplying on the left by the polarization field $\phi^{\dagger}(\xv)$ and integrating over space, $\int {\rm d}\xv$. One can generalize this approach further by \emph{not integrating} over space. Further, instead of taking the inner product against the entire polarization field, one can select to a particular overlap of particular field components, both to the left and the right.

The abstract vector notation description of these generalizations is achieved by multiplying by an (almost) arbitrary matrix $\DD_i$ before multiplying by $\phi^{\dagger}$:
\begin{align}
    \phi^{\dagger} \DD_i \left( \Gamma_0 + \xi \right) \phi = -\phi^{\dagger} \DD_i \psi_{\rm inc}.
    \label{eq:GenConstraint}
\end{align}
for many possible choices of matrix, indexed by $i$. In the space-dependent notation, this is equivalent to
\begin{align}
    \int \int \phi^{\dagger}(\xv) \DD_i(\xv) \left( \Gamma_0(\xv,\xv') + \xi(\xv)\delta(\xv-\xv') \right) \phi(\xv')\,{\rm }d\xv d\xv' = -\int\phi^{\dagger}(\xv) \DD_i(\xv) \psi_{\rm inc}(\xv){\rm }d\xv.
    \label{eq:GenConstraint2}
\end{align}
Notice that in the latter expression, we have made the matrix $\DD_i$ to have a single position argument. This is not the most general analog of matrix $\DD_i$ in \eqref{eq:GenConstraint}; instead, it could have two position arguments (e.g., as a convolution operator). Of course, one can multiply the volume integral equation by an arbitrary linear operator such as a convolution. For general bounds, however, it is crucial to multiply by a ``local'' operator, localized to a single position. The source of this is a subtle point that was, in fact, already implicit in the bounds of Sec.~\ref{sec:SingleFreq}.

Consider a scatterer with volume $V_s$, which means the susceptibility $\chi$ is nonzero (and the parameter $\xi$ does not diverge) only on $V_s$. Assume a designable volume $V_d$ that contains all possible scatterer volumes. The volume integral equation at the heart of the quadratic constraints is defined \emph{only} on the domain of the scatterer: the operators/matrices $\Gamma_0$ and $\xi$ are only evaluated at points in $V_s$. From this starting point, then, the integrals in \eqref{eq:GenConstraint2} should both be evaluated over the domain of a specific scatterer, $V_s$. This would inhibit any possible bounds, however, which should apply over the designable domain $V_d$, encompassing many possible scattering volumes. After multiplying to the left by $\phi^{\dagger}(\xv)$, where $\xv$ is the \emph{same} location at which the Green's function is evaluated, a new possibility emerges. Now, at points $\xv$ outside of $V_s$, the quadratic constraint can still hold, because $\phi(\xv)$ is zero at those points. The same is true on the right-hand side of \eqref{eq:GenConstraint2}. The same argument applies to $\xv'$: it can be extended to the entire designable domain, because at points outside $V_s$, $\phi(\xv')=0$. Hence, with a spatially local $\DD$ matrix, \eqref{eq:GenConstraint} and \eqref{eq:GenConstraint2} can both be extended to the designable domain, thereby become ``\textbf{domain-oblivious}'' constraints that apply to all possible designs within $V_d$.

Suppose we have many constraints from many distinct $\DD_i$ matrices. (We discuss below how one might choose how many.) How can bounds be computed? One approach is to extend the Lagrangian duality concepts discussed above. Equivalently, one can connect the duality approach for quadratic constraints to a ``lift-and-relax'' procedure. We discuss the mathematical concept here and refer the interested reader to Refs.~\cite{Molesky2020,Kuang2020_computational} for granular details that arise in photonic design problems. For a general optimization of a quadratic objective $\xv^{\dagger} \AA \xv$ subject to constraint $\xv^{\dagger} \DD_i \xv=c_i$, for many $i$, one can apply two tricks~\cite{Luo2010-pa}: take the trace of each quadratic form, use the cyclic permutation invariance of the trace operator, and then define a matrix variable $X = \xv\xv^{\dagger}$. With these tricks, one can transform the \textbf{quadratically constrained quadratic program}:
\begin{equation}
\begin{aligned}
\max_{\xv} \quad & \xv^{\dagger} \AA \xv = \Tr\left( \AA \xv\xv^{\dagger} \right)\\
\textrm{s.t.} \quad & \xv^{\dagger} \DD_i \xv = \Tr\left( \DD_i \xv \xv^{\dagger}\right) = c_i
\end{aligned}
\quad
\Rightarrow
\quad 
\begin{aligned}
\max_{X} \quad &\Tr\left(\AA X\right) \\
\textrm{s.t.} \quad &  \Tr\left(\DD_i X\right) = c_i \\
\quad &X \geq 0, \operatorname{rank}\left(X\right) = 1,
\end{aligned}
\end{equation}
where the final two conditions on the right-hand side correspond to the condition $X=\xv\xv^{\dagger}$, which is equivalent to enforcing that $X$ is positive semidefinite and has rank one. The objective and all constraints involving $\DD_i$ in the right-hand side are all \emph{linear} (and thereby convex) functions of the new matrix variable $X$. This is the ``lifting'' part of the procedure: if we started with $N\times1$ vectors $\xv$, the new matrix variables has $N^2$ elements, occupying a much higher-dimensional space. The benefit of moving to this higher-dimensional space is that the quadratic functions have been converted to linear functions. The positive semidefinite condition on $X$ is still a convex constraint, but the rank-one constraint is not. This is where we ``relax:'' one can simply drop the rank-one constraint, leaving a \textbf{semidefinite program (SDP)}, which is now truly a convex problem. Removing the rank-one constraint can only increase the optimal solution, meaning that the optimum of the SDP is a fundamental limit to the original quadratically constrained quadratic program. The promise of a convex problem is that its global optimal can be computed ``efficiently,'' i.e., in polynomial time, though the value of the polynomial is not ideal. For a design problem with $N$ degrees of freedom, the computational runtime of standard SDP solver scales as $N^{3.5}$ or $N^4$, which inhibits application at large scale~\cite{Luo2010-pa}. There is promising recent research on this front, with specialized solvers showing lower-order scaling and much larger problem sizes than previously possible~\cite{Burer2003-hs,Arora2007-vj,Boumal2016-kc,Ding2021-hd}.

\begin{figure*}[tb]
\centering
\includegraphics[width=0.99\linewidth]{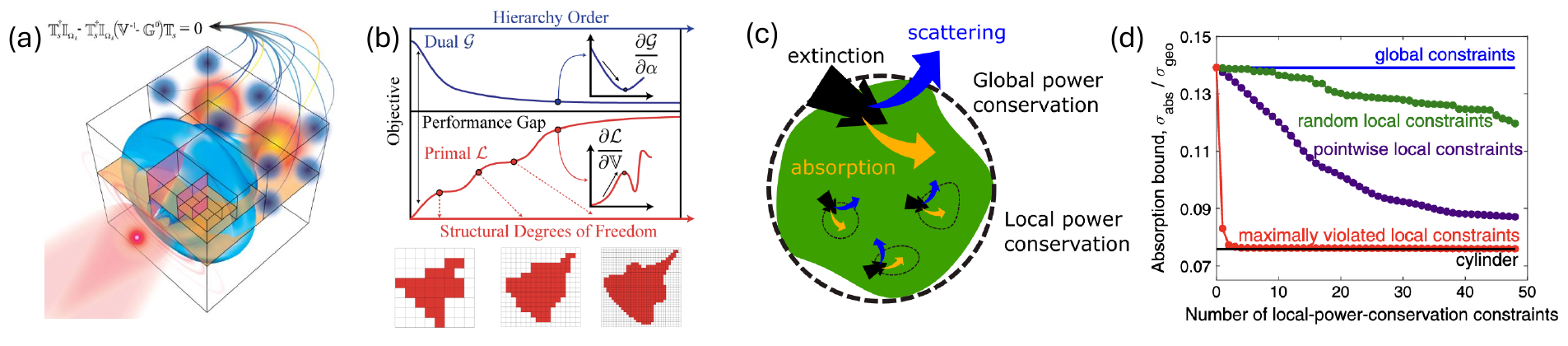}
\caption{\textbf{Computational bounds via local conservation laws.} (a) The hierarchical mean-field $\TT$-operator viewpoint: a structured scatterer inside a designable cubic domain (light blue), partitioned into nested subdomains. Constraints on the scattering $\TT$ operator are imposed on each subdomain rather than only globally. (b) As the hierarchy order increases (more, finer clusters), the convex dual bound tightens from above while the achievable primal performance improves from below, closing the performance gap. (c) Local-conservation law viewpoint, via polarization currents: at every point within a freeform scatterer in a designable region, extinction must equal local absorption plus local scattering, yielding an infinite family of domain-oblivious constraints. (d) Convergence of an absorption bound versus the number of imposed local constraints. Iteratively selecting the maximally violated constraint (red) converges to the cylinder-saturated optimum (black) within a few iterations; random (green) or pointwise (purple) selection converges far more slowly, and global constraints alone (blue) yield a much looser bound. (Panels (a,b) adapted from Ref.~\cite{Molesky2020}; panels (c,d) adapted from Ref.~\cite{Kuang2020_computational}.)}
\label{fig:CompBounds}
\end{figure*}
An important consideration regarding computational complexity is how many quadratic constraints are formed? In other words, how many $\DD_i$ matrices should be chosen? A simple ``default'' is as follows. For $N_s$ spatial degrees of freedom (DOFs), the $\DD_i$ matrices must be diagonal in space, and hence there are only $N_s$ independent $\DD_i$ matrices that can be imposed. (For example, one each with a single 1 at a unique diagonal, and all other entries 0.) For $N_{ns}$ non-spatial DOFs (e.g. polarization), the $\DD_i$ matrix need not be diagonal, and hence there are at most $N_{ns}^2$ matrices for each spatial DOF. Together, then, the number of independent $\DD_i$ matrices is $N_s N_{ns}^2$~\footnote{A closer consideration of the scaling exponents for the constraint count appears especially important for quantum control problems. See Ref.~\cite{zhang2021conservation} for an introduction to quantum control bounds using this ``$\DD$-matrix'' approach. For photonic design problems, the number of constraints scales linearly with the number of space points but quadratically with non-spatial DOFs such as polarization. For a quantum control problem, say, transforming quantum state $A$ to quantum state $B$ in the least amount of time, the number of analogous constraints would be linear in the number of time points but quadratic in the size of the Hilbert space. Quantum control problems with large Hilbert spaces would therefore have a constraint count that grows quite rapidly. In theory, this can be ameliorated by only taking a minimal subset of the required constraints, restoring a linear scaling with Hilbert space size. But equivalent quadratic constraint sets can lead to different semidefinite relaxations~\cite{Anstreicher2000-da}, so it remains to be seen whether the constraint count can be kept reasonable while maintaining the quantitative utility of the bounds.}. These can be collectively imposed and bounded using the lift-and-relax approach discussed above. To reduce the computational burden, and increase the size of the systems that can be analyzed with this approach, it is beneficial to reduce the number of constraints without further relaxations of the bound. One approach to do so is simply to enforce the constraints on average over spatial regions. One can do this in a hierarchical manner, starting with coarse resolution and slowly increasing it until bound convergence is achieved, as depicted in Fig.~\ref{fig:CompBounds}(a,b). From this perspective, one is successively tightening both the bounds (upper limits, blue) as well as the performance of the optimal design (red), with increasing spatial resolution. Successful examples of this approach are given in Ref.~\cite{Molesky2020}. Another heuristic approach starts with the two global conservation-law constraints, identifying the non-imposed space-diagonal $\DD_i$ matrix whose corresponding constraint is \emph{maximally violated}, impose that constraint, and iterate. In numerical examples such as those given in Ref.~\cite{Kuang2020_computational} and depicted in Fig.~\ref{fig:CompBounds}(c,d), this technique has been shown to converge surprisingly quickly. Yet, it is not even clear that enforcing local conservation at the finest spatial resolution is sufficient. There is an at-first paradoxical result known in semidefinite programming: adding ``redundant'' quadratic constraints can produce inequivalent semidefinite relaxations, and hence inequivalent bounds. We discuss an example below, in the differential-equation formulation, showing evidence of this phenomenon in photonic design. But there is no general understanding of how to formulate an ``optimal'' set of constraints (for various definitions of ``optimal''), leaving the open questions:
\begin{openq}
    What minimal set of quadratic constraints, possibly including ``redundant'' variations~\cite{Anstreicher2000-da}, leads to the tightest bounds? Can these sets be further reduced in size with minimal change to the bounds, for computational efficiency?
\end{openq}

\begin{figure*}[tb]
\centering
\includegraphics[width=0.99\linewidth]{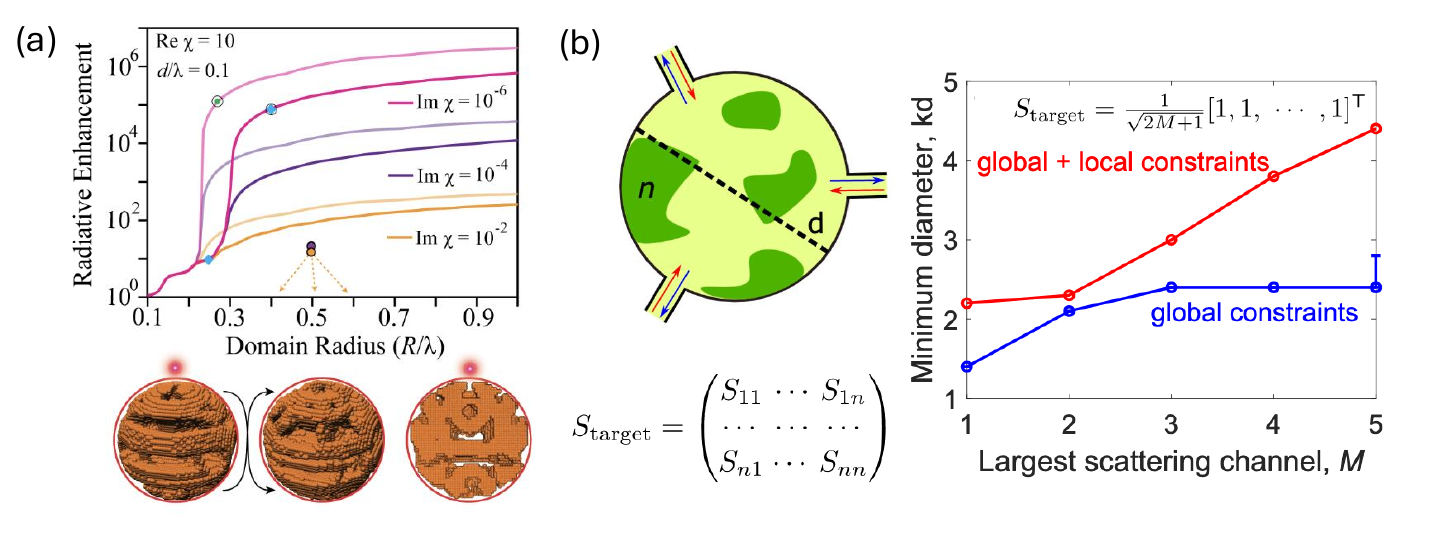}
\caption{\textbf{Applications of the computational-bound framework.} (a) Bounds on the near-field radiative Purcell enhancement of a dipole emitter coupled to a spherical, freeform-patternable dielectric domain, plotted against the domain radius $R/\lambda$ for several values of the material loss $\Im\chi$. Light curves apply only global conservation; dark curves apply the hierarchical local constraints and tighten the bounds by more than an order of magnitude. Markers are inverse-designed structures whose performance approaches the bounds. (b) Minimum scatterer diameter $kd$ for a 2D freeform dielectric region to implement a target scattering matrix $S_{\rm target}$ that couples up to the $M$-th angular-momentum channel. Global constraints alone (blue) yield an unphysically small minimum size that saturates with $M$; adding local-power-conservation constraints (red) gives bounds that grow with $M$, consistent with channel-counting intuition. (Panel (a) adapted from Ref.~\cite{Molesky2020}; panel (b) adapted from Ref.~\cite{Kuang2020_computational}.)}
\label{fig:ExCompBounds}
\end{figure*}
Examples of the successful application of the computational bound approach described above are shown in Fig.~\ref{fig:ExCompBounds}. This figure shows two scenarios. In the first, Fig.~\ref{fig:ExCompBounds}(a), from Ref.~\cite{Molesky2020}, bounds and inverse designs are computed for the enhancement in local density of states (LDOS), related to the Purcell factor, at a single point close to a spherical domain comprising a nearly lossless dielectric that can be arbitrarily patterned. The success of the computational-bound approach can be seen in the distinction between the light-shaded lines and their darker counterparts. The former are bounds only applied across the entire domain, whereas the latter use more spatial resolution to refine the bounds and come close to--or even coincide with--inverse designs, given by the markers. The second example, in Fig.~\ref{fig:ExCompBounds}(b), from Ref.~\cite{Kuang2020_computational}, considers a circular-shaped scattering domain that can be arbitrarily patterned with a lossless dielectric, and identifies the smallest domain over which a targeted scattering matrix can be achieved with high fidelity. The bound approach cannot directly identify a minimal domain size, but one can instead identify bounds on the maximum fidelity of achieving the target matrix, for a fixed domain size, and then perform, e.g., a binary search to discover the smallest domain allowing for a given fidelity.

Although the discussion above centers around integral equations and a volume-integral formulation of scattering theory, the use of quadratic constraints for computational bounds can also be directly applied to the (more common) differential equation formulation of photonic design problems, an approach developed extensively in Ref.~\cite{Gertler2025-ri}. Consider a designable domain in which the material parameter $\nu$ (e.g., the permittivity) can take one of two values, $\nu_1$ or $\nu_2$, as well as a nondesignable domain with a single parameter value at each space point $\xv$: $\nu_{\rm nd}(\xv)$, which may be spatially varying. The nondesignable domain may consist of free space, other background or nondesignable structures (perhaps an interior object to be cloaked), and/or perfectly matched layers (PMLs, which can be thought of as absorbing materials). We assume the use of PMLs or similar absorbing regions, to simplify the boundary conditions (typically to perfect conductors). We can write Maxwell's equations across the two domains as:
\begin{align}
    M_0 \psi + \nu\psi &= s, \quad \nu(\xv) = \nu_1 \textrm{ or } \nu_2, \quad &\xv \textrm{ in designable region}, \nonumber \\
    M_0 \psi + \nu_{\rm bg}\psi &= s, \quad \quad &\xv \textrm{ in nondesignable region},
    \label{eq:DiffEqConstraints}
\end{align}
where $s$ denotes the sources (e.g., producing the incident fields). At every point $\xv$ in the designable region, we know that $\psi(\xv)$ satisfies \emph{either} $(M_0+\nu_1)\psi-s=0$ or $(M_0+\nu_2)\psi-s=0$. We can enforce this either-or constraint by taking the inner product of the two, sandwiched around arbitrary matrices $\DD_i$:
\begin{align}
    \left[(M_0+\nu_1)\psi-s\right]^{\dagger}\DD_i\left[(M_0+\nu_2)\psi-s\right] = 0, \quad \textrm{ in the designable region},
\end{align}
over all independent $\DD_i$. These quadratic constraints are exactly the differential-equation analogs of the integral-equation constraints in \eqref{eq:GenConstraint2}. A version of these constraints was proposed in Ref.~\cite{Angeris2021}, but a crucial cross-constraint from the complex-valued nature of these equations was missed, as was an additional subtlety regarding the nondesignable domain~\footnote{This subtlety does not arise in the integral-equation approach because the nondesignable domain is embedded in the Green's function and absent from the polarization field; the degrees of freedom \emph{only} arise within the designable region.}. The correct formulation --- including both the cross-constraint and the redundant constraints over the nondesignable region introduced below --- was developed in Ref.~\cite{Gertler2025-ri}. In theory, one could simply leave the linear constraint equation of \eqref{eq:DiffEqConstraints}. However, it turns out to substantially improve the resulting bounds if one adds ``redundant'' quadratic constraints formed by multiplying the linear equation ``by itself:''
\begin{align}
\left[(M_0+\nu_{\rm bg})\psi-s\right]^{\dagger}\DD_i\left[(M_0+\nu_{\rm bg})\psi-s\right] = 0, \quad \textrm{ in the nondesignable region}.
\end{align}
This exemplifies the idea that ``equivalent'' quadratic constraints can lead to different constraints after relaxation, as discussed above. There is another subtlety regarding the differential equation formulation. In particular, the either-or formulation above relies on the \textbf{local} nature of the constitutive relation; nonlocal constitutive relations would potentially require a different approach, though they are relatively rare in the landscape of photonic design problems. What is far more common, however, is a \emph{nonlocal numerical implementation of local constitutive laws}. Typical finite element methods~\cite{Jin2011}, for example, use overlapping basis functions that create nearest-neighbor couplings even for local permittivities/materials. Nonlocal implementations offer no obstacle for simulation---the differential operators are similarly nonlocal, and matrix inversion can still be done quickly, but they do present an obstacle for local constraint formation. Hence an important question to study is:
\begin{openq}
    Is there an efficient, general computational framework for forming quadratic conservation-law constraints when local constitutive relations are numerically approximated with nonlocal matrix equations, as is common in finite element formulations?
\end{openq}
A conceptual advance in Ref.~\cite{Gertler2025-ri} is the recognition that, when properly formulated, photonic design problems become \textbf{sparse} QCQPs. The key reframing is to treat the field $\psi(\xv)$ (rather than the material distribution $\nu(\xv)$) as the primary degree of freedom---a \textbf{field-optimization} perspective complementary to the conventional \textbf{structure-optimization} viewpoint. The binary-material requirement is enforced through the quadratic field constraints above, and the resulting QCQP inherits the sparsity of the underlying Maxwell operator, in stark contrast to the dense matrices that arise from integral-equation formulations. SDP relaxations that would be intractable for dense matrices remain tractable at sizes relevant to large-area metasurface design, thanks to the sparsity property. Furthermore, the SDP solutions were observed to have bounded rank, so that the same SDP that produces an upper bound also yields a near-globally-optimal design through a ``lift-and-recover'' procedure. The construction extends naturally to anisotropic, magnetic, and bianisotropic materials, and the underlying mathematical structure suggests applicability to the broader class of bilinear differential-equation design problems, from structural mechanics and fluid dynamics to quantum control. A technical detail of importance for practical implementation---namely, the construction of local quadratic constraints when local constitutive laws are discretized with overlapping (nonlocal) finite-element bases---is addressed in the Supplementary Materials of the same reference, providing one resolution to the open question above.

\begin{figure*}[tb]
\centering
\includegraphics[width=0.99\linewidth]{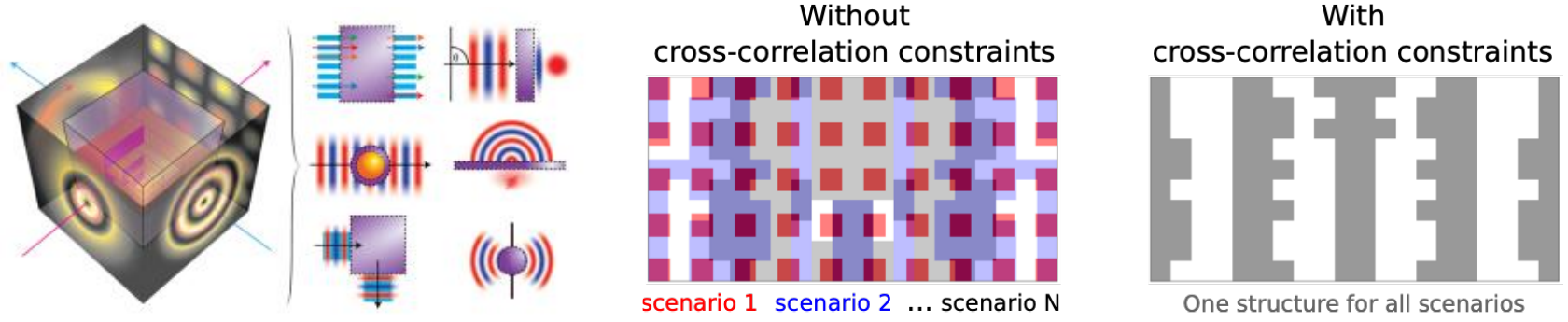}
\caption{\textbf{Multi-functional design via cross-correlation constraints.} (Left) A multi-functional device must produce distinct responses---different focal spots, scattering patterns, transmission angles, etc.---across multiple incident fields, frequencies, or material configurations, while comprising a single physical structure. (Middle) Without cross-correlation constraints, the bound formulation treats each scenario independently, and the SDP relaxations recover incompatible per-scenario patterns with no single physical realization. (Right) Adding the cross-correlation constraints of \eqref{eq:MultiScenarioConstraints} enforces a single shared structure across all scenarios, yielding a physically realizable multi-functional design. (Left panel adapted from Ref.~\cite{Molesky2022}; middle and right panels adapted from Ref.~\cite{Shim2021}.)}
\label{fig:MultifunctionalIllustration}
\end{figure*}

An important aspect of the computational bound approach is that it can be applied to \textbf{multi-functional and tunable designs}: a single physical structure that must produce distinct responses across multiple excitations, frequencies, or material configurations---for example, a liquid-crystal metasurface that steers a beam differently across two voltage states (Fig.~\ref{fig:MultifunctionalIllustration}, left). The framework for bounding such designs was developed in Ref.~\cite{Shim2021}. Going back to the volume-integral formulation (but equally applicable to differential equations), consider two scenarios, with two background Green's functions, $\Gamma_0^{(1)}$ and $\Gamma_0^{(2)}$, two material tensors, $\xi^{(1)}$ and $\xi^{(2)}$, two incident fields, $\psi_{\textrm{inc},1}$ and $\psi_{\textrm{inc},2}$, and two polarization fields $\phi_1$ and $\phi_2$, satisfying the equations
\begin{align}
    \left(\Gamma_0^{(1)}+\xi^{(1)}\right) \phi_1 &= -\psi_{\textrm{inc},1} \nonumber \\
    \left(\Gamma_0^{(2)}+\xi^{(2)}\right) \phi_2 &= -\psi_{\textrm{inc},2}.
    \label{eq:IE2S}
\end{align}
Of course, one can multiply sub-equation $i$ on the left by $\phi_i^{\dagger} \DD_k$, for arbitrary block-diagonal $\DD_k$, to form independent constraints for each scenario $i$. Without added constraints, however, this would correspond to forcing binary design constraints independently for each scenario. One can force a single design pattern, possibly with different material tensors across that pattern (as in the liquid-crystal example), by forming the additional constraints
\begin{align}
    \phi_2^{\dagger} \DD_{\ell} \left(\Gamma_0^{(1)}+\xi^{(1)}\right) \phi_1 &= -\phi_2^{\dagger} \DD_{\ell} \psi_{\textrm{inc},1} \nonumber \\
    \phi_1^{\dagger} \DD_{m} \left(\Gamma_0^{(2)}+\xi^{(2)}\right) \phi_2 &= -\phi_1^{\dagger} \DD_{m} \psi_{\textrm{inc},2}.
    \label{eq:MultiScenarioConstraints}
\end{align}
These \textbf{cross-correlation constraints} have a clean physical interpretation. The first constraint forces, at each design point, \emph{either} the first VIE \emph{or} that the second polarization field is zero, and vice versa with the second constraint. When combined with the usual ``self'' constraints, this implies either the presence or absence of the design at each point across all scenarios, enforcing a single shared structure (Fig.~\ref{fig:MultifunctionalIllustration}, right) rather than the inconsistent per-scenario patchwork that would arise without them (Fig.~\ref{fig:MultifunctionalIllustration}, middle). These constraints can be loosened if there is a more complex prescribed correlation between design patterns across scenarios. Ref.~\cite{Shim2021} demonstrates the framework on two prototypical problems: bounds on the maximum reflectivity contrast achievable in a tunable optical filter (a figure of merit relevant for sensing), and bounds on the maximum switching efficiency of a liquid-crystal-based beam-deflection metagrating between two voltage states.

The culmination of the analytical, semi-analytical, and computational bounds developed over the past decade, and described above, offers a tantalizing promise: for any photonic design problem, there should be a prescriptive, automated way to compute bounds. Just as electromagnetic software increasingly offers automated inverse-design capabilities, one can envision specifying a design space, the set of allowed materials, and the objective of interest, with the software automatically producing an unsurpassable limit. Such limits could \emph{certify} inverse designs that meet them, \emph{trigger} further optimization when a gap remains, or---most ambitiously---\emph{guide} the design itself, blurring the historical separation between top-down limits and bottom-up design. Two complementary realizations of this idea have already appeared. In Ref.~\cite{Gertler2025-ri}, the same SDP that produces the bound also yields a near-globally-optimal design directly via a ``lift-and-recover'' procedure, a single-shot capability convincingly demonstrated in one-dimensional problems. In Ref.~\cite{Chao2026}, by contrast, the SDP solution is used to construct an informed initialization for conventional gradient-based inverse design, allowing the optimizer to escape the poor local optima that plague random initializations and to approach the bounds (within a factor of two, in the Purcell-enhancement examples therein) in higher-dimensional settings. Three open challenges stand between this vision and routine practice. First, \textbf{scale}: SDP relaxations remain expensive, and the sparsity exploited in Ref.~\cite{Gertler2025-ri} is one step among many needed to reach industrial-scale design problems. Second, \textbf{tightness}: the bounds described above match inverse designs in many cases (Figs.~\ref{fig:CompBounds}, \ref{fig:ExCompBounds}), but a general characterization of when relaxations are tight---and a systematic procedure for tightening them when they are not---is missing. Third, \textbf{scope}: the framework as developed here applies to linear, time-invariant Maxwell systems with local constitutive relations. Extending it to nonlinear susceptibilities, time-varying media, and quantum-mechanical degrees of freedom would substantially broaden the class of problems amenable to general, automated bounds.

\subsection{Limits on the polarizability of small scatterers} \label{sec:polarizabilities}
While our discussion in this chapter so far has concerned general scatterers of arbitrary size, in this section we focus on electrically small scatterers, whose response can be characterized by a single parameter, i.e., their dominant polarizability, typically the electric dipole polarizability $\alpha_e$. In this context, one may ask a similar set of questions as those considered in Chapter \ref{sec:bounds_materials} (namely, how large or small can the real and imaginary parts of the susceptibility be?) but now applied instead to the polarizability. Indeed, the electric susceptibility and electric dipole polarizability have somewhat analogous definitions, with the former relating the induced polarization density to the total macroscopic field at a point in space, $\Pv=\varepsilon_0\chi_e\Ev$, and the latter relating the induced dipole moment to the \emph{local} electric field at the center of the scatterer, $\pv=\alpha_e\Ev_{\mathrm{loc}}$ (the local field is equal to the incident field at the center of the scatterer if the object is isolated \cite{tretyakov2003analytical}). This analogy is not unexpected, as the polarizability of atoms, molecules, or generic scatterers is the fundamental quantity entering homogenization and mixing formulas used to determine the effective macroscopic constitutive parameters, such as $\chi_e$, of a (meta)material composed of an array of these elements (the relationship between the local field acting on each inclusion and the macroscopic averaged field is at the core of homogenization theories\cite{tretyakov2003analytical}). There are, however, some deep differences between the physical and analytical properties of $\alpha_e$ and $\chi_e$, which make the problem of establishing constraints and bounds on $\alpha_e$ quite different from what we discussed in Chapter \ref{sec:bounds_materials} for $\chi_e$. In particular, we should mention relevant differences in the role of passivity and causality.

    \textbf{Passivity}. Different from the problem of wave propagation in a homogeneous medium, in a scattering problem, energy can be removed from an incident field not only through absorption, but also re-radiation, i.e., scattering. This fact has implications for the conditions that a passive polarizability must satisfy. To see this, we apply the optical theorem, Eq. (\ref{eq:PextOT}), to an isolated particle described by its electric dipole polarizability, whose polarization density is $\Pv=\delta(\xv) \pv=\delta(\xv) \alpha_e \Einc(\xv)$. The requirement that the extinction power $P_{\mathrm{ext}}$ (scattered plus absorbed power) be nonnegative immediately yields a passivity condition on the polarizability: $\Im \alpha_e \ge 0$, directly analogous to the standard passivity conditions on the constitutive parameters. However, one can derive a stronger condition by additionally requiring that the scattered power, $P_s \propto  |\alpha_e|^2$ (power radiated by the induced dipole), does not exceed the extinction power, $P_{ext} \propto \Im\alpha_e$, which yields the so-called \textbf{Sipe-Kranendonk condition} (or radiation condition) on the electric dipole polarizability (similar conditions can also  be derived for magnetic and higher-order multipolar polarizabilities): 
    \begin{align}
    -\Im [\alpha_e^{-1}] \ge \frac{k_0^3}{6\pi \varepsilon_0}.
    \end{align}
     This result directly implies that the polarizability $\alpha_e$ is complex even in the absence of material absorption, due to radiation damping. It also highlights an important limitation of the standard quasi-static expression for small particles, $\alpha_{e,\mathrm{qs}}$, which is derived by neglecting retardation effects (i.e., by solving Laplace’s equation instead of the full wave equation). Because $\alpha_{e,\mathrm{qs}}$ is purely real, it predicts nonzero scattered power but zero extinction power, thereby violating passivity. The quasi-static polarizability can be corrected to respect passivity by explicitly adding a radiation correction~\footnote{If a metamaterial is composed of a periodic array of \emph{lossless} elements characterized by this complex polarizability, a proper homogenization procedure would still yield a purely real effective permittivity, as the imaginary part of the interaction constant (proportional, again, to the imaginary part of the Green's function of the host medium) exactly cancels, in the periodic case, the imaginary contribution arising from the radiation correction term \cite{tretyakov2003analytical}. See also footnote \ref{footnote_Ewald_Oseen}.}:
    \begin{align}
    \tilde\alpha_{e,qs}= \left( \alpha_{e,qs}^{-1}-i\frac{k_0^3}{6\pi \varepsilon_0} \right)^{-1}.
    \label{eq:polar_corr}
    \end{align}
    Interestingly, the radiation correction term only depends on the wave frequency, the properties of the surrounding medium, and the electric-dipolar character of the radiation (it is universal for any electric-dipolar scatterer in free space). This term is in fact proportional to the imaginary part of the free-space electromagnetic Green's function and, therefore, to the local density of photonic states, Eq. (\ref{eq:ldos}), multiplied by $\omega$. Hence, the radiation correction can be modified by placing the scatterer in a different environment, for example near a substrate, inside a cavity, or within a waveguide, exactly as the spontaneous emission rate of an emitter (classical or quantum) depends on its photonic surroundings (Purcell effect \cite{Novotny2012-py}).
    
    \textbf{Causality}. In contrast to the electric susceptibility $\chi_e$ of a bulk material, the polarizability of a scatterer does not, in general, satisfy simple KK relations. This difference ultimately originates from the point-dipole (or point-multipole) approximation used to model a finite object through its polarizability. In this description, the scatterer is assumed to respond to the local field at its center, where the dipole moment is evaluated; however, in reality, a scattered wave can be generated as soon as the incident wave reaches the outer boundary of the object, namely, up to a time $a/c$ \emph{before} the incident field reaches the center point, with $a$ being the radius of the smallest sphere, centered at that point, that encloses the scatterer. Overall, compared to an ideal dipole radiating at the origin, the scattered wave exhibit a time advance of $2a/c$, which leads to an apparent violation of causality in the point-dipole description (effect preceding the cause). This fact has important implications for the frequency dispersion of the polarizability $\alpha_e(\omega)$, which is not constrained by KK relations in the same way as $\chi_e(\omega)$. Notably, derivative bounds that apply to the susceptibility or permittivity of passive media, Eq. (\ref{eq:derivative_bounds}), do not apply to the polarizability $\alpha_e(\omega)$ of passive scatterers. These issues, and their implications for the homogenization of metamaterials composed of small polarizable elements, are discussed in detail in \cite{alu2011causality}.

We also note that many of the properties of the polarizability, and the causality issues described above, are directly related to the properties of the scattering matrix elements and of the Mie scattering coefficients (similar difficulties in encoding causality and deriving KK relations for scattering parameters were alluded to in Section \ref{sec:passivity} and will be discussed in greater detail in Section \ref{sec:cloaking}). For example, for a spherically symmetric scatterer, the Mie scattering coefficient for the first-order transverse-magnetic (TM) vector spherical harmonic can be written as
    \begin{align}
    c_1^{\mathrm{TM}}=-\frac{U_1^{\mathrm{TM}}}{U_1^{\mathrm{TM}}+i\,V_1^{\mathrm{TM}}}.
    \label{eq:c1TM}
    \end{align}
where the quantities $U_1^{\mathrm{TM}}$ and $V_1^{\mathrm{TM}}$ (real in the lossless case) are the determinant of matrices determined by imposing boundary conditions at the spherical surfaces and are expressed as a combination of spherical Bessel and Neumann functions \cite{alu2005polarizabilities,bohren2008absorption}. The coefficient $c_1^{\mathrm{TM}}$ can be shown to be directly proportional to the electric dipole polarizability \cite{alu2005polarizabilities,monticone2014embedded}:
    \begin{align}
    \alpha_e(\omega)=-i\frac{6\pi \varepsilon_0}{k_0^3}c_1^{\mathrm{TM}}=\frac{6\pi \varepsilon_0}{k_0^3}\left(\frac{V_1^{\mathrm{TM}}}{U_1^{\mathrm{TM}}}-i\right)^{-1},
    \label{eq:alpha_e}
    \end{align}
which provides an exact expression for the polarizability of a spherically symmetric scatterer of any size and converges to the radiation-corrected quasi-static expression, Eq. (\ref{eq:polar_corr}), in the long-wavelength limit. Similar relations can also be derived between higher-order transverse-magnetic (transverse-electric) Mie scattering coefficients and electric (magnetic) multipolar polarizabilities.

\begin{figure*}[tb]
\centering
\includegraphics[width=0.6\linewidth]{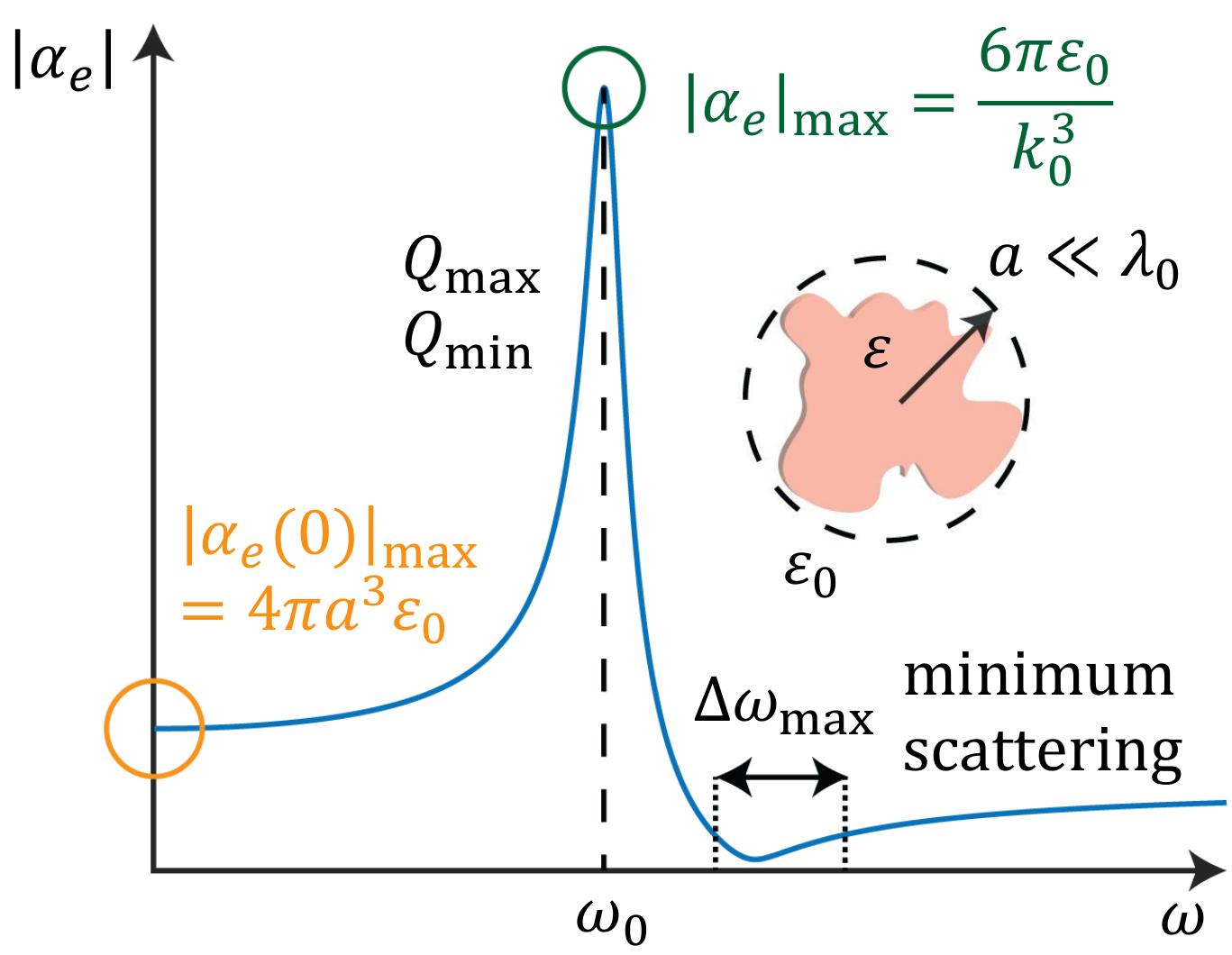}
\caption{\textbf{Limits on the polarizability of small scatterers}. Magnitude of the electric-dipole polarizability for a representative small scatterer (a subwavelength sphere composed of a lossy Drude metal with $\varepsilon/\varepsilon_0=\varepsilon_\infty-\omega_p^2/(\omega^2+i\gamma_m\omega)$ and $\varepsilon_\infty>1$), illustrating some of the limits discussed in Sections \ref{sec:polarizabilities}, \ref{sec:plasmonicQ}.}
\label{fig:polarizability}
\end{figure*}

Given these considerations, we can then address some fundamental questions regarding the ultimate limits of the electric dipole polarizability (Fig. \ref{fig:polarizability}).

\paragraph{How large can the magnitude of $\alpha_e$ be?} Unlike the susceptibility, which can diverge at resonance in absence of material loss, the polarizability cannot, as it is intrinsically limited by radiation damping. In particular, while the quasi-static polarizability $\alpha_{e,qs}$ diverges at resonance when absorption is neglected, the radiation-corrected polarizability $\tilde\alpha_{e,qs}$ in Eq. (\ref{eq:polar_corr}) remains finite, taking the value $\tilde \alpha_{e,qs}=i 6\pi \varepsilon_0/k_0^3$. At resonance, the polarizability is therefore purely imaginary and attains its maximum possible magnitude, corresponding to the maximum achievable extinction and scattered power, consistent with the optical theorem (\ref{eq:PextOT}) (incident field and induced polarization current are in phase). This conclusion holds more generally, beyond the quasi-static approximation, which can be shown, for instance, by using the relation between the Mie scattering coefficient and the polarizability [Eq. (\ref{eq:alpha_e})], together with the fact that, for a passive system, the scattering coefficients reach unity at resonance.

\paragraph{How large can the magnitude of $\alpha_e$ be, at zero frequency?} Given the importance of the static polarizability for scattering sum rules (e.g., Eq.~(\ref{eq:SRExt0})), it is natural to ask how large this quantity can be. This question can be answered using a monotonicity theorem established in \cite{jones1985scattering,sjoberg2009variational} for both isotropic and anisotropic materials, which shows that, for a fixed arbitrary volume, an increase (decrease) in the static permittivity or permeability in any portion of this volume cannot result in a decrease (increase) of the static electric or magnetic polarizability (the diagonal elements of the polarizability tensors in the anisotropic case). This result implies that the maximum possible static electric polarizability is achieved by homogeneously filling the available volume with a material of maximal static permittivity, i.e., in the limit $\left|\varepsilon(0)\right| \to \infty$~\footnote{\label{footnote_static_epsilon}What values can the static permittivity take? Consider the KK relation (\ref{eq:KKRe}) for the permittivity, $\mathop{{\rm Re}} \varepsilon(\omega) = \varepsilon_0+ \frac{2}{\pi } ~ \Pvl\!\int_0^\infty  \frac{\Omega \mathop{{\rm Im}} \varepsilon(\Omega) } {{\Omega^2} - {\omega ^2}}d\Omega$. If $\operatorname{Im}\varepsilon(\omega)$ is regular at zero frequency, one can take the limit $\omega\to 0$ to obtain the static-permittivity sum rule: $\mathop{{\rm Re}} \varepsilon(0) = \varepsilon_0+ \frac{2}{\pi } \int_0^\infty  \frac{\mathop{{\rm Im}} \varepsilon(\Omega) } {\Omega}d\Omega$. Since $\operatorname{Im}\varepsilon(\Omega)\ge0$ for passive media, this sum rule implies $\operatorname{Re}\varepsilon(0)\ge\varepsilon_0$. The case of metals, which have a singularity of $\operatorname{Im}\varepsilon(\Omega)$ at $\omega =0$ (see Section \ref{sec:causality}) must be treated with care, but it leads to the same conclusion, as shown in Ref. \cite[Sec. 82]{landau2013electrodynamics}, for any finite conductivity. A notable exception is the case of infinite conductivity (perfect conductors), for which $\mathop{{\rm Re}} \varepsilon(0) \to -\infty$, as in a ``lossless'' Drude model with the loss concentrated as a delta function of $\operatorname{Im}\varepsilon(\Omega)$ at zero frequency. In summary, at zero frequency, causality and passivity imply that the static permittivity satisfies $\mathop{{\rm Re}} \varepsilon(0) \ge \varepsilon_0$, or diverges as $\mathop{{\rm Re}} \varepsilon(0) \to -\infty$ for perfect conductors.}. For any scatterer contained within a sphere of radius $a$, its static polarizability is therefore bounded by 
    \begin{align}
\alpha_e(0)_{\max}= \lim_{\left|\varepsilon(0)\right| \to \infty} 4\pi \varepsilon_0 a^3 \frac{\varepsilon(0)-1}{\varepsilon(0)+2}=4\pi a^3 \varepsilon_0.
    \label{eq:alpha_e_DC}
    \end{align}
No amount of geometric structuring can yield a static electric polarizability exceeding this value. 

\paragraph{How small can the magnitude of $\alpha_e$ be?} In the absence of absorption, the polarizability can become identically zero at a desired frequency [in Eq. (\ref{eq:polar_corr}), if $\alpha_{e,qs}=0$, then the radiation-corrected polarizability $\tilde \alpha_{e,qs}$ vanishes, even though the radiation correction term itself is nonzero]. This corresponds to a scatterer that is invisible at that frequency, at least with respect to its electric dipole response (which typically dominates the scattering of electrically small particles). More interesting questions are related to (i) how small the magnitude of the polarizability (and therefore the scattered power) can be in the presence of a given level of absorption (the problem of ``invisible sensors''), and (ii) how small the magnitude of the polarizability can remain over a desired bandwidth (the problem of broadband invisibility; bandwidth-related questions are particularly challenging due to the causality-related considerations discussed above). Both questions will be addressed, in greater generality, in Section \ref{sec:cloaking}. 

\paragraph{How sharp can a polarizability resonance be? (and the Chu limit for electrically small antennas)} Due to radiation damping, one might expect that polarizability resonances are always broadened to some extent, even in the absence of absorption. This, however, is not necessarily true. Consider, for example, a lossless electrically small scatterer contained within a spherical volume $V$. If such a scatterer exhibits a resonance, it must behave as a quasi-static resonance, analogous to a LC circuit resonator (the polarizability multiplied by $i \omega$ is effectively an admittance, as it relates the electric field to a current, i.e., the polarization current). We therefore assume that the scatterer has a quasi-static polarizability of the form $\alpha_{e,qs}=3 \varepsilon_0 V \omega_0^2/(\omega_0^2-\omega^2)$, so that the resonance lineshape and strength are consistent with those of an LC resonance, and, at zero frequency, this expression converges to the largest possible static polarizability, namely, that of a sphere of volume $V$ filled with a material of infinite permittivity, $\alpha_{e}(0)=3 \varepsilon_0 V$, consistent with Eq. (\ref{eq:alpha_e_DC}). Using the radiation-corrected polarizability in Eq. (\ref{eq:polar_corr}), we can then compute the full-width-at-half-maximum bandwidth of $|\tilde \alpha_{e,qs}|^2$, which is proportional to the scattered power. Assuming $k_0^3V \ll 1$ and recalling that the peak value is fixed by radiation damping, $|\tilde \alpha_{e,qs}|_{\mathrm{max}}=6\pi \varepsilon_0/k_0^3$, we obtain: $\Delta \omega=\omega_0 k_0^3V/2\pi$. Finally, we can calculate the $Q$ factor of the resonance, defined as the inverse of the fractional bandwidth, $Q=\omega_0/\Delta\omega$. Substituting $V=4\pi a^3/3$ for a sphere of radius $a$, we find:
    \begin{align}
    Q=\frac{3}{2}\frac{1}{(k_0a)^3}.
    \label{eq:Q_Thal}
    \end{align}
Importantly, this represents a fundamental \emph{lower} bound on the $Q$ factor of the resonance of an electrically small scatterer, corresponding to an \emph{upper} bound on its fractional bandwidth. If the assumed static polarizability were reduced by a factor $A$ relative to the maximum value used above (as if the sphere were smaller), the FWHM bandwidth would decrease by the same factor. Interestingly, this bound coincides exactly with the so-called \textbf{Thal limit}, \cite{thal2006new,gustafsson2009illustrations,yaghjian2018overcoming} a stricter version (by a factor of precisely 3/2) of the well-established \textbf{Chu limit} \cite{chu1948physical,harrington1960effect,mclean1996reexamination}, one of the most important theoretical limitations in antenna theory, which establishes the minimum possible $Q$ (or equivalently, maximum possible bandwidth) of lossless electrically small antennas. A substantial body of work refines and generalizes these classical $Q$ bounds---expressing stored energy in terms of fields, currents, and input impedance, optimizing $Q$ via convex optimization, and trading $Q$ against efficiency, gain, and bandwidth~\cite{gustafsson2006bandwidth_Q,gustafsson2015antenna_Q_fields,gustafsson2015stored_energy_antenna_Q,jonsson2015stored_energies_currents,capek2017minimization_Q,schab2018energy_stored,gustafsson2019tradeoff_efficiency_Q,gustafsson2024modes_bounds_synthesis}. The derivation above is different from the standard one for the Chu-Thal limit, but it leads to the same result, as expected, since scatterers are effectively antennas driven by an incident field. Returning to the original question of this paragraph, this result shows that, in the absence of absorption, the $Q$ factor of a polarizability resonance can be arbitrarily large despite radiation damping, especially as the size of the object decreases (in the presence of absorption, we will see in Section \ref{sec:plasmonicQ} that the $Q$ factor of an important class of subwavelength scatterers, namely, plasmonic nanostructures, is governed by the loss rate of the material itself independently of size and geometry). A notable example of scatterers exhibiting arbitrarily sharp resonances are those supporting so-called \textbf{bound states in the continuum (BICs) or embedded eigenstates} \cite{hsu2016bound,monticone2014embedded,silveirinha2014trapping,doeleman2018experimental}. In these systems, a scattering resonance and a scattering zero move arbitrarily close, leading to a vanishing linewidth (corresponding to a complex-frequency pole of the polarizability approaching the real axis and landing exactly on a real-frequency zero). For isolated three-dimensional objects, BICs can arise if the structure incorporates media with vanishing permittivity or permeability \cite{hsu2016bound,monticone2014embedded,silveirinha2014trapping} and they correspond to ``nonradiating'' eigenmodes of the scatterer \cite{monticone2014embedded,monticone2018trapping,monticone2019can}, for example eigenmodes with purely radial polarization currents on a spherical shell, a current distribution that cannot radiate to the far field. Light can couple into \emph{quasi}-BICs, i.e., BICs with small but nonzero linewidth, leading to very large internal fields at steady state, which may be useful to enhance weak light-matter interactions, provided that absorption losses remains sufficiently low. However, as we will see in Section \ref{sec:sensing}, in the presence of absorption, the radiation-limited portion of the $Q$-factor, as is typically done with BICs, is not necessarily optimal for applications in which the resonance amplitude is also important, for example sensing, as the amplitude decreases more rapidly with material loss if $Q$ is high.

\paragraph{Magneto-electric polarizability}
Finally, to conclude this brief discussion of the fundamental limits of the polarizability of small scatterers, we note that for small objects with complex geometries, an incident field may induce not only an electric dipole moment but also a magnetic dipole moment. In general, such a scatterer must therefore be described by electric and magnetic polarizabilities, as well as by magneto-electric coupling coefficients: $\pv=\alpha_e\Ev_{\mathrm{loc}}+\alpha_{em}\Hv_{\mathrm{loc}}$, $\mathbf{m}=\alpha_m\Hv_{\mathrm{loc}}+\alpha_{me}\Ev_{\mathrm{loc}}$ in direct analogy with the constitutive relations of bianisotropic media [Eq. (\ref{eq:const_bian})]. One can derive passivity constraints analogous to those discussed in Chapter \ref{sec:bounds_materials} for bianisotropic materials, including bounds on the strength of the magneto-electric (chiral) polarizability \cite{sersic2011magnetoelectric,sersic2012ubiquity}.

\subsection{Plasmonic field confinement, enhancement, and quality factor}\label{sec:plasmonicQ}

While the previous section discussed the general properties of electrically small scatterers and their resonances, one may ask how a resonance can arise within a subwavelength volume, where retardation effects are negligible. In this context, an important class of systems is provided by plasmonic structures and scatterers, which can indeed support resonances even in the deep quasi-static regime (that is, resonances not associated with wave retardation effects, as in lumped-element circuits) owing to their negative permittivity. This property enables significantly stronger field confinement and enhancement than is achievable in dielectric structures of the same size (see also Section \ref{sec:mode_volume}), physically originating from the coupling between electromagnetic fields and collective electron motion. However, regardless of how a plasmonic structure is engineered, the achievable field enhancement is ultimately constrained by material absorption losses \cite{khurgin2012reflecting,khurgin2015deal}. As we will discuss in this section, this limitation is intrinsically tied to the fundamental physics of quasi-static resonances in plasmonic systems.

In general, an \textbf{electromagnetic resonance} occurs when the electric and magnetic stored energies are equal, $U_H=U_E$ (corresponding to equal and opposite inductive and capacitive reactances in a circuit), such that the energy of the electromagnetic field can oscillate indefinitely between the two forms in the lossless case (with no need for additional work to be performed). This condition corresponds to a self-sustained oscillation of the electromagnetic field, namely, an eigenmode.
However, if the electric field is confined to a small volume---for example, a sphere of radius $a$, filled with a transparent material of refractive index $n$ and zero conductivity---the spatial derivatives in Maxwell's equations imply that the magnetic field magnitude scales as $n^2a/\lambda_0$. One can then estimate the ratio between magnetic and electric energies as $U_H/U_E \sim (\pi n a/\lambda_0)^2$ \cite{khurgin2012reflecting}. For subwavelength dielectric objects, this ratio is much smaller than unity, which implies that the electromagnetic resonance condition, $U_H=U_E$, cannot be satisfied. Deeply subwavelength lumped-element circuits, the type familiar to any undergraduate engineering or physics student, circumvent this limitation by relying on metals, which have very large (approximately real) conductivity, $\sigma(\omega)\approx \sigma_0 = Ne^2/m \gamma_m$, at the relevant frequencies (microwave and lower frequencies, where $\omega\ll \gamma_m$, with $\gamma_m$ being the metal damping rate). This property can be used to create loops of large conduction current and, therefore, generate strong magnetic fields and large inductances, which, in turn, can restore the energy balance necessary for resonance. 

\begin{figure}[tbhp]
\centering
\includegraphics[width=0.9\linewidth]{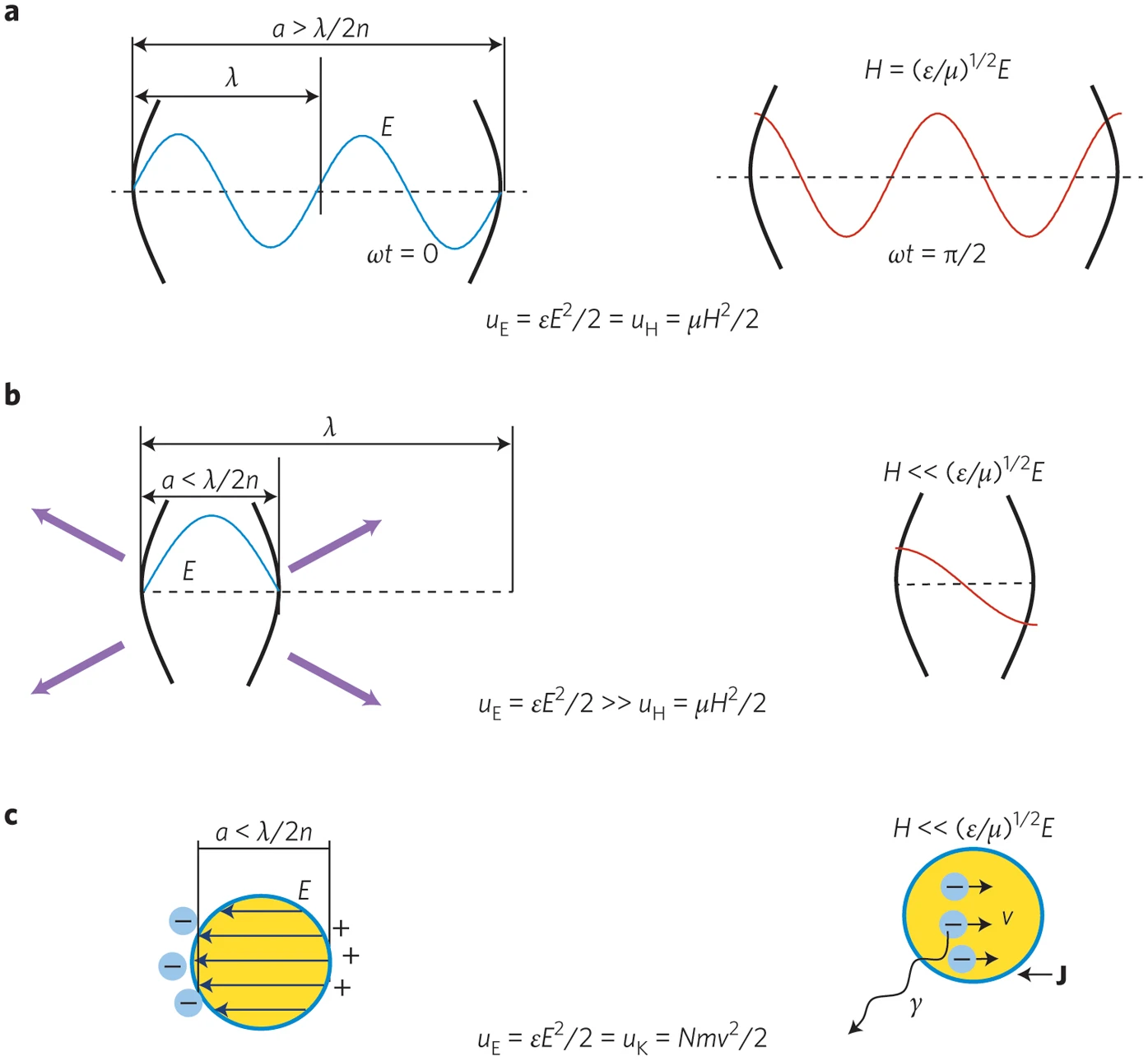}
\caption{\textbf{Energy requirements for resonance in photonic and plasmonic structures.} (a) In a conventional optical cavity with dimensions larger than approximately half a wavelength, the stored energy oscillates (indefinitely, in the lossless case) between electric-field energy (left) and magnetic-field energy (right). (b) In a subwavelength dielectric cavity, the magnetic energy is too small to support a self-sustained oscillation (a resonance) within such a confined volume, and the energy is radiated outward. (c) In a subwavelength plasmonic structure, the presence of free carriers allows the stored energy to oscillate between electric-field energy (left) and the kinetic energy of the carriers (right). This enables resonance even when the magnetic field is nearly negligible, but it is inherently accompanied by loss due to the strong damping of carrier motion in metals. (Figure adapted from Ref.~\cite{khurgin2015deal}.)}
\label{fig:energy_balance}
\end{figure}

This mechanism, however, no longer applies at optical frequencies, where the metal conductivity drastically changes, becoming complex and frequency dependent (Drude model \cite{bohren2008absorption}). Fortunately, however, the electromagnetic response of metals has another trick up its sleeve. While the real part of the Drude conductivity governs ohmic resistance, the imaginary part, which becomes significant for $\omega > \gamma_m$, yields an inductive reactance known as \textbf{kinetic inductance}. This effect originates from the inertia of the charged carriers moving inside the metal and is fundamentally distinct from the magnetic (``Faraday'') inductance of conventional circuits. The associated carrier kinetic energy adds to the magnetic stored energy because the oscillations of the carrier velocity are 90 degrees out of phase relative to the harmonic electric field (in the lossless case), analogously to how, in the \emph{quasi-static} regime, the magnetic field is 90 degrees out of phase relative to the electric field according to Ampere's Law. Consequently, the kinetic energy can help rebalance the energy relation, $U_K+U_H=U_E$, thereby enabling resonances in the deeply subwavelength regime also at optical frequencies. The time-averaged kinetic energy in a volume $V$ is $U_K=\frac{1}{4}m|v|^2NV$, where the velocity $v$ can be determined from the microscopic equation of motion of the carriers. Using the standard Drude model for free electrons, one finds 
\begin{equation}
U_K=\frac{1}{4} \varepsilon_0 a^3\frac{\omega_p^2}{\gamma_m^2+\omega^2} |\Ev|^2,
\label{eq:kin_energy}
\end{equation}
for a volume $V=a^3$.

The fact that deeply subdiffractive resonances rely on carrier kinetic energy comes with an unavoidable sensitivity to loss.
Dissipation is directly linked to the damping of the carrier velocity, characterized by the scattering rate $\gamma_m$, an intrinsic property of the material (see also Section \ref{sec:small_imaginary}). Because the total energy of a mode is stored half of the time \emph{inside} the metal, in the form of kinetic energy dissipated at a rate $2\gamma_m$, it is intuitively clear that the overall electromagnetic energy decays at a rate on the order of $\gamma_m$, independent of the geometry or size of the scatterer. This intuitive explanation can be formalized following the elegant argument in \cite{wang2006general}, slightly modified here in order to more clearly highlight the role of carrier kinetic energy. 

Consider a generic subwavelength structure composed of two materials: a low-loss plasmonic material characterized by a complex frequency-dispersive permittivity $\varepsilon_m=\varepsilon_m'+i\varepsilon_m''$, with $\varepsilon_m''\ll|\varepsilon_m'|$, and a dielectric with real permittivity $\varepsilon_d$, assumed non-dispersive. In the quasi-static limit, valid for electrically small structures, we have $\nabla \times \Ev = 0$ and $\Ev = -\nabla \Phi$, where here, and in the following, we consider real-valued fields, not phasors. From this, one can show that the following integral must vanish for any spatial distribution of the permittivity $\varepsilon$:
\begin{align*}
\int_{\Omega}  \varepsilon(\xv) \Ev \cdot \Ev\, dV 
&= \int_{\Omega} \Ev \cdot \Dv\, dV 
= - \int_{\Omega} \nabla \Phi \cdot \Dv\, dV \\
&= - \int_{\Omega} \left[ \nabla \cdot (\Phi \Dv) - \Phi (\nabla \cdot \Dv) \right] dV \\
&= - \int_{S_{\Omega}} \Phi (\Dv \cdot \vec{n})\, ds = 0 .
\end{align*}
Here, $\Omega$ is a large volume enclosing the structure, with boundary surface $S_\Omega$ chosen sufficiently far away such that, in the quasi-static limit, we may assume $\Dv = 0$ at $S_\Omega$ for any localized resonant field supported by the structure. A first important observation that follows from the condition $\int_{\Omega} \varepsilon(\xv) \Ev \cdot \Ev\, dV = 0$, is that, for a non-trivial (i.e., nonzero) self-sustained mode to exist in a deeply subwavelength structure, the permittivity $\varepsilon$ 
cannot be positive everywhere. This confirms that materials with negative permittivity are strictly necessary to support deeply subwavelength resonances. Another important implications of the vanishing of the integral above is that, if $\Omega$ is divided into sub-volumes containing only metal and dielectric regions, $\Omega_{\text{metal}}$ and $\Omega_{\text{dielectric}}$, one obtains the identity
\begin{equation}
\int_{\Omega_{\text{dielectric}}} \varepsilon_d \overline{E^2} \, dV 
= \int_{\Omega_{\text{metal}}} -\varepsilon_m \overline{E^2} \, dV \approx \int_{\Omega_{\text{metal}}} -\varepsilon_m' \overline{E^2},
\label{eq:dielectric_metal}
\end{equation}
where we have also performed time averaging (denoted by the overbar), since this identity will be used to simplify the expression of the stored energy and the $Q$-factor of a plasmonic resonance. 

In the dielectric region, the total time-averaged stored energy is
\begin{equation}
U_{d}=\frac{1}{2}\int_{\Omega_{\text{dielectric}}} \varepsilon_d \overline{E^2} \, dV,
\end{equation}
where we ignored the magnetic contribution, which is negligible for deeply subwavelength structures, as discussed above. In the metal region, we separate the stored energy into two contributions: the energy stored in the material and the electric-field energy as if the field were in free space \cite{askne1970energy,ramakrishna2008physics} (again neglecting the magnetic energy)\footnote{Different from \cite{wang2006general}, we adopt here a ``microscopic perspective'' on the stored energy in the dispersive metal, while ultimately arriving at the same conclusion.}:
\begin{equation}
U_{m}=U_{m,mat}+U_{m,E}=U_{m,mat}+\frac{1}{2}\int_{\Omega_{\text{metal}}} \varepsilon_0 \overline{E^2} \, dV.
\end{equation}
In a material with free electrons (neglecting any bound-electron contribution for simplicity\footnote{If the electromagnetic response arises from bound electrons, as in a Lorentz model, both the kinetic and potential energies of the dipole oscillators must be included \cite{askne1970energy,ramakrishna2008physics}.}), the energy stored in the medium is the kinetic energy of the electrons driven by the harmonic electric field, as discussed above. Thus, the material contribution is $U_{m,mat}=U_K$, given by Eq. (\ref{eq:kin_energy}), and the total time-averaged stored energy in the metal is
\begin{equation}
U_{m}=\frac{1}{2}\int_{\Omega_{\text{metal}}} \varepsilon_0 \left(1+\frac{\omega_p^2}{\gamma_m^2+\omega^2}\right) \overline{E^2} \, dV.
\label{eq:energy_metal}
\end{equation}
Note that the expression in parentheses is exactly equal to $\frac{d(\omega \varepsilon_m')}{d \omega}$ for a lossless Drude metal with $\gamma=0$. However, using $\frac{d(\omega \varepsilon_m')}{d \omega}$ to calculate the stored energy in a dispersive material is strictly valid only in the lossless limit (see also Chapter \ref{sec:bounds_materials} and footnote \ref{footnote_stored_energy}) and may incorrectly predict negative energies at frequencies where $\varepsilon_m''$ is large. Instead, separating the stored energy into field and material contributions, as in Eq. (\ref{eq:energy_metal}), yields an exact and positive expression, even in strongly lossy and dispersive media.

As discussed above, a resonance is possible when energy can oscillate between two forms. In the quasistatic plasmonic case, where magnetic energy is negligible, these two forms are the electron kinetic energy and the electric-field energy. If the system is a homogeneous metal, the resonance condition reduces to $U_{K}=U_{m,E}$, which can be satisfied at the real frequency $\omega=\omega_p$, if $\gamma=0$. This corresponds to the intrinsic volume plasmon resonance of the plasmonic material. If, instead, the system is a structure consisting of both metallic and dielectric regions, the resonance condition is $U_{K}=U_{m,E}+U_d$, yielding a resonance frequency that depends on the geometry of the structure (but is independent of its size in the quasi-static limit), as shown in \cite{wang2006general}. This corresponds to a localized surface plasmon resonance of the nanostructure. Finally, one can calculate the $Q$ factor of a plasmonic resonance of this type, for a generic subwavelength structure consisting of only one type of Drude metal,
\begin{align}
Q &\equiv \omega\frac{ \text{ Stored energy}}{\text{Power loss}} 
 = \omega\frac{ U_m+U_d}{\text{Power loss}}\\
&= \omega\frac{ \frac{1}{2}\int_{\Omega_{\text{metal}}} \varepsilon_0 \left(1+\frac{\omega_p^2}{\gamma_m^2+\omega^2}\right) \overline{E^2} \, dV 
   + \frac{1}{2}\int_{\Omega_{\text{dielectric}}} \varepsilon_d \overline{E^2} \, dV}
  {\omega \int_{\Omega_{\text{metal}}} \varepsilon_m'' \, \overline{E^2} \, dV }
\end{align}

\begin{figure*}[tb]
\centering
\includegraphics[width=0.5\linewidth]{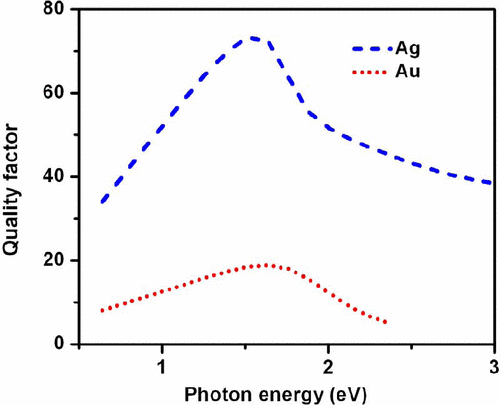}
\caption{\textbf{Quality factor of plasmonic resonances in subwavelength metallic nanostructures}, as a function of resonance frequency, for gold (dashed line) and silver (dotted line). In the quasi-static regime, once the resonance frequency and the plasmonic material are specified, the $Q$-factor is uniquely determined and cannot be improved by structuring. (Figure adapted from Ref.~\cite{wang2006general}.)}
\label{fig:PlasmonicQ}
\end{figure*}

We can simplify this expression by invoking Eq. (\ref{eq:dielectric_metal}) (a consequence of the quasi-static nature of the problem), which allows all integrals to be evaluated solely over the metallic region. Substituting the Drude expressions for $\varepsilon_m'$ and $\varepsilon_m''$, we find, after some algebra,
\begin{align}
Q = \frac{ \int_{\Omega_{\text{metal}}} \varepsilon_0 \left(1+\frac{\omega_p^2}{\gamma_m^2+\omega^2}\right) \overline{E^2} \, dV 
   - \int_{\Omega_{\text{metal}}} \varepsilon_m' \overline{E^2} \, dV}
  {2 \int_{\Omega_{\text{metal}}} \varepsilon_m'' \, \overline{E^2} \, dV }= \frac{\omega}{\gamma_m}
\end{align}
which directly shows that the $Q$-factor of any subwavelength plasmonic nanostructure depends only on the resonance frequency and the intrinsic loss rate of the plasmonic material. In other words, once the resonance frequency and the plasmonic material are chosen, the $Q$-factor is independent of geometry, (subwavelength) size, or the combination with any dielectric material (Fig. \ref{fig:PlasmonicQ})~\footnote{The original derivation in \cite{wang2006general}, which does not rely on any microscopic argument but uses $\frac{d(\omega \varepsilon_m')}{d \omega}$ to compute the stored energy, leads to an equivalent final expression that depends only on the permittivity function of the material. For a Drude metal, this coincides with our result in the low-loss regime.}.

Maximizing the $Q$-factor of nano-plasmonic systems is, therefore, a materials problem rather than a structural-design problem. Indeed, significant research efforts have focused on reducing intrinsic losses in plasmonic materials \cite{khurgin2012reflecting,boltasseva2011low}. However, even in the hypothetical case of a perfect metal with no intrinsic bulk losses, field confinement and enhancement is ultimately limited by surface-collision induced damping (Landau damping), unless a lossless metal could be realized with no available electronic states for collision-induced transitions, as discussed in Section \ref{sec:small_imaginary}\footnote{If a truly lossless metal were possible, a spherical plasmonic scatterer of radius $a$ would operate exactly at the Chu-Thal limit [Eq.~(\ref{eq:Q_Thal})] (for a Drude permittivity with unit high-frequency limit, $\varepsilon_\infty=\varepsilon_0$), which can be verified by analyzing the polarizability resonance of a subwavelength sphere made of a lossless Drude material.}. When surface-collision damping dominates, $\gamma_m$, and therefore $Q$, would become geometry dependent; however, tighter field confinement in geometries with sharper or smaller features can only increase this type of loss. Hence, also in this regime, geometrical structuring would not be useful to increase the $Q$-factor of a plasmonic resonance at a given frequency.

\subsection{Minimum mode volume and minimum physical volume for a mode} \label{sec:mode_volume}
\begin{figure*}[tb]
\centering
\includegraphics[width=0.99\linewidth]{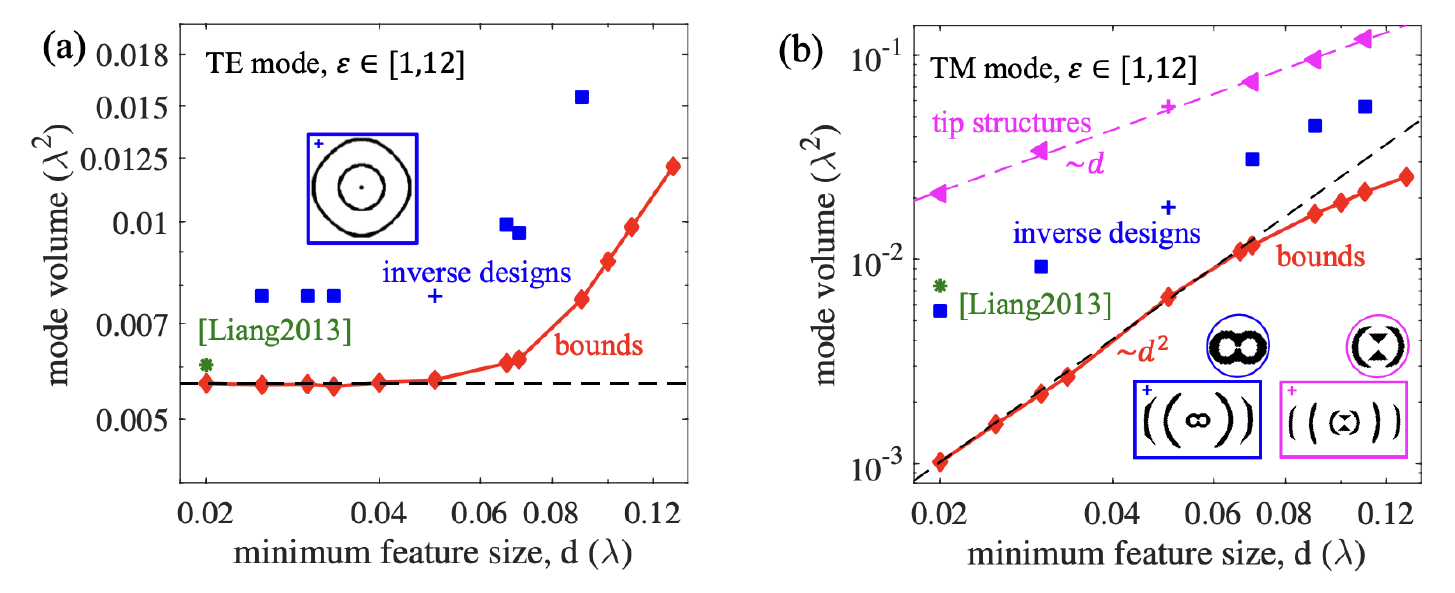}
\caption{\textbf{Minimum mode volumes of 2D dielectric cavities.} (a) For scalar waves (or vector waves with out-of-plane polarization), there can be no singular field response, and the minimum mode volume scales with $\lambda^2$ and cannot be further reduced by fine features of size $d\ll \lambda$. (b) For vector waves (in-plane polarization), the mode volume can be decreased (at a fixed wavelength $\lambda$) with increasingly fine features. A key open question is the ultimate feasible scaling. Included are simple sharp-tip designs (pink), more complex inverse designs (blue), and computational bounds (red). (Figure adapted from Ref.~\cite{Zhao2020minimum}.)}
\label{fig:ModeVolumeScaling}
\end{figure*}

In contrast to the plasmonic structures discussed in the previous section, achieving tight field confinement is significantly more challenging using dielectrics. In this context, two fundamental questions arise in a wide range of photonics problems, particularly for systems based on dielectric media: What is the minimum possible ``mode volume''? And what is the minimum physical volume required to support a mode (or multiple orthogonal modes)? As we will see, although closely related, these two questions have fundamentally different answers. 

The mode volume $V$ of a high-$Q$ modal field $\Ev$ in a non-magnetic structure with permittivity distribution $\varepsilon(\xv)$ is defined as \cite{Novotny2012-py}
\begin{equation}
    V = \frac{\int \varepsilon(\xv)\, |\Ev(\xv)|^{2}\, {\rm d}\xv}
         {\varepsilon(\xv_{0})\, |\Ev(\xv_{0})|^{2}},
         \label{eq:mode_vol}
\end{equation}
which, in the absence of frequency dispersion, can be interpreted as the ratio of the total electric field energy to the energy density at a maximum intensity point $\xv_0$~\footnote{See, for example, \cite{kristensen2011effective,lalanne2018light,wu2021nanoscale}, for possible generalizations to open, leaky optical cavities with radiation-limited $Q$, based on quasi-normal mode theory.}. This concept of mode volume is relevant for both linear and nonlinear light-matter interactions, as many effects scale as $1/V$ or higher powers thereof \cite{Novotny2012-py}. For instance, in the Purcell effect, the enhancement of spontaneous emission rate is proportional to the local density of states, which scales as $Q/V$ for a single high-$Q$ resonator. 

If no constraints are imposed on the permittivity distribution, the mode volume has no fundamental lower bound, even for purely dielectric structures. Indeed, it has long been known \cite{andersen1978field} that perfectly sharp wedges and tips, whether dielectric or metallic, support integrable field singularities [i.e., the integral in the numerator of Eq. (\ref{eq:mode_vol}) remains finite], yielding a vanishing mode volume. In the dielectric case, the electric field enhancement near such singular structures can be understood by discretizing the wedge and iteratively applying the boundary conditions at successive interfaces approaching the tip. In particular, the normal component of the electric field at the interface between two dielectrics is enhanced in the lower-permittivity medium (e.g., air outside the wedge) by the ratio of the higher to lower permittivities ($E_{l\perp}=E_{h\perp}\varepsilon_h/\varepsilon_l$), as discussed, for example, in \cite{choi2017self}. Even when practical constraints are imposed on the permittivity distribution, such as a minimum feature size and/or a minimum distance between edges and the field maximum, sharp all-dielectric structures have been shown to support mode volumes several orders of magnitude smaller than the conventional ``diffraction limit'', $(\lambda_0/2n)^2$. These structures are typically designed through computational optimization \cite{liang2013formulation,hu2016design,hu2018experimental,wang2018maximizing,albrechtsen2022nanometer,ouyang2024singular}, although rational design strategies have also been proposed \cite{choi2017self,mao2025singulonics}. Furthermore, Ref.~\cite{Zhao2020minimum} demonstrated that global lower bounds on the mode volume of dielectric resonators can be computed using the optimization frameworks discussed in Section \ref{sec:comput_bounds} (specifically, Lagrangian duality). For two-dimensional structures, this revealed an approximately quadratic scaling of the minimum achievable mode volume with the minimum feature size $d$ for TM modes (Fig. \ref{fig:ModeVolumeScaling}), i.e., modes with in-plane polarization, for which the electric field can be enhanced through boundary conditions, as discussed above. 

\begin{figure*}[tb]
\centering
\includegraphics[width=0.99\linewidth]{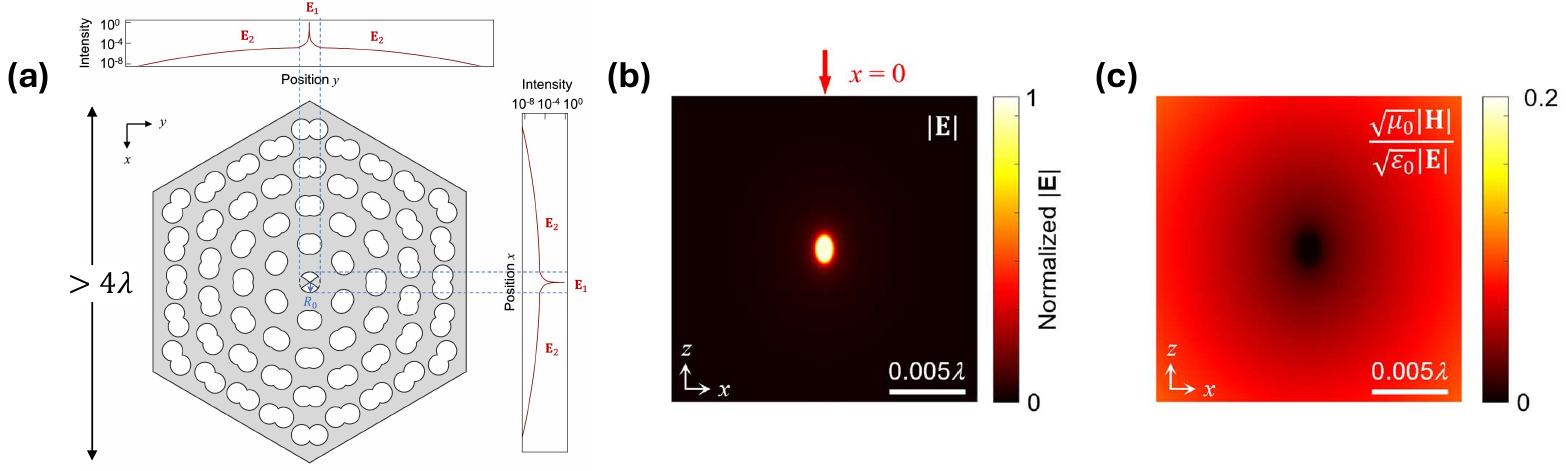}
\caption{\textbf{Light confinement beyond the diffraction limit in all-dielectric structures.} (a) Example of a structure supporting a strongly confined electric field distribution associated with a deeply subwavelength mode volume. The structure consists of a 3D biconical antenna embedded within a twisted photonic crystal cavity. Although the mode volume, as defined in Eq. (\ref{eq:mode_vol}), is subwavelength, the overall physical size of the supporting structure must be greater than a wavelength in order to sustain the resonant mode and reduce radiation leakage. (b) Simulated electric field distribution, and (c) normalized magnetic-to-electric field amplitude ratio, in the vicinity of the maximum field point. These results show that the magnetic-field energy becomes negligible in regions where the electric field is strongly confined, preventing the energy-balance condition necessary for resonance from being satisfied in a subwavelength dielectric structure. (Figure adapted from Ref.~\cite{mao2025singulonics}.)}
\label{fig:confinement}
\end{figure*}

Interestingly, all reported implementations of dielectric resonators exhibiting deeply subdiffractive mode volumes appear to rely on structures whose overall physical size is comparable to, or larger than, the wavelength. In other words, it does not appear possible to realize a subdiffractive mode volume within a dielectric structure that is, itself, entirely subwavelength. This is, in fact, a fundamental limitation of dielectric resonator, which can be understood qualitatively from the same general arguments discussed in Section \ref{sec:plasmonicQ}. An electromagnetic mode is a self-sustained oscillation in which energy can oscillate between two forms, usually electric $U_E$ and magnetic $U_H$ energy, indefinitely in the lossless case. However, as shown in Section \ref{sec:plasmonicQ}, if the electric field is highly confined within a small dielectric volume of characteristic size $a$, the associated magnetic energy in that region becomes smaller than the electric energy by a factor proportional to $(a/\lambda_0)^2$, which implies that the resonance condition for a self-sustained mode, $U_E=U_H$, cannot be satisfied if the magnetic field is confined to the same small volume. A much larger surrounding region is therefore always required, one in which the magnetic and electric fields are not strongly confined, to generate sufficient magnetic energy to balance the total electric energy, as in a conventional dielectric cavity. Such a region must extend over a distance on the order of $\lambda_0/2n$ in at least one direction, and an even larger confining structure, such as a photonic crystal, is then typically needed to suppress radiative losses (Fig. \ref{fig:confinement}) \cite{khurgin2025research,mao2025singulonics}. In summary, in all-dielectric structures it is fundamentally impossible to confine both electric and magnetic fields of a mode beyond the diffraction limit simultaneously. In contrast, there is no fundamental obstacle to achieving deeply subdiffractive confinement of the electric field alone---and hence ultra-small mode volumes as defined by Eq. (\ref{eq:mode_vol})---provided that this confinement is supported by a much larger, wavelength-scale structure \cite{khurgin2025research}. 

As discussed in Section \ref{sec:plasmonicQ}, deeply subwavelength modes can be supported in deeply subwavelength structures only if the  energy balance required for resonance can be restored. This can be achieved either (i) through highly conducting metals at low frequencies, such that conduction currents can be shaped to create large magnetic fields, and therefore large magnetic stored energy, as in standard quasi-static circuit resonators, or (ii) through materials with negative permittivity (plasmonic materials or polar insulators), where the kinetic energy of carriers can balance the electric energy and enable a resonance even if the magnetic field is nearly absent. What is the minimum volume required to support a resonance in these systems? Or, how must the system's size scale if the target resonant frequency is lowered (hence, the resonant wavelength $\lambda_0$ is increased)?
Using negative-permittivity materials, the size of a small (subdiffractive) resonator in the quasi-static regime becomes essentially independent of the resonant frequency (which is uniquely determined by the material parameters and the shape of the structure, rather than its size). Instead, in the lumped-element circuit case (again operating in the quasi-static regime), lowering the resonant frequency requires increasing either the capacitance $C$ or the inductance $L$, since $\omega_0=1/\sqrt{LC}$. If perfect conductors are available, arbitrarily large inductance and capacitance values could, in principle, be realized within a fixed volume by using thinner wires and more closely spaced plates, thereby reducing the resonant frequency without enlarging the structure.
This is no longer true if the available materials are imperfect (lossy), in which case a few open questions remain: 
\begin{openq}
What scaling laws govern the minimum size of a resonant circuit with imperfect conductors and/or imperfect dielectrics? How must the system's size scale with resonant frequency/wavelength and loss? Or, along similar lines, can one derive global bounds on the $Q$-factor of a lumped-element circuit as a function of available volume and material dissipation? 
\end{openq}
These questions are highly relevant for the continued miniaturization of integrated electronic components, such as on-chip inductors and resonators \cite{kang2018chip}. The arguments used in Section \ref{sec:plasmonicQ} for the quasi-static plasmonic case do not carry over to the circuit case, since they neglect the magnetic field energy altogether. Plasmonics and electronics represent fundamentally distinct quasistatic regimes, as also emphasized in \cite{khurgin2012reflecting}. For the specific case of planar inductors, upper bounds on their $Q$-factor as a function of size and dissipation were recently derived \cite{abdelrahman2026maximum} using the computational-bound framework (specifically, semidefinite programming) discussed in Section \ref{sec:comput_bounds}. 

Going back to the all-dielectric case, another intriguing question one could ask is the following: if a structure is engineered to produce \emph{multiple} points of ultra-confined electric fields, each associated with a very small mode volume $V$, and if these hot spots are separated by deeply subwavelength distances over some area, can they be used to realize orthogonal information/energy-carrying channels with a density beyond the diffraction limit? As a more concrete example, could such a subdiffractive arrangement of near-field hot spots enable super-resolution near-field imaging with a single small probe? If all of these hot-spots are supported by the same wavelength-scale mode, i.e., if the fields must ``flow'' through the same mode
, the answer is clearly no: one would have access to only a single independent channel. 

One might instead imagine that each ultra-confined field maximum corresponds to a distinct orthogonal mode. However, in the dielectric case, each such mode must be supported by a wavelength-scale resonant structure, as discussed above, which suggests that the overall device may need to grow in size as $\lambda_0/2n$ times the number of channels, at least in one direction.
More broadly, these considerations motivate the following conjectures for all-dielectric systems: (i) there is no fundamental obstacle to ``packing'' an arbitrary number of electric-field hot-spots within an arbitrarily small area through sufficiently intricate structuring (only limited, in practice, by the minimum feature size), (ii) despite each hot spot being associated with a subdiffractive mode volume as defined in Eq. (\ref{eq:mode_vol}), the number of power-carrying orthogonal modes/channels that can be supported within a given area remains constrained by the diffraction limit (set by the maximum available refractive index), regardless of structuring. We stress that, to the best of our knowledge, a rigorous proof of these statements is still lacking. We also refer the reader to Section~\ref{sec:thickness} for a discussion of the physical size (in particular the thickness) needed to accommodate a given number of channels for a specified optical functionality, and related open questions.

\begin{openq}
Can the conjectures stated above for all-dielectric systems---(i) that any number of subdiffractive electric-field hot spots can be packed into an arbitrarily small area through structuring, and (ii) that the number of power-carrying orthogonal channels in a given area remains constrained by the diffraction limit regardless of structuring---be rigorously proven (or disproven)?
\end{openq}

\subsection{Scattering reduction: Invisibility, cloaking, and impedance matching} \label{sec:cloaking}
In many application domains, ranging from energy harvesting and photovoltaics to stealth, photolithography, microwave circuits, and vision systems, one is interested in reducing electromagnetic scattering from an object, either the total scattering cross section or scattering in specific directions, most commonly backward scattering or reflection. In this section, we briefly review known fundamental constraints and limits for the general problem of scattering reduction, with a particular focus on cloaking and invisibility.

We begin with a simple question: \textbf{is perfect invisibility possible}? More precisely, can the total scattering cross section of an object be completely suppressed by surrounding it with additional material (a ``cloak'')?\footnote{Here we only consider the case of an isolated object in free space, and not cases in which the scatterer is embedded in high-index, lossy, or diffusive media \cite{alu2008effects,schittny2014invisibility}, or placed on a ground plane \cite{li2008hiding}.} The answer is yes, but only at a \emph{single} frequency and in the \emph{absence} of absorption, with these restrictions discussed in the next paragraphs. A notable theoretical solution for a perfect single-frequency invisibility cloak is based on the concept of ``transformation optics,'' and we refer the reader to \cite{pendry2006controlling,fleury2015invisibility} for further details.

\paragraph{Tradeoffs between scattering and absorption}
One may then ask: \emph{Is perfect invisibility possible at a single frequency in the presence of nonzero absorption?} Or, can the total scattering cross section of an object be completely suppressed by surrounding it with \emph{lossy} material? The answer is no, as follows directly from the optical theorem for plane-wave scattering, Eq. (\ref{eq:PextOT}), which relates extinction power to the forward-scattering amplitude. If the scatterer absorbs energy, its absorption cross section $\sigma_\mathrm{abs}$ is nonzero, and therefore its extinction cross section $\sigma_\mathrm{ext}=\sigma_\mathrm{abs}+\sigma_\mathrm{scat}$ (sum of absorption and scattering) is also nonzero. The optical theorem then implies that the forward scattering amplitude cannot vanish: the object must cast a shadow. Thus, no passive absorbing object can be perfectly invisible. This also implies that any passive sensor or receiving antenna---any system that must absorb energy in order to detect a signal---necessarily produces nonzero scattering (although in principle this scattering may be entirely confined to the forward direction). In different scenarios, this general observation can be formalized into fundamental tradeoffs between absorption and scattering at a fixed frequency, derived from power conservation in passive systems~\cite{Fleury2014-cd}. 

\begin{figure*}[tb]
\centering
\includegraphics[width=0.99\linewidth]{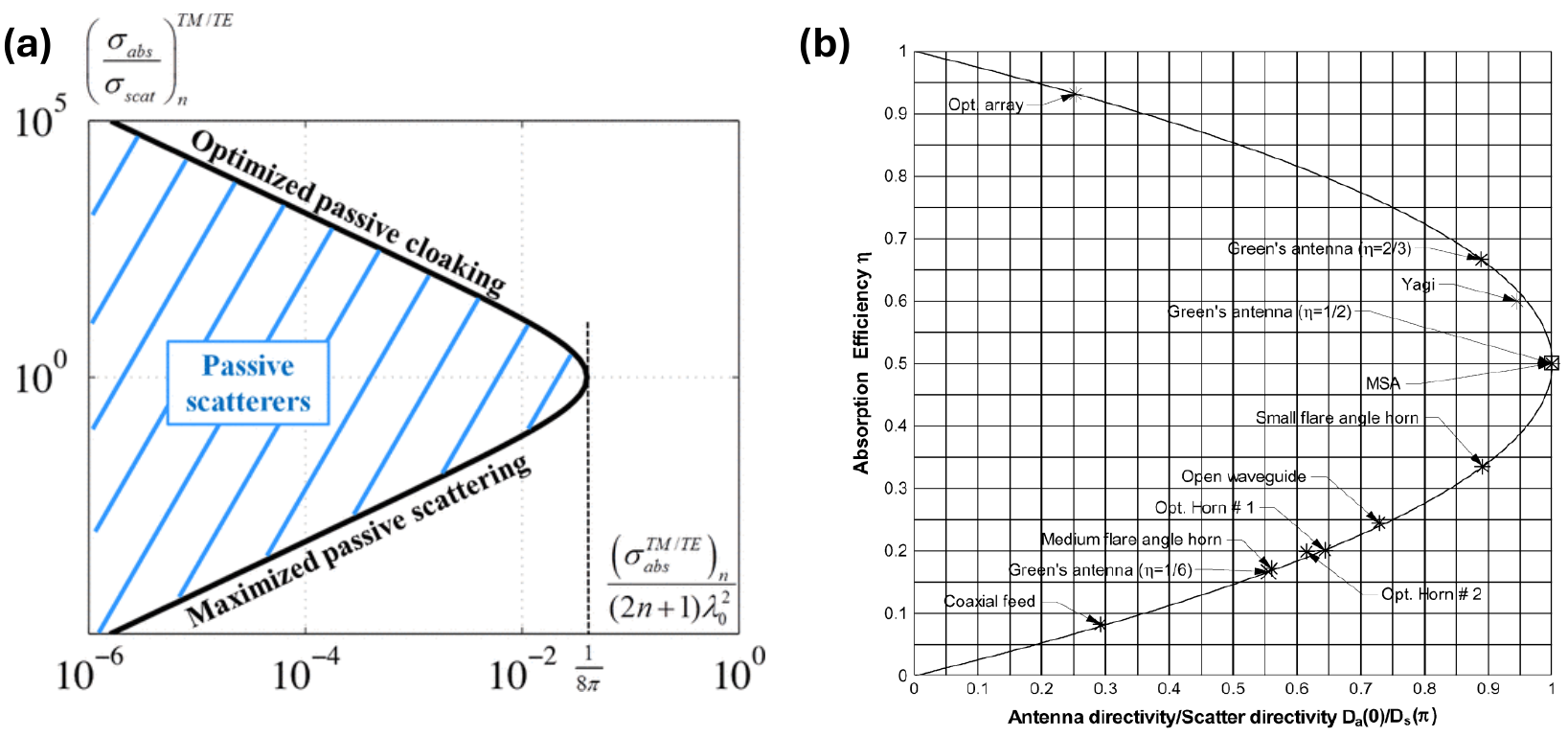}
\caption{\textbf{Fundamental tradeoffs among scattering, absorption, and directivity.} (a) Admissible values of the absorption-to-scattering ratio for a given absorption level, as dictated by passivity for spherically symmetric scatterers. The boundary of the blue shaded region represents the optimal scattering-absorption tradeoff achievable by passive objects. (b) Optimal tradeoff between absorption efficiency and antenna/scatterer directivity, shown together with several representative examples. (Panel (a) adapted from Ref.~\cite{Fleury2014-cd}; panel (b) adapted from Ref.~\cite{andersen2005absorption}.)}
\label{fig:tradeoff_abs_scat}
\end{figure*}

For the problem of scattering from an isolated object, assumed here to be spherically symmetric for simplicity, the spherical scattering coefficients $s_n$\footnote{The spherical scattering coefficients $s_n$ are related to the Mie scattering coefficients $c_n$ used in Section ~\ref{sec:polarizabilities} by $c_n=(s_n-1)/2$, as discussed later in this section. Accordingly, the Mie coefficients of a passive object are restricted, in the complex plane, to the closed disk of radius $1/2$ centered at $-1/2$.}, which relate incoming and outgoing spherical waves, are constrained by passivity to satisfy $|s_n|\leq 1$. They are therefore restricted, in the complex plane, to the closed unit disk. For plane-wave incidence, the scattering, extinction, and absorption cross sections can be calculated using Mie theory and expressed in terms of these coefficients, for both TE and TM vector spherical harmonics, as
\begin{equation}
\sigma_{\mathrm{scat}}
= \sum_{n=1}^{+\infty} (\sigma_{\mathrm{scat}})_n
= \frac{\pi}{2 k_0^{2}} \sum_{n=1}^{+\infty} (2n+1)
\left( \left| s_n^{\mathrm{TM}} - 1 \right|^{2}
     + \left| s_n^{\mathrm{TE}} - 1 \right|^{2} \right) ,
\end{equation}
\begin{equation}
\sigma_{\mathrm{ext}}
= \sum_{n=1}^{+\infty} (\sigma_{\mathrm{ext}})_n
= \frac{\pi}{k_0^{2}} \sum_{n=1}^{+\infty} (2n+1)
\left( 1 - \operatorname{Re}s_n^{\mathrm{TM}}
     + 1 - \operatorname{Re}s_n^{\mathrm{TE}} \right) ,
\end{equation}
\begin{equation}
\sigma_{\mathrm{abs}}
= \sum_{n=1}^{+\infty} (\sigma_{\mathrm{abs}})_n
= \frac{\pi}{2 k_0^{2}} \sum_{n=1}^{+\infty} (2n+1)
\left( 1 - \left| s_n^{\mathrm{TE}} \right|^{2}
     + 1 - \left| s_n^{\mathrm{TM}} \right|^{2} \right) .
\end{equation}
For each spherical harmonic, one can then define an absorption efficiency, or absorption-to-scattering ratio, as
\begin{equation}
\left( \frac{\sigma_{\mathrm{abs}}}{\sigma_{\mathrm{scat}}} \right)_{n}^{\mathrm{TE/TM}}
= 
\frac{\,1 - \left| s_{n}^{\mathrm{TE/TM}} \right|^{2}\,}
     {\left|\, s_{n}^{\mathrm{TE/TM}} - 1 \,\right|^{2}} \, .
\end{equation}
By plotting this quantity together with the normalized absorption cross section $\frac{(\sigma_{\mathrm{abs}})_{n}^{\mathrm{TE/TM}}}{ (2n+1)\,\lambda_{0}^{2}}$ for all possible complex values of $s_n$ allowed by passivity, the graph in Fig. \ref{fig:tradeoff_abs_scat}(a) is obtained (blue shaded region), which reveals a fundamental limit on the absorption efficiency of each harmonic as a function of the absorption level. Maximum absorption is achieved when scattering and absorption are equal, a condition known as \textbf{conjugate matching} in microwave engineering \cite{pozar2011microwave}. More generally, for any given level of absorption (a vertical line in the plot), one can identify the maximum and minimum scattering allowed by passivity. In particular, the plot shows that, in the presence of nonzero absorption, scattering cannot perfectly vanish. However, for small absorption levels, the absorption-to-scattering ratio can become very large, a regime that enables the realization of low-scattering sensors or \textbf{cloaked sensors} \cite{alu2009sensor,Fleury2014-cd,bernal2022cloaked}. 

We also note that similar fundamental tradeoffs between  scattering cross section, absorption cross section, and the directivity of the scattered field have been identified in generic antenna systems~\cite{andersen2005absorption} (Fig. \ref{fig:tradeoff_abs_scat}(b)). These tradeoffs can be derived starting from the following inequality, which follows from the optical theorem~\cite{andersen2005absorption}:
\begin{equation}
\sigma_{\mathrm{abs}}+\sigma_{\mathrm{scat}}\le \lambda\sqrt{\sigma_{\mathrm{scat}}\frac{D_s(\hat{\mathbf{k}}_i)}{\pi}},
\end{equation}
where $D_s(\hat{\mathbf{k}}_i)$ is the directivity of the scattered field in the forward direction $\hat{\mathbf{k}}_i$. This inequality again shows that perfectly invisible absorbing objects cannot exist: the absorption cross section must vanish if the scattered power is zero, unless the forward directivity diverges, which, however, is not possible for finite objects with any nonzero level of loss (see also footnote \ref{footnote_discontinuous_scattering}). Finally, beyond the analytical approaches discussed here, fundamental trade-offs between the absorption and scattering cross sections of lossy objects confined to arbitrarily shaped volumes have recently been studied \cite{schab2020tradeoffs_absorption_scattering,jelinek2021fundamental_cloaks} as multi-objective optimization problems using convex relaxation methods, as those discussed in Section \ref{sec:comput_bounds}.

\paragraph{Perfect broadband invisibility} Going back to the lossless case, another question one may ask is \emph{whether perfect invisibility is possible over a nonzero bandwidth}. The answer is no, provided that the object is finite and surrounded by free space, as can be understood from causality arguments. Intuitively, perfect invisibility requires a cloaking device to reroute waves around an obstacle, reconstructing the wavefront behind it without imperfections, exactly as if the object were not there. This would require the wave to travel faster than the speed of light in vacuum, $c$, in order to compensate for the longer path around the object. While this is not a problem for monochromatic waves, as the phase velocity can take any value (including infinity), this requirement would violate relativistic causality\footnote{The signal velocity, equal to the group velocity in the lossless case, must not exceed $c$.} for non-monochromatic signals unless the response is frequency dispersive, which implies bandwidth constraints. 

\begin{figure*}[tb]
\centering
\includegraphics[width=0.99\linewidth]{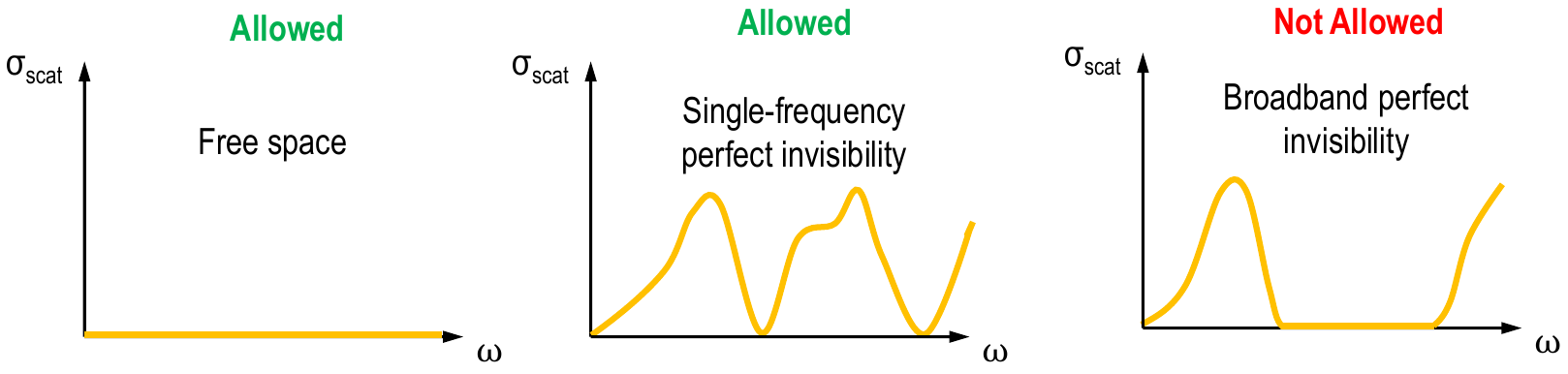}
\caption{\textbf{Perfect invisibility?} Scattering cross sections, as a function of frequency, that are allowed and disallowed for finite, causal, stable scattering systems in free space. Nontrivial perfect invisibility is only possible on a discrete set of frequencies.
}
\label{fig:SCS_allowed}
\end{figure*}

An alternative and more general way to see the impossibility of broadband perfect invisibility (not just cloaking, but any invisibility effect\footnote{Cloaking is a special type of invisibility effect in which invisibility is achieved by wrapping additional material around an otherwise unmodified object. More generally, however, one may consider a given volume and a fixed amount of matter (defined for example by the total number of electrons) and ask whether invisibility can be achieved by rearranging that matter, or by adding more material.}) can be formulated in terms of the analyticity of the forward scattering amplitude (a consequence of causality, cf. Section \ref{sec:Plane_wave_extinction}), together with stability and the optical theorem. The argument goes as follows: Suppose that the total scattering cross section $\sigma_\mathrm{scat}(\omega)$ vanishes over a finite bandwidth. Since $\sigma_\mathrm{scat}$ is proportional to the angular integral of the scattering amplitude $|s(\omega,\theta,\phi)|^2$, this implies that the scattering amplitude vanishes in all directions, including the forward direction, for every frequency in that bandwidth~\footnote{\label{footnote_discontinuous_scattering}This statement is subtler than it may appear. Strictly speaking, the vanishing of the angular integral implies that the positive integrand is zero \emph{almost everywhere} in angle, i.e., except on a set of zero measure, as the integrand may be discontinuous. For a finite scatterer, however, the far-field scattering amplitude is a smooth function of angle, since the scattered field outside the smallest sphere enclosing the object admits a convergent spherical-wave expansion. Therefore, $s$ must vanish \emph{everywhere} in angle, including in the forward direction. This argument can fail for scatterers that are infinite in one or more directions, for which the angular scattering pattern may become infinitely sharp, hence discontinuous.}. Then, consider that the forward scattering amplitude $s(\omega,0,\phi)$ is analytic in the upper half of the complex-frequency plane, $\operatorname{Im}\omega>0$, as a consequence of causality and stability. We further assume that $s(\omega,0,\phi)$ has no poles or other singularities on the real-frequency axis, so that it can be analytically continued to an open neighborhood of the real-frequency interval under consideration~\footnote{In other words, we allow the system to be non-passive (active), but exclude cases with scattering poles, or other singularities, on the real axis and above.}.
Under these weak assumptions, the identity theorem of complex analysis implies that, if the forward amplitude vanishes on a real-frequency interval of nonzero measure (nonzero bandwidth), then it must vanish identically throughout the connected domain of analyticity, including the upper half-plane and the real axis. By the optical theorem, Eq. (\ref{eq:PextOT}), the extinction cross section $\sigma_\mathrm{ext}(\omega)$ must then also vanish over the entire real-frequency spectrum.
Finally, since $\sigma_\mathrm{ext}=0$ implies $\sigma_\mathrm{scat}=0$~\footnote{In a non-passive scattering system, one could in principle have a nonzero scattering cross section even when the extinction cross section vanishes, because negative absorption (gain) could exactly compensate scattering. However, if this requirement were imposed over the entire frequency spectrum, such a hypothetical scatterer would need to be made of a medium with net gain at every frequency, which is prohibited by the positivity of the $f$-sum rule, Eq. (\ref{eq:f-sum rule}). Two other loopholes are worth noting. First, with prior knowledge of the incident field and its time of arrival, an active scatterer can be designed to radiate a field that perfectly cancels the scattered field over any nonzero bandwidth \cite{selvanayagam2013experimental}. Second, in time-varying and/or nonlinear systems, the forward scattering amplitude becomes a function of multiple frequency variables, which can alter the relevant analyticity properties depending on the relationships among the involved frequencies (certain nonlinear susceptibilities and reflectivities are meromorphic \cite{peiponen2002dispersion,Lucarini2005}; see also Section \ref{sec:nonlinear}), potentially invalidating the argument above.}, we conclude that, if scattering vanishes over \emph{any} nonzero bandwidth, then it must also vanish on the entire real-frequency axis. Since any object made of ordinary matter necessarily interacts with electromagnetic fields over some nonzero portion of the spectrum, this confirms the impossibility of broadband perfect invisibility for any finite, causal, stable scattering system (Fig. \ref{fig:SCS_allowed}).

\paragraph{Global bounds on cloaking} The discussion in the previous paragraphs shows that, over any nonzero bandwidth, invisibility and cloaking effects are necessarily imperfect, even in the lossless case. Quantifying the relationship between this imperfection, i.e., the residual scattering, and bandwidth has been one of the central questions in this area. We first note that the all-frequency integrated extinction sum rules, Eqs.~(\ref{eq:SRExt0}) and~(\ref{eq:SRExtInf}), provide only a partial answer. These sum rules still allow $\sigma_\mathrm{scat}$ to be arbitrarily close to zero over very broad frequency ranges, provided this reduction is compensated by larger scattering elsewhere in the spectrum. Nevertheless, this observation already points to several interesting implications for invisibility cloaks.

\begin{figure*}[tb]
\centering
\includegraphics[width=0.7\linewidth]{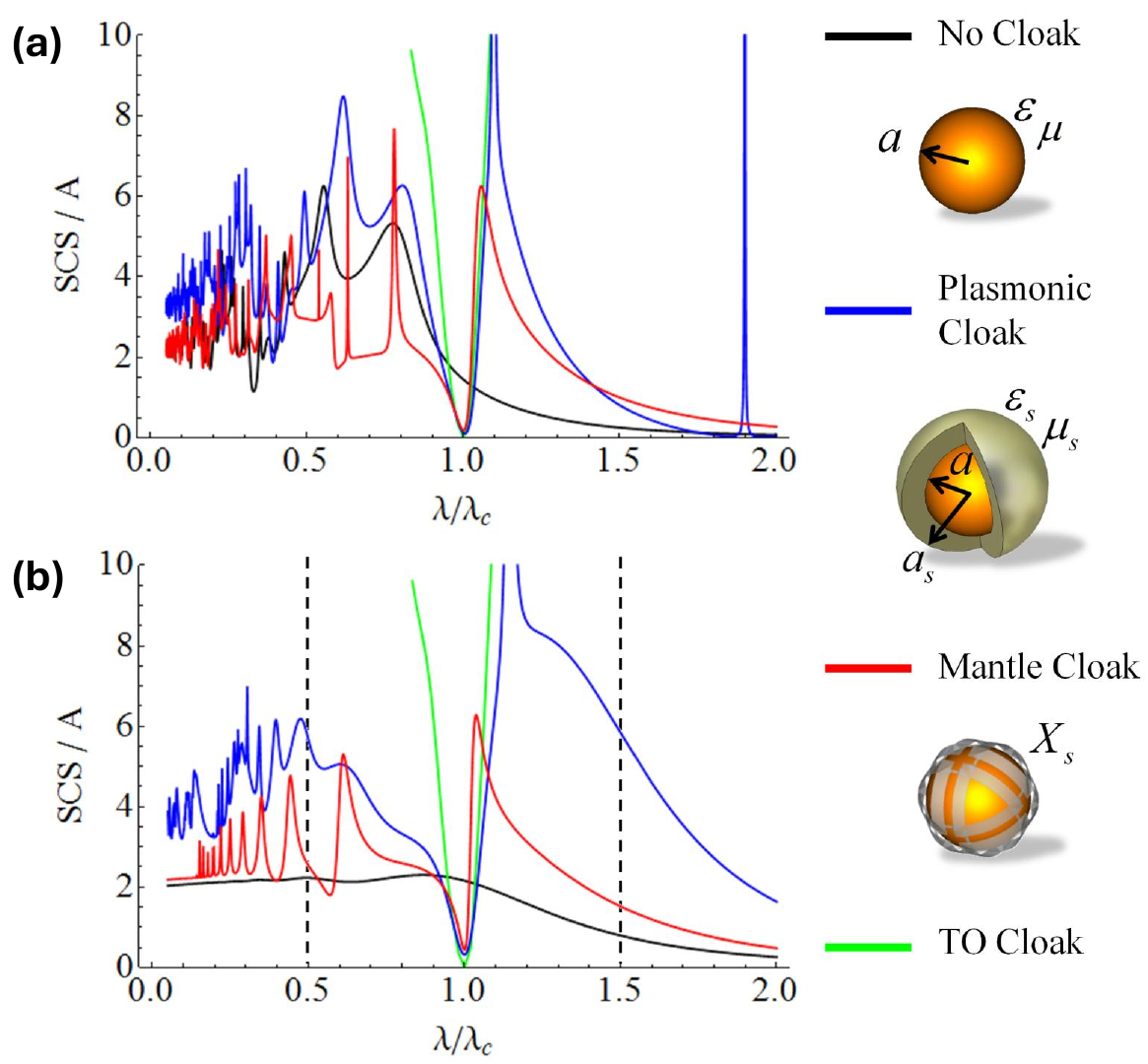}
\caption{\textbf{A global bound on cloaking.} Normalized scattering cross section over a broad wavelength range for different cloaking techniques. The cloaks are designed to suppress the scattering of (a) a dielectric sphere and (b) a perfectly conducting sphere at the central wavelength $\lambda_c$. In both panels, the scattering spectrum of the uncloaked object (black curve) is compared with those of several passive cloaks: a plasmonic cloak (blue curve), a mantle cloak (red curve), and a transformation-optics cloak (green curve). In all cases, as a consequence of all-frequency sum rules and relevant monotonicity theorems on the static polarizability, the passive cloaks increase the scattering cross section integrated over the entire frequency spectrum. (Figure adapted from Ref.~\cite{monticone2013cloaked}.)}
\label{fig:global_cloaking}
\end{figure*}

Consider the low-frequency sum rule, Eq. (\ref{eq:SRExt0}), which relates the integrated scattering cross section of an object (in the lossless case) to its static polarizability. One may then ask whether adding a cloak, designed to suppress scattering within a specified frequency window, can also reduce the integrated scattering. Over the entire electromagnetic spectrum, does a cloaked object scatter less or more than the original object? Following the argument in Ref.~\cite{monticone2013cloaked}, this question can be answered using Eq.~(\ref{eq:SRExt0}) together with the \textbf{monotonicity of the static electric and magnetic polarizabilities} with respect to the static permittivity and permeability. This monotonicity property, valid for both isotropic~\cite{jones1985scattering} and anisotropic~\cite{sjoberg2009variational} materials, was already briefly discussed in Section~\ref{sec:polarizabilities}. Focusing on the lossless case for simplicity, it implies that adding material around an object in free space necessarily increases the polarizability of the cloaked object, compared with the original uncloaked object, provided that $\varepsilon(0)>\varepsilon_0$ and $\mu(0)>\mu_0$ at every point in the cloak volume. The condition $\varepsilon(0)>\varepsilon_0$ is guaranteed for passive materials by the static-permittivity sum rule obtained from the zero-frequency limit of the KK relations (see footnote \ref{footnote_static_epsilon}). By contrast, $\mu(0)$ is not generally guaranteed to exceed $\mu_0$ (indeed, static diamagnetic media do exist), which is related to the fact that the KK relations for the permeability must be applied with a finite upper limit of integration~\cite[Sec.~82]{landau2013electrodynamics}. From these observations, it follows that the integrated extinction sum rule implies that any cloak made of linear, passive, nondiamagnetic media necessarily increases the integrated scattering of the original uncloaked object. Thus, under these assumptions, any cloaked object scatters more than the original object for sufficiently broadband incident signals (Fig. \ref{fig:global_cloaking}).

This result provides a global bound on cloaking: no passive, nondiamagnetic cloak can reduce the integrated scattering of the original uncloaked object over the entire spectrum. It does not, however, directly constrain the continuous bandwidth over which scattering can be reduced below a specified nonzero level, nor does it reveal how such bandwidth limits may depend on general properties of the system, such as its size. Before reviewing a few approaches that have been developed to address this question, it is useful to explain why this problem is significantly more difficult than the corresponding problem of deriving bounds on plane-wave reflection in planar structures, or on guided-wave reflection in wave-guiding structures, such as transmission-line circuits, for which fundamental limits have been known for decades. The fundamental reason for this difference lies in the different implications of causality in these settings, as also alluded to in Sections \ref{sec:passivity}, \ref{sec:polarizabilities}, \ref{sec:T-matrix-response}. This distinction goes to the heart of the problem of formulating causality constraints on the scattering matrix. 

\paragraph{Bounds on spherical-wave scattering}
Consider a propagating planar wave packet with a sharp front, such that the field vanishes for $t<z/c$, incident on a finite object of radius $a$, assumed here to be spherically symmetric for simplicity. This planar wave packet can be decomposed into spherical multipole wave packets, i.e., spherical waves of fixed angular momentum (often referred to as partial waves). Each of these spherical waves vanishes for $t<-a/c$, that is, for all times before the planar wavefront first touches the surface of the scatterer. The standard causality condition for spherical waves requires that each scattered spherical wave packet also vanish at the surface up to this time. Although this condition appears natural in the spherical-wave basis, it does not exclude the following unphysical situation. At $t=-a/c$, when the incident planar wavefront first touches the leftmost point of the scatterer, a scattered spherical wave could in principle emerge from the entire surface, including the rightmost point. A scattered wave would then be generated instantaneously across a distance $2a$, violating relativistic causality. A stronger causality condition can be formulated for the total scattered field (without expanding it into spherical waves) for fixed scattering angles, by requiring causal propagation also inside the scatterer, as discussed, for example, in Ref.~\cite[Sec.~3.2]{Nussenzveig1972}. This is the strategy needed to derive KK relations for the forward scattering amplitude, \eqref{eq:KKExt}. Here, however, we are specifically interested in KK relations for the spherical-wave scattering coefficients, and the associated fundamental limits~\footnote{\label{footnote_quantum_scat}It is worth noting that none of the causality conditions mentioned in this section, which require defining the time at which the wave first reaches the scatterer, can be directly applied to quantum scattering of particles. In this context, propagating wave packets cannot be constructed to have sharp wavefronts for any nonzero time length: a sharp front would require arbitrarily high frequencies, and hence arbitrarily large energies, $E= \hslash   \omega$, which in turn imply arbitrarily large particle velocities and instantaneous spreading through space. Causality in quantum scattering must therefore be formulated differently; see Ref.~\cite[p.~75]{Nussenzveig1972}. Incidentally, another (related) major difference from the classical case is the absence of negative frequencies, since energy is nonnegative, which has important implications for quantum scattering~\cite{Nussenzveig1972} and, among other things, for the generation of single-photon states~\cite{gulla2021approaching}.}. 

Given an incoming spherical wave $a_n(t+r/c)$, the outgoing spherical wave $b_n(t-r/c)$ can appear as soon as the input wavefront reaches $r=a$, but not earlier, by causality. This implies that the output of the system can appear a time $2a/c$ earlier than it would have been possible in free space (i.e., in the absence of the scatterer). Thus, the causal impulse response of this system must be $s_n(t-2a/c)$ rather than $s_n(t)$, in order to account for this time shift relative to free space (only with this time shift is the output correctly obtained from the input through convolution with the impulse response.). One can then show that the Fourier transform of this shifted causal impulse response, $e^{i k_02a}s_n(\omega)$, has the required properties to derive KK relations, whereas the spherical scattering coefficient $s(\omega)$ itself does not \cite[Sec. 2.4]{Nussenzveig1972},\cite{bernland2011physical}. It then follows that
\begin{align}
    \Re [s_n(\omega)e^{2i\omega a/c}] = 1+\frac{\omega}{\pi} \Pvl\!\int_{-\infty}^{\infty} \frac{\Im[s_n(\omega')e^{2i\omega' a/c}]}{\omega'(\omega' - \omega)}\,{\rm d}\omega'.
    \label{eq:KK_scattering}
\end{align}
This relation is less useful than standard KK relations because it explicitly depends on the radius of the scatterer, and because the real and imaginary parts of the scattering coefficient are mixed by the exponential factor. This fact is at the origin of the ``causality issues'' of the scattering matrix alluded to in previous sections, as a scattering matrix defined in terms of ports/channels associated with spherical waves of different multipolar order $n,m$ (corresponding to fixed input and output angular moments) inherits these problems.
\begin{openq}
Is it possible to use the constraints of \eqref{eq:KK_scattering} to identify useful bounds on scattering processes, for example on directional/multipolar scattering, especially as a function of bandwidth?
\end{openq}
While this remains an open question in the general case, some useful sum rules and fundamental bandwidth limits can still be derived for the reduction of individual spherical scattering coefficients, as discussed below.

Intuitively, given the form of the function $f(\omega)= s_n(\omega)e^{2i\omega a/c}$, it is natural to consider its complex logarithm, $\log z = \ln|z|+i \arg z$ (principal value), in order to try to derive dispersion relations and sum rules for $|s_n(\omega)|$. This idea does indeed work, but it involves several subtleties. The first is that $\log f_n(\omega)$ is not, in general, holomorphic in $\mathbb{C}^+$, because $s_n(\omega)$ may have zeros in the upper half-plane. These zeros, denoted by $\omega_{n,i}$ and counted according to their multiplicity, can be removed by introducing the Blaschke product 
\begin{equation}
\Bl_n(\omega)=\prod_{i} \frac{1-\omega/\omega_{n,i}}{1-\omega/\omega_{n,i}^*}.
\end{equation}
Importantly, on the real-frequency axis, $\left| f'(\omega) \right|=\left| f(\omega)/\Bl(\omega) \right|=\left| f(\omega) \right|=\left| s_n(\omega) \right|$. One can then use complex-frequency integration, together with the low- or high-frequency asymptotic behavior, to derive all-frequency sum rules. This is similar to the approach used to derive other sum rules we have encountered in previous sections, but with the main difference that the resulting expressions will depend on the locations of the upper half-plane zeros. 

More generally, for passive scattering systems, these sum rules can be obtained more directly using the theory of Herglotz--Nevanlinna functions, which we briefly described in Section \ref{sec:passivity}. Recall that a Herglotz function is a function $h$ that is holomorphic in $\mathbb{C}^+$ and satisfies $\operatorname{Im}h(z)\ge 0$ there. For the spherical reflection coefficient $s_n$, one can define the Herglotz function\footnote{One can also define corresponding Herglotz functions for non-diagonal elements of the scattering matrix, $s_{n,m}$ with $n\ne m$, but additional technical difficulties arise in deriving the associated sum rules~\cite{bernland2011physical}.} \cite{bernland2011physical}
\begin{equation}
h_n(\omega)=-i\log\left( \frac{s_n(\omega)e^{2i\omega a/c}}{\Bl_n(\omega)} \right)
\label{eq:herglotz_spherical_s}
\end{equation}
Sum rules of the form given in Eq.~(\ref{eq:h_sum_rules}) can then be derived from the asymptotic expansions of $h_n(\omega)$. For example, the low-frequency expansion of Eq.~(\ref{eq:herglotz_spherical_s}) gives $h_n(\omega)=a_1\omega+O(\omega^2)=2a\,\omega/c+2\sum_i\omega\,\Im(1/\omega_{n,i})+O(\omega^2)$. Then, using Eq.~(\ref{eq:h_sum_rule_p1}), one obtains
\begin{equation}
\frac{2}{\pi} 
\int_{0}^{\infty} 
\frac{1}{\omega^{2}} 
\ln \frac{1}{|s_n(\omega)|} 
\, d\omega
\le \frac{2a}{c} 
+ 2\sum_{i} \Im\!\left( \frac{1}{\omega_{n,i}} \right),
\label{eq:sum_rule_sn}
\end{equation}
Note that $\Im \frac{1}{\omega_{n,i}}= -\frac{\Im \omega_{n,i}}{|\omega_{n,i}|^2}$ is always non-positive since $\Im \omega_{n,i}>0$ by construction (we removed the upper-half-plane zeros using a Blaschke product). Moreover, $|s_n(\omega)|\le 1$ in passive systems. From these observations, it follows that
\begin{equation}
\left| 
\int_{0}^{\infty} 
\frac{1}{\omega^{2}} 
\ln |s_n(\omega)| 
\, d\omega \right|
\le \frac{\pi a}{c}.
\end{equation}
Equivalently, after a simple change of variables, this result can be written as an all-wavelength integral $\left| \int_{0}^{\infty} \ln |s_n(\lambda)| \, d\lambda \right|
\le 2 \pi^2 a$. 

\begin{figure*}[tb]
\centering
\includegraphics[width=0.6\linewidth]{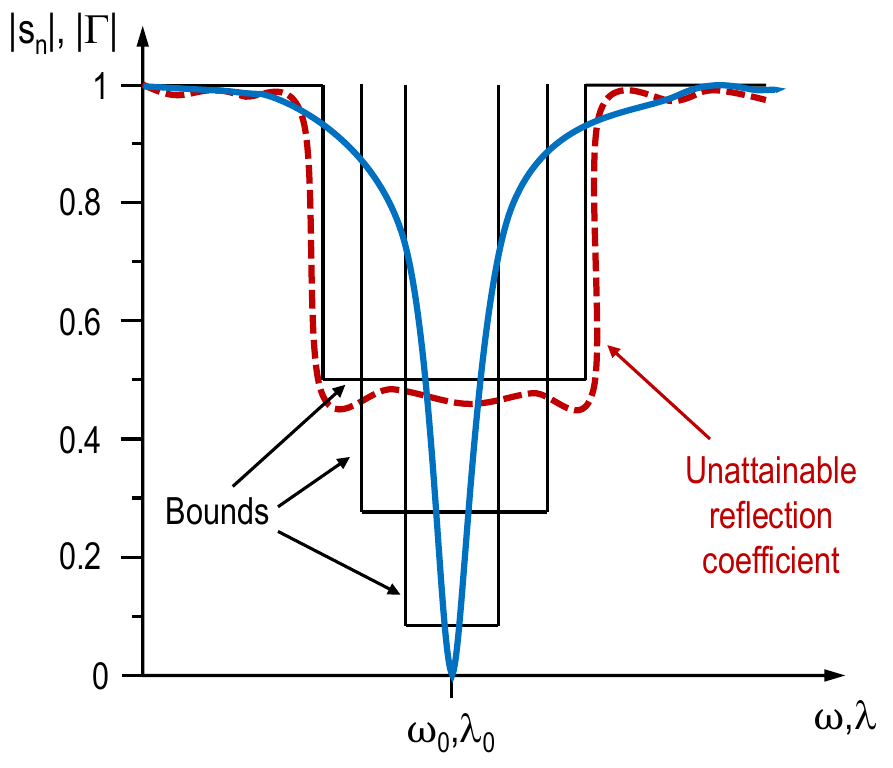}
\caption{\textbf{Fundamental bandwidth bounds on spherical-wave, plane-wave, and guided-wave reflection.} Interpretation of Eqs.~(\ref{eq:s_bound}) (spherical-wave absorption), (\ref{eq:Gamma_bound}) (Rozanov bound), and (\ref{eq:Gamma_bound_BF}) (Bode--Fano limit) as fundamental tradeoffs between in-band reflection magnitude and bandwidth. These tradeoffs are governed by physical size in Eqs.~(\ref{eq:s_bound}) and (\ref{eq:Gamma_bound}), and by the properties of the load impedance in Eq.~(\ref{eq:Gamma_bound_BF}). For a given center frequency or wavelength, the bounds are illustrated for three different values of the in-band reflection or scattering magnitude, each corresponding to a different maximum achievable bandwidth. The figure shows representative examples of attainable (blue) and unattainable (red) reflection coefficients. (Figure adapted from Ref.~\cite{bernland2011physical}.)}
\label{fig:Reflection_Bounds}
\end{figure*}

This sum rule can then be converted into a fundamental tradeoff between scattering suppression, bandwidth, and size. To do this, assume that $|s_n(\omega)|=1$ everywhere, except on a frequency range $\left[\omega_1,\omega_2\right]$ where $|s_n(\omega)|\le S_0<1$. Then, 
\begin{equation}
\frac{\pi a}{c}\ge \left| \int_{0}^{\infty} \frac{1}{\omega^{2}} \ln |s_n(\omega)| \, d\omega \right| \ge \left| \ln S_0 \right| \int_{\omega_1}^{\omega_2} \frac{1}{\omega^{2}} \, d\omega \ge \left| \ln S_0 \right| \frac{B}{\omega_0},
\label{eq:s_bound_0}
\end{equation}
where $B$ is the fractional bandwidth around the center frequency $\omega_0$, and we used the fact that $\int_{\omega_1}^{\omega_2} \frac{1}{\omega^{2}} \, d\omega=\frac{B}{\omega_0}\frac{1}{1-B^2/4}\ge\frac{B}{\omega_0}$. Finally, we obtain the desired fundamental tradeoff:
\begin{equation}
\frac{\left| \ln S_0 \right| B}{\pi} \le k_0 a.
\label{eq:s_bound}
\end{equation}
This result shows that reducing the magnitude of a spherical scattering coefficient to increasingly small values can only be achieved over increasingly narrow bandwidths, as illustrated in Fig.~\ref{fig:Reflection_Bounds}. The tradeoff is governed by the size of the smallest sphere, centered on the origin of the spherical-wave expansion, that encloses the scatterer ~\footnote{More stringent tradeoffs that depend on the static polarizability of the object can also be derived, as shown in Ref.~\cite{bernland2011physical}.}. 

Is Eq. (\ref{eq:s_bound}) a bound on cloaking and invisibility? Unfortunately, no. Since the magnitude of the spherical scattering coefficient $s_n(\omega)$ is equal to unity at real frequencies for a \emph{lossless} spherically symmetric scatterer, this result should instead be interpreted as a bandwidth limit on the \emph{absorption} of spherical waves by a lossy scatterer. For a given incoming spherical wave, absorption reduces the amplitude of the corresponding outgoing spherical wave. (For a nonspherical object, $|s_n|<1$ is possible even without loss, through conversion into other spherical-wave channels; Eq.~(\ref{eq:s_bound}) then bounds absorption plus conversion out of the channel.) This is distinct from the problem of invisibility: by absorbing incident light, the object generally becomes more easily detectable.

\paragraph{Bounds on plane-wave or guided-wave reflection and impedance matching} The approach described in the previous section for deriving sum rules and bounds is also relevant to the cases of plane-wave reflection from planar structures and of guided-wave reflection from waveguide terminations or circuit loads. There are, however, some subtle differences between the planar and spherical cases. 

Consider, for example, a plane wave normally incident on a planar structure, assumed for simplicity to be transversely invariant. Causality implies that, given an incident wave $a(t-x/c)$, the reflected wave $b(t+x/c)$ can appear as soon as the incident wavefront reaches the first interface at $x=0$, but not earlier. Here, in contrast to the spherical or cylindrical case, there is no problem in defining the coordinate origin on the surface of the planar scatterer. Moreover, in the absence of the scatterer, the output reflected wave is simply zero, whereas in the spherical-wave problem an outgoing wave is always present, with a delay relative to the case in which the scatterer is present. Thus, the causal impulse response of the system is simply $\Gamma(t)$ and its Fourier transform $\Gamma(\omega)$ is a good candidate for deriving KK relations, without the additional phase shift required in the spherical case. In particular, one may consider the logarithm of the reflection coefficient, $\log \Gamma=\ln|\Gamma|+i\arg\Gamma$, in order to obtain relations between its amplitude and phase. Unfortunately, however, this function is problematic in the high-frequency limit: since the material response vanishes as $|\omega|\to\infty$ (see Section~\ref{sec:causality}), the reflection coefficient also vanishes, and therefore $\log \Gamma$ diverges. As a result, the function is not square integrable. Different strategies can be used to regularize this divergence and derive KK-like relations between the reflection phase and amplitude; we refer the reader to Ref.~\cite[Sec.~4.6.2]{Lucarini2005} for further details. In particular, in the absence of zeros of the reflection coefficient in $\mathbb{C}^+$ (corresponding to branch points of the logarithm)---which is true, for example, for normal incidence at a single vacuum-matter interface \cite{smith1977dispersion,Lucarini2005}---the following well-known relation for the phase of the reflection coefficient can be derived \cite{smith1977dispersion}:
\begin{align}
    \arg \Gamma (\omega) = -\frac{2\omega}{\pi} \Pvl\!\int_{0}^{\infty} \frac{\ln|\Gamma(\omega')|}{\omega'^2 - \omega^2}\,{\rm d}\omega'.
    \label{eq:KK_reflection}
\end{align}
With knowledge of the high- or low-frequency asymptotic behavior of $\Gamma(\omega)$, this relation can then be used to derive all-frequency sum rules and, potentially, bounds on the magnitude of $\Gamma$ in passive systems, for which $|\Gamma|\le 1$. 

Alternatively, and more generally, one can use essentially the same approach as for the spherical coefficients $s_n$ in the previous section. First, one constructs the Herglotz function $h(\omega)=-i\log\left(\Gamma(\omega)/\Bl(\omega)\right)$ (where $\Bl(\omega)$ is again a Blaschke product constructed from the upper-half-plane zeros $\omega_{i}$ of $\Gamma(\omega)$); second, the associated sum rules are bounded using the coefficients of the low- or high-frequency asymptotic expansion of $\Gamma$ for the structure under consideration. This leads to some of the most important and well-established fundamental limits in antenna theory and microwave engineering, which we briefly review below.

(i) Consider a transversely invariant structure, such as a multilayered slab, placed above a perfectly reflecting plane, or ground plane, and illuminated by a normally incident plane wave. In this case, the low-frequency asymptotic expansion of the reflection coefficient is $\Gamma(\omega)=-1-i\,\omega/c\,2\sum_i\mu_{s,i}d_i+O(\omega^2)$, where $\mu_{s,i}$ is the static relative magnetic permeability of the $i$th layer of thickness $d_i$~\cite{rozanov2002ultimate,gustafsson2011physical_high_impedance}. The corresponding low-frequency expansion of the Herglotz function is $h(\omega)=\pi+\omega/c\,2\sum_i\mu_{s,i}d_i+2\sum_i\omega\,\Im(1/\omega_{i})+O(\omega^2)$. From this expansion, one can derive the reflection-coefficient analogue of the sum rule in Eq.~(\ref{eq:sum_rule_sn}). Following the same steps as in the previous section, this leads to the so-called \textbf{Rozanov bound} on electromagnetic absorbers, commonly written as an all-wavelength integral~\cite{rozanov2002ultimate}:
\begin{equation}
\left| \int_{0}^{\infty} \ln |\Gamma(\lambda)| \, d\lambda \right|
\le 2\pi^{2} \sum_i\mu_{s,i}d_i.
\end{equation}
The Rozanov bound can also be expressed as a tradeoff between bandwidth, maximum in-band reflection magnitude $\Gamma_0$, and absorber thickness. Evaluating the integral as done in Eq. (\ref{eq:s_bound_0}) for the spherical case, assuming $|\Gamma (\omega)|=1$ outside the finite bandwidth of interest, one obtains 
\begin{equation}
\frac{\left| \ln \Gamma_0 \right| B}{\pi} \le k_0 \sum_i\mu_{s,i}d_i,
\label{eq:Gamma_bound}
\end{equation}
where $B=(\lambda_2-\lambda_1)/\lambda_0$ is the fractional bandwidth around the center wavelength $\lambda_0$ (see Fig.~\ref{fig:Reflection_Bounds}). The similarity between Eqs. (\ref{eq:s_bound}) and (\ref{eq:Gamma_bound}) indicates that these two bounds have the same physical interpretation as fundamental reflection-bandwidth-size tradeoffs for spherical-wave and plane-wave absorption, respectively~\footnote{These bounds can be generalized to oblique incidence and to planar absorbers that are not transversely invariant but periodic. In the general case, the low-frequency asymptotic expansion of the specular reflection coefficient is $\Gamma(\omega)=-1-i\omega/c(2d\cos\theta+\alpha/A)+O(\omega^2)$, where $\theta$ is the angle of incidence $\alpha$ is the static polarizability of the unit cell subject to periodic boundary conditions and $A$ is the area of the unit cell \cite{gustafsson2011physical_high_impedance}. Moreover, Ref.~\cite{firestein2023sum} extended the Rozanov bound to cases in which the perfectly conducting termination is replaced by various types of penetrable impedance sheets.}.

\begin{figure*}[tb]
\centering
\includegraphics[width=0.8\linewidth]{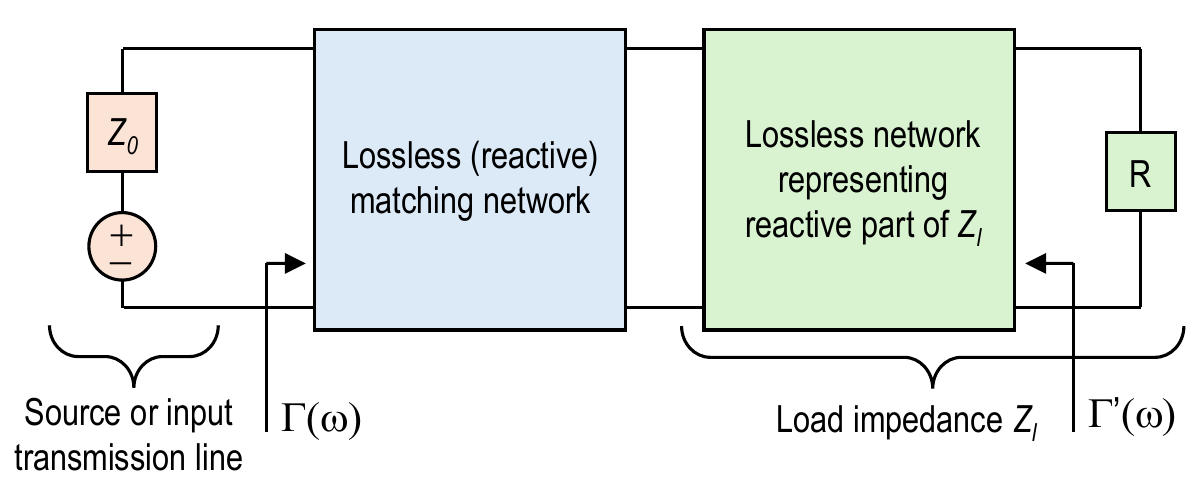}
\caption{\textbf{The impedance-matching problem.} Circuit topology used in the derivation of the Bode-Fano limit on broadband impedance matching.}
\label{fig:Bode_Fano_Circuit}
\end{figure*}

(ii) Consider now a transmission line terminated by a load impedance. This system may represent either an actual electrical circuit or an equivalent-circuit model of a plane-wave propagation problem, since both plane waves and guided waves in transmission lines are TEM waves. Following arguments similar to those above, one can derive general limits on the reflection coefficient between the load and the transmission line. These limits constrain the extent to which reflections can be suppressed, achieving impedance matching, using an arbitrary, lossless, i.e., reactive, impedance-matching network. Here we discuss only one illustrative example: a parallel resistor-capacitor (RC) load. This is the example originally studied by Bode in 1945~\cite{bode1945network}, which initiated the systematic study of matching networks and their fundamental limits. For arbitrary load impedances, we refer the reader to the classic paper by Fano~\cite{fano1950theoretical} and to more recent generalizations~\cite{carlin1998wideband,bernland2011sum}. In a transmission-line problem, the reflection coefficient at the location of the load is given by $\Gamma(\omega)=(Z_l(\omega)/Z_0-1)/(Z_l(\omega)/Z_0+1)$, where $Z_l$ is the load impedance and $Z_0$ is the characteristic impedance of the transmission line (i.e., the wave impedance of the guided TEM wave), here assumed real for a lossless transmission line. For a parallel RC load, the load impedance is $Z(\omega)=R/(1-i\omega R C)$. We slightly modify the problem by assuming $Z_0=R$, which does not change the intrinsic difficulty of the matching problem~\footnote{These two resistances can always be made equal by introducing an ideal transformer, which scales any impedance by $a^2$, where $a$ is the turns ratio. Apart from the need to introduce this ideal transformer, the original and modified problems are equally difficult to match; the reason for this modification will become clear below.}. Since any physically realizable impedance function can be represented as the input impedance of a reactive two-port network terminated by a pure resistance~\cite{fano1950theoretical}, the overall system can be represented as in Fig.~\ref{fig:Bode_Fano_Circuit}. For the RC load, a sum rule can be derived from the high-frequency asymptotic expansion of the reflection coefficient, $\Gamma(\omega)=-1+2i/(RC)\,\omega^{-1} +O(\omega^{-2})$ (which is independent of the matching network, as further discussed below). The corresponding high-frequency expansion of the Herglotz function is $h(\omega)=A-2/(RC)\,\omega^{-1}+2\sum_i\omega^{-1}\Im\omega_{i}+O(\omega^{-2})$, where $A$ is a constant. Using Eq.~(\ref{eq:h_sum_rule_p0}), one then obtains the so-called \textbf{Bode-Fano limit} for a parallel RC load,
\begin{equation}
\left| \int_{0}^{\infty} \ln |\Gamma(\omega)| \, d\omega \right|
\le \frac{\pi}{RC}.
\label{eq:Bode_Fano}
\end{equation}
This result can again be expressed as a reflection-bandwidth tradeoff (Fig.~\ref{fig:Reflection_Bounds}), analogous to the bounds discussed in the previous sections: 
\begin{equation}
\frac{\left| \ln \Gamma_0 \right| \Delta\omega}{\pi} \le \frac{1}{R C}.
\label{eq:Gamma_bound_BF}
\end{equation}
where $\Gamma_0$ is the maximum allowed in-band reflection magnitude over a bandwidth $\Delta\omega$. Unlike the Rozanov bound, however, the Bode--Fano limit constrains the ability to achieve broadband reflection suppression not through absorption, but through reactive impedance matching, i.e., through interference effects produced by a lossless matching network.

Note that, because of the symmetry of the resistive terminations in Fig.~\ref{fig:Bode_Fano_Circuit} after the transformation described above, and because everything between the two terminating resistances is reactive and therefore lossless, the magnitude of the reflection coefficient is the same whether looking into the structure from the source side or from the load side, i.e., $|\Gamma|=|\Gamma'|$. Thus, since the reactive part of the load, here the parallel capacitance, has a transmission zero as $\omega\to\infty$ (its reactance vanishes and it becomes a short circuit), the high-frequency asymptotic expansion of the load-side reflection coefficient $\Gamma'$, and hence the sum rule for $\ln|\Gamma|=\ln|\Gamma'|$, is independent of the reactive matching network. The properties of the matching network only affect the locations of the upper-half-plane zeros of $\Gamma'(\omega)$, which however do not affect the resulting sum-rule bound in Eq.~(\ref{eq:Bode_Fano}), but only determine how closely a given matching network approaches the bound. The Bode-Fano limit is therefore uniquely determined by the properties of the load. This is a distinctive feature of this bound, valid for arbitrary load impedances, and is what makes it particularly general and useful.
 
Finally, it is worth stressing that both the Rozanov bound and the Bode--Fano limit, although historically developed and applied in the context of microwave systems, are also directly relevant to photonic systems. Indeed, these limits apply to any transversely invariant structure (but extensions to multi-channel cases exist \cite{nie2015broadband}) designed to suppress reflection either through absorption (Rozanov) or interference effects (Bode-Fano), such as thin-film absorbers or anti-reflection coatings, respectively.

\paragraph{Bandwidth bounds on invisibility and cloaking} Returning to the three-dimensional scattering problem, a scattering matrix constructed from the spherical scattering coefficients $s_{n,m}$ (also known as spherical reflection coefficients), which relates incoming and outgoing spherical waves, is unitary in the lossless case (cf. Section \ref{sec:tr_symmetry}). Its eigenvalues therefore have unit magnitude on the real-frequency axis and can be written as phase factors, representing the \emph{scattering phase shifts} of the corresponding channels (see also \cite[Sec. 10.4]{Jackson1999}). Importantly, this shows that the coefficients $s_{n,m}$ are distinct from the Mie scattering coefficients that appear in the problem of plane-wave scattering by an object, and that are related to the object’s multipolar polarizabilities; see Section~\ref{sec:polarizabilities}. While the matrix elements $s_{n,m}$ relate incoming spherical waves to outgoing spherical waves, the Mie coefficients relate outgoing spherical waves to a standing-wave excitation. This is because a plane wave can be expanded into \emph{standing} spherical waves, represented by spherical Bessel function $j_n$ of order $n$, according to the frequency-domain identity (also known as partial-wave expansion; see, e.g., \cite[Sec. 2.2]{Nussenzveig1972}):
\begin{equation}
e^{ikz} = e^{ikr\cos\theta} = \sum_{n=0}^{\infty} (2n+1)i^{n} j_{n}(kr) P_{n}(\cos\theta),
\label{eq:plane_wave_expansion}
\end{equation}
where $P_n$ is the $n$th Legendre polynomial. In the relevant case of spherically symmetric objects, the Mie coefficients $c_n$ are related to the elements $s_n$ of the diagonal scattering matrix as $s_n=2c_n+1$ \footnote{This follows from the fact that, while spherical Hankel functions of the first and second kind, $h_n^{(1,2)}$, represent outgoing and incoming spherical waves (for the $e^{-i\omega t}$ convention), $j_n$ represents a spherical standing wave, and $2j_n(z) = h_n^{(1)}(z) + h_n^{(2)}(z)$.}. Different from $s_n$, in the lossless case the magnitude of the Mie coefficients on the real axis can vary between zero ($c_n=0$, zero scattering in that channel) and unity ($c_n=-1$, maximum resonant scattering), consistent with their connection to the polarizabilities of a scatterer (Section \ref{sec:polarizabilities}). Similar to $s_n$, however, the Mie coefficients do not satisfy standard KK relations, as expected from the causality issues associated with spherical-wave scattering discussed above. Instead, the relevant relations take the form of Eq.~(\ref{eq:KK_scattering}) with the replacement $s_n=2c_n+1$. Moreover, this relationship between $s_n$ and $c_n$ makes it even more difficult to derive useful bounds on the Mie coefficients and polarizabilities by using the procedure derived above: a sum rule for $\ln|s_n|=\ln|2c_n+1|$ does not generally imply a useful constraint on $|c_n|$.

Given these difficulties, dispersion and bandwidth bounds on invisibility and cloaking, and more broadly, on scattering minimization in the lossless case, remain sparse. We briefly mention a few different approaches below, all of which lead to similar qualitative conclusions. Considering a standard cloaking configuration, in which an object is surrounded by a cloak that re-routes waves around it with minimal distortion, Ref. \cite{craeye2012rule} derived a rule of thumb for the cloaking bandwidth based on the following argument. For an incident broadband light pulse, the time delay $\Delta T$ due to the (average) extra path length $\Delta R$ traveled inside the cloak, around the concealed object, must be small compared to the pulse duration $\tau$ in order to minimize waveform distortions. $\tau$ can be taken as the characteristic pulse duration for the available bandwidth $\Delta f$, $\tau\simeq1/\Delta f=2\pi/\Delta\omega$ (the convention of Ref.~\cite{craeye2012rule}). Since the time delay is equal to the group delay in the lossless case, we are essentially requiring that the peak of the pulse not be delayed by more than the width of the pulse envelope. Ref. \cite{craeye2012rule} also defines a ``quality factor'' for imperfect cloaking, $Q_c$, as the ratio of pulse width and delay, i.e., $Q_c=\tau/\Delta T\ge1$, which characterizes the pulse distortion and, ultimately, the quality of the cloaking effect. Numerical evaluations show that $Q_c$ on the order of 5-10 corresponds to good-quality cloaking (a relation between $Q_c$ and the scattering cross section should clearly exist, but it was not investigated in \cite{craeye2012rule}). Then, since $\Delta T>\Delta R/c$, because the free-space speed of light is the maximum possible velocity of the wave packet, and since $\Delta T=\tau/Q_c=2\pi/(\Delta\omega Q_c)$, one obtains the following rule of thumb for the maximum fractional bandwidth of imperfect cloaking:
\begin{equation}
B<\frac{\lambda_0}{\Delta R\,Q_c}<\frac{\lambda_0}{a\,Q_c}
\label{eq:cloak_bound_1}
\end{equation}
where $\lambda_0$ is the wavelength at the center frequency and $a$ is the radius of the object to be concealed, with $a<\Delta R$. As an example, for a human-scale object, assuming an extra path length on the order of one meter and $Q_c=5$, at a center wavelength $\lambda_0=500~\mathrm{nm}$, corresponding to green light, the fractional bandwidth would be approximately $B=10^{-5}\%$. Using different arguments, but still based on causality (specifically, on the requirement that the group velocity in the cloak remains subluminal, $v_g\le c$), Ref.~\cite{chen2007extending} derived another upper bound on the cloaking bandwidth, of the form: $B\le \Delta r_0/a$. This result again shows that the fractional cloaking bandwidth must scale inversely with the radius of the object. Here, $\Delta r_0$ is another parameter that indirectly quantifies the quality of imperfect cloaking (see \cite{chen2007extending} for details).

Different approaches further confirm this scaling of cloaking bandwidth with object size and performance. For example, in \cite{monticone2016invisibility}, approximate bandwidth bounds on cloaking were derived by drawing inspiration from the Bode-Fano limit on broadband impedance matching discussed above. First, the Bode-Fano limit was applied directly to the problem of minimizing backscattering, i.e., reflection, from a planar dielectric slab, which was modeled, around one of its resonances, as an RLC circuit within a transmission-line model for wave propagation. This approach yields a bandwidth bound for a slab of thickness $d$ and permittivity $\varepsilon$: $B\le \frac{\lambda_0}{d}\frac{1}{(\varepsilon/\varepsilon_0-1)|\ln \Gamma_0|}$, where $\Gamma_0$ is the maximum allowed reflection magnitude within the desired bandwidth (for an electrically thin slab, $k_0 d\ll1$, the slab reduces to a shunt capacitance, twice that of the resonant RLC model, $C=\varepsilon_0(\varepsilon/\varepsilon_0-1)d$, and the bound becomes twice as tight). Owing to the properties of the Bode-Fano limit discussed above, this bandwidth bound is determined entirely by the ``load'' (the slab and the space behind it) and is independent of the employed lossless ``matching network'' or cloak. In this planar case, such a cloak may be an antireflection coating of arbitrary thickness and number of layers.
An analogous approach was then applied heuristically to circuit models for the $n$th-order spherical scattering from a spherically symmetric object, defined so that the circuit reflection coefficient $\Gamma$ approximates the corresponding Mie scattering coefficient around the frequency of interest, while enforcing causality and passivity. For the case of optically large, impenetrable, spherical objects (i.e., any object enclosed within a large perfectly reflecting spherical shell), surrounded by a generic lossless cloak, summing the bounds on the magnitudes of these modified Mie coefficients ultimately leads to a bandwidth bound on the suppression of the total scattering cross section $\sigma_\mathrm{scat}$: 
\begin{equation}
B<\frac{\lambda_0}{a}\frac{1}{\left|\ln\left( \frac{1}{2} \frac{\sigma_\mathrm{scat}}{\pi a^2} \right)\right|}.
\end{equation}
This bound again scales inversely with the object size, $a/\lambda_0$, and depends on a measure of the quality of imperfect cloaking, here the logarithm of the maximum allowed normalized scattering cross section within the bandwidth of interest. If the scattering cross section is required to vanish, the fractional bandwidth over which this can be achieved also vanishes, as expected. As an example of imperfect cloaking, consistent with the discussion following Eq.~(\ref{eq:cloak_bound_1}), consider again a human-scale object enclosed in a sphere of radius $a=1~\mathrm{m}$. This bound predicts that a ten-fold suppression of the scattering cross section at a center wavelength $\lambda_0=500~\mathrm{nm}$ (green light) could be achieved only over a fractional bandwidth of approximately $B=2\times10^{-5}\%$.
Finally, Ref. \cite{hashemi2012diameter} also showed that the bandwidth of transformation-optics-based cloaks must scale inversely with the object diameter, assuming the normalized scattering cross section $\sigma_s/\pi a^2$ is small at the design frequency $\omega_0$. This scaling law follows from the more general result that the normalized scattering cross section averaged over a bandwidth around $\omega_0$ must scale linearly with the diameter as a consequence of causality constraints. The derivation of this result relies on a sophisticated approach showing that the calculation of the frequency-averaged scattering cross section is equivalent to solving an appropriate scattering problem at a single complex frequency, which, in turn, can be mapped to a scattering problem at a single real frequency with transformed complex materials parameters. We refer the reader to \cite{hashemi2012diameter} for further details. More recently, the same complex-frequency evaluation of bandwidth averages has been combined with the computational-bound framework of Section~\ref{sec:comput_bounds} to obtain structure-agnostic bounds on the bandwidth-averaged extinction of passive cloaks~\cite{strekha2024limitations}.

\begin{figure*}[tb]
\centering
\includegraphics[width=1.0\linewidth]{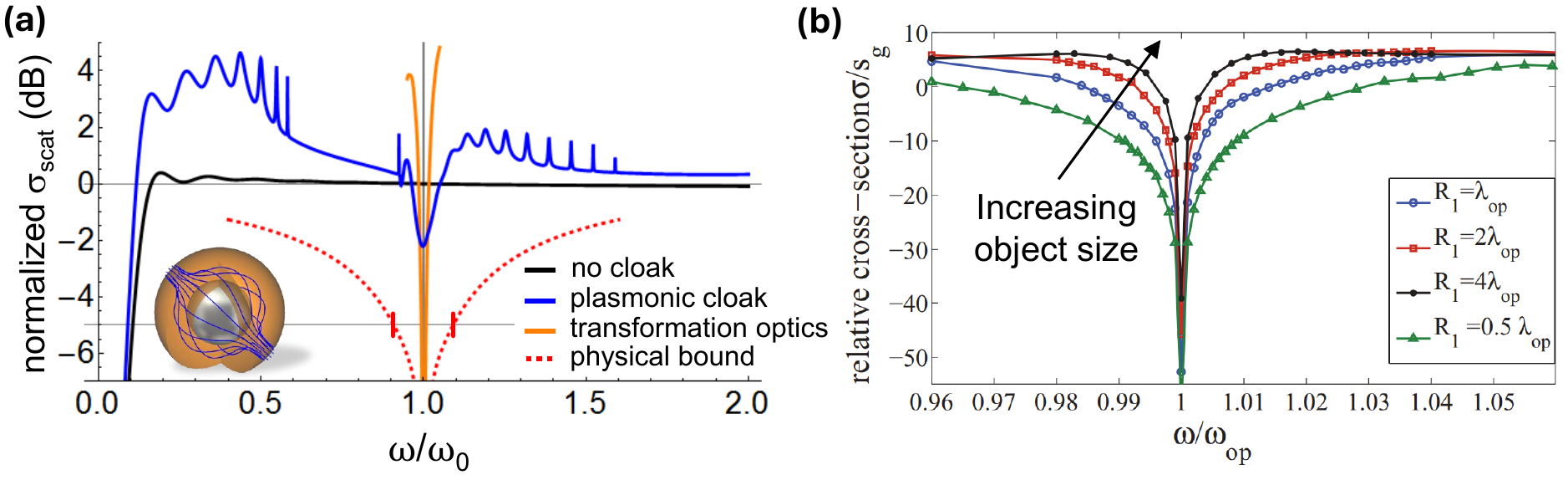}
\caption{\textbf{Physical bounds on cloaking.} (a) Normalized scattering cross section of an impenetrable sphere of radius $a=\lambda_0$ (black curve) and of two cloaking devices, compared with the approximate physical bound from Ref.~\cite{monticone2016invisibility} (dashed red line). Existing bounds on cloaking are essentially limits on the size--bandwidth--performance product, showing that, as the maximum allowed in-band scattering is reduced, the maximum achievable fractional bandwidth becomes increasingly small, with the tradeoff becoming more severe as the object size increases.
(b) Normalized scattering cross sections for a transformation-optics cloak, shown for different object radii. The cloaking bandwidth decreases inversely with object size.
(Panel (a) adapted from Ref.~\cite{monticone2016invisibility}; panel (b) adapted from Ref.~\cite{hashemi2012diameter}.)}
\label{fig:Cloaking}
\end{figure*}

All these approaches lead to essentially the same causality-based scaling limitations for invisibility cloaks, in the form of a bound on the size-bandwidth (or size-bandwidth-performance) product. A few open questions remain, however. In particular, the approaches discussed above apply primarily to the standard cloaking configuration in which light is rerouted around a shielded object, and they rely on various assumptions, approximations, or heuristic arguments. We therefore conclude this section with the following open question:

\begin{openq}
    Are there more general causality- and passivity-based limitations on the size-bandwidth-performance product for any type of invisibility cloaks? More broadly, can one derive rigorous limitations for any imperfect invisibility effect, not only cloaking, given a specified amount of matter within a prescribed volume?
\end{openq}

\subsection{Sensing with scattering resonances} \label{sec:sensing}
An important application of both dielectric and plasmonic photonic structures is sensing, since the optical response can change significantly in response to local variations in the surrounding environment, especially for guided/localized fields in resonant configurations. The key figure of merit in this context is the so-called \textbf{limit of detection (LOD)}, which is the minimum change in the measurand that can be detected by the sensor \cite{white2008performance,molina2019fundamental,conteduca2022beyond}. Are there general, fundamental, lower bounds on the LOD? 

\begin{figure*}[tb]
\centering
\includegraphics[width=0.99\linewidth]{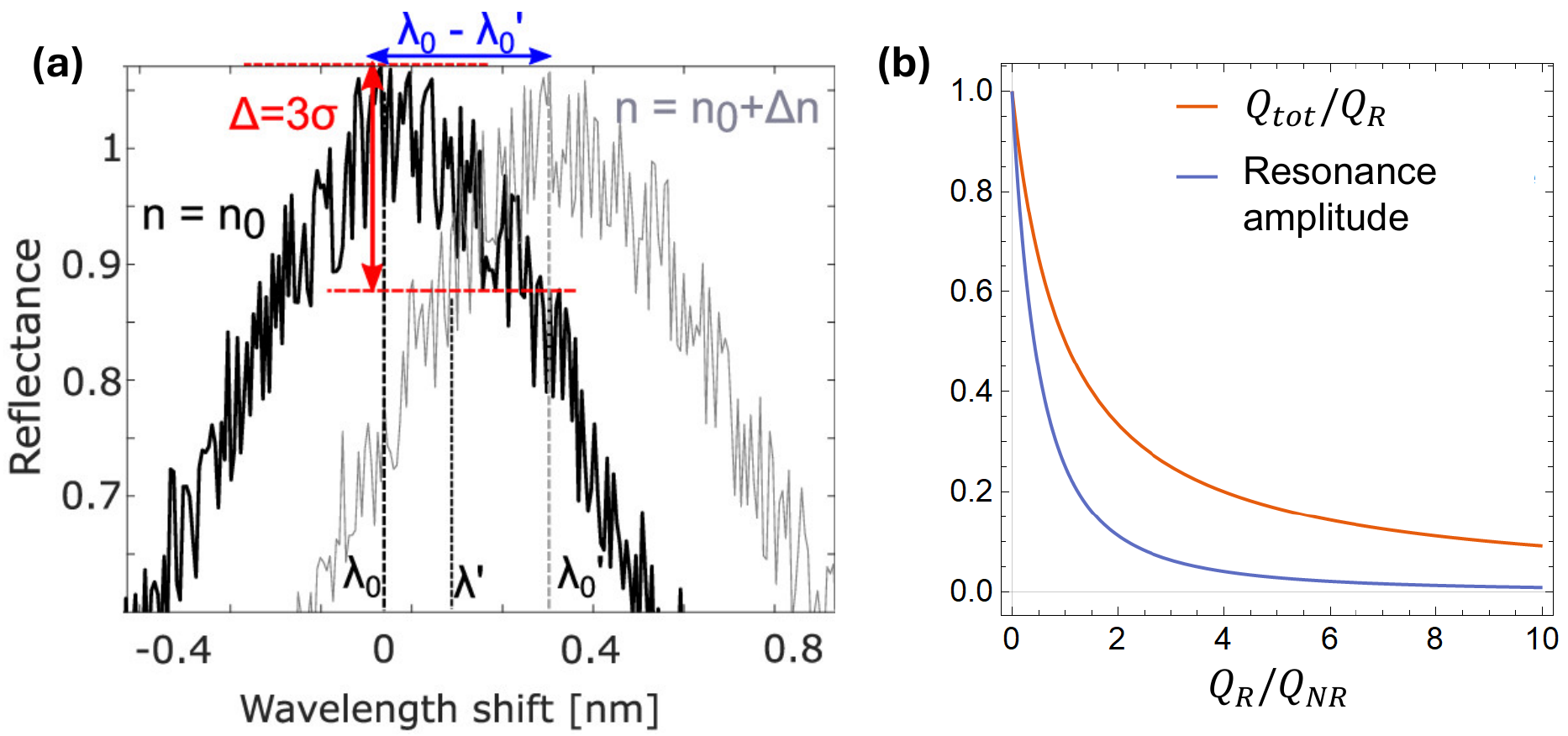}
\caption{\textbf{Sensing with scattering resonances.} (a) Typical experimental reflectance spectra for a refractive index sensor with noise, before (black curve) and after (gray curve) a perturbation. (b) Dependence of the total normalized $Q$-factor (orange curve) and resonance amplitude $A(\lambda_0)$ (blue curve) on the ratio $Q_R/Q_{NR}$. (Figure adapted from Ref.~\cite{conteduca2022beyond}.)}
\label{fig:sensing}
\end{figure*}

In the typical resonant photonic sensing configuration, one measures an amplitude variation due to a shift in resonance wavelength, from $\lambda_0$ to $\lambda_0'$, induced by a perturbation in the environment (refractive index change, geometry change, etc.). The ultimate limit of any sensing scheme is set by noise. It is generally accepted that for a signal to be considered detectable, it must be at least three times greater than the noise standard deviation $\sigma$ \cite{white2008performance,molina2019fundamental,conteduca2022beyond}. Thus, the minimum detectable amplitude variation satisfies: $A(\lambda_0)-A(\lambda')=3\sigma$, where $A(\lambda)$ is the wavelength-dependent signal amplitude and $\lambda'$ is the wavelength at which the amplitude has changed by $3\sigma$, as illustrated in Fig. \ref{fig:sensing}. Given a model for the resonance lineshape, this condition allows one to determine the wavelength offset $\lambda'-\lambda_0$. Since the resonance position is then known only within $\pm(\lambda'-\lambda_0)$, we follow Ref.~\cite{conteduca2022beyond} and define the minimum detectable wavelength shift as $\Delta \lambda_{\mathrm{min}}=2(\lambda'-\lambda_0)$, and, hence, obtain the LOD through the sensitivity $S$, defined as the ratio of wavelength shift to the change in the measurand (for example, in a refractive index sensor, $S=\Delta \lambda/\Delta n$). As a representative case, consider a single Lorentzian resonance coupled to two symmetric power-carrying channels (input/reflection and transmission channels). Using temporal coupled mode theory \cite{fan2003temporal,hsu2013observation}, one obtains the following wavelength-dependent expression for the signal amplitude: 
\begin{equation}
A(\lambda) \approx 
\left| 
\frac{Q_R^{-1}}{2i \left( \tfrac{\lambda - \lambda_0}{\lambda_0} \right) + Q_R^{-1} + Q_{NR}^{-1}} 
\right|^2
\label{eq:res_amplitude}
\end{equation}
where $Q_R$ and $Q_{NR}$ denote the radiative and non-radiative $Q$-factors, associated, respectively, with leakage into the input/output channels and absorption within the resonator (or losses through other channels, as in the case, for example, of scattering losses due to fabrication imperfections).
The LOD can then be obtained, after some algebra and to leading order in $3\sigma/A(\lambda_0)\ll1$, as \cite{conteduca2022beyond}:
\begin{equation}
\mathrm{LOD} \approx \frac{\lambda_0}{S Q_R A(\lambda_0)}\sqrt{3 \sigma}.
\label{eq:lod}
\end{equation}

The LOD can be minimized through what may be termed \emph{intrinsic} approaches, namely, minimizing the noise $\sigma$ or maximizing the sensitivity $S$, or through \emph{extrinsic} approaches, namely, optimizing the resonance properties encoded by the product $Q_R A$. Independently of the $Q$-factor or amplitude of the resonance employed for sensing~\footnote{Our distinction between intrinsic and extrinsic approaches to LOD optimization is only approximately valid under the assumption that the eigenfrequencies of the system are independent of the power-carrying channels coupled to the resonator. This corresponds to the weak-coupling, perturbative regime appropriate for a coupled mode theory description.}, the ``bulk'' sensitivity $S$ is determined by how the relevant eigenfrequency varies with the quantity of interest \cite{yu2011extraordinarily}. The investigation of fundamental limits on $S$ remains an active area of research, particularly in light of recent work on so-called exceptional-point sensors, in which eigenfrequencies depend on system parameters through a square-root branch-point dependence, whose derivative, in principle, diverges in the vicinity of the exceptional point \cite{wiersig2020review}\footnote{Exceptional points are modal degeneracies of non-Hermitian systems, where two or more eigenstates (both their eigenvalue and eigenvector) coalesce \cite{heiss2012physics} Their non-Hermitian character may arise from absorption or radiation losses.}. An increase in $S$, however, also makes the system more sensitive to noise itself, leading to a reduction in the effective sensitivity of the system \cite{duggan2022limitations}. Whether such sensors can truly outperform their conventional counterparts remains a matter of debate \cite{lau2018fundamental,wiersig2026fundamental}.

Regarding \emph{extrinsic} approaches, it is important to emphasize that the relevant figure of merit is $Q_R A(\lambda_0)$, not simply $Q_R$ or the total $Q$-factor ($Q_{tot}^{-1}=Q_{R}^{-1}+Q_{NR}^{-1}$). As argued in \cite{conteduca2022beyond}, in the presence of nonzero dissipation, $Q_{NR}^{-1} \ne 0$, arbitrarily increasing $Q_R$, as often done, for example, in sensors based on bound states in the continuum (see Section \ref{sec:polarizabilities}), is not necessarily optimal. This is because, for $Q_{NR}^{-1} \ne 0$, an increase in $Q_R$ suppresses the resonance amplitude $A(\lambda_0)$ strongly, as shown in Fig. \ref{fig:sensing}. Instead, Eq. (\ref{eq:res_amplitude}) shows that the figure of merit $Q_R A(\lambda_0)$ is maximized when 
\begin{equation}
Q_R=Q_{NR}. 
\end{equation}
This condition is known as \textbf{critical coupling} (or \textbf{conjugate matching} in the electromagnetics literature \cite{pozar2011microwave}), and it ensures maximum power coupled into the resonator (cf. the concept of cloaked sensors in Section \ref{sec:cloaking}). Under critical coupling, one finds $Q_R A(\lambda_0)=Q_R/4$. More broadly, this discussion highlights that when the amplitude of a resonance is itself of interest, in addition to its $Q$-factor, the presence of dissipation must be treated with care, as losses suppress the resonance amplitude more rapidly than they reduce the total $Q$-factor (Fig. \ref{fig:sensing}).

Finally, we note that, as shown in \cite{conteduca2022beyond}, replacing a Lorentzian resonance with a Fano resonance---arising from the interference between a sharp resonant mode and a broader resonance or background within the same channel \cite{limonov2017fano}---can improve the LOD by up to a factor of two. This improvement is achieved by optimizing the background phase to produce a more asymmetric lineshape. To the best of our knowledge, no further enhancement of the LOD through resonance engineering has been demonstrated beyond what is predicted by this model. It should be noted, however, that the derivation outlined above relies on temporal coupled-mode theory, which assumes high-$Q$ modes in the weak-coupling limit. This raises questions about its generality and broader applicability. In particular, the following questions remain open:

\begin{openq}
        Can more general analytical expressions for the LOD be derived in regimes where coupled-mode theory is not accurate, for example using quasi-normal-mode frameworks \cite{yang2015simple,lalanne2018light,tao2020coupling,zhang2020quasinormal}? Can rigorous lower bounds on the LOD be established, given a specified noise level, material loss, and optimizable sensor volume, based on more general physical principles?
\end{openq}

Moreover, one may wonder more broadly what an optimal scattering-based sensing scheme looks like, beyond conventional approaches that rely on measuring frequency shifts in scattering spectra. This also raises interesting questions about optimal sensing modalities from the viewpoint of information-theoretic metrics. For example, Ref.~\cite{feldman2025information} recently showed that angle-resolved measurements of scattering from a Fano-resonant structure can be quantitatively more informative than measurements of spectral shifts, by comparing the intrinsic Fisher information content of the two approaches.

\subsection{Fundamental limits on the delay-bandwidth product} \label{sec:delay-bandwidth}
The ability to delay light is of significant importance in many applications in photonics and electromagnetism \cite{tucker2005slow,krauss2008we,khurgin2018slow}. Light can be delayed by increasing the propagation length or reducing the group velocity in waveguiding structures, or by temporarily storing energy in resonators. Limitations on the product between the pulse bandwidth $\Delta \omega$ and the achievable delay $\Delta T$ over that bandwidth arise naturally in these contexts. We emphasize that such delay-bandwidth product is distinct from the time-bandwidth product associated with the ``uncertainty principle'' of Fourier analysis, which states that the intensity-weighted root-mean-square widths of a pulse in time and frequency satisfy $\Delta t \Delta \omega \ge 1/2$ \cite{Folland1997} (if the widths are instead measured as intensity full widths at half maximum, the product is $\approx 2$ for common transform-limited pulse shapes, e.g., $4\ln 2\approx 2.8$ for a Gaussian pulse, but this is not a strict bound). To avoid confusion, we therefore use different symbols for the time delay, $\Delta T$, and the temporal width of a pulse, $\Delta t$. 

In the case of a single, linear, time-invariant resonator coupled to an input and an output port, the time evolution of the complex mode amplitude $a$ can be derived, within temporal coupled mode theory, from the first-order differential equation \cite{Haus1984,fan2003temporal}:
\begin{align}
    \frac{da}{dt}=-i\omega_0 a-\gamma a + \kappa s_+,
\end{align}
where $\omega_0$ is the resonant frequency, $\gamma$ is the decay rate (including both dissipative loss inside the cavity and radiation loss into the ports), $\kappa$ is the in-coupling coefficient, and $s_+$ is the complex amplitude of an incident wave. The quantities $a$ and $s_+$ are normalized such that $|a|^2$ represents the stored \emph{energy} in the resonator and $s_+$ the \emph{power} carried by the incident wave. For a time-harmonic excitation at frequency $\omega$, the stored energy in the resonator is
\begin{align}
    |a(\omega)|^2= \frac{|\kappa s_+|^2}{(\omega-\omega_0)^2+\gamma^2}.
\end{align}
The bandwidth of the resonance, and therefore the bandwidth of pulses that can be stored in the resonator, is given by the full width at half maximum of this Lorentzian lineshape, which yields $2\gamma$. For how long can a pulse be stored in the resonator? With no external source, a nonzero amplitude in the resonator decays exponentially as $\text{exp}(-\gamma t)$. The achievable time delay can then be identified with the resonance lifetime, i.e., the time required for the amplitude to decrease by a factor of $e$, yielding $\Delta T = 1/\gamma$ (this coincides with the group delay at resonance of the wave transmitted to the output port, $\propto a$). Importantly, this shows that \emph{both} the bandwidth and the time delay depend solely on the total decay rate $\gamma$, and not on the in-coupling coefficient $\kappa$. The delay-bandwidth product for a single LTI resonator in transmission is therefore: $\Delta T \Delta \omega = 2$ (or $\Delta T = 2Q/\omega_c$, in terms of the resonance $Q$-factor and central frequency $\omega_c)$. If multiple resonators are employed, the total achievable delay for a given bandwidth can be increased, with the delay-bandwidth product scaling with the number of resonators that one could fit within the physical length of the structure \cite{wang2003compact,tucker2005slow} (see also Section \ref{sec:mode_volume}). These results do not rely on reciprocity or time-reversal symmetry, and therefore remain valid even in strongly nonreciprocal configurations, such as cavities coupled to topologically protected one-way waveguides \cite{mann2019nonreciprocal,hassani2019truly,monticone2020truly}. Further details on the role of reciprocity in this context are discussed in \cite{mann2019nonreciprocal}. By contrast, if the system is made nonlinear or time-varying, the delay-bandwidth product can be modified, as the parameters of the system (resonant frequency, decay rate, coupling coefficients) could become functions of the mode amplitude itself or of time, respectively, changing the fundamental equation of motion of the system. For example, dynamically opening and closing a cavity to trap and release a pulse can certainly surpass the delay-bandwidth limit of LTI systems. More sophisticated approaches include schemes based on group velocity reduction through a process of adiabatic bandwidth compression \cite{yanik2004stopping,tucker2005slow} or more general forms of dynamic modulations \cite{minkov2018localization}.

\begin{figure*}[tb]
\centering
\includegraphics[width=\linewidth]{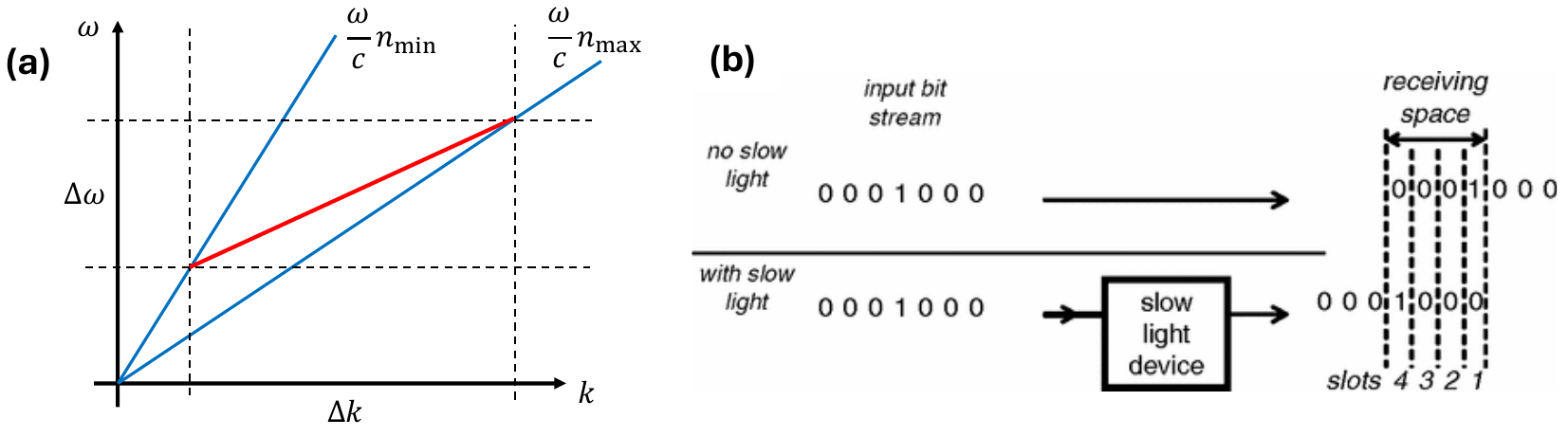}
\caption{\textbf{Limits to the delay-bandwidth product.} (a) Dispersion diagram of a guided mode with linear dispersion (red line, shown only within the bandwidth $\Delta\omega$) bounded by the light lines corresponding to the maximum and minimum available refractive indices (blue lines). This construction is used to derive the delay-bandwidth-product limit for waveguiding structures made of transparent dielectric materials, Eq.~(\ref{eq:Tucker_delay_band}). (b) Illustration of a bit pattern delayed by $S$ bit periods (here, $S=3$) through scattering in a slow-light device, used to derive a general delay-bandwidth-product limit for one-dimensional LTI devices, Eq.~(\ref{eq:Miller_delay_band}). In the device design, scattering into a total of $S+1$ bit slots must be controlled, hence requiring the control of $S+1$ orthogonal functions in the receiving space.
(Panel (b) adapted from Ref.~\cite{miller2007fundamentalPRL}.)}
\label{fig:Delay_bandwidth}
\end{figure*}

Another LTI system for which the delay-bandwidth product admits an intuitive derivation is a waveguide made of transparent, non-dispersive dielectric materials \cite{tucker2005slow}. The dispersion curves of the guided (bound) modes in such structures are bounded by the light lines corresponding to plane-wave propagation in homogeneous materials with the highest and lowest refractive index in the system~\footnote{This can be proven, for any lossless waveguide with arbitrary cross section, using standard perturbation-theory arguments for Hermitian eigenproblems (here, lossless waveguiding). One can derive \cite[p.~18]{Joannopoulos2011} the relation $\Delta \omega_m(k) = -\frac{\omega_m(k)}{2} \frac{\int d^{3}\xv \, \Delta \varepsilon(\xv)\, |\mathbf{E}(\xv)|^{2}} {\int d^{3}\xv \, \varepsilon(\xv)\, |\mathbf{E}(\xv)|^{2}}$, which implies that any increase (decrease) in permittivity  must result in a decrease (increase) in the eigenfrequency of a guided bound mode $\omega_m$, thereby confining the dispersion curves of the bound modes to the region between the light lines associated with homogeneous media having the highest and lowest permittivity.}. Consider a hypothetical mode with optimal linear dispersion (constant group velocity) bounded by these two light lines, as illustrated in Fig. \ref{fig:Delay_bandwidth}(a). One can achieve smaller values of non-dispersive group velocity (i.e., flatter linear dispersion) and therefore larger time delay for a given physical length only at the expense of reduced bandwidth. Assuming linear dispersion, the delay is $\Delta T = L/v_g=L \Delta k/\Delta \omega$. Then, since the maximum achievable $\Delta k$ is determined by the separation between the light lines, we obtain
\begin{align}
\Delta T \Delta \omega \le \frac{L}{c}\left(n_{\text{max}}\omega_2 - n_{\text{min}}\omega_1\right) = 2\pi \frac{L}{\lambda_c} (n_{\text{max}}-n_{\text{min}}) + \frac{L\,\Delta\omega}{2c}(n_{\text{max}}+n_{\text{min}})
\label{eq:Tucker_delay_band}
\end{align}
where $\omega_{1,2}=\omega_c\mp\Delta\omega/2$ and $\lambda_c$ is the free-space wavelength at the central frequency $\omega_c$. The second term can be neglected only for $\Delta \omega/\omega_c \ll 2(n_{\text{max}}-n_{\text{min}})/(n_{\text{max}}+n_{\text{min}})$; at weak index contrast it can dominate unless the bandwidth is correspondingly reduced.

Both cases considered above (resonators and waveguides) indicate that increasing the delay-bandwidth product for a given device length requires increasing the maximum refractive-index contrast within the structure, either to enable flatter linear dispersion over wider bandwidths or to realize smaller resonators that can be packed in larger numbers within a fixed length. These considerations suggest that the limits discussed above must originate from some more fundamental constraint of LTI systems, intuitively related to the (spatial and spectral) density of resonances/modes or orthogonal functions that can be realized in a physical electromagnetic system. A fundamental bound of this type was derived by David Miller \cite{miller2007fundamentalPRL} for generic one-dimensional LTI structures composed of arbitrary distributions of frequency-dispersive refractive index, including materials with loss and/or gain. The bound follows from a more general result by the same author \cite{miller2007fundamentalJOSAB}, which establishes an upper limit on the number of orthogonal functions that can be generated in a ``receiving space'' through scattering of an incident wave by an LTI structure. In brief, the problem of delaying a pulse can be mapped into this more general framework by asking whether a structure can delay a pulse by a certain number of time slots $S$. Achieving such a delay requires the pulse in the receiving/output space to appear in one specific slot among the $S+1$ available slots up to and including the desired one. This, in turn, requires the ability to control the amplitudes of the fields in these independent time slots, i.e., to control $S+1$ orthogonal functions generated through LTI scattering (Fig. \ref{fig:Delay_bandwidth}(b)). Omitting several technical details, the general results of Ref. \cite{miller2007fundamentalJOSAB} imply that, for narrowband pulses and structures much longer than the wavelength, a pulse interacting with a generic one-dimensional scattering structure can be delayed by a maximum number of time slots $S_{\text{max}} \le \frac{\pi}{2\sqrt{3}} \frac{L}{\lambda_c} \eta_{\text{max}}$, where $\eta_{\text{max}}$ is the maximum value of $|\varepsilon(z,\omega)/\varepsilon_b-1|$, with $\varepsilon$ and $\varepsilon_b$ denoting the permittivity of the device and background medium, respectively, at \emph{any} point within the structure and \emph{any} frequency within the bandwidth of interest. This bound can be recast as a delay-bandwidth constraint by considering how many time slots are required to impart a delay $\Delta T$ to the shortest possible pulse for a given bandwidth. Since the minimum pulse duration is approximately $\Delta t_{\text{min}}\approx 2/\Delta \omega$, the number of slots required is $S=\Delta T/\Delta t_{\text{min}} \approx \Delta T\Delta \omega/2$. The limit to the delay-bandwidth product is then given by
\begin{align}
\Delta T \Delta \omega \le \frac{\pi}{\sqrt{3}} \frac{L}{\lambda_c} \eta_{\text{max}}.
\label{eq:Miller_delay_band}
\end{align}
The slot bound $S_{\text{max}}$ is independent of the specific one-dimensional device design and does not rely on the simplifying assumptions used above (such as transparent materials, linear dispersion, or a well-defined group velocity). Its recast as Eq.~(\ref{eq:Miller_delay_band}), however, is approximate: as noted in Ref.~\cite{miller2007fundamentalPRL}, it rests on a heuristic relation between bandwidth and pulse width (here $\Delta\omega\,\Delta t \approx 2$). Moreover, $\eta_{\text{max}}$ constrains only the value of the permittivity within the band, not its dispersion: a uniform lossless slab in vacuum whose relative permittivity is swept from $1-\eta_{\text{max}}$ to $1+\eta_{\text{max}}$ across the band by resonances outside it reaches a group-delay--bandwidth product $\Delta T\Delta\omega \approx \eta_{\text{max}}\,\omega_c L/c$, i.e., $2\sqrt{3}$ times the right-hand side of Eq.~(\ref{eq:Miller_delay_band}). We also note that, when both limits apply, Eq.~(\ref{eq:Miller_delay_band}), which refers to the delay relative to propagation through the background medium, is close to the first (leading) term of Eq.~(\ref{eq:Tucker_delay_band}), and exceeds it if $\varepsilon_\mathrm{max}\gtrapprox6$.

Despite the usefulness and generality of these limits for one-dimensional delay lines, it is somewhat surprising that their derivation does not explicitly invoke causality-based arguments, sum rules, or constraints on scattering operators or on the local density of states, such as those discussed earlier in this chapter. Moreover, their extension to two- or three-dimensional structures is far from obvious. These observations motivate the following open question:

\begin{openq}
        Can rigorous fundamental limits on the delay-bandwidth product for generic structures be derived directly from causality arguments and sum rules, as done in previous sections for other scattering-related quantities?
\end{openq}

\begin{figure*}[tb]
\centering
\includegraphics[width=0.99\linewidth]{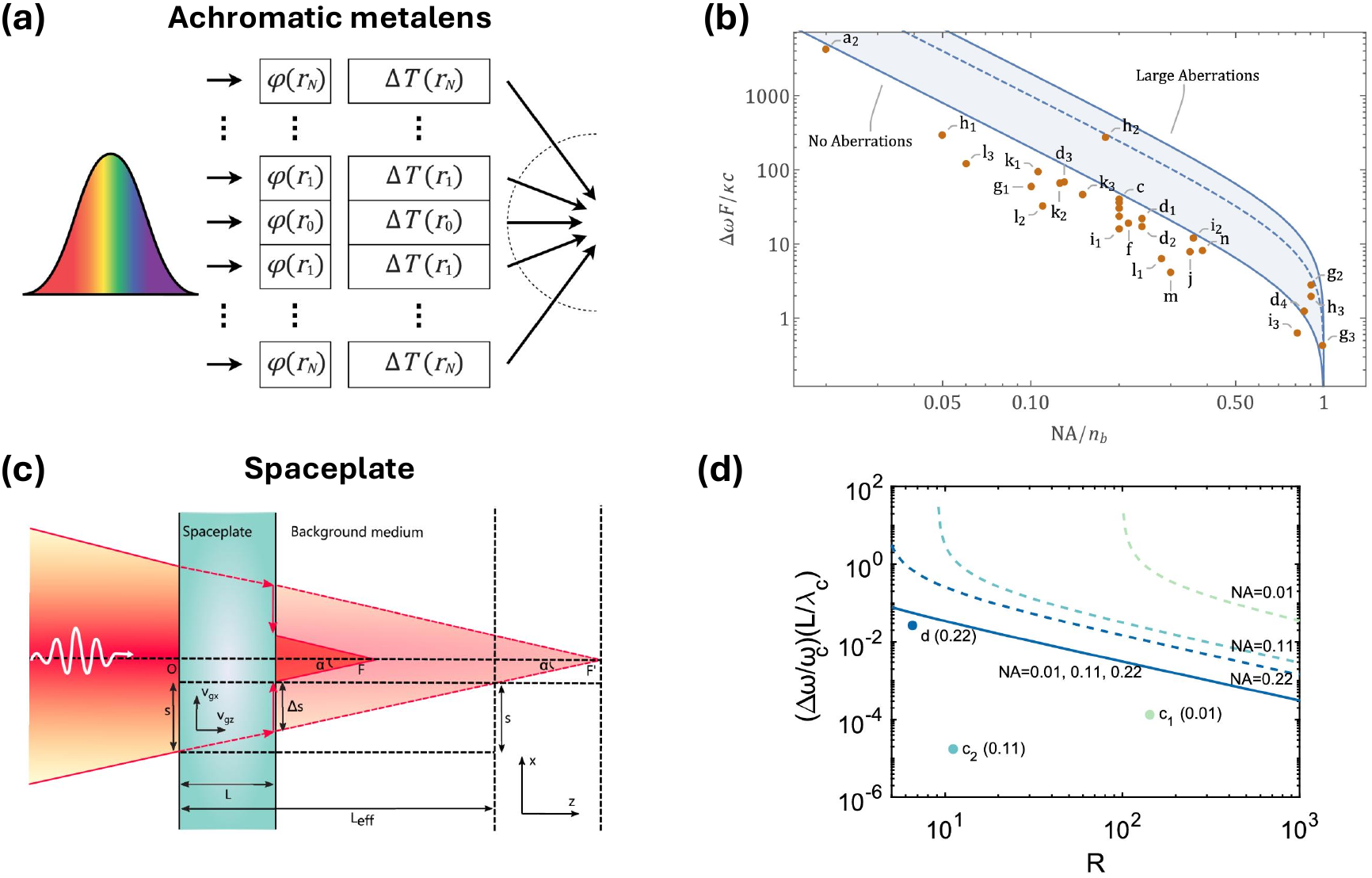}
\caption{\textbf{Application of delay-bandwidth limits to meta-optics.} (a) Delay-line model of a thin, dispersion-engineered achromatic metalens. (b) Comparison of published achromatic metalens designs against the derived (normalized) bandwidth limits, shown as a function of numerical aperture (NA) and for different levels of monochromatic aberrations (wavefront deformations). (c) Schematic of a spaceplate, which effectively replaces a region of free space of length $L_\mathrm{eff}$ by imparting an angle-dependent phase shift $\Delta s$. Achromatic broadband operation requires the spaceplate to impart an angle-dependent time delay, i.e., to function as an angle-dependent delay line. (d) Comparison of published spaceplate designs against the derived (normalized) bandwidth limits, plotted as a function of compression ratio $R$, for different values of NA and under different assumptions for the transverse group velocity $v_{gx}$ (dashed lines $v_{gx}\to c$; solid lines $v_{gx}\to c \sin \alpha$). (Panels (a,b) adapted from Ref.~\cite{presutti2020focusing}; panels (c,d) adapted from Ref.~\cite{shastri2022extent}.)}
\label{fig:meta-optic}
\end{figure*}
Finally, we note that, beyond their obvious relevance to optical delay lines and slow-light-enhanced light-matter interactions \cite{tucker2005slow,krauss2008we,khurgin2018slow,khurgin2010slow}, limits on the delay--bandwidth product have become particularly relevant for establishing bandwidth bounds on meta-optical devices, whose performance is often constrained by the narrow bandwidths associated with their typical resonant nature. We briefly discuss two important classes of meta-optical devices below.

\textbf{Dispersion-engineered achromatic metalenses}, which are flat metasurface-based lenses designed to provide the group delay required to compensate for differences in arrival time at the focal point \cite{chen2018broadband,shrestha2018broadband,presutti2020focusing}, as illustrated in Fig. \ref{fig:meta-optic}(a). These devices can therefore be treated as \emph{space-dependent} one-dimensional delay lines\footnote{Not all flat lenses are constrained by these bandwidth limits. In dispersion-engineered achromatic metalenses, the achromatic behavior relies on the dispersive response of meta-atoms acting as truncated waveguides, which justifies the analogy with local delay lines. By contrast, in achromatic \emph{diffractive} lenses \cite{meem2021imaging,engelberg2021achromatic}, the achromatic response arises from wavelength-dependent redistribution of energy among diffraction orders due to the overall microscopic surface structure, making them potentially more broadband, but introducing other limitations. See Ref.~\cite{engelberg2021achromatic} for a detailed discussion of the different (and somewhat complementary) limitations of these two main classes of achromatic flat lenses.}, yielding the following bandwidth bound for a metalens in free space \cite{presutti2020focusing}
\begin{equation}
\Delta \omega \le \frac{\kappa c}
{F} \frac{\sqrt{1-\mathrm{NA}^2}}
{\left(1-\sqrt{1-\mathrm{NA}^2}\right)},
\end{equation}
where $F$ is the focal length of the metalens, NA is the numerical aperture, and $\kappa$ is the upper bound on the delay-bandwidth product ($\Delta T \Delta \omega \le \kappa$), which may take different forms, as discussed above, for different classes of devices or under different assumptions. A comparison between the bandwidth bound and previously reported achromatic metalens designs is shown in Fig. \ref{fig:meta-optic}(b).
    
\textbf{Broadband spaceplates}, which are space-compression devices that impart a wavevector-dependent (i.e., nonlocal) lateral shift and longitudinal group delay to emulate the broadband response of a free-space volume within a shorter physical length, as illustrated in Fig. \ref{fig:meta-optic}(c) \cite{shastri2023nonlocal,shastri2022extent}. Under suitable assumptions, these structures can therefore be viewed as \emph{angle-dependent} one-dimensional delay lines, yielding the bandwidth bound \cite{shastri2022extent}
\begin{equation}
\frac{\Delta \omega}{\omega_c}
\le
\frac{1}{2\pi}
\frac{\kappa}{L/\lambda_c}
\frac{\sqrt{1-\mathrm{NA}^2}\,\, v_{gx}/c}
{R\cdot \mathrm{NA} - v_{gx}/c},
\end{equation}
where $\lambda_c$ is the center wavelength, $\omega_c = 2\pi c / \lambda_c$ is the corresponding center angular frequency, $L$ is the physical length of the spaceplate, $R=L_\mathrm{eff}/L$ is the compression ratio, and $v_{gx}$ is the transverse group velocity (different assumptions about $v_{gx}$ can be made; see \cite{shastri2022extent}). A comparison between this bound and previously reported spaceplate designs is shown in Fig. \ref{fig:meta-optic}(d).

\section{Discussion and conclusions}
\begin{figure*}[tb]
\centering
\includegraphics[width=0.9\linewidth]{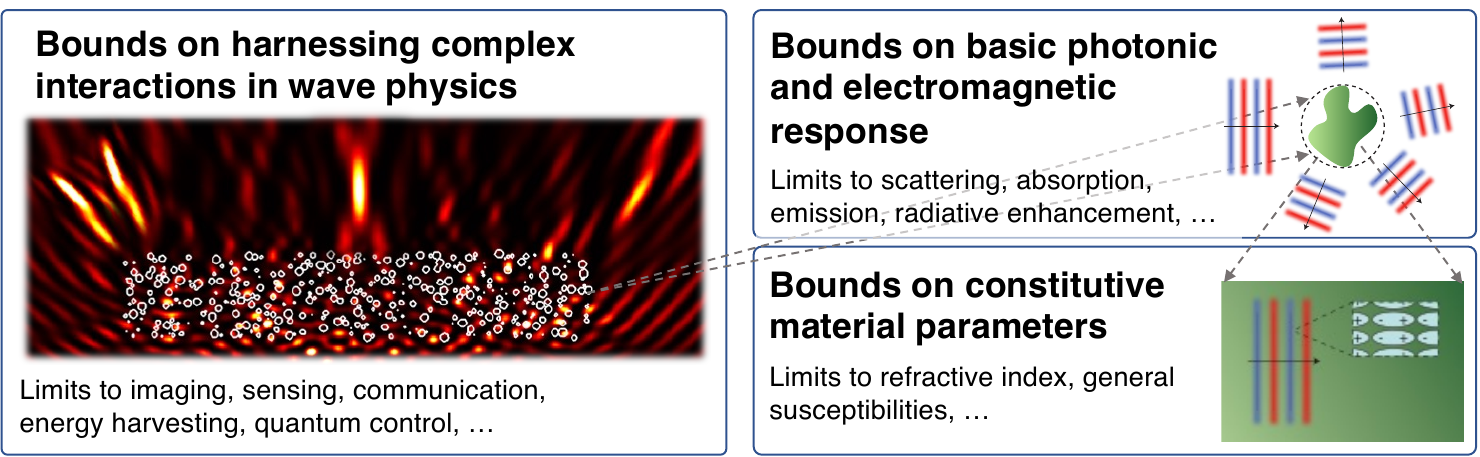}
\caption{\textbf{Toward a universal framework for fundamental limits of light interactions with matter,} organized as a three-tier hierarchy: bounds on complex applications, bounds on basic photonic and electromagnetic response, and bounds on the constitutive material parameters themselves.}
\label{fig:future_vision}
\end{figure*}

A major goal of fundamental limits in photonics and electromagnetics is to characterize what is possible in increasingly --- and ideally arbitrarily --- complex interactions of light and matter. As depicted in Fig.~\ref{fig:future_vision}, this aim sits at the apex of a hierarchy whose lower levels include bounds on basic single-body electromagnetic response (scattering, absorption, emission, radiative enhancement) and, at the foundation, bounds on the constitutive material parameters themselves (refractive index, loss and gain, nonlinear susceptibilities). In this Tutorial we have traced the rapid progress at each level from both directions: ``bottom up,'' starting from material-loss figures of merit like $|\chi|^2/\Im\chi$ and proceeding through plasmonic and dipolar scatterers up to multi-channel scatterers and metasurfaces; and ``top down,'' starting from symmetries, conservation laws, and optical-thickness arguments and proceeding to specific design objectives. Even within the linear, time-invariant scope set in the Introduction, many active areas could not be covered here: bounds on waveguide modes (existence~\cite{Lee2008}, sum rules~\cite{Haakestad2005}, minimum mode volumes~\cite{Arbabi2014}); bounds on photonic-crystal modes, including quality factors~\cite{Osting2013} and band gaps~\cite{Rechtsman2009}; topological-gap bounds~\cite{Onishi2024}; speed-of-light limitations in passive linear media~\cite{welters2014speedoflight}; and the few but growing results pushing beyond LTI, including causal-but-not-passive media~\cite{ivanenko2020quasi_herglotz}. More generally, broad fields with different or modified effective governing equations, such as ultrafast optics~\cite{Weiner2009}, with strong time dependencies, or disordered media~\cite{Carminati2021-zb} (with diffusive or radiative-transfer models), are ripe for general bounds.

An ideal vision might be a single unified theory that automatically takes in any photonic problem and returns a tight, computationally efficient bound. We doubt such a single theory will form, but instead expect a foundational framework accompanied by a flowering of problem-dependent paths. An instructive precedent is convex optimization: a field where, in principle, every local optimum is global, strong duality holds under mild conditions, and many problems are polynomial-time solvable. Even there, a working practitioner must pick among formulations (LP, QP, SOCP, SDP), among solvers (interior-point methods are standard; first-order methods scale to very large problems), and engineer problem-specific reformulations and warm-starts~\cite{boyd2004convex}. The foundational principles are few; the techniques built on them are many. We expect the same shape for photonic fundamental limits: the underlying principles (linearity, causality, reciprocity, conservation laws) compress to the short list captured in our Chapter~1, but the catalog of bound-derivation techniques will continue to grow problem by problem, with each new application area contributing its own combinations of relaxations and bespoke constructions.

Looking forward, our long-term vision is simple to state and hard to achieve: for any well-posed electromagnetic question --- focusing efficiency, scattering cross-section, near-field enhancement, communication capacity, anything else --- given a designable volume and a palette of available materials, characterize what is possible. The pursuit cuts across an unusually wide swath of adjacent fields: classical scattering theory and antenna design supply foundational physics models and intuition; modern materials science fixes available constitutive parameters (or offers unprecedented new ones); convex and nonconvex optimization theory provides the relaxations and duality structures from which most bounds are derived; inverse-design and machine-learning techniques serve as a tool for finding bound-saturating structures or identifying theoretical gaps; and information theory increasingly supplies the language in which questions are most naturally posed. We hope the open questions threaded through this Tutorial provide concrete starting points for the community --- particularly if formalized into a public, growing collection and attacked with both human and artificial intelligence. We intend to revisit this landscape a decade hence and reflect on how it has shifted. We close with a phrase that has guided much of our own thinking on this subject: \emph{there is freedom in constraints}.

\section*{Acknowledgments}
\addcontentsline{toc}{section}{Acknowledgments}
We gratefully acknowledge the support and patience of the inviting Editor, Prof. Guifang Li. We also thank Matteo Ciabattoni for his assistance in preparing some of the figures. ODM acknowledges support from the Simons Foundation through the Simons Collaboration on Extreme Wave Phenomena Based on Symmetries (award no.\ SFI-MPS-EWP-00008530-09), the Air Force Office of Scientific Research under Grant No.\ FA9550-22-1-0393, and the National Science Foundation under Award No.\ 2522005. FM acknowledges support from the Air Force Office of Scientific Research under Grant No. FA9550-22-1-0204 through Dr. Arje Nachman, and the National Science Foundation under Award No.\ 2522004. The importance of sustained federal investment in basic scientific and engineering research cannot be overstated. "Without scientific progress no amount of achievement in other directions can insure our health, prosperity, and security as a nation in the modern world." (Vannevar Bush, ``Science, the Endless Frontier", 1945)

\section*{Data availability}
\addcontentsline{toc}{section}{Data availability}
Code that reproduces the numerical checks and several of the figures in this Tutorial is available in Ref.~\cite{CompanionRepo}. Other data underlying the results presented in this paper are not publicly available at this time but may be obtained from the authors upon reasonable request.

\clearpage
\section*{Bibliography}
\renewcommand{\sectionmark}[1]{\markright{\normalsize\hspace{5pt}#1}{}}
\sectionmark{Bibliography}
\addcontentsline{toc}{section}{Bibliography}
\printbibliography[heading=bibempty]

\end{document}